%% file: sampling-tkdd.tex
\documentclass[prodmode,acmtkdd]{acmsmall}
\usepackage{setspace}
\usepackage[ruled,linesnumbered,vlined]{algorithm2e}
\usepackage{times,graphicx,xspace,comment,subfigure,algpseudocode,multirow,url,amsmath}
\usepackage{tabularx}
\usepackage{hyphenat}
\usepackage{url}
\usepackage{microtype}
\DisableLigatures{encoding=*,family=*}
\usepackage{booktabs}
\usepackage{balance}  
\usepackage{caption}

\newcommand{\remove}[1]{}
\newfont{\cour}{cmtt10}
\newcommand{\ol}{\setlength{\itemsep}{0pt.}\begin{enumerate}}
\newcommand{\eol}{\end{enumerate}\setlength{\itemsep}{-\parsep}}
\newcommand{\eg}{\emph{e.g.}\xspace}
\newcommand{\ie}{\emph{i.e.}\xspace}

\newcommand{\algo}{{ES-i}\xspace}

\renewcommand{\paragraph}[1]{\vspace{0.1in}\noindent{\bf #1.}}

\usepackage{pifont}

\def\fivepc{$^{\ast}$}

\renewcommand{\algorithmcfname}{ALGORITHM}
\SetAlFnt{\small}
\SetAlCapFnt{\small}
\SetAlCapNameFnt{\small}
\SetAlCapHSkip{0pt}
\IncMargin{0.5mm}

\begin{document}

\markboth{N.K. Ahmed et al.}{Network Sampling: From Static to Streaming Graphs}

\title{Network Sampling: From Static to Streaming Graphs}
\author{NESREEN K. AHMED, JENNIFER NEVILLE, and RAMANA KOMPELLA 
\affil{Purdue University}
}

\input{abstract}

\category{}{Database Management}{Database Applications---\textit{Data mining}}

\terms{Design, Algorithms, Experimentation, Performance}


\maketitle

\input{introduction}
\input{foundations}

\input{computational-models}

\input{related}

\input{sampling-stat}
\input{evaluation-static}    
\input{sampling-stream}
\input{evaluation-stream}
\input{applications}
\input{conclusions}

\emergencystretch 1.5em
\bibliographystyle{acmsmall}
\bibliography{sampling-tkdd}

\input{appendix}

\end{document}

%% file: abstract.tex
\begin{abstract}
Network sampling is integral to the analysis of social, information, and biological networks. Since many real-world networks 
are massive in size, continuously evolving, and/or distributed in nature, the network structure is often sampled in order to facilitate study.
For these reasons, a more thorough and complete understanding of network sampling is critical to support the field of network science. In this paper, we outline a framework for the general problem of network sampling, by highlighting the different objectives, population and units of interest, and classes of network sampling methods. In addition, we propose a spectrum of computational models for network sampling methods, ranging from the traditionally studied model based on the assumption of a static domain to a more challenging model that is appropriate for streaming domains.
We design a family of sampling methods based on the concept of graph induction that generalize across the full spectrum of computational models (from static to streaming) while efficiently preserving many of the topological properties of the input graphs.
Furthermore, we demonstrate how traditional static sampling algorithms can be modified for graph streams for each of the three main classes of sampling methods: node, edge, and topology-based sampling.
Our experimental results indicate that our proposed family of sampling methods more accurately preserves the underlying properties of the graph for both static and streaming graphs. Finally, we study the impact of network sampling algorithms on the parameter estimation and performance evaluation of relational classification algorithms.
\end{abstract}

%% file: introduction.tex
\section{Introduction}
\label{sec:intro}

Networks arise as a natural representation of data in various domains, ranging from social to biological to information domains.
However, the majority of these real-world networks are massive and continuously evolving over time (\ie, streaming).
As an example, consider online activity and interaction networks formed from electronic communication (e.g., email, IMs, SMS), social media (e.g., Twitter, blogs, web pages), and content sharing (e.g., Facebook, Flicker, Youtube).
These social processes provide a prolific amount of continuous streaming data that is naturally represented as a network where the nodes are people or objects and the edges are the interactions among them (\eg, Facebook users posting 3.2 billion likes and comments every day~\cite{allfacebook}).
Modeling and analyzing these large dynamic networks have become increasingly important for many applications, such as identifying the behavior and interests of individuals (\eg, viral marketing, online advertising) and investigating how the structure and dynamics of human-formed groups evolve over time.

Unfortunately, many factors make it difficult, if not impossible, to study these networks in their entirety.
First and foremost, the sheer size of many networks makes it computationally infeasible to study the entire network.
Moreover, some networks are not completely visible to the public (\eg, Facebook) or can only be accessed through crawling (\eg, Web). In other cases, the size of the network may not be as large but the measurements required to observe the underlying network are costly (\eg, experiments in biological networks).
Thus, network sampling is at the heart and foundation of our study to understand network structure---since researchers typically need to select a (tractable) subset of the nodes and edges to make inferences about the full network. 

From peer-to-peer to social networks, sampling arises across many different settings. For example, sampled networks may be used in simulations and experimentation, to measure performance before deploying new protocols and systems in the field---such as new Internet protocols, social/viral marketing schemes, and/or fraud detection algorithms. 
In fact, many of the network datasets currently being analyzed as complete networks are themselves samples due to the above limitations in data collection. This means it is critical that researchers understand the impact of sampling methods on the structure of the constructed networks.
All of these factors motivate the need for a more refined and complete understanding of \textit{network sampling}.
In this paper, we outline a spectrum of computational models for network sampling and investigate methods of sampling that generalize across this spectrum, going from the simplest and least constrained model focused on sampling from static graphs to the more difficult and most constrained model of sampling from graph streams. 

Traditionally, network sampling has been studied in the case of simple static graphs (\eg~\cite{leskovec2006slg}). 
These works typically make the simplifying assumption that the graphs are of moderate size and have static structure. Specifically, it is assumed that 
the graphs fit in the main memory (\ie, algorithms assume the full neighborhood of each node can be explored in a constant time) and many of  
the intrinsic complexities of realistic networks, such as the time-evolving nature of these systems,
are totally ignored.
For domains that meet these assumptions, we propose a family of sampling methods based on the concept of graph induction and evaluate our methods against state-of-the-art sampling methods from each of the three classes of network sampling algorithms (node, edge, and topology-based sampling).
More importantly, we show that our family of methods preserve the properties of different graphs more accurately than the other sampling methods.

While studying static graphs is indeed important, the assumption that the graph fits in memory is not realistic for many real world domains (\eg, online social networks).
When the network is too large to fit in memory, sampling requires random disk accesses that incur large I/O costs.
Naturally, this raises the question: how can we sample from these large networks {\em sequentially}, one edge at a time, while \textit{minimizing} the number of {\em passes} over the edges? 
In this context, most of the topology-based sampling procedures such as breadth-first search, random walks, or forest-fire sampling are not appropriate as they require random exploration of a node's neighbors (which requires many passes over the edges).
In contrast, we demonstrate that our sampling methods naturally applies to large graphs, requiring only {\em two passes} over the edges.
Moreover, our proposed sampling algorithm is still able to accurately preserve the properties of the large graph while minimizing the number of passes over the edges (more accurately than alternative algorithms).

Finally, in addition to their massive size, many real-world networks are also likely to be \textit{streaming} over time. 
A {\em streaming graph} is a continuous, unbounded, rapid, time-varying 
stream of {\em edges} that is clearly too large to fit in memory except for probably short windows of time (\eg, a single day).
Streaming graphs occur frequently in the real-world and can be found in many modern online and communication applications such as: Twitter posts, Facebook likes/comments, email communications, network monitoring, sensor networks, among many other applications. 
Although these domains are quite prevalent, there has been little focus on developing network sampling algorithms that address the complexities of streaming domains. Graph streams differ from static graphs in three main aspects: 
(i) the massive volume of edges is far too large to fit in the main memory,  
(ii) the graph structure is not fully observable at any point in time (\ie, only sequential access is feasible, not random access),  and 
(iii) efficient, real-time processing is of critical importance. 

\begin{figure}
\centering
\includegraphics[width=5.5in]{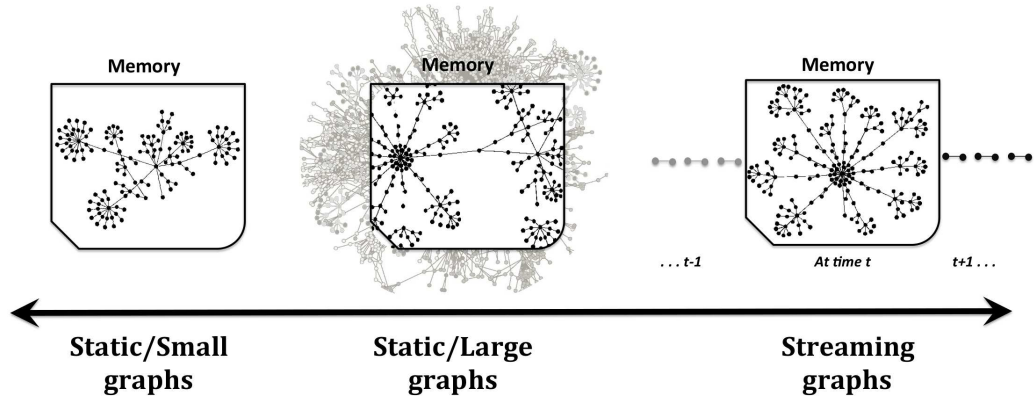}
\vspace{-6.mm}
\caption{Spectrum of Computational Models for network sampling: from static to streaming.} \label{model-spectrum}
\vspace{-2.mm}
\end{figure}

The above discussion shows a natural progression of computational models for sampling---from static to streaming. The majority of previous work has focused on sampling from static graphs, which is the simplest and least restrictive problem setup. In this paper, we also focus on the more challenging issues of sampling from disk-resident graphs and graph streams.
This leads us to propose a spectrum of computational models for network sampling as shown in Figure~\ref{model-spectrum} where we clearly outline the three computational models for sampling from: (1) static graphs, (2) large graphs, and (3) streaming graphs. 
This spectrum not only provides insights into the complexity of the computational models (\ie, static vs. streaming), but also the complexity of the algorithms that are designed for each scenario.
More complex algorithms are more suitable for the simplest computational model of sampling from static graphs. In contrast, as the complexity of the computational model increases to a streaming scenario, efficient algorithms become necessary. 
Thus, there is a trade-off between the complexity of the sampling algorithm and the complexity of the computational model (static $\rightarrow$ streaming).
A subtle but important consequence is that any algorithm designed to work over graph streams is also applicable in the simpler computational models (\ie, static graphs). 
However, the converse is not true, algorithms designed for sampling a static graph that can fit in memory, may not be generally applicable for graph streams (as they may require an intractable number of passes to implement).

Within the spectrum of computational models for network sampling, we formally discuss in Section~\ref{sec:foundations} the various objectives of network sampling (\eg, sampling to estimate network parameters). We provide insights on how conventional objectives in static domains can be naturally adapted to the more challenging scenario of streaming graphs. In Sections~\ref{sec:sampling-stat} and~\ref{sec:sampling-stream}, we primarily focus on the task of representative subgraph sampling from both static and streaming graphs. As an example problem definition, we consider the case of sampling representative subgraphs from graph streams. 
Formally, the input is assumed to be a graph $G\!=\!(V,E)$, presented as a stream of edges $E$ in no particular order. Then, the goal is to sample a subgraph $G_s\!=\! (V_s, E_s)$ with a subset of the nodes ($V_s\!\subset\! V$) and/or edges ($E_s \!\subset\! E$) from the population graph stream $G$. The objective is to ensure that $G_s$ is a {\em representative} subgraph that matches many of the topological properties of $G$. In addition to the {\em sample representativeness} requirement, a graph stream sampling algorithm is also required to be {\em efficient}---and needs to decide whether to include an edge $e \in E$ in the sample or not as the edge is {\em streamed} in.  The sampling algorithm may maintain a state $|\Psi|$ and consult the state to determine whether to sample the edge or not, but the total storage associated with the algorithm is preferred to be of the order the size of the output sampled subgraph $G_s$, \ie, $|\Psi|$ = $O(|G_s|)$.

In this paper, we formalize the problem of sampling from graph streams.
We show how to extend traditional sampling algorithms from each of the three classes of sampling methods (node, edge, and topology-based sampling) for use on graph streams. For edge sampling from graph streams, we use the approach in~\cite{aggarwal2011outlier}. 
Furthermore, we propose our graph stream sampling method (from the family of methods based on the concept of graph induction), which is space-efficient (uses space in the order of the sampled subgraph) and runs in a {\em single pass} over the edges. Our family of sampling methods based on the concept of graph induction generalize across the full spectrum of computational models (from static to streaming) while efficiently preserving many of the graph properties for streaming and static graphs.
In addition, our proposed family of sampling methods offers a good balance between algorithm complexity and sample representativeness while remaining general enough for any computational model.
Notably, our family of algorithms, while less complex, preserve the properties of {\em static} graphs even better than the more complex algorithms that do not generalize to the streaming model, over a broad set of real-world networks.

\vspace{3mm}
\noindent In conclusion, we summarize the contributions of this paper as follows:
\begin{list}{$-$}{\leftmargin=1em}
\item A detailed framework outlining the general problem of network sampling, highlighting the different goals, population and units, and classes of network sampling methods (see Section~\ref{sec:foundations}).
\item A further elaboration of the above framework to include a spectrum of computational models within which to design network sampling methods (\ie, going from static to streaming graphs) (see Section~\ref{sec:computational-models}).
\item Introduction of a family of sampling methods based on the concept of \textit{graph induction} that has the following properties (see Sections~\ref{sec:sampling-stat} and ~\ref{sec:sampling-stream}):
\begin{enumerate}[(a)]
\item Preserve many key graph characteristics more accurately than alternative state-of-the-art algorithms.
\item Generalize to a streaming computational model with a minimum storage space (\ie, space complexity is on the order of the sample size).
\item Run efficiently in a few number of passes over the edges (\ie, runtime complexity is on the order of the number of edges).
\end{enumerate}
\item Systematic investigation of the three classes of static sampling methods on a variety of datasets and extension of these algorithms for application to graph streams (see Sections~\ref{sec:sampling-stat} and ~\ref{sec:sampling-stream}).
\item Empirical evaluation that shows our sampling methods are applicable to large graphs that don't fit in the main memory (see Sections~\ref{sec:sampling-stat} and ~\ref{sec:sampling-stream}).
\item Further task-based evaluation of sampling algorithm performance in the context of relational classification. This investigation illustrates the impact of network sampling on the parameter estimation and evaluation of classification algorithms overlaid on the sampled networks  (see Section~\ref{sec:applications}).
\end{list}{}{}

%% file: foundations.tex
\section{Foundations of Network Sampling}
\label{sec:foundations}

In the context of statistical data analysis, a number of issues arise and need to be considered carefully before collecting data and making inferences based on them. 
At first, we need to identify the relevant {\em population} to be studied. 
Then, if sampling is necessary then we need to decide how to sample from that population. 
Generally, the term {\em population} is defined as the full set of representative units that one wishes to study (\eg, units may be individuals in a particular city).
In some instances, the population may be relatively small and bounded, and is therefore easy to study in its entirety (\ie, without sampling). 
For instance, it is relatively easy to study the set of graduate students in an academic department. 
Conversely, in other situations the population may be large, unbounded, or difficult and/or costly to access in its entirety. 
For instance, the complete set of Facebook users. 
In this case, a sample of units should be collected and characteristics of the population can be estimated from the sampled units.

Network sampling is of interest to a variety of researchers from many distinct fields (\eg statistics, social science, databases, data mining, machine learning) due to the range of complex datasets that can be represented as graphs.
While each area may investigate different types of networks, they all have focused primarily on {\em how} to sample.

For example, in social science, snowball sampling is used extensively to run survey sampling in populations that are difficult-to-access (\eg, sampling the set of drug users in a city)~\cite{watters1989targeted}. Similarly, in Internet topology measurements, breadth first search is used to {\em crawl} distributed, large-scale Online social networks (\eg Facebook)~\cite{mislove07imc}. 
Moreover, in structured data mining and machine learning, the focus has been on developing algorithms to sample small(er) subgraphs from a single large network~\cite{leskovec2006slg}. These sampled subgraphs are further used to learn models (\eg, relational classification models~\cite{Friedman99learningprobabilistic}), evaluate and compare the performance of algorithms (\eg, different classification methods~\cite{rossi2012time}), and study complex network processes (\eg, information diffusion~\cite{bakshy2012role}). We provide a detailed discussion of the related work in Section~\ref{sec:related}.

While this large body of research has developed methods to sample from networks, much of the work is problem-specific and there has been less work focused on developing a broader foundation for network sampling.
More specifically, it is often not clear {\em when} and {\em why} particularly sampling methods are appropriate. This is because the goals and population are often not explicitly defined or stated up front, which makes it difficult to evaluate the quality of the recovered samples for other applications. 
One of the primary aims of this paper is to define and discuss the foundations of network sampling more explicitly, such as:  objectives/goals, population of interest, units, classes of sampling algorithms (\ie, node, edge, and topology-based), and techniques to evaluate a sample (\eg, network statistics and distance metrics). In this section, we will outline a solid methodological framework for network sampling. 
The framework will facilitate the comparison of various network sampling algorithms, and help to understand their relative strengths and weaknesses with respect to particular sampling goals. 

\subsection{Notation}
\label{notation}

Formally, we consider an input network represented as a graph $G\!=\!(V,E)$ with the node set $V=\{v_1,v_2,...,v_N\}$ and edge set $E=\{e_1,e_2,...,e_M\}$, such that $N=|V|$ is the number of nodes, and $M=|E|$ is the number of edges. We denote $\eta(.)$ as any topological graph property. Therefore, $\eta(G)$ could be {\em a point statistic} (\eg, average degree of nodes in $V$) or {\em a distribution} (\eg, degree distribution of $V$ in $G$). 

Further, we define $\Lambda = \{a_1,a_2,...,a_k\}$ as the set of $k$ attributes associated with the nodes describing their properties. Each node $v_i \in V$ is associated with an attribute vector $[a_1(v_i),a_2(v_i),...,a_k(v_i)]$ where $a_j(v_i)$ is the $j^{th} $ attribute value of node $v_i$. For instance, in a Facebook network where nodes represent users and edges represent friendships, the node attributes may include age, political view, and relationship status of the user.  

Similarly, we denote $\beta=\{b_1,b_2,...,b_l\}$ as the set of $l$ attributes associated with the edges describing their properties. Each edge $e_{ij}=(v_i,v_j) \in E$ is associated with an attribute vector $[b_1(e_{ij}),b_2(e_{ij}),...,b_l(e_{ij})]$. In the Facebook example, edge attributes may include relationship type (\eg, friends, married), relationship strength, and type of communication (\eg, wall post, picture tagging).

Now, we define the network sampling process. Let $\sigma$ be any sampling algorithm that selects a random sample $S$ from $G$ (\ie, $S=\sigma(G)$). The sampled set $S$ could be a subset of the nodes ($S=V_s\!\subset\! V$) , or edges ($S=E_s \!\subset\! E$), or a subgraph ($S\!=\! (V_s, E_s)$ where $V_s\!\subset\! V$ and $E_s \!\subset\! E$).
The size of the sample $S$ is defined relative to the graph size with respect to a sampling fraction $\phi$ ($0 \leq \phi \leq 1$). In most cases the sample size is defined as a fraction of the nodes in the input graph, \eg, $|S|=\phi \cdot |V|$. But in some cases, the sample size is defined relative to the number of edges ($|S|=\phi \cdot |E|$).

\subsection{Goals, Units, and Population} 
\label{sec:goals}

While the explicit aim of many network sampling algorithms is to select a smaller subgraph from $G$, there are often other more implicit goals of the process that are left unstated. 
Here, we formally outline a range of possible goals for network sampling: 

\begin{enumerate}[({Goal} 1)]
\item \textsc{Estimate network parameters}\\
Used to select a subset $S$ of the nodes (or edges) from $G$, to estimate properties of $G$.
Thus, $S$ is a good sample of $G$ if,
\begin{eqnarray*} 
\eta(S) \approx \eta(G)
\end{eqnarray*}

For example, let $S=V_s \subset V$ be the subset of sampled nodes, we can estimate the average degree of nodes $G$ using $S$ as
\begin{eqnarray*} 
\hat{deg}_{avg}={1 \over {|S|}} {\sum_{v_i \in S}{deg(v_i \in G)}}
\end{eqnarray*}

where $deg(v_i \in G)$ is the degree of node $v_i$ as it appears in $G$, and a direct application of statistical estimators helps to correct the sampling bias of $\hat{deg}_{avg}$ \cite{hansen1943theory}.

\vspace{2mm}
\item \textsc{Sample a representative subgraph}\\
Used to select a small subgraph $S=G_s=(V_s,E_s)$ from $G$, such that $S$ preserves some topological properties of $G$. 
Let $\eta_{\mathcal{A}}$ be a set of topological properties, then $S$ is a good sample of $G$ if, 

\begin{eqnarray*} 
\eta_{\mathcal{A}}(S) \approx \eta_{\mathcal{A}}(G)
\end{eqnarray*}

Generally, the subgraph representativeness is evaluated by picking a set of graph topological properties that are important for a wide range of applications. This ensures that the sample subgraph $S$ can be used instead of $G$ for testing algorithms, systems, and/or models in an application. For example, \cite{leskovec2006slg} uses topological properties like degree, clustering, and eigenvalues to evaluate the samples.

\vspace{2mm}
\item \textsc{Estimate node attributes}\\
Used to select a subset $S=V_s$ of the nodes from $G$ to study node attributes.
Let $f_a$ be a function involving node attribute $a$, then, $S \subset V$ is a good sample of $V$ if, 

\begin{eqnarray*} 
f_a(S) \approx f_a(V)
\end{eqnarray*}

For example, if $a$ represents the age of users, we can estimate the average age in $G$ using $S$ as
\begin{eqnarray*} 
\hat{a}_{avg}={1 \over {|S|}} {\sum_{v_i \in S}{a(v_i)}}
\end{eqnarray*}

Similar to goal $1$, statistical estimators can be used to correct the bias.

\vspace{2mm}
\item \textsc{Estimate edge attributes}\\
Used to select a subset $S=E_s$ of the edges from $G$ to study edge attributes.
Let $f_b$ be a function involving edge attribute $b$, then, $S \subset E$ is a good sample of $E$ if, 

\begin{eqnarray*} 
f_b(S) \approx f_b(E)
\end{eqnarray*}

For example, if $b$ represents the relationship type of friends (\eg, married, coworkers), we can estimate the proportion of married relationships in $G$ using $S$ as
\begin{eqnarray*} 
\hat{p}_{married}={1 \over {|S|}} {\sum_{e_{ij} \in S}{1_{(b(e_{ij})=married)}}}
\end{eqnarray*}
\end{enumerate}

Clearly, the first two goals ($1$ and $2$) focus on characteristics of entire networks, while the last two goals ($3$ and $4$) focus on characteristics of nodes or edges in isolation. 
Therefore, these goals maybe difficult to satisfy simultaneously---i.e., if the sampled data enable accurate study of one, it may not allow accurate study of the others. 
For instance, a representative subgraph sample could be a biased estimate of global graph parameters (\eg, density).

Once the goal is outlined, the population of interest can be defined relative to the goal. In many cases, the definition of the population may be obvious.
The main challenge is then to select a representative subset of units in the population in order to make the study cost efficient and feasible.
Other times, the population may be less tangible and difficult to define. For example, if one wishes to study the characteristics of a system or {\em process}, there is not a clearly defined set of items to study. Instead, one is often interested in the overall behavior of the system. In this case, the population can be defined as the set of possible outcomes from the system (e.g., measurements over all settings) and these {\em units} should be sampled according to their underlying probability distribution.

In the first two goals outlined above, the objective of study is an entire network (either for structure or parameter estimation). In goal $1$, if the objective is to estimate local properties from the nodes (\eg degree distribution of $G$), then the elementary units are the nodes, and then the population would be the set of all nodes $V$ in $G$. However, if the objective is to estimate global properties (\eg diameter of $G$), then the elementary units correspond to subgraphs (any $G_s \subset G$) rather than nodes and the population should be defined as the set of subgraphs of a particular size that could be drawn from $G$. In goal $2$, the objective is to select a subgraph $G_s$, thus the elementary units correspond to subgraphs, rather than nodes or edges (goal $3$ and $4$). As such, the population should also be defined as the set of subgraphs  of a particular size  that could be drawn from $G$. 

\subsection{Classes of Sampling Methods}
\label{sec:sampling-classes}
Once the population has been defined, a sampling
algorithm $\sigma$ must be chosen to sample from $G$. Sampling algorithms can be categorized as node, edge, and topology-based sampling, based on whether nodes or edges are locally selected from $G$ (node and edge-based sampling) or if the selection of nodes and edges depends more on the existing topology of $G$ (topology-based sampling).

\noindent
Graph sampling algorithms have two basic steps: 
\begin{enumerate}[(1)]
\item \textit{Node selection:} used to sample a subset of nodes $S=V_s$ from $G$, (\ie, $V_s \subset V$). 
\item \textit{Edge selection:} used to sample a subset of edges $S=E_s$ from $G$, (\ie, $E_s \subset E$)
\end{enumerate}

When the objective is to sample only nodes or edges (\ie, goals $1,3,4$), then either step $1$ or step $2$ is used to form the sample $S$. When the objective is to sample a subgraph $G_s$ from $G$ (\ie, goal 2), then both step $1$ and $2$ from above are used to form $S$, (\ie, $S\!=\!(V_s,E_s$)). In this case, the edge selection is often conditioned on the selected node set in order to form an {\em induced subgraph} by sampling a subset of the edges incident to $V_s$ (\ie $E_s=\{e_{ij}=(v_i,v_j), e_{ij} \in E | v_i,v_j \in V_s\}$. 
We distinguish between two different approaches to graph induction---{\em total} and {\em partial} graph induction---which differ by whether {\em all} or {\em some} of the edges incident on $V_s$ are selected.
The resulting sampled graphs are referred to as the {\em induced subgraph} and {\em partially induced subgraph} respectively. 

While the discussion of the algorithms in the next sections focuses more on sampling a subgraph $G_s$ from $G$, they can easily generalize to sampling only nodes or edges.

\paragraph{Node sampling (NS)} In classic node sampling,
nodes are chosen independently and uniformly at random from 
$G$ for inclusion in the sampled graph $G_s$. For a
target fraction $\phi$ of nodes required, each node is
simply sampled with a probability of $\phi$.  Once the nodes
are selected for $V_s$, the sampled subgraph is constructed to be the {\em induced
subgraph} over the nodes $V_s$, \ie, all edges among the
$V_s \in G$ are added to $E_s$.
While node sampling is intuitive and relatively
straightforward, the work in \cite{stumpf2005ssf} shows that
it does not accurately capture properties of graphs with
power-law degree distributions.  Similarly, \cite{lee:06}
shows that although node sampling appears to capture nodes
of different degrees well, due to its inclusion of all edges
for a chosen node set only, the original level of connectivity is
not likely to be preserved.  

\paragraph{Edge sampling (ES)} In classic edge sampling,
edges are chosen independently and uniformly at random from 
$G$ for inclusion in the sampled graph $G_s$.
Since edge sampling focuses on the
selection of edges rather than nodes to populate the sample, the node set is constructed by including both incident nodes in $V_s$ when a particular edge is sampled (and added to $E_s$).  The resulting subgraph is partially induced, which means
no extra edges are added over and above those that were chosen during
the random edge selection process. 
Unfortunately, ES fails to preserve many desired graph properties. Due to the
independent sampling of edges, it does not preserve clustering and connectivity.
It is however more likely to capture path lengths, due to its bias towards high
degree nodes and the inclusion of both end points of selected edges. 

\paragraph{Topology-based sampling} 
Due to the known
limitations of NS (\cite{stumpf2005ssf,lee:06}) and ES (bias
toward high degree nodes), researchers have also considered
many other topology-based sampling methods, which use breadth-first search (\ie sampling without replacement) or random walks (\ie sampling with replacement) over the graph to construct a sample. 

One example is
snowball sampling, which adds nodes and edges using breadth-first
search (but with only a fraction of neighbors explored) from a randomly selected seed node.  Snowball
sampling accurately maintains the network connectivity
within the snowball, however it suffers from
\emph{boundary bias} in that many peripheral nodes (\ie,
those sampled on the last round) will be missing a large
number of neighbors~\cite{lee:06}.

Another example is the Forest Fire Sampling (FFS) method~\cite{leskovec2006slg}, which uses {\em partial} breadth-first search where only a fraction of neighbors are followed for each node.  The algorithm starts by picking a node uniformly at random and adding
it to the sample. It then "burns'' a random proportion of its outgoing
links, and adds those edges, along with the incident nodes, to the sample.  The fraction is determined by sampling from a geometric distribution with mean
$p_f/(1-p_f)$).  The authors recommend setting $p_f=0.7$, which
results in an average burn of $2.33$ edges per node. The process is repeated recursively  for
each burned neighbor until no new node is selected, then a new
random node is chosen to continue the process until the desired sample size is obtained. Also, there are other examples such as respondent-driven sampling~\cite{heckathorn1997respondent} and expansion sampling~\cite{Maiya2010www}, we give more details in Section~\ref{sec:related}.

In general, such topology-based sampling approaches form the sampled graph out of the explored nodes and edges, and usually perform better than simple algorithms such as NS and ES. 

\subsection{Evaluation of Sampling Methods} \label{sec:evaluation}

When the goal is to study the entire input network---either for measuring the quality of parameter estimates (goal $1$), or measuring the representativeness of the sampled subgraph structure (goal $2$)---the accuracy of network sampling methods is often evaluated by comparing network statistics (\eg, degree). We first define a suite of common network statistics and then discuss how they can be used more quantitatively to compare sampling methods.

\paragraph{Network Statistics}
The commonly considered network statistics  can be compared along two dimensions: {\em local} vs. {\em global} statistics, and {\em point} statistic vs. {\em distribution}. A local statistic is is used to describe a characteristic of a local graph element (\eg, node, edge, subgraph). For example, node degree and node clustering coefficient. On the other hand, a global statistic is used to describe a characteristic of the entire graph. For example, global clustering coefficient and graph diameter. Similarly, there is also the distinction between point statistics and distributions. A point-statistic is a single value statistic (\eg, diameter) while a distribution is a multi-valued statistic (\eg, distribution of path length for all pairs of nodes). Clearly, a range of network statistics are important to understand the full graph structure. 
 
In this work, we focus on the goal of sampling a representative subgraph $G_s$ from $G$, by using distributions of network characteristics calculated on the level of nodes, edges, sets of nodes or edges, and subgraphs. Table~\ref{tab:gprops} provides a summary for the six network statistics we use and we formally define the statistics below:

\begin{table*}
\small
\centering
\caption{Description of Network Statistics}
\begin{tabularx}{\linewidth}{lXc}
\toprule
\textbf{{Network Statistic}} & \textbf{{Description}} \\
\midrule
\textsc{Degree dist.} & Distribution of nodes degrees in the network \\
\\
\textsc{Path length dist.} & Distribution of all shortest paths \\
\\
\textsc{Clustering coefficient dist.} & Distribution of local clustering per node \\
\\
\textsc{K-core dist.} & Distribution of sizes of the largest subgraphs where nodes have at least k interconnections \\
\\
\textsc{Eigenvalues} & Distribution of the eigenvalues of the network adjacency matrix vs. their rank \\
\\
\textsc{Network values} & Distribution of  eigenvector components of the largest eigenvalue of the network adjacency matrix vs. their rank \\
\bottomrule
\end{tabularx}
\label{tab:gprops}
\end{table*}

\begin{enumerate}[(1)]
\item \textit{Degree distribution}: The fraction of nodes with degree $k$, for all $k > 0$ 
\begin{center}
$p_k = {{|\{v \in V | deg(v)=k \}|}\over {N}}$ 
\end{center} 
Degree distribution has been widely studied by many researchers to understand the connectivity in graphs. Many real-world networks were shown to have a power-law degree distribution, for example in the Web  \cite{kleinberg1999web}, citation graphs \cite{redner1998popular}, and online social networks \cite{chakrabarti2004r}.    

\vspace{2mm}
\item \textit{Path length distribution}: Also known as the {\em hop} distribution and denotes the fraction of pairs $(u,v) \in V$ with a shortest-path distance ($dist(u,v)$) of $h$, for all $h > 0$
\begin{center}
$p_h = {{|\{(u,v) \in V | dist(u,v) = h\}|}\over {N^2}}$ 
\end{center}

The path length distribution is essential to know how the number of paths between nodes expands as a function of distance (\ie, number of hops).

\vspace{2mm}
\item \textit{Clustering coefficient distribution}: The fraction of nodes with clustering coefficient ($cc(v)$) $c$, for all $0 \leq c \leq 1$
\begin{center}
$p_c = {{|\{v \in V' | cc(v) = c\}|}\over {|V'|}}$, where $V'=\{v \in V | deg(v) >1\}$ 
\end{center}
Here the clustering coefficient of a node $v$ is calculated as the number of triangles centered on $v$ divided by the number of pairs of neighbors of $v$ (\eg, the proportion of $v$'s neighbor that are linked). 
In social networks and many other real networks, nodes tend to cluster. Thus, the clustering coefficient is an important measure to capture the transitivity of the graph \cite{watts1998small}.  

\vspace{2mm}
\item \textit{K-core distribution}: The fraction of nodes in graph $G$ participating in a {\em k-core} of order $k$. The {\em k-core} of $G$ is the largest induced subgraph with minimum degree $k$.
Formally, let $U \subseteq V$, and $G_{[U]}\!=\!(U,E')$ where $E'\!=\!\{e_{u,v} \in E | u,v \in U\}$. Then $G_{[U]}$ is a {\em k-core} of order $k$ if $\forall v \! \in \! U \; deg_{G_{[U]}}(v) \geq k$.

Studying {\em k-cores} is an essential part of social network analysis as they demonstrate the connectivity and community structure of the graph~\cite{carmi2007model,alvarez2005k,kumar2010structure}. We denote the {\em maximum core number} as the maximum value of $k$ in the {\em k-core} distribution. The {\em maximum core number} can be used as a lower bound on the degree of the nodes that participate in the largest induced subgraph of $G$. Also, the core sizes can be used to demonstrate the localized density of subgraphs in $G$ \cite{seshadhri2011depth}.   

\vspace{2mm}
\item \textit{Eigenvalues}:  The set of real eigenvalues $\lambda_1 \geq ... \geq \lambda_N$ of the corresponding adjacency matrix $A$ of $G$.
Since {\em eigenvalues} are the basis of spectral graph analysis, we compare the largest $25$ eigenvalues of the sampled graphs to their real counterparts.

\vspace{2mm}
\item \textit{Network values}: The  distribution of the eigenvector components associated with the largest eigenvalue $\lambda_{max}$.
We compare the largest $100$ network values of the sampled graphs to their real counterparts.
 
\end{enumerate}
Next, we describe the use of these statistics for comparing sampling methods quantitatively.

\paragraph{Distance Measures for Quantitatively Comparing Sampling Methods}

The goal is to select a {\em representative} sample
that minimizes the distance between the property in $G$ and
the property in $G_s$: $dist[\eta(G),\eta(G_S)]$. When the goal is to provide estimates of global network parameters  (\eg, average degree), then $\eta(.)$ may measure {\em point statistics}. However, when the goal is to provide a representative subgraph sample, then $\eta(.)$ may measure {\em distributions} of network properties (\eg, degree distribution). These distributions reflects how the graph structure is distributed across nodes and edges.

The $dist$ function could be typically any distance measure (\eg, absolute difference). In this paper, since we focus on using distributions to characterize graph structure, we use four different distributional distance measures for evaluation.
\begin{enumerate}[(1)]
\item \textit{Kolmogorov-Smirnov (KS) statistic:} Used to assess the distance between two cumulative distribution functions (CDF). The KS-statistic is a widely used measure of the
agreement between two distributions, including in \cite{leskovec2006slg} where it is used to illustrate the accuracy of FFS.
It is computed as the maximum vertical distance between the two distributions, where $x$ represents the range of the random
variable and $F_1$ and $F_2$ represent two CDFs: 
\begin{center}
$KS(F_1,F_2) = max_x{|F_1(x) - F_2(x)|}$
\end{center}

\vspace{2mm}
\item \textit{Skew divergence (SD):} Used to assess the difference between two probability density functions (PDF)~\cite{lee:01}. Skew divergence is used to measure the Kullback-Leibler (KL) divergence between two PDFs $P_1$ and $P_2$ that do not have continuous support over the full range of values (\eg, skewed degree). KL measures the average number of extra bits required to represent samples from the original distribution when using the sampled distribution. However, since KL divergence is not defined for distributions with different areas of support, skew divergence {\em smooths} the two PDFs before computing the KL divergence:  

\begin{center}
$SD(P_1, P_2, \alpha) = KL[\alpha P_1 + (1-\alpha) P_2\: ||\: \alpha P_2 + (1-\alpha) P_1 ]$
\end{center}

The results shown in \cite{lee:01} indicate that using SD yields better results than other methods to approximate KL divergence on non-smoothed distributions. In this paper, as in \cite{lee:01}, we use $\alpha=0.99$.
 
\vspace{2mm}
\item \textit{Normalized $L_1$ distance:} In some cases, for evaluation we will need to measure the distance between two positive m-dimensional real vectors $p$ and $q$ such that $p$ is the true vector and $q$ is the estimated vector. For example, to compute the distance between two vectors of eigenvalues. In this case,  we use the normalized $L_1$ distance:

\begin{center}
$L_1(p, q) =  {1 \over{m}} \sum_{i=1}^{m}{|p_i-q_i|\over{p_i}}$
\end{center}

\vspace{2mm}
\item \textit{Normalized $L_2$ distance:} In other cases, when the vector components are fractions (less than one), we use the normalized euclidean distance $L_2$ distance (\eg, to compute the distance between two vectors of network values): 

\begin{center}
$L_2(p, q) = { ||p-q|| \over ||p||}$
\end{center}
\end{enumerate}

%% file: computational-models.tex
\section{Models of Computation}
\label{sec:computational-models}

In this section, we discuss the different models of computation that can be used to implement network sampling methods. At first, let us assume the network $G=(V,E)$ is given (\eg, stored on a large storage device). Then, the goal is to select a sample $S$ from $G$.

Traditionally, network sampling has been explored in the context of a {\em static model of computation}. 
This simple model makes the fundamental assumption that it is easy and fast (\ie, constant time) to randomly access any location of the graph $G$. For example, random access may be used to query the entire set of nodes $V$ or to query the neighbors $\mathcal{N}(v_i)$ of a particular node $v_i$ (where $\mathcal{N}(v_i)=\{v_j \in V | e_{ij}\!=\!(v_i,v_j)  \in E\}$). However, random accesses on disks are much slower than random accesses in  main memory. A key disadvantage of the static model of computation is that it does not differentiate between a graph that can fit entirely in the main memory and a graph that cannot. Conversely, the primary advantage of the static model is that, since it is the natural extension of how we {\em understand and view} the graph, it is a simple framework within which to design algorithms.

Although design sampling algorithms with a static model of computation in mind is indeed appropriate for some applications, it
assumes the input graphs are relatively small, can fit entirely into main memory, and have static structure (\ie, not changing over the time). This is unrealistic for many domains. 
For instance, many social, communication, and information networks naturally change over time and are massive in size (\eg, Facebook, Twitter, Flickr).
The sheer size and dynamic nature of these networks make it difficult to load the full graph entirely in the main memory.
Therefore, the static model of computation cannot realistically capture all the intricacies of graphs as we understand them today.

Many real-world networks that are currently of interest are \textit{too large} to fit into memory. In this case, sampling methods that require random disk accesses can incur large I/O costs for loading and reading the data.
Naturally, this raises a question as to how we can sample from large networks sequentially rather than assuming random access (\eg, representing the graph as a stream of edges that is accessed in sequence)? In this context, most of the topology based sampling procedures such as breadth-first search and random-walk sampling are no longer appropriate as they require the ability to randomly access a node's neighbors $\mathcal{N}(v_i)$. If access is restricted to sequential passes over the edges, a large number of passes over the edges would be needed to repeatedly select $\mathcal{N}(\cdot)$. In a similar way, node sampling would no longer be appropriate as it not only requires random access for querying a node's neighbors but it also requires random access to the entire node set $V$ in order to obtain a uniform random sample.

\begin{figure*}[!h]
\centering
\subfigure{\label{}\includegraphics[width=0.3\textwidth]{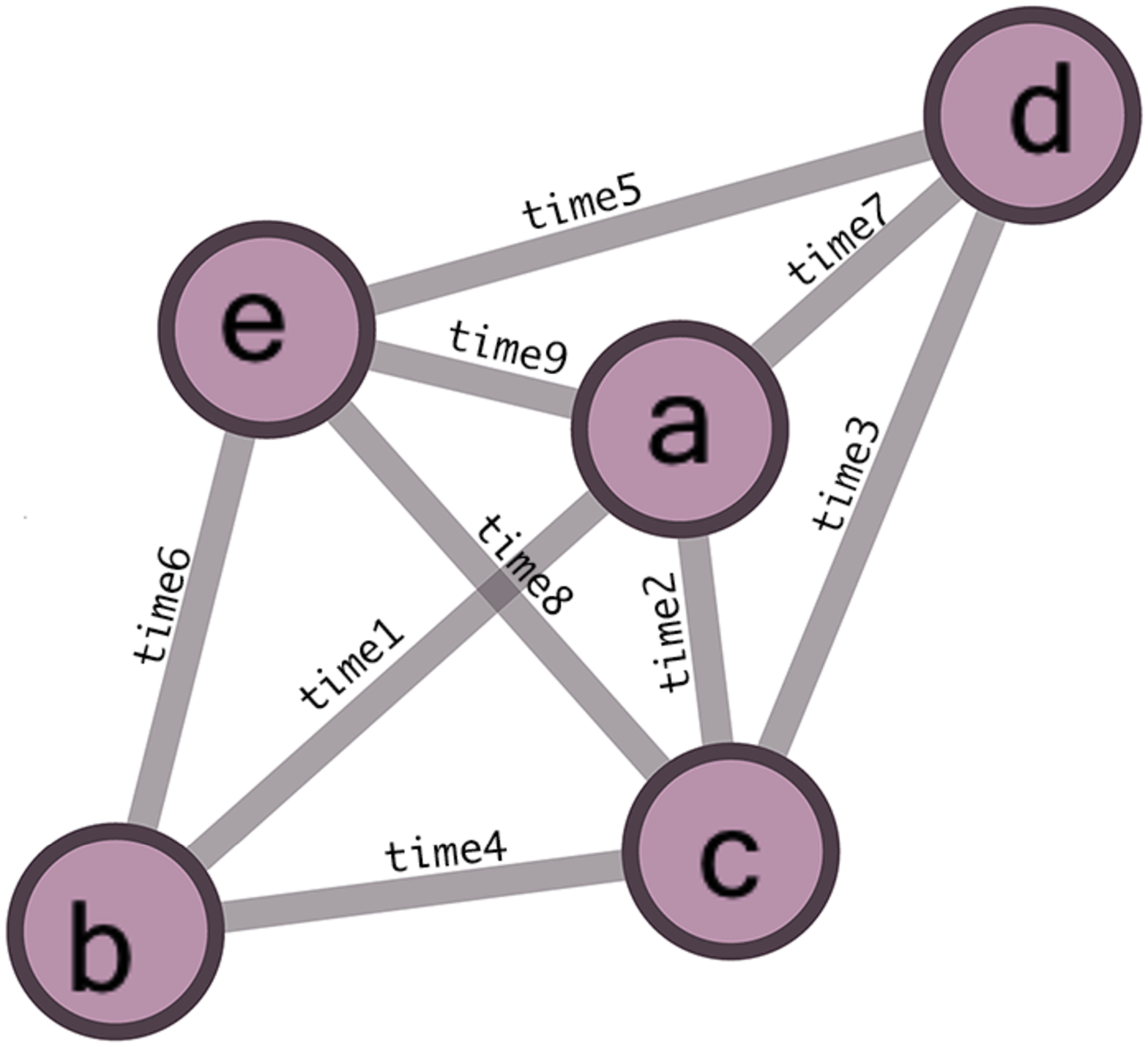}}
\hspace{4.mm}
\subfigure{\label{}\includegraphics[width=0.35\textwidth]{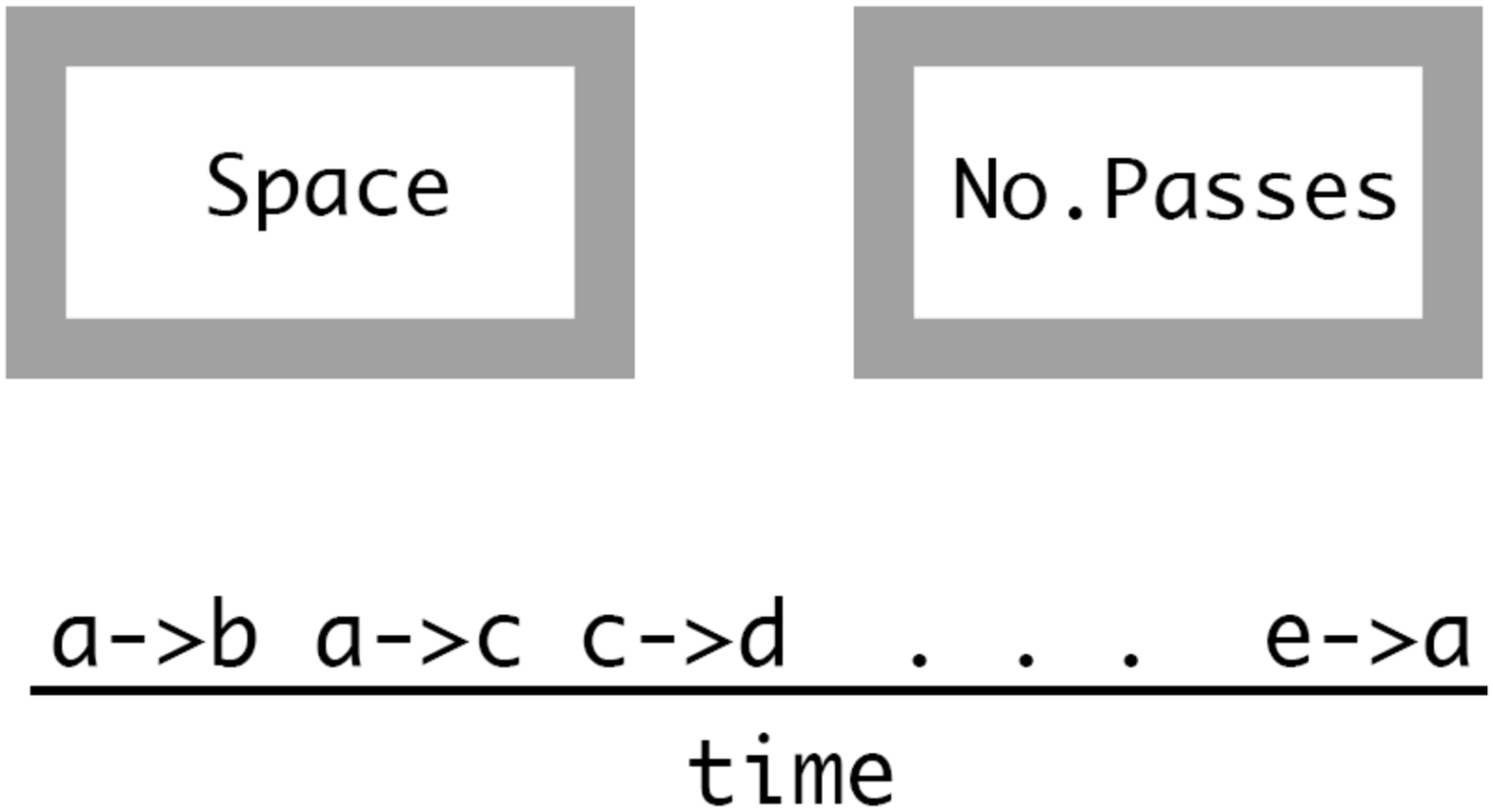}}
\vspace{-2mm}
\caption{Illustration of a {\em graph stream}---a sequence of edges ordered over time.}
\label{fig:gstream1}
\vspace{-2mm}
\end{figure*}

A {\em streaming model of computation}  in which the graph can only be accessed sequentially as a stream of edges, is therefore more preferable for these situations \cite{zhang2010survey}. The streaming model completely discards the possibility of random access to $G$ and the graph can only be accessed through an ordered scan of the edge stream. The sampling algorithm may use the main memory for holding a portion of the edges temporarily and perform random accesses on that subset. In addition, the sampling algorithm may access the edges repeatedly by making multiple passes over the graph stream. 
Formally, for any input network $G$, we assume $G$ arrives as a graph stream (as shown in Figure~\ref{fig:gstream1}).

\noindent 
\begin{definition}[Graph Stream]  A \emph{graph stream} is an ordered sequence of edges 
$e_{\pi(1)},e_{\pi(2)},...,e_{\pi(M)}$, where $\pi$ is any random permutation on the edge indices $[M]=\{1,2,...,M\}$, 
$\pi : [M] \rightarrow [M]$.   
\label{def:gstream_defn}
\end{definition}

Definition~\ref{def:gstream_defn} is usually called the "adjacency stream'' model in which the graph is presented as a stream of edges in any arbitrary order. In contrast, the "incidence stream'' model assumes all edges incident to a vertex are presented in order successively~\cite{buriol2006counting}. In this paper, we assume the adjacency stream model.

While most real-world networks are too large to fit into main memory, many are also likely to be naturally \textit{streaming}. 
A streaming graph is a continuous, unbounded, time-varying, transient stream of edges that is both too large and too dynamic to fit into memory.
These types of streaming graphs occur frequently in real-world communication and information domains. For example real-time tweets between users in Twitter, email logs, IP traffic, sensor networks, web search traffic, and many other applications. 
While sampling from these streaming networks is clearly challenging, their characteristics preclude the use of static models of computation thus a more systematic investigation is warranted.
This naturally raises the second question: how can we sample from large graph streams in a {\em single pass} over the edges? 
Generally graph streams differ from static graphs in three main aspects: 
\begin{enumerate}[(1)]
\item The massive volume of edges streaming over the time is far too large to fit in main memory.
\item The graph can only be accessed  sequentially in a single pass (\ie, no random access to neighboring nodes or to the entire graph).
\item Efficient, real-time processing is of critical importance.
\end{enumerate}

In a streaming model, as each edge $e \in E$ arrives, the sampling algorithm $\sigma$ needs to decide whether
to include the edge or not as the edge {\em streams} by.  The sampling
algorithm $\sigma$ may also maintain state $\Psi$ and consult the state to
determine whether to sample $e$ or not. 

\noindent
The complexity of a streaming sampling algorithm is measured by:
\begin{enumerate}[(1)]
\item Number of passes over the stream $\omega$.
\item Space required to store the state $\Psi$ and the output.
\item Representativeness of the output sample $S$.
\end{enumerate}

Multiple passes over the stream (\ie, $\omega>1$) may be allowed for massive disk-resident graphs but 
multiple passes are not realistic for datasets where the graph is continuously streaming over time. In this case, a requirement of a single pass is more suitable (\ie, $\omega=1$).
The total storage space (\ie, $\Psi$) is usually of the order of the size of the output (\ie, $G_s$): $|\Psi|$ = $O(|G_s|)$.  Note that this requirement is
potentially larger than the $o(N,t)$ (and preferably $polylog(N,t)$) that streaming
algorithms typically require~\cite{muthu}. But, since any algorithm cannot
require less space than its output, we relax this requirement in our definition as follows.

\begin{definition}[Streaming Graph Sampling] A \emph{streaming graph
sampling algorithm} is any sampling algorithm $\sigma$ that
produces a sampled graph $G_s$ by sampling edges of
the input graph $G$ in a sequential order, preferably in {\em one pass} (\ie, $\omega=1$), while preferably maintaining state
$\Psi$ such that $|\Psi| \leq O(|G_s|)$. 
 \label{stream_defn}
\end{definition}

Clearly, it is more difficult to design sampling algorithms for the {\em streaming graph model}, but it is critical to address the fundamental intricacies of graphs as we understand them today.

We now have what can be viewed as a complete spectrum of computational models for network sampling, which ranges from the simple, yet least realistic, static graph model to the more complex, but more realistic, streaming model (as in Figure~\ref{model-spectrum}). In the next sections, we evaluate algorithms for representative subgraph sampling in each computation model across the spectrum.

We note that our assumption in this work is that the population graph $G$ is visible in its entirety ({\em collected and stored on disk}). In many analysis domains this assumption is valid, but in some cases the full structure of the population graph may be unknown prior to the sampling process (\eg, the deep Web or distributed information in peer-to-peer networks).
Web/network crawling is used extensively to sample from graphs that are not fully visible to the public but naturally allow methods to explore the neighbors of a given node (\eg, hyperlinks in a web page). Topology-based sampling methods (\eg, breadth-first search, random walk) have been widely used in this context. However, many of these methods assume the graph $G$ is {\em well connected} and remains {\em static} during crawling, as discussed in~\cite{gjoka2011practical}. 

%% file: related.tex
\section{Related work}
\label{sec:related}

Generally speaking, there are two bodies of work related to this paper: (i) network sampling methods, investigating and evaluating sampling methods with different goals of collecting a sample and (ii) graph steams, mining graph streams. In this section, we describe and put the related work in perspective of the framework we discussed in Section~\ref{sec:foundations}. 

The problem of sampling graphs has been of interest in many different fields of
research. Most of this body of research has focused on {\em how} to sample and evaluate the goodness of the sample relative to the specific goal of the research.

\paragraph{Network sampling in social science} In social science, the classic work done by Frank in~\cite{frank1977survey} and his review papers in~\cite{frank1980sampling} and~\cite{frank1981survey} provide the basic solutions to the first problems that arise when only a sample of the actors in a social network is available. Also, in~\cite{goodman1961snowball}, the concept of ``chain-referral'' sampling originated when Goodman introduced the snowball sampling method. Further, Granovetter introduced the network community to the problem of making inferences about the entire population from a sample (\eg, estimation of network density)~\cite{granovetter1976network}. And then later, respondent-driven sampling was proposed in~\cite{heckathorn1997respondent} and analyzed in~\cite{gile2010respondent} to reduce the biases associated with chain referral sampling of hidden populations. For an excellent survey about estimation of sample properties, we refer the reader to the work in~\cite{kolaczyk2009sampling}.
Generally, the work in this area focuses on either the estimation of global network parameters (\eg, density) or the estimation of actors (node) attributes, \ie, goals $1$ and $3$.

\paragraph{Statistical properties of network sampling} Another important trend of research focused on analyzing the statistical properties of sampled subgraphs. For example, the work in \cite{lee:06,yoon08} studied the statistical
properties of sampled subgraphs produced by the classical node, edge and random walk sampling and discussed the bias in estimates of topological properties. Similarly, the work done in~\cite{stumpf2005ssf} show that the sampled subgraph of a scale free network is far from being scale free. Conversely, the work done in~\cite{lakhina2003sampling} shows that under traceroute sampling, the degree distribution is a Power law even when the actual distribution is a Poisson. 
Clearly, the work in this area has focused on representative subgraph sampling (\ie, goal $2$) considering how sampling changes the topological properties of the original network. 

\paragraph{Network sampling in network systems research} A large body of research in network systems focused on Internet measurement, which targets the problem of topology measurements in large-scale online networks, such as peer-to-peer networks (P2P), world wide web (WWW), and online social networks (OSN). The sheer size, and distributed structure of these networks make it hard to measure the properties of the entire network. Network sampling, via {\em crawling}, has been used extensively in this context. In OSNs, sampling methods that don't allow nodes to be revisited are widely used (\eg, breadth-first search~\cite{ahn2007analysis,mislove07imc,wilson2009userint}). Breadth-first search, however, has been shown to be biased towards high degree nodes~\cite{ye2010crawling} but the work in~\cite{kurant2011towards} suggested analytical solutions to correct the bias. Random walk sampling has also been used, such as the work in~\cite{gjoka2010walking} to sample a uniform sample from users in Facebook, and Last.fm. For a recent survey covering assumptions and comparing different methods of crawling, we refer the reader to~\cite{gjoka2011practical}. Similar to OSNs, random walk sampling and its variants were used extensively to sample the WWW~\cite{baykan2009comparison,henzinger2000near}, and P2P networks~\cite{gkantsidis2004rwp}. Since the classical random walk is biased towards high degree nodes, some improvements were applied to correct the bias. For example, the work in~\cite{stutzbach2006usu} applied metropolis-hastings random walk (MHRW) to sample peers in Gnutella network, and the work in~\cite{rasti2009respondent} applied re-weighted random walk (RWRW) to sample P2P networks. Other work used m-dependent random walk and random walks with jumps~\cite{ribeiro10imc,avrachenkov2010improving}. 

Overall, the work done in this area has focused extensively on sampling a uniform subset of nodes from the graph, to estimate topological properties of the entire network from the set of sampled nodes (\ie, goal 1).  

\paragraph{Network sampling in structured data mining} Network sampling is a core part of data mining research. Representative subgraph sampling was first defined in ~\cite{leskovec2006slg}. Instead of sampling, the work in~\cite{krishnamurthy2007sli} explored reductive methods to shrink the existing topology of the graph. Further, the work in~\cite{hubler08icdm} proposed a generic metropolis algorithm to optimize the representativeness of a sampled subgraph (by minimizing the distance of several graph properties). Unfortunately, the number of steps until convergence is not known in advance (and usually large), and each step requires the computation of a distance function, which may be costly. However, other work discussed the difficulty of getting a  "universal representative'' subgraph that preserves {\em all} properties of the target network. For example, our work in~\cite{ahmed2010reconsidering} discussed the possible correlation between properties, where accurately preserving some properties leads to underestimate/overestimate other properties (\eg, preserving average degree of the target network leads to overestimating its density). Also, the work done in~\cite{Maiya2011kdd} investigated the connection between the biases of topology-based sampling methods (\eg, breadth-first search) and some topological properties (\eg, degree). Therefore, some work focused on obtaining samples for specific applications, and to satisfy specific properties of the target network, such as to preserve the community structure~\cite{Maiya2010www}, to preserve the pagerank between all pairs of sampled nodes~\cite{vattani2011preserving}, and to visualize the graph~\cite{jia2008visualization}.

Other network sampling goals have been considered as well. For example, sampling nodes to perform A/B testing of social features~\cite{backstrom2011network}, sampling nodes to analyze estimators of the fraction of users with a certain property~\cite{dasgupta2012social}, (\ie, goal $3$), and sampling tweets (edges) to analyze the language used in twitter (\ie, goal $4$). In addition,~\cite{al2009output} samples the output space of graph mining algorithms,~\cite{papagelis2011sampling} collects information from social peers for enhancing the information needs of users, and~\cite{de2010does} studies the impact of sampling on the discovery of information diffusion.   

Much of this work has focused on sampling in the static model of computation, where it is assumed that the graph can be loaded entirely in main memory, or the graph is distributed and allows exploring the neighborhood of nodes in a crawling fashion.
   
\paragraph{Graph Streams} Data stream querying and mining has gained a lot of interest in the past years~\cite{babcock2002models,golab2003issues,muthu,aggarwal2006data}. For example, for sequence sampling (\eg, reservoir sampling)~\cite{Vitter:85,babcock2002sampling,agrwal}, for computing frequency counts~\cite{manku,charikar2002finding} and load shedding~\cite{tatbul2003load}, for mining concept drifting data streams~\cite{wang2003mining,gao2007general,fan2004systematic,fan2004streamminer}, clustering evolving data streams~\cite{guha2003clustering,aggarwal2003framework}, active mining and learning in data streams~\cite{fan2004active,li2009positive}, and other related mining tasks~\cite{domingos2000mining,hulten2001mining,gaber2005mining,wang2005mining}.  

Recently, there has been an increasing interest in mining and querying {\em graph streams} as a result of the proliferation of graph data (\eg, social networks, emails, IP traffic, Twitter hashtags). Following the earliest work on graph streams~\cite{raghavan1999computing}, various problems were explored in the field of mining graph streams. For example, to count triangles~\cite{yossef,buriol2006counting}, finding common neighborhoods~\cite{buchsbaum2003finding}, estimating pagerank values~\cite{atish}, and characterizing degree sequences in multi-graph streams~\cite{cormode2005space}. More recently, there is the work done on clustering graph streams~\cite{aggarwal2010clustering}, outlier detection~\cite{aggarwal2011outlier}, searching for subgraph patterns~\cite{chen2010continuous}, and mining dense structural patterns~\cite{aggarwal2010dense}.   

Graph stream sampling was utilized in some of the work mentioned above. For example, the work in~\cite{atish} performs short random walks from uniformly sampled nodes to estimate pagerank scores. Also,~\cite{buriol2006counting} used sampling to estimate number of triangles in the graph stream. Moreover, the work in~\cite{cormode2005space} used a min-wise hash function to sample nearly uniformly from the set of all edges that has been at any time in the stream. The sampled edges were later used to maintain cascaded summaries of the graph stream. More recently,~\cite{aggarwal2011outlier} designed a structural reservoir sampling approach (based on min-wise hash sampling of edges) for structural summarization. For an excellent survey on mining graph streams, we refer the reader to~\cite{mcgregor2009graph,zhang2010survey}.

The majority of this work has focused on sampling a subset of nodes uniformly from the stream to estimate parameters such as the number of  triangles or pagerank scores of the graph stream (\ie goal $1$). Also as we discussed above, other work has focused on sampling a subset of edges uniformly from the graph stream to maintain summaries (\ie goal $2$). These summaries can be further pruned (in the decreasing order of the hash value~\cite{aggarwal2011outlier}) to satisfy a specific stopping constraint (\eg specific number of nodes in the summary). In this paper, since we focus primarily on sampling a representative subgraph $G_s \subset G$ from the graph stream, we compare to some of these methods in Section~\ref{sec:sampling-stream}.

%% file: sampling-stat.tex
\section{Sampling from Static Graphs}
\label{sec:sampling-stat}

In this section, we focus on how to sample a representative subgraph $G_s\!=\!(V_s,E_s)$ from $G\!=\!(V,E)$ (\ie, goal $2$ from Section~\ref{sec:goals}). A representative sample $G_s$ is essential for many applications in machine learning, data mining, and network simulations. As an example, it can be used to drive realistic simulations and experimentation before deploying new protocols and systems in the field~\cite{krishnamurthy2007sli}. We evaluate the representativeness of $G_s$ relative to $G$, by comparing distributions of six topological properties calculated over nodes, edges, and subgraphs (as summarized in Table~\ref{tab:gprops}). 

First, we distinguish between the degree of the sampled nodes before and after sampling. For any node $v_i \in V_s$, we denote $k_i$ to be the node degree of $v_i$ in the input graph $G$. Similarly, we denote $k_{i}^s$ to be the node degree of $v_i$ in the sampled subgraph $G_s$. Note that $k_i=|\mathcal{N}(v_i)|$, where $\mathcal{N}(v_i)=\{v_j \in V | e_{ij}\!=\!(v_i,v_j) \in E\}$) is the set of neighbors of node $v_i$. Clearly, when a node is sampled, it is not necessarily the case that all its neighbors are sampled as well, and therefore $0 \leq k_i^s \leq k_i$.

In this section, we propose a simple and efficient sampling algorithm based on the concept of graph-induction: {\em induced edge sampling} (for brevity \algo).  
\algo has several advantages over current sampling methods as we show later in this section:
\begin{enumerate}[(1)]
\item \algo preserves the topological properties of $G$ better than many of current sampling algorithms.
\item \algo can be easily implemented as a {\em streaming} sampling algorithm using only two passes over the edges of $G$ (\ie, $\omega = 2$).
\item \algo is suitable for sampling large graphs that cannot fit into main memory.
\end{enumerate}

We compare our proposed algorithm \algo to state-of-the-art sampling algorithms from each of the three classes of network sampling (node, edge, and topology-based). More specifically, we compare to node (NS), edge (ES), and forest fire (FFS) sampling methods. Note that all the baseline methods are implemented under the assumption of a static model of computation. However, we show how \algo can be implemented as a streaming algorithm that takes only two passes over the edges of $G$ (\ie $\omega = 2$). Thus its computational complexity is $O(2E)$.   

\subsection{Algorithm}
\label{sec:algo_static}

We formally specify \algo in Algorithm~\ref{algo:ties}. Initially, \algo selects the nodes in pairs by sampling edges uniformly (\ie, $p(e_{ij} \mbox{ is selected})= 1/|E|$) and adds them to the sample ($V_s$). Then, \algo augments the sample with all edges that exist between any of the sampled nodes ($E_s=\{e_{ij}\!=\!(v_i,v_j) \in E | v_i,v_j \in V_S\}$).  These two steps together form the sample subgraph $G_s=(V_s,E_s)$.
For example, suppose edges $e_{12} = (v_1, v_2)$ and $e_{34} = (v_3, v_4)$ are sampled in the first step, that leads
to the addition of the vertices $v_1, ..., v_4$ into the sampled graph. In the second step, \algo adds all the edges that exist between the sampled nodes---for example, edges $e_{12} = (v_1, v_2)$, $e_{34} = (v_3, v_4)$, $e_{13} =(v_1,v_3)$, $e_{24} = (v_2, v_4)$, and any other possible combinations involving $v_1, ..., v_4$ that appear in $G$.  

\begin{algorithm}[h]
\DontPrintSemicolon
\SetKwInOut{Input}{Input}
\SetKwInOut{Output}{Output}
\Input{Sample fraction $\phi$, Edge set $E$}
\Output{Sampled Subgraph $G_s = (V_s, E_s)$}
\BlankLine
$V_s = \emptyset, E_s = \emptyset$\;
// Node selection step\;
\While {$|V_s|$ $<$ $\phi \times |V|$}{
	$r$ = random $(1,|E|)$\; 
	// uniformly random\; 
	$e_r = (u, v)$\;
	$V_s = V_s \cup \{u, v\}$\;
}

// Edge selection step\;
\For{$k =1:|E|$} {
	$e_k = (u, v)$\;
	\If {$u \in V_s$ AND $v \in V_s$} {
		$E_s = E_s \cup \{e_k\}$	\;
	}
}
\caption{ES-i($\phi$, $E$)}
\label{algo:ties}
\end{algorithm}

\paragraph{Downward bias caused by sampling} Since any sampling algorithm (by definition) selects only a subset of the nodes/edges in the graph $G$, it naturally produces subgraphs with underestimated degrees in the degree distribution of $G_s$. 
We refer to this as a {\em downward bias} and note that it is a property of all network sampling methods, since only a fraction of a node's neighbors may be selected for inclusion in the sample (\ie $k_i^s \leq k_i$ for any sampled node $v_i$).  
 
Our proposed sampling methods exploits two key observations. First, by selecting nodes via edge sampling the method  
is inherently biased towards the selection of nodes with high degrees, resulting
in an {\em upward bias} in the (original) degree distribution if only observed from the sampled nodes (\ie, using the degree of the sampled nodes as observed in $G$).
The upward bias resulting from edge sampling can help offset the downward bias of the sampled degree distribution of $G_s$. 

In addition to improving estimates of the sampled degree distribution, selecting high degree nodes also helps to produce a more connected sample subgraph that preserves the topological properties of the graph $G$. This is due to the fact that high degree nodes often represent {\em hubs} in the graph, which serve as good navigators through the graph (\eg, many shortest paths usually pass through these hub nodes).  

However, while the upward bias of edge sampling can help offset some issues of sample selection bias, it is not sufficient to use it in isolation to construct a good sampled subgraph. Specifically, since the edges are each sampled independently, edge sampling is unlikely to preserve much structure 
{\em surrounding} each of the selected nodes.  This leads us to our second observation, that a simple {\em graph induction} step over the sampled nodes  (where we sample all the edges between any sampled nodes from $G$) is a key to recover much of the connectivity in the sampled subgraph---offsetting the downward degree bias as well as increasing local
clustering in the sampled graph. More specifically, graph induction increases the likelihood that triangles will be sampled among the set of selected nodes, resulting in higher clustering coefficients and shorter path lengths in $G_s$.
 
These observations, while simple, make the sample subgraph $G_s$ approximate the
characteristics of the original graph $G$ more accurately, even better than
topology-based sampling methods. 

\paragraph{Sampling very large graphs} As we discussed before, many real networks are now too large to fit into main memory. This raises the question: how can we sample from $G$ sequentially, one edge at a time, while minimizing the number of passes over the edges? As we discussed in Section~\ref{sec:computational-models}, most of the topology-based sampling methods would no longer be applicable, since they require many passes over the edges. Conversely, \algo runs sequentially and requires only two passes over the edges of $G$ (\ie, $\omega=2$). In the first pass, \algo samples edges uniformly from the stream of edges $E$ and adding their incident nodes to the sample. Then, it proceeds in a second pass by adding all edges which have both end-points already in the sample. This makes \algo suitable for sampling large graphs that cannot fit into main memory. 

Next, we analyze the characteristics of \algo and after that, in the evaluation, we show how it accurately preserves many of the properties of the graph $G$.

\subsection{Analysis of \algo}
\label{sec:theory}
\input{theory-analysis.tex}

%% file: theory-analysis.tex
In this section, we analyze the bias of \algo's node selection analytically by comparing to the unbiased case of uniform sampling in which all nodes are sampled with a uniform probability (\ie, $p={1\over N}$).
First, we denote $f_D$ to be the degree sequence of $G$ where $f_D(k)$ is the number of nodes with
degree $k$ in graph $G$. Let $n=|V_s|$ be the target number of nodes in
the sample subgraph $G_s$ (\ie., $\phi={n \over N}$). 

\paragraph{Upward bias to high degree nodes}
We start by analyzing the upward bias to select high degree nodes by calculating the expected value of the number of nodes with original degree $k$ that are added to the sample set of nodes $V_s$. 
Let $E[f_D(k)]$ be the expected value of $f_D(k)$ for the sampled set $V_s$ when $n$ nodes are sampled uniformly with probability $p= {1\over N}$:

\begin{eqnarray*}
E[f_D(k)] &=& f_D(k) \cdot n \cdot p \\
&=&  f_D(k) \cdot {n\over{N}} \\
\end{eqnarray*}

Since, \algo selects nodes proportional to their degree, the probability of sampling a node $v_i$ with degree $k_i\!=\!k$ is $p' \!=\! {k\over{\sum_{j=1}^{N}{k_j}}}$. Note that we can also express the probability as $p' \!=\! {k\over{2 \cdot |E|}}$. 
Then we let $E'[f_D(k)]$ denote the expected value of $f_D(k)$ for the sampled set $V_s$ when nodes are sampled with \algo:
\begin{eqnarray*}
E'[f_D(k)] &=&  f_D(k) \cdot n \cdot p' \\
&=&  f_D(k) \cdot n \cdot {k\over{2 \cdot |E|}}\\
\end{eqnarray*}

\noindent
This leads us to Lemma~\ref{lemma1}.
\noindent
\begin{lemma}
\algo is biased to high degree nodes. \\
$E'[f_D(k)] \geq E[f_D(k)]$ if $k \geq k_{avg}$, where $k_{avg}={{2 \cdot |E|}\over{N}}$ is the average degree in $G$.
\label{lemma1}
\end{lemma}

\noindent \textsc{Proof:}
Consider the threshold $k$ at which the expected value of $f_D(k)$ using \algo sampling is greater than the expected value of $f_D(k)$ using uniform random sampling:
\allowdisplaybreaks{
\begin{align*}
{E'[f_D(k)] - E[f_D(k)]} & \geq 0 \\
{{f_D(k) \cdot n \cdot {k\over{2 \cdot |E|}}} - {f_D(k) \cdot {n\over{N}}}} & \geq 0\\
{{f_D(k) \cdot {n \over N} \cdot {k\over{k_{avg}}}} - {f_D(k) \cdot {n\over{N}}}} & \geq 0\\
{{k\over{k_{avg}}} -1} & \geq 0  \\
{k\over{k_{avg}}} & \geq 1  \\
\end{align*}
}
\vspace{-2.mm}
The above statement is true when $k \geq k_{avg}$.
\hfill $\Box$
\vspace{2.mm}

\paragraph{Downward bias caused by sampling} 
Next, instead of focusing on the {\em original} degree $k$ as observed in the graph $G$, we focus on the {\em sampled} degree $k^s$ as observed in the sample subgraph $G_s$, where  $0 \leq k^s \leq k$.
Let $k_i^s$ be a random variable that represent the sampled degree of node $v_i$ in $G_s$, given that the original degree of node $v_i$ in $G$ was $k_i$.
We compare the expected value of $k_i^s$ when using uniform sampling to the expected value of $k_i^s$ when using \algo. 
Generally, the degree of the node $v_i$ in $G_s$ depends on how many of its neighbors in $G$ are sampled. When using uniform sampling, the probability of sampling one of the node's neighbors is $p= {1\over N}$: 
\begin{eqnarray*} 
E[k_i^s] &=& {\sum_{j=1}^{k_i} p \cdot n} = k_i \cdot  {n \over N} \\
\end{eqnarray*}

When using \algo, the probability of sampling any of the node's neighbors is proportional to the degree of the neighbor. Let $v_j$ be a neighbor of $v_i$ (\ie, $e_{ij}\!=\!(v_i,v_j) \in E$), then the probability of sampling $v_j$ is $p' = {k_j\over{2 \cdot |E|}}$:

\begin{eqnarray*} 
E'[k_i^s] &=&  \sum_{j=1}^{k_i} p' \cdot n \\
&=& {\sum_{j=1}^{k_i} {k_j \over {2 \cdot |E|}}  \cdot n  } \\
&=& {n \over N} \cdot {{\sum_{j=1}^{k_i} k_j} \over {k_{avg}}} \\
\end{eqnarray*}

Now, let us define the variable ${k_{\mathcal{N}}}={\sum_{k'} {k' \cdot P(k'|k)}}$, where $k_{\mathcal{N}}$ represents the average degree of the neighbors of a node with degree $k$ as observed in $G$. The function $k_{\mathcal{N}}$ has been widely used as a global measure of the {\em assortativity} in a network~\cite{newman2002assortative}. If ${k_{\mathcal{N}}}$ is increasing with $k$, then the network is assortative---indicating that nodes with high degree connect to, on average, other nodes with high degree. Alternatively, if ${k_{\mathcal{N}}}$ is decreasing with $k$, then the network is dissortative---indicating that nodes of high degree tend to connect to nodes of low degree. 

Note that here, we define ${k_{\mathcal{N}i}}={{\sum_{j=1}^{k_i} k_j} \over {k_i}}$ as the average degree of the neighbors of node $v_i$. Note that  ${k_{\mathcal{N}i}} \geq 1$. In this context, ${k_{\mathcal{N}i}}$ represents the local assortativity of node $v_i$, so then:
\begin{eqnarray*} 
E'[k_i^s] &=& {n \over N} \cdot {{\sum_{j=1}^{k_i} k_j} \over {k_{avg}}} \\
&=& {n \over N} \cdot {k_i \over {k_{avg}}} \cdot {{\sum_{j=1}^{k_i} k_j} \over {k_i}} \\ 
&=& k_i \cdot {n \over N}  \cdot {{k_{\mathcal{N}i}} \over {k_{avg}}} \\  
\end{eqnarray*}

This leads us to Lemma~\ref{lemma2}.

\begin{lemma}
The expected sampled degree in $V_s$ using \algo is greater than the expected sampled degree based on uniform node sampling.
For any node $v_i \in V_s$, $E'[k_i^s] \geq E[k_i^s]$ if ${k_{\mathcal{N}i}} \geq k_{avg}$, where the average degree in $G$ is $k_{avg}={{2\cdot |E|}\over{N}}$ and the average degree of $v_i$'s neighbors in $G$ is ${k_{\mathcal{N}i}}={{\sum_{j=1}^{k_i} k_j} \over {k_i}}$.
\label{lemma2}
\end{lemma}

\noindent \textsc{Proof:}
Consider the threshold $k$ at which the expected value of $k^s$ using \algo sampling is greater than the expected value of $k^s$ using uniform random sampling:

\allowdisplaybreaks{
\begin{align*}
{E'[k_i^s] - E[k_i^s]} & \geq 0 \\
{k_i \cdot {n \over N}  \cdot {{k_{\mathcal{N}i}} \over {k_{avg}}}} - {k_i \cdot  {n \over N}}  & \geq 0\\
{{k_{\mathcal{N}i} \over {k_{avg}}} - 1} & \geq 0 \\
{{k_{\mathcal{N}i} \over {k_{avg}}} } & \geq 1
\end{align*}
}
\vspace{-2.mm}
The above statement is true when $k_{\mathcal{N}i} \geq k_{avg}$.
\hfill $\Box$
\vspace{2.mm}

Generally, for any sampled node $v_i$ with degree $k_i$, if the average degree of $v_i$'s neighbors is greater than the average degree of $G$, then the expected sampled degree $k_i^s$ under \algo is higher than if uniform sampling is used. This typically the case in many real networks where high degree nodes are connected with other high degree nodes.

\begin{figure*}[h!t!]
\centering
\vspace{-5.mm}
\subfigure[Original degree in $G$, $k$]{\label{fig:arxiv deg dist}\includegraphics[width=0.45\textwidth]{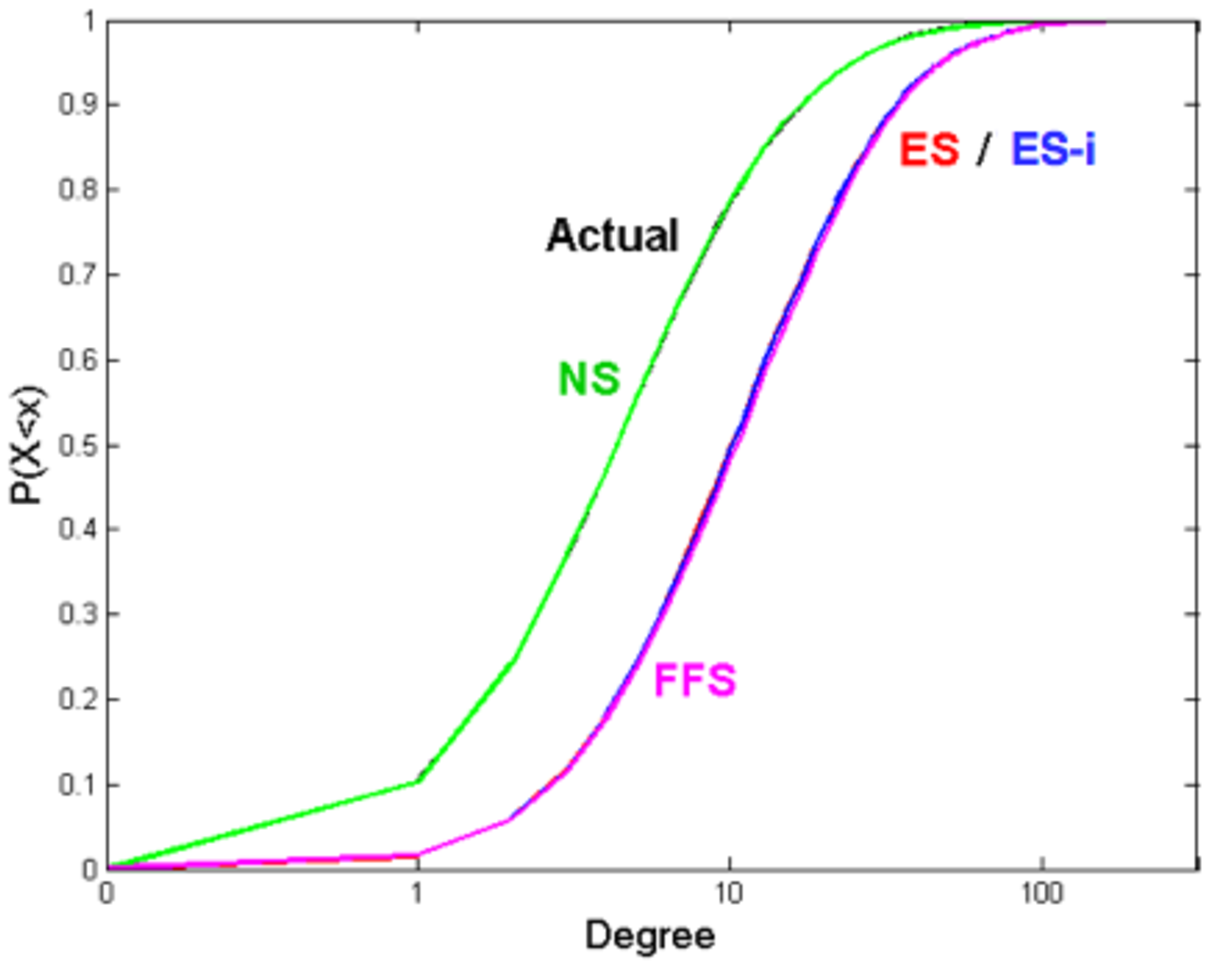}}
\hspace{-5.mm}
\subfigure[Sampled degree in $G_s$, $k^s$]{\label{fig:condmat deg dist}\includegraphics[width=0.45\textwidth]{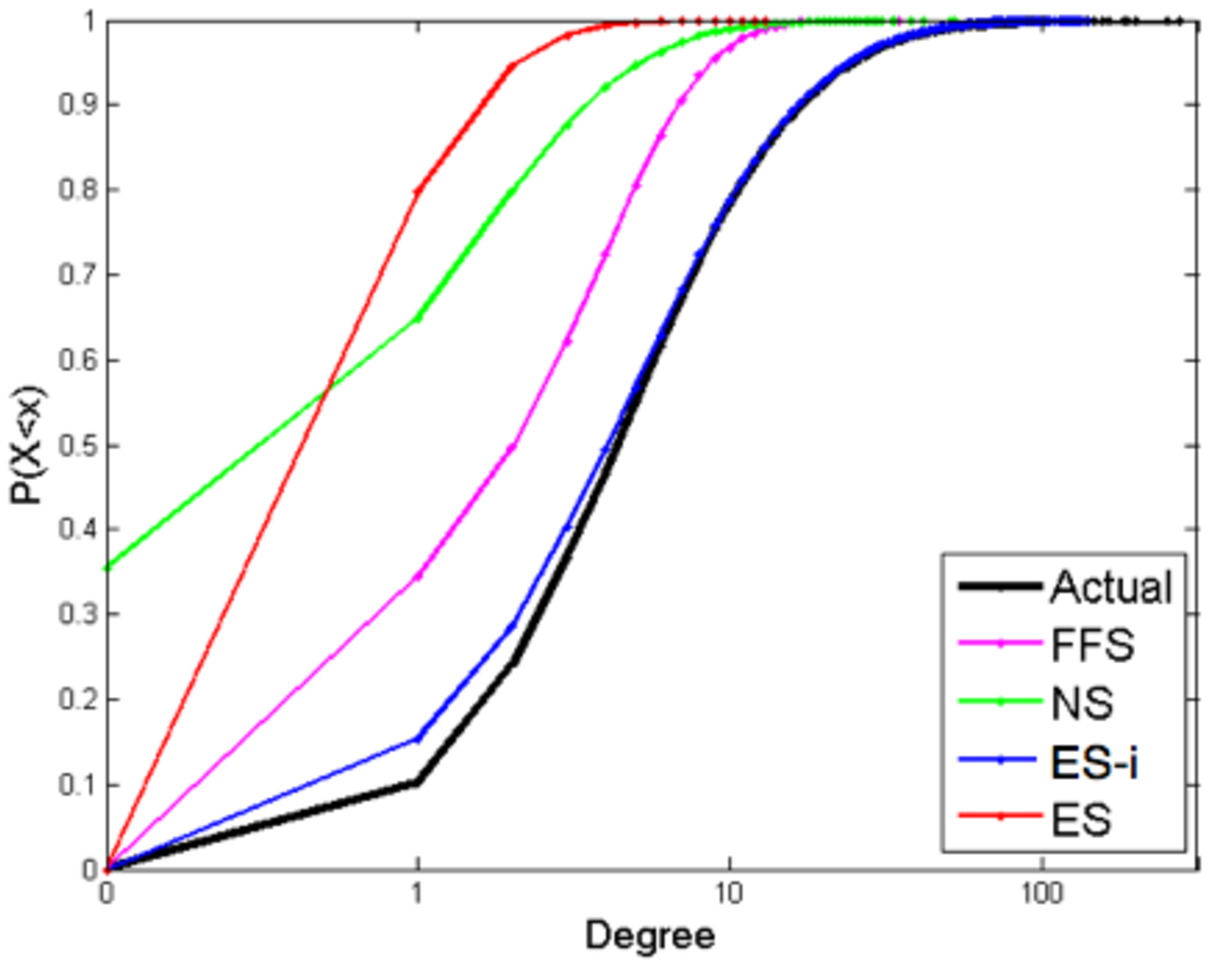}}
\vspace{-3mm}
\caption{Illustration of original degrees (in $G$) vs. sampled degrees (in $G_s$) for subgraphs selected by NS, ES, \algo, and FFS on the CondMAT network.}
\label{fig:deg-example}
\vspace{-1mm}
\end{figure*}

In Figure~\ref{fig:deg-example}, we empirically investigate the difference between the
sampled degrees $k^s$ and original degrees $k$ for an example network---the \textsc{CondMAT} graph. Specifically we compare the degree distribution of $G_s$ (in Figure~\ref{fig:deg-example}b) to the degree distribution of $G$ (in Figure~\ref{fig:deg-example}a) as observed only from the set of sampled nodes $V_s$.
Furthermore, we compare to the {\em full} degree distribution of $G$, \ie, as observed over the set of all nodes $V$.

In Figure~\ref{fig:deg-example}a, NS accurately estimates the actual degree distribution as observed in $G$. However, in Figure~\ref{fig:deg-example}b, NS underestimates the sampled degree distribution in $G_s$, \ie, the NS curve
is skewed upwards. On the other hand, in Figure~\ref{fig:deg-example}a, ES, FFS, and \algo overestimate the original degree distribution in $G$, since they are biased to selecting higher degree nodes. In Figure~\ref{fig:deg-example}b, ES and FFS both underestimate the sampled degree distribution as observed in $G_s$. In contrast to other sampling methods, \algo comes closer to replicating the original degree distribution of $G$ in $G_s$. We conjecture that the combination of high degree bias together with the graph induction helps \algo to compensate for the underestimation caused by sampling subgraphs.

%% file: evaluation-static.tex
\subsection{Experimental Evaluation}
\label{sec:evaluation-stat}
In this section, we present results evaluating the various sampling methods on static graphs. We compare the performance of our proposed algorithm \algo to other algorithms from each class (as discussed in section~\ref{sec:sampling-classes}): node sampling (NS), edge sampling (ES) and forest fire sampling (FFS). We compare the algorithms on seven different real-world networks.  We use online social networks from \textsc{Facebook} from New Orleans City \cite{viswanath-2009-activity} and \textsc{Flickr} \cite{PURR1002}. Social media networks drawn from \textsc{Twitter}, corresponding to users tweets surrounding the United Nations Climate Change Conference in Copenhagen, December 2009 ($\#cop15$) \cite{ahmed2010time}. Also, we use a citation graph from ArXiv \textsc{HepPh}, and a collaboration graph from \textsc{CondMAT}~\cite{leskovecRepository}.  Additionally, we use an email network \textsc{Email-Univ} that corresponds to a month of email communication collected from Purdue university mail-servers \cite{ahmed2012space}. Finally, we compare the methods on a large social network from \textsc{LiveJournal}~\cite{leskovecRepository} with $4$ million nodes (included only at the $20\%$ sample size). Table~\ref{tab:point stats} provides a summary of the global statistics of these six network datasets.

\begin{table*} 
\small
\centering 
\caption{Characteristics of Network Datasets}
\begin{tabularx}{\textwidth}{lXXXXXX}
\toprule 
\textbf{{Graph}} & \textbf{{Nodes}} & \textbf{{Edges}} & \textbf{{Weak Comps.}} & \textbf{{Avg.  Path}} & \textbf{{Density}} & \textbf{{Global Clust.}}\\
\midrule
\textsc{HepPH} & 34,546 & 420,877 & 61 & 4.33 & $7\times10^{-4}$ & 0.146\\
\textsc{CondMAT} & 23,133 & 93,439 & 567 & 5.35 & $4\times10^{-4}$ & 0.264\\
\midrule
\textsc{Twitter} & 8,581 & 27,889 & 162 & 4.17 & $7\times10^{-4}$ & 0.061\\
\textsc{Facebook} & 46,952 & 183,412 & 842 & 5.6  & $2\times10^{-4}$ & 0.085\\
\textsc{Flickr} & 820,878 & 6,625,280 & 1 & 6.5 & $1.9\times10^{-5}$& 0.116\\
\textsc{LiveJournal} & 4,847,571 & 68,993,773 & 1876 & 6.5 & $5.8\times10^{-6}$& 0.2882\\
\midrule
\textsc{Email-PU Univ} & 214,893 & 1,270,285 & 24 & 3.91 & $5.5\times10^{-5}$ & 0.0018\\
\bottomrule
\end{tabularx}
\vspace{-2mm}
\label{tab:point stats}
\end{table*} 

Next, we outline the experimental methodology and results. For each experiment, we apply to the full network and sample subgraphs over a range of sampling fractions $\phi=[5\%,40\%]$. For each sampling fraction, we report the average results over ten different trials.   

\begin{figure*}[!h!t]
\centering
\subfigure[Degree]{\label{fig:avg ks deg}\includegraphics[width=0.268\textwidth]{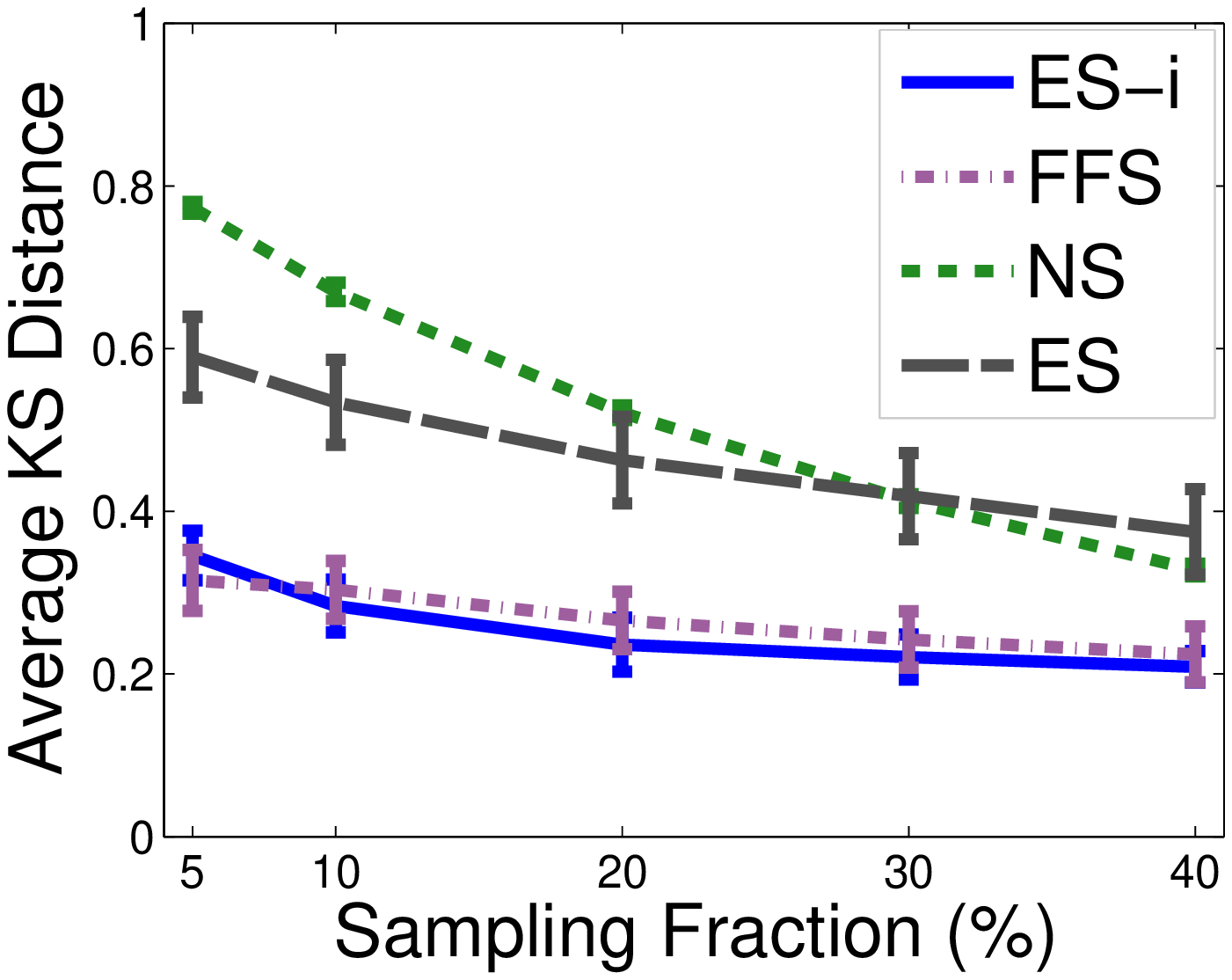}}
\hspace{-5.mm}
\subfigure[Path length]{\label{fig:avg ks plen}\includegraphics[width=0.268\textwidth]{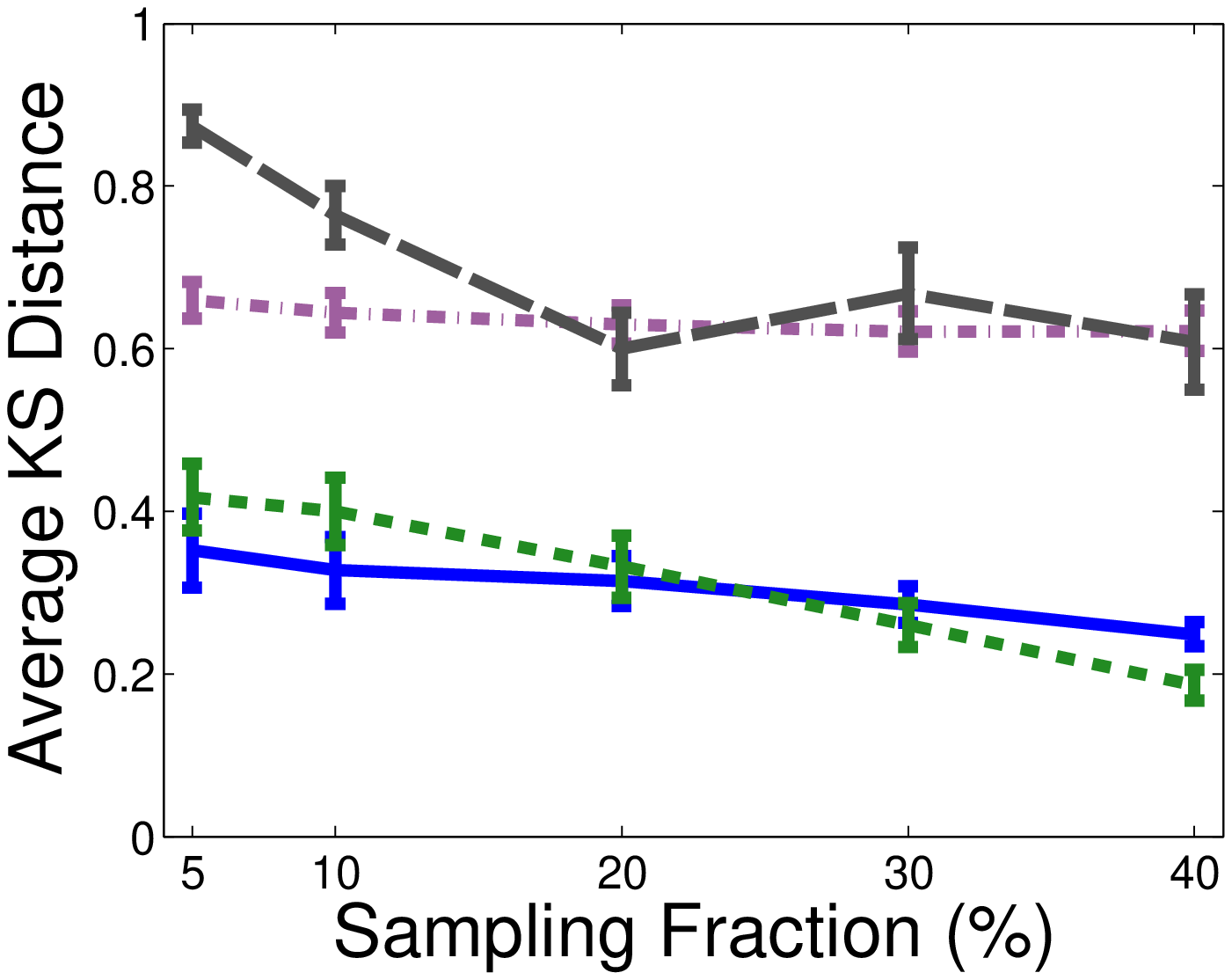}}
\hspace{-5.mm}
\subfigure[Clustering Coefficient]{\label{fig:avg ks cc}\includegraphics[width=0.268\textwidth]{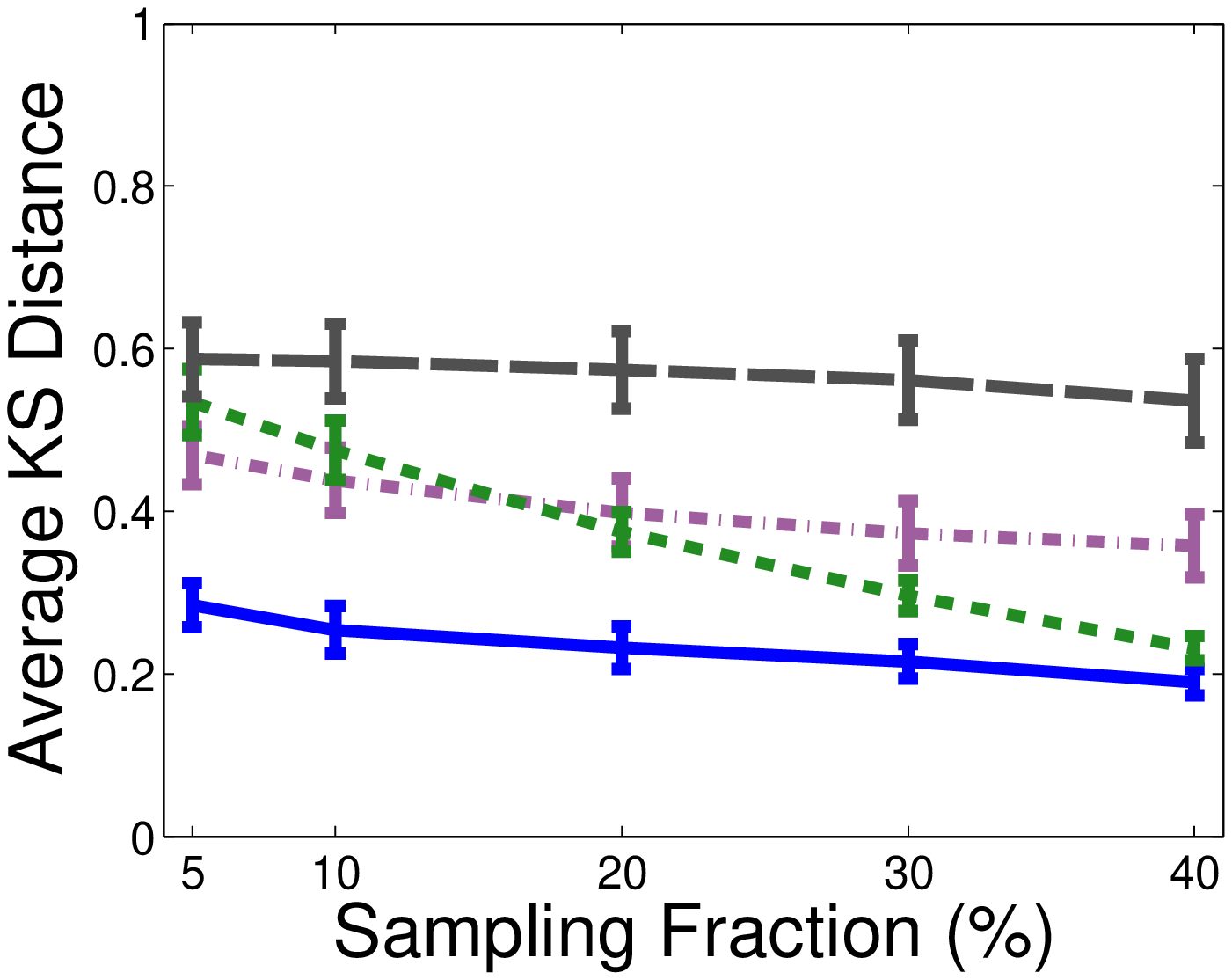}}
\hspace{-5.mm}
\subfigure[Kcore decomposition]{\label{fig:avg ks core}\includegraphics[width=0.268\textwidth]{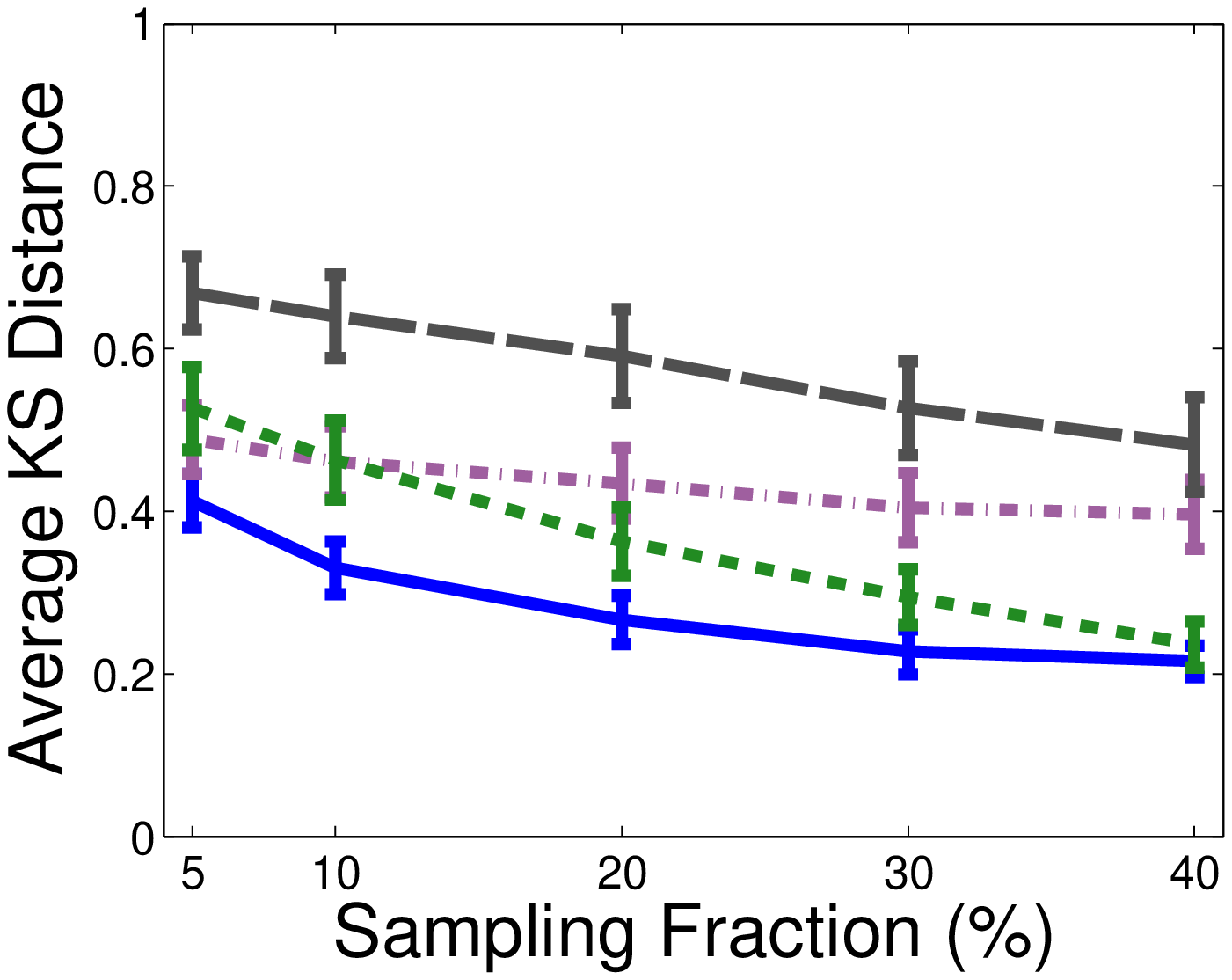}}\\
\vspace{-2mm}
\hspace{-2.mm}
\subfigure[Degree]{\label{fig:avg skl deg}\includegraphics[width=0.268\textwidth]{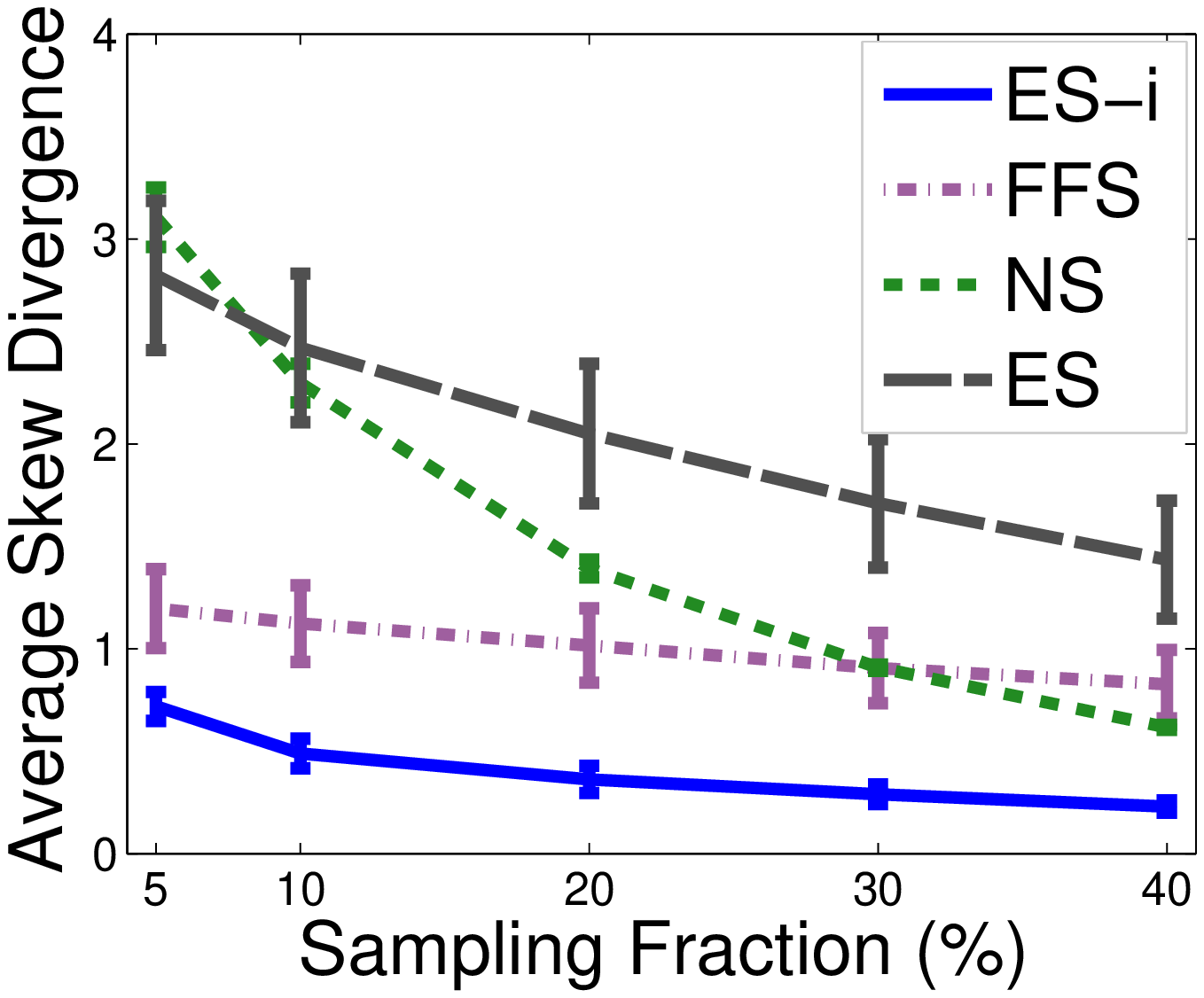}}
\hspace{-5.mm}
\subfigure[Path length]{\label{fig:avg skl plen}\includegraphics[width=0.268\textwidth]{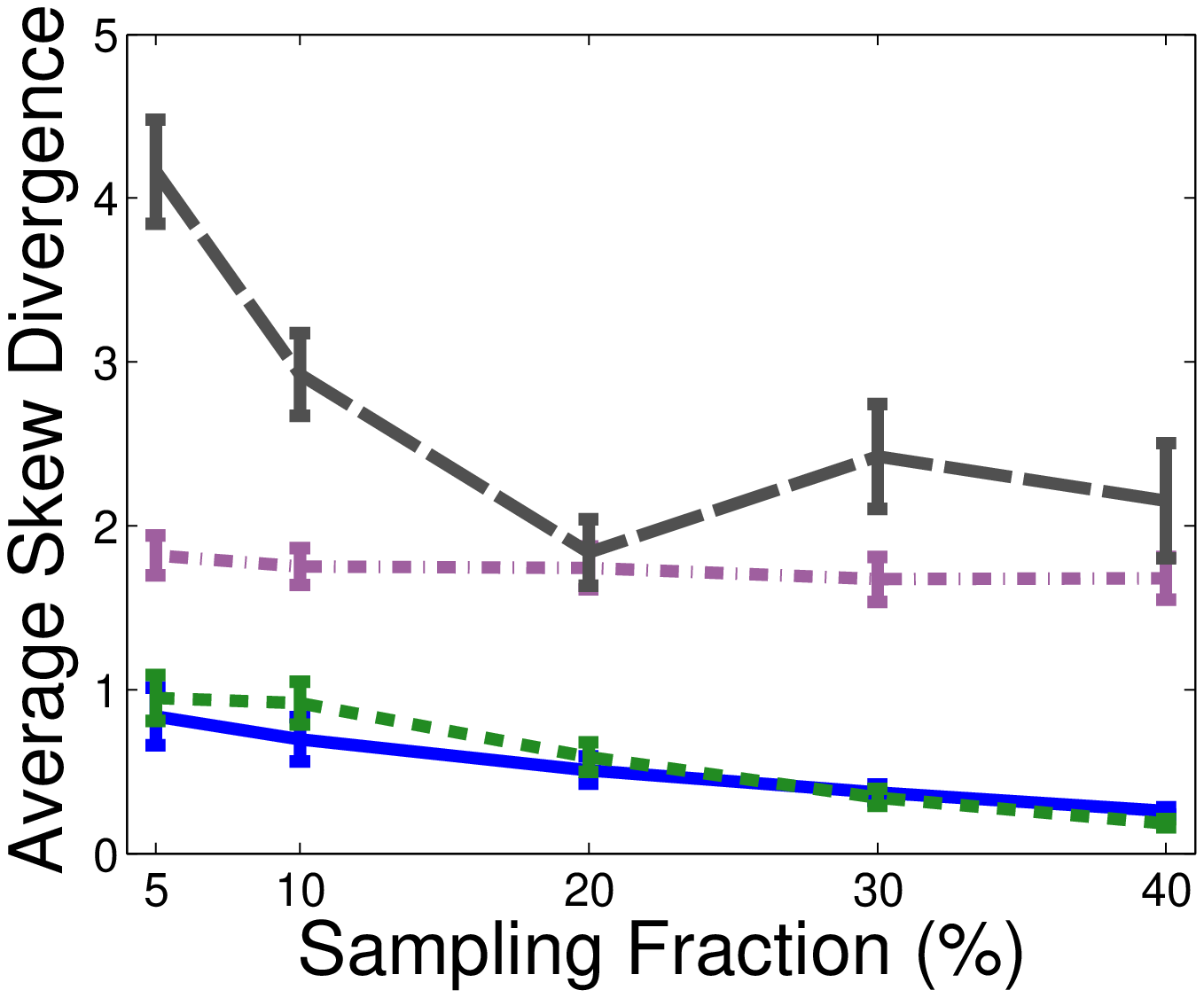}}
\hspace{-5.mm}
\subfigure[Clustering Coefficient]{\label{fig:avg skl cc}\includegraphics[width=0.268\textwidth]{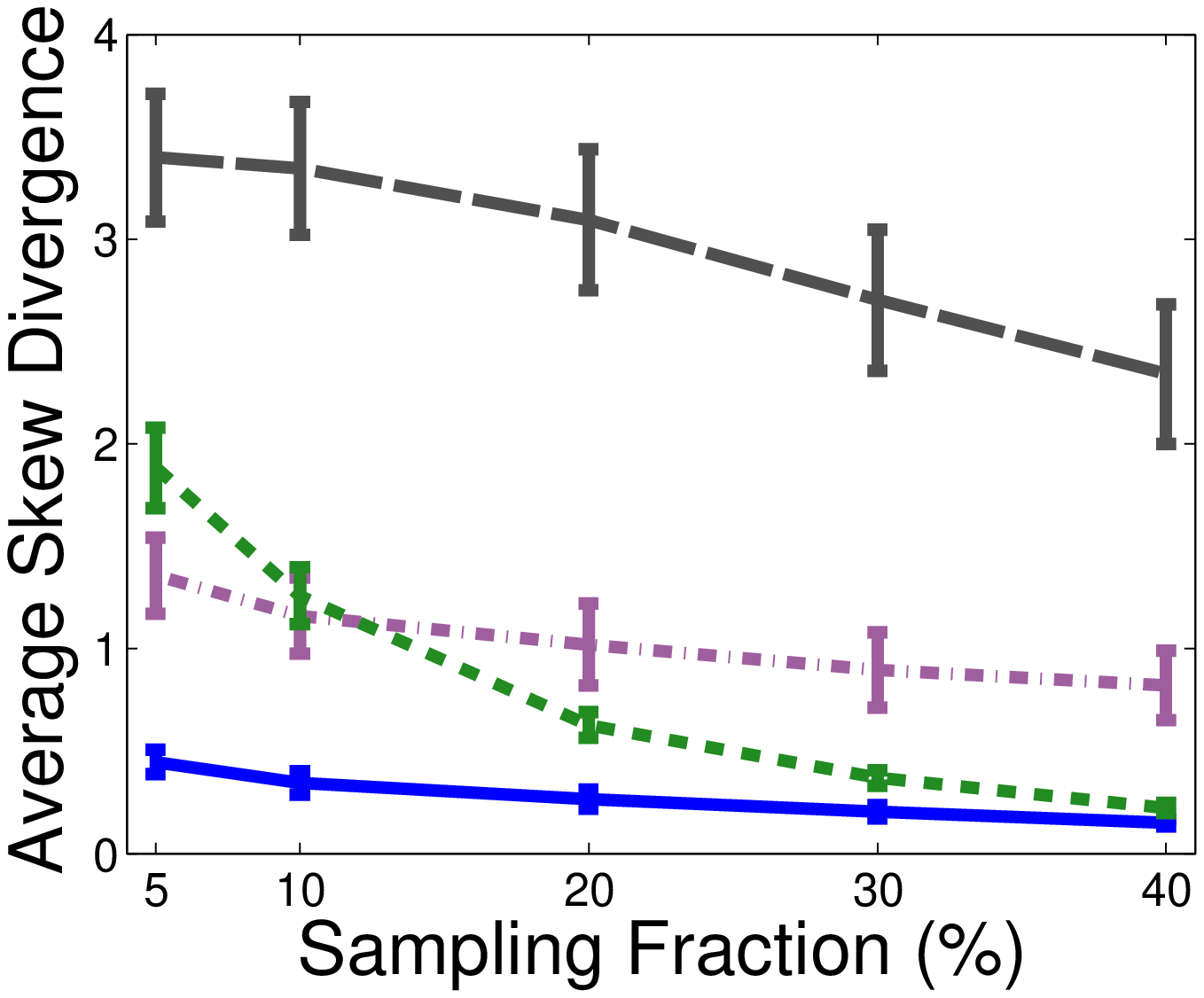}}
\hspace{-5.mm}
\subfigure[Kcore decomposition]{\label{fig:avg skl core}\includegraphics[width=0.268\textwidth]{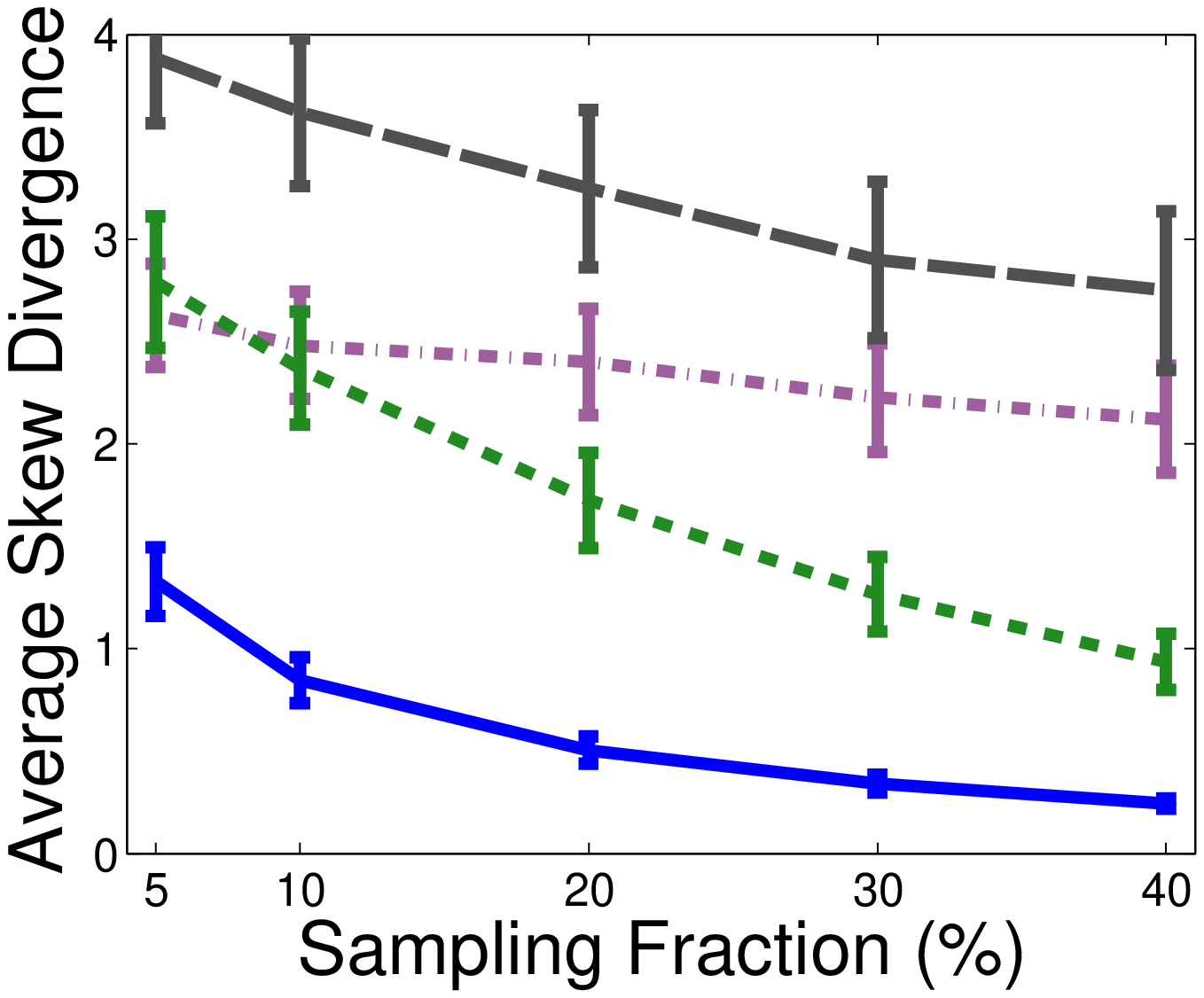}}
\vspace{-2mm}
\hspace{-2.mm}
\subfigure[Eigen Values]{\label{fig:avg evals}\includegraphics[width=0.268\textwidth]{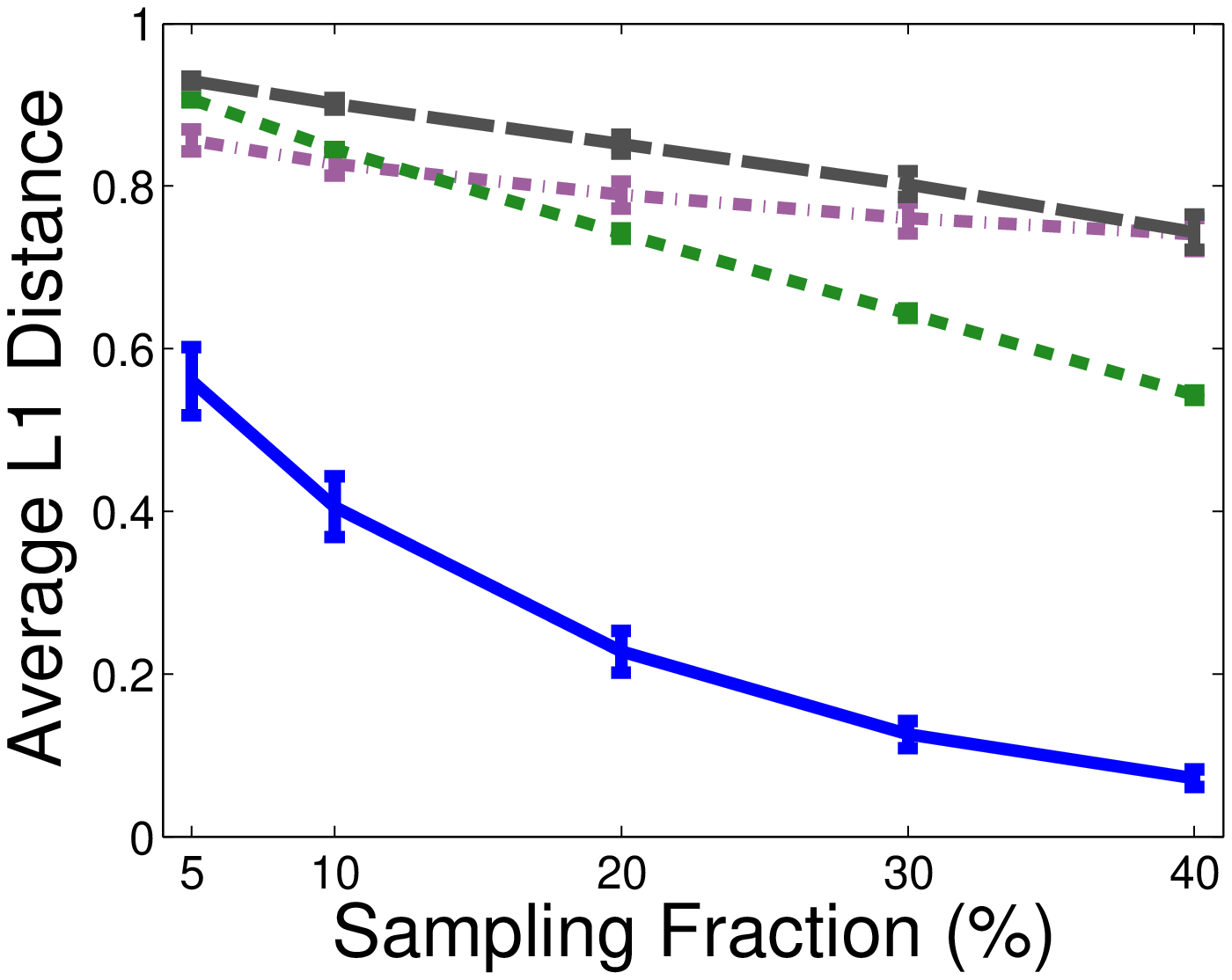}}
\hspace{-5.mm}
\subfigure[Network Value]{\label{fig:avg netvals}\includegraphics[width=0.268\textwidth]{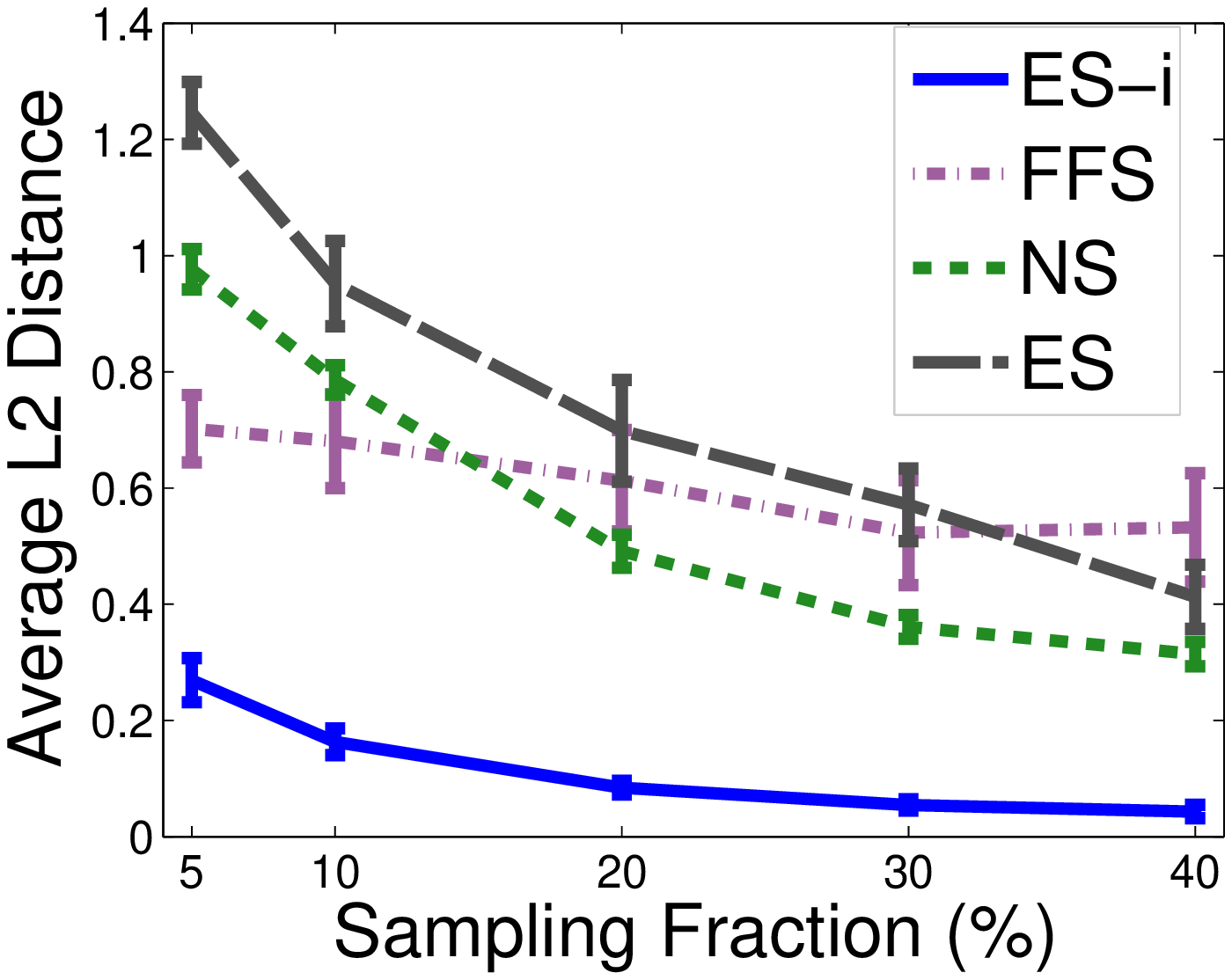}}
\vspace{-2mm}
\caption{(a-d) Average KS distance, (e-h) average skew divergence, and (i-j) average L1 and L2 distance respectively, across 6 datasets.
}\label{fig:all_distances}
\vspace{-2mm}
\end{figure*}

\paragraph{Distance metrics} Figures~\ref{fig:avg ks deg}--\ref{fig:avg ks core} show the average KS statistic for degree, path length, clustering coefficient, and k-core distributions on the six datasets. Generally, \algo outperforms the other methods for each of the four distributions. FFS performs similar to \algo in the degree distribution, however, it does not perform well for path length, clustering coefficient, and k-core distributions. This implies that FFS can capture the degree distribution but not connectivity between the sampled nodes. NS performs better than FFS and ES for path length, clustering coefficient, and k-core statistics but not for the degree statistics. This is due to the uniform sampling of the nodes that makes NS is more likely to sample low degree nodes and miss the high degree nodes (as discussed in~\ref{sec:theory}). Clearly, as the sample size increases, NS is able to select more nodes and thus the KS statistic decreases. \algo and NS perform similarly for path length distribution. This is because they both form a fully induced subgraph out of the sampled nodes.  Since induced subgraphs are more connected, the distance between pairs of nodes is generally smaller in induced subgraphs. 

In addition, we also used skew divergence as a second measure. Figures~\ref{fig:avg skl deg}--\ref{fig:avg skl core} show the average skew divergence statistic for degree, path length, clustering coefficient, and k-core distributions on the six datasets. Note that skew divergence computes the divergence between the sampled and the real distributions on the entire support of the distributions. While the skew divergence shows that \algo outperforms the other methods similar to KS statistic, it also shows that \algo performs significantly better across the entire support of the distributions. 

Finally, Figures~\ref{fig:avg evals} and ~\ref{fig:avg netvals} show the L1 and L2 distances for eigenvalues and network values respectively. Clearly, \algo outperforms all the other methods that fail to improve their performance even when the sample size is increased up to $40\%$ of the full graph.  

\paragraph{Distributions} While the distance measures are important to quantify the divergence between the sampled and the real distributions, by analyzing only the distance measures it is unclear whether the sampled statistics are an over-estimate or under-estimate of the original statistics. Therefore, we plot the distributions of for all networks at the $20\%$ sample size. We choose the $20\%$ as a representative sample size, however, we note that same conclusions hold for the other sample sizes. Note that we plot the $CCDF$ for degree and k-core distributions, $CDF$ for path length and clustering coefficient distribution, and we plot eigenvalues and network values versus the rank. 
Figures~\ref{fig:dist_comp_fbor},~\ref{fig:dist_comp_condmat},~\ref{fig:dist_comp_arxiv},~\ref{fig:dist_comp_twcop},~\ref{fig:dist_comp_email}, ~\ref{fig:dist_comp_flickr}, and~\ref{fig:dist_comp_socjor} (Appendix~\ref{appendix}) show the plots for all the distributions across all networks. 

\begin{description}
\item[\textbf{Degree Distribution}] Across all networks, \algo captures the tail of the degree distribution (high degree nodes) better than NS, ES, and FFS. However, \algo under-estimates the low degree nodes for \textsc{Twitter}, \textsc{Email-Univ}, and \textsc{Flickr}. FFS and NS capture a large fraction of low degree nodes but they fail to capture the high degree nodes. 
\vspace{2mm}
\item[\textbf{Path length Distribution}] \algo preserves the path length distribution of \textsc{HepPH}, \textsc{CondMAT}, and \textsc{LiveJournal}, however, it underestimates the distributions of \textsc{Twitter}, \textsc{Email-Univ}, and \textsc{Flickr}. Conversely, NS over-estimates the distributions of \textsc{HepPH}, \textsc{CondMAT}, and \textsc{LiveJournal} but successfully preserves the distributions of the other datasets.
\vspace{2mm}
\item[\textbf{Clustering Coefficient Distribution}] \algo generally captures the clustering coefficient more accurately than other methods. While \algo under-estimates the low clustering coefficients specially in \textsc{Email-Univ}, and \textsc{Flickr}, the other methods fail to capture the clustered structures in almost all the datasets.
\vspace{2mm}
\item[\textbf{K-Core Distribution}] Similar to the previous statistics, \algo nicely preserves the distribution of core sizes for \textsc{HepPH}, \textsc{CondMAT}, and \textsc{Facebook}, but it over-estimates the core structures of the other datasets. On the other hand, NS, ES, and FFS generally fail to capture the core structures for the majority of the datasets (except \textsc{Flickr}).   
In addition to the distribution of the core sizes, we compared the {\em max-core} number in the sampled graphs to their real counterparts for the $20\%$ sample size (Table~\ref{table:kcore_static}).  Note that the {\em max-core number} is the maximum value of $k$ in the k-cor distribution. In contrast to \algo, the {\em max-core number} of NS, ES, and FFS is consistently an order of magnitude smaller than the real {\em max-core number}. This indicates that NS, ES, and FFS do not preserve the local density in the sampled subgraph structures. 
\vspace{2mm}
\item[\textbf{Eigenvalues}] The NS, ES, and FFS methods generally fail to approximate the eigenvalues of the original graph in the sample. Conversely, \algo accurately approximates the eigenvalues of \textsc{Twitter}, \textsc{Email-Univ}, \textsc{Flickr}, and \textsc{LiveJournal} and closely approximates the eigenvalues of \textsc{HepPH}, \textsc{CondMAT}, and \textsc{Facebook} (at $20\%$ sample size). By the {\em interlacing} theorem of the eigenvalues of induced subgraphs~\cite{haemers1995interlacing}, the eigenvalues of \algo in $G_s$ can be used to estimate bounds on their counterparts in the input graph $G$: $\lambda_i \leq \mu_i \leq \lambda_{i+(N-n)}$ such that $\mu_i$ is the $i^{th}$ eigenvalue of $G_s$, and $\lambda_i$ is the $i^{th}$ eigenvalue of $G$.
\vspace{2mm}
\item[\textbf{Network values}] Similar to the other graph measures, \algo accurately approximates the network values of the graph compared to other methods.
\end{description}
\vspace{-2mm}

\begin{table*}[h!]
\centering\small
\begin{minipage}[c]{0.65\textwidth}
\caption{Comparison of {\em max-core-number} at the 20\% sample size for \algo, NS, ES, FFS versus Real value of $G$.}
\begin{tabularx}{1.0\textwidth}{ l ccccc }
\toprule
\textbf{Graph} & \textbf{Real max core no.} & \textbf{ES-i} & \textbf{NS} & \textbf{ES} & \textbf{FFS} \\
\midrule
\small
\textsc{HepPH}
&30&23\fivepc &8&2&4\\
\textsc{CondMAT}
&25&20\fivepc &7&2&6\\
\midrule
\textsc{Twitter}
&18&18\fivepc &5&2&3\\
\textsc{Facebook}
&16&14\fivepc &4&2&3\\
\textsc{Flickr}
&406&406\fivepc &83&21&7\\
\textsc{LiveJournal}
&372&372\fivepc &84&6&7\\
\midrule
\textsc{Email-UNIV}
&47&46\fivepc &15&3&7\\
\bottomrule
\end{tabularx}
\vspace{-1mm}
\label{table:kcore_static}
\end{minipage}
\end{table*}

\noindent
\textit{\textbf{Summary}---We summarize the main empirical conclusions in this section:}

\textit{
\begin{enumerate}[(1)]
\vspace{-4mm}
\item Sampled subgraphs collected and constructed by \algo accurately preserve a range of network statistics that capture both local and global distributions of the graph structure.
\item Due to its bias to selecting high degree nodes, \algo generally favors dense and clustered areas of the graph, which results in connected sample subgraphs---in contrast with other methods.
\item NS, ES and FFS generally construct more sparsely connected sample subgraphs.
\end{enumerate}
}

%% file: sampling-stream.tex
\section{Sampling from Streaming Graphs}
\label{sec:sampling-stream}
In this section, we focus on how to sample a representative subgraph $G_s$ from $G$, such that $G$ is presented as a stream of edges in no particular order. Note that in this paper we focus on space-efficient sampling methods.
Using the definition of a streaming graph
sampling algorithm as discussed in Section~\ref{sec:computational-models}, we now present streaming variants of
different sampling algorithms as discussed in Section~\ref{sec:sampling-classes}.

\paragraph{Streaming Node Sampling} One key problem with
traditional node sampling we discussed in~\ref{sec:sampling-classes} is that the algorithms assume that nodes can be accessed at
random. In our stream setting, new nodes arrive into the
system only when an edge that contains the new node is added
into the system; it is therefore difficult to identify which $n$
nodes to select {\em a priori}. To address this, we utilize the idea of reservoir sampling~\cite{Vitter:85} to implement a streaming variant of node sampling (see Algorithm~\ref{algo:NS}).

The main idea is to select nodes uniformly at random with
the help of a uniform random hash function. A uniform random hash function defines a true random permutation on the nodes in the graph, meaning that any node is equally likely to be the minimum. Specifically, we keep track of
nodes with $n$ smallest hash values in the graph; nodes are
only added if their hash values represent the top-$n$
minimum hashes among all nodes seen thus far in the stream.
Any edge that has both vertices already in the reservoir is
automatically added to the original graph. 

Since the reservoir is finite, a node with smaller hash value may arrive late in the stream and replace a node that was sampled earlier. 
In this case, all edges incident to the node that was dropped will be removed from the sampled subgraph. 
Once the reservoir is filled up to $n$ nodes, it will remain at $n$ nodes, but since the selection is based on the hash value, nodes will be dropped and added as the algorithm samples from all portions of the stream (not just
the front). Therefore, it guarantees a uniformly sampled set of nodes from the graph stream. 

\begin{algorithm}[h]
\DontPrintSemicolon
\SetKwInOut{Input}{Input}
\SetKwInOut{Output}{Output}
\Input{Sample Size $n$, Graph Stream $S$}
\Output{Sampled Subgraph $G_s = (V_s, E_s)$}
\BlankLine
$V_s = \emptyset, E_s = \emptyset$\;
$h$ is fixed uniform random hash function\;
$t=1$\;
\For{$e_t$ in the graph stream $S$}{
		$(u, v) = e_t$\;
       \If {$u \notin V_s \&$ $h(u)$ is top-n min hash}{
		$V_s = V_s \cup u$\;
		Remove all edges incident on replaced node\;
       }
       \If {$v \notin V_s \&$ $h(v)$ is top-n min hash}{
		$V_s = V_s \cup v$\;
		Remove all edges incident on replaced node \;
       }
       \If {$u,v \in V_s$}{
  		$E_s = E_s \cup e$\;
	}
	$t=t+1$ \;
}
\caption{Streaming Node Sampling NS($n$,$S$)}
\label{algo:NS}
\end{algorithm}

\paragraph{Streaming Edge Sampling} 
Streaming edge sampling can be implemented similar to streaming node sampling. Instead of
hashing individual nodes, we focus on using hash-based
selection of edges (as shown in Algorithm~\ref{algo:ES}). We use the approach that was first proposed in ~\cite{aggarwal2011outlier}. 
More precisely, if we are interested in obtaining $m$ edges
at random from the stream, we can simply keep a reservoir of
$m$ edges with the minimum hash value.  Thus, if a new edge
streams into the system, we check if its hash value is
within top-$m$ minimum hash values. If it is not, then we do
not select that edge, otherwise we add it to the reservoir
while replacing the edge with the previous highest top-$m$
minimum hash value.
One problem with this approach is that our goal is often in
terms of sampling a certain number of nodes $n$. Since we use a
reservoir of edges, finding the right $m$ that provides $n$
nodes is hard. It also keeps varying depending on
which edges the algorithm ends up selecting. Note that
sampling fraction could also be specified in terms of
fraction of edges; the choice of defining it in terms of
nodes is somewhat arbitrary in that sense. For our
comparison purposes, we ensured that we choose a large
enough $m$ such that the number of nodes was much higher
than $n$, but later iteratively pruned out sampled edges
with the maximum hash values until the target number of
nodes $n$ was reached.

\begin{algorithm}[h!]
\DontPrintSemicolon
\SetKwInOut{Input}{Input}
\SetKwInOut{Output}{Output}
\Input{Sample Size $n$, Graph Stream $S$}
\Output{Sampled Subgraph $G_s = (V_s, E_s)$}
\BlankLine
$V_s = \emptyset, E_s = \emptyset$\;
$h$ is fixed uniform random hash function\;
$t=1$\;
\For{$e_t$ in the graph stream $S$}{
       $(u, v) = e_t$\;
	\If {$h(e_t)$ is in top-$m$ min hash}{
           $E_s= E_s \cup e_t$\; 
	     $V_s = V_s \cup \{u,v\}$\;
        }
        Iteratively remove edges in $E_s$ in decreasing order such that $|V_s|=n$ nodes\;  
        $t=t+1$\; 
}
\caption{Streaming ES($n$, $S$)}
\label{algo:ES}
\end{algorithm}

\paragraph{Streaming Topology-Based Sampling} We also
consider a streaming variant of a topology-based sampling
algorithm. Specifically, we consider a simple BFS-based
algorithm (shown in Algorithm~\ref{algo:BFS}) that works as follows. This algorithm essentially
implements a simple breadth-first search on a sliding window
of $w$ edges in the stream. In many respects, this algorithm
is similar to the forest-fire sampling (FFS) algorithm. Just
as in FFS, it starts at a random node in the
graph and selects an edge to burn (as in FFS parlance) among
all edges incident on that node within the sliding window. 
For every edge burned, let $v$ be the incident node at the other end of the burned edge. 
We enqueue $v$ onto a queue $Q$ in order to get a chance to burn its incident edges within the window.
For every new streaming edge, the sliding window moves one step,
which means the oldest edge in the window is dropped and a
new edge is added. (If that oldest edge was sampled, it will
still be part of the sampled graph.) 
If as a result of the sliding window moving one step, the
node has no more edges left to burn, then the burning
process will dequeue a new node from $Q$. If the queue is empty, the process jumps to a random node within the sliding window
(just as in FFS). This way, it does BFS as much as possible
within a sliding window, with random jumps if there is no
more edges left to explore.  Note that there may be other
possible implementation of a streaming variant for topology-based sampling, but since
to our knowledge, there are no streaming algorithms in the literature, we
include this as a reasonable approximation for comparison. 
This algorithm has a similar problem as the edge sampling
variant in that it is difficult to control the exact number of
sampled nodes, and hence some additional pruning needs to be
done at the end (see Algorithm~\ref{algo:BFS}).

\begin{algorithm}[h!]
\DontPrintSemicolon
\SetKwInOut{Input}{Input}
\SetKwInOut{Output}{Output}
\Input{Sample Size $n$, Graph Stream $S$,Window Size=$wsize$}
\Output{Sampled Subgraph $G_s = (V_s, E_s)$}
\BlankLine
$V_s = \emptyset, E_s = \emptyset$\;
$W = \emptyset$\;
Add the first $wsize$ edges to $W$\;
$t=wsize$\;
Create a queue $Q$\;
// uniformly sample a node from $W$\;
$u=Uniform(V_W)$\;
\For{$e_t$ in the graph stream $S$}{
	\textbf{if} $u \notin V_s$ \textbf{then} add $u$ to $V_s$\;
	\eIf {$W.incident\_edges(u) \neq \emptyset$}{
		Sample $e$ from $W.incident_edges(u)$\;
		Add $e=(u, v)$ to $E_s$\;
		Remove $e$ from $W$\;
		Add  $v$ to $V_s$\;
		Enqueue $v$ onto $Q$\;
		}
		{ 
		\lIf{$Q = \emptyset$}{$u=Uniform(V_W)$}\;
		\lElse{$u = Q.dequeue()$}
		}
		Move the window $W$\;       
       \If {$|V_s|$ \textgreater $ n$}{
  		Retain $[e] \subset E_s$ such that $[e]$ has $n$ nodes\;
		Output $G_s = (V_s, E_s)$\;
	}	
       $t=t+1$\;
}
\caption{Streaming BFS($n$, $S$,$wsize$)}
\label{algo:BFS}
\end{algorithm}

\paragraph{Partially-Induced Edge Sampling (PIES)}
We finally present our main algorithm called PIES that
outperforms the above implementation of stream sampling algorithms.  
Our approach discussed in section~\ref{sec:sampling-stat} outlines a sampling algorithm based on edge sampling concepts. A key advantage of using edge sampling is its bias towards high degree nodes. 
This upward bias helps offset the downward bias (caused by subgraph sampling) to some extent. Afterwards, forming the induced graph will help capture the connectivity among the sampled nodes. 

Unfortunately, full graph induction in a streaming fashion
is difficult, since node selection and graph induction requires
at least two passes (when implemented in the obvious,
straightforward way).  Thus, instead of full induction of
the edges between the sampled nodes, we can utilize {\em
partial} induction and combine edge-based node sampling
with the graph induction (as shown in
Algorithm~\ref{algo:pies}) in a single step. The partial
induction step induces the sample in the forward direction only. 
in other words, it adds any edge among a pair of sampled nodes if it
occurs {\em after} both the two nodes were added to the sample.

PIES aims to maintain a dynamic sample while the graph is
streaming by utilizing the same reservoir sampling idea we have
used before.  In brief, we add the first $m$ edges of
the stream to a {\em reservoir} such that the reservoir contains $n$ nodes and then the rest of the
stream is processed randomly by replacing existing records
in the reservoir.  Specifically, PIES runs over the edges
in a single pass, and adds deterministically the first $m$
edges incident to $n$ nodes of the stream to the sample.  Once it achieves
the target sample size, then for any streaming edge, it adds
the incident nodes to the sample (probabilistically) by
replacing other sampled nodes from the node set
(selected uniformly at random).  At each step, it will also add the
edge to the sample if its two incident nodes are already in the sampled
node set---producing a partial induction effect). 

\vspace{3mm}
\noindent
Now, we discuss the properties of PIES to illustrate its characteristics.
\begin{enumerate}
\item \textit{PIES is a two-phase sampling method.}
\noindent 
A two-phase sampling method is a method in which an initial sample of units is selected from the population (\eg, the graph stream), and then a second sample is selected as a subsample of the first. PIES can be analyzed as a two-phase sampling method. The first phase samples edges (\ie, edge sampling) from the graph stream with probability $p_{e}=\frac{m}{t}$ if the edge is incident to at least one node that does not belong to the reservoir (here $m$ is the number of initial edges in the reservoir). Also, an edge is  sampled with probability $p_e=1$ if the edge is incident to two nodes that belong to the reservoir. After that, the second phase samples a subgraph uniformly (\ie, node sampling) to maintain only $n$ nodes in the reservoir (\ie, all nodes in the reservoir are equally likely to be sampled). 
\vspace{1mm}
\item \textit{PIES has a selection bias to high degree nodes.}
PIES is biased to high degree nodes due to its first phase that relies on edge sampling. Naturally, edge sampling is biased towards high degree nodes since they tend to have more edges compared to lower degree nodes.
\vspace{1mm}
\item \textit{PIES samples an induced subgraph uniformly from the sub-sampled edge stream $E'(t)$ at any time $t$ in the stream.}
At any time $t$ in the graph stream $E$, PIES subsamples $E(t)$ to $E'(t)$ (such that $|E'(t)| \leq |E(t)|$). Then, PIES samples a uniform induced subgraph from $E'(t)$, such that all nodes in $E'(t)$ have an equal chance to be selected. Now, that we analyzed PIES, it can be easily adapted depending on a specific choice of the network sampling goal.
\end{enumerate}

\begin{algorithm}[h]
\DontPrintSemicolon
\SetKwInOut{Input}{Input}
\SetKwInOut{Output}{Output}
\Input{Sample Size $n$, Graph Stream $S$}
\Output{Sampled Subgraph $G_s = (V_s, E_s)$}
\BlankLine
$V_s = \emptyset, E_s = \emptyset$\;
$t=1$\;
\While{graph is streaming}{
       $(u, v) = e_t$\;
        \eIf {$|V_s|$ \textless $ n$}{
        \lIf{$u \notin V_s$} {$V_s = V_s \cup \{u\}$}\;
        \lIf{$v \notin V_s$} {$V_s = V_s \cup \{v\}$}\;
        $E_s = E_s \cup \{e_t\}$\;
        $m=|E_s|$\;
        }
        {
        $p_{e}=\frac{m}{t}$\;
        draw $r$ from continuous Uniform(0,1)\;
        \If {$r \leq p_{e}$}{
        draw $i$ and $j$ from discrete Uniform[1,$|V_s|$]\;
        \lIf{$u \notin V_s$} {$V_s = V_s \cup \{u\}$ , drop node $V_s[i]$ with all its incident edges}\;
        \lIf{$v \notin V_s$}{$V_s = V_s \cup \{v\}$ , drop node $V_s[j]$ with all its incident edges}}
        \lIf{$u \in V_s$ AND $v \in V_s$}{$E_s = E_s \cup \{e_t\}$}
        }
        $t=t+1$ \;
}
\caption{PIES(Sample Size $n$, Stream $S$)}
\label{algo:pies}
\end{algorithm}

%% file: evaluation-stream.tex
\subsection{Experimental Evaluation}
\label{sec:evaluation-stream}
In this section, we present results of sampling from streaming graphs presented as an arbitrarily ordered sequence of edges. The experimental setup is similar to what we used in section~\ref{sec:evaluation-stat}. We compare the performance of our proposed algorithm PIES to the proposed streaming implementation of node (NS), edge (ES), and breadth-first search sampling (BFS) methods. Note that we implement breadth first search using a sliding window of $100$ edges.

Similar to the experiments in section~\ref{sec:evaluation-stat}, we use ten different runs. To assess algorithm variation based on the edge sequence ordering, we randomly permute the edges in each run (while ensuring that all sampling methods use the same sequential order). Note that we compare the methods on a large social network from \textsc{LiveJournal}~\cite{leskovecRepository} with $4$ million nodes and $68$ million edges (included only at the $20\%$ sample size).  

\begin{figure*}[!h!t]
\centering
\subfigure[Degree]{\label{fig:str_avg ks deg}\includegraphics[width=0.268\textwidth]{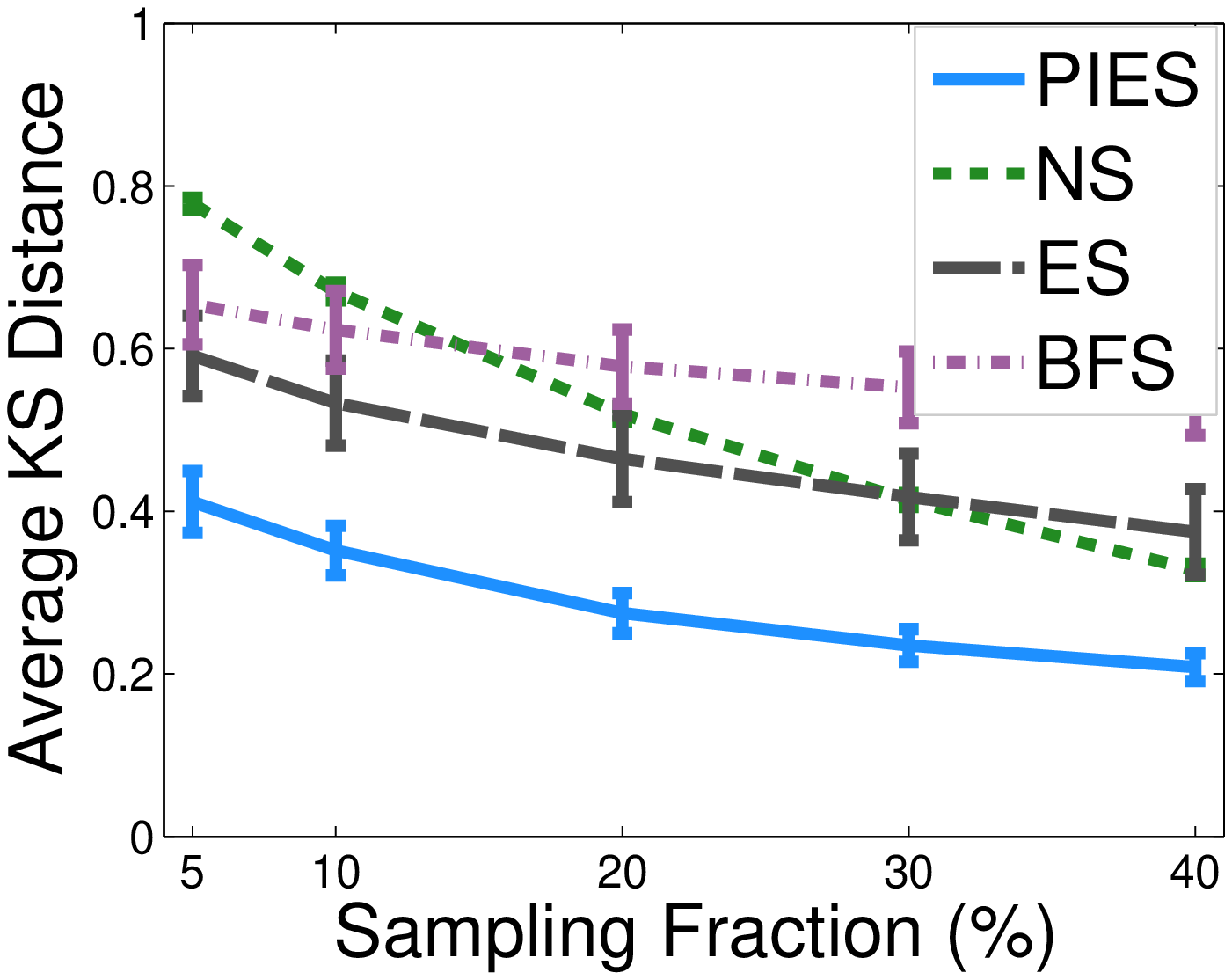}}
\hspace{-5.mm}
\subfigure[Path length]{\label{fig:str_avg ks plen}\includegraphics[width=0.268\textwidth]{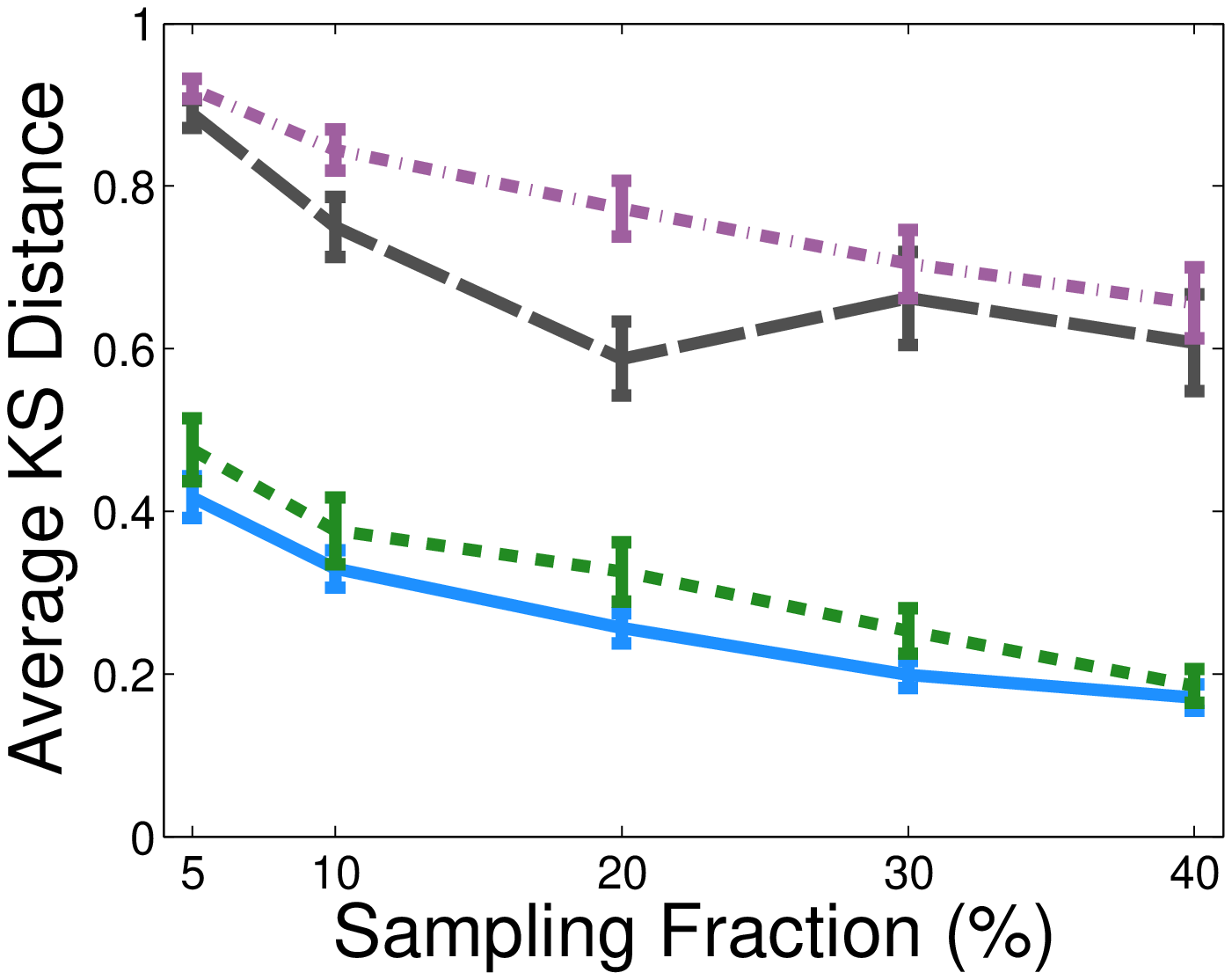}}
\hspace{-5.mm}
\subfigure[Clustering Coefficient]{\label{fig:str_avg ks cc}\includegraphics[width=0.268\textwidth]{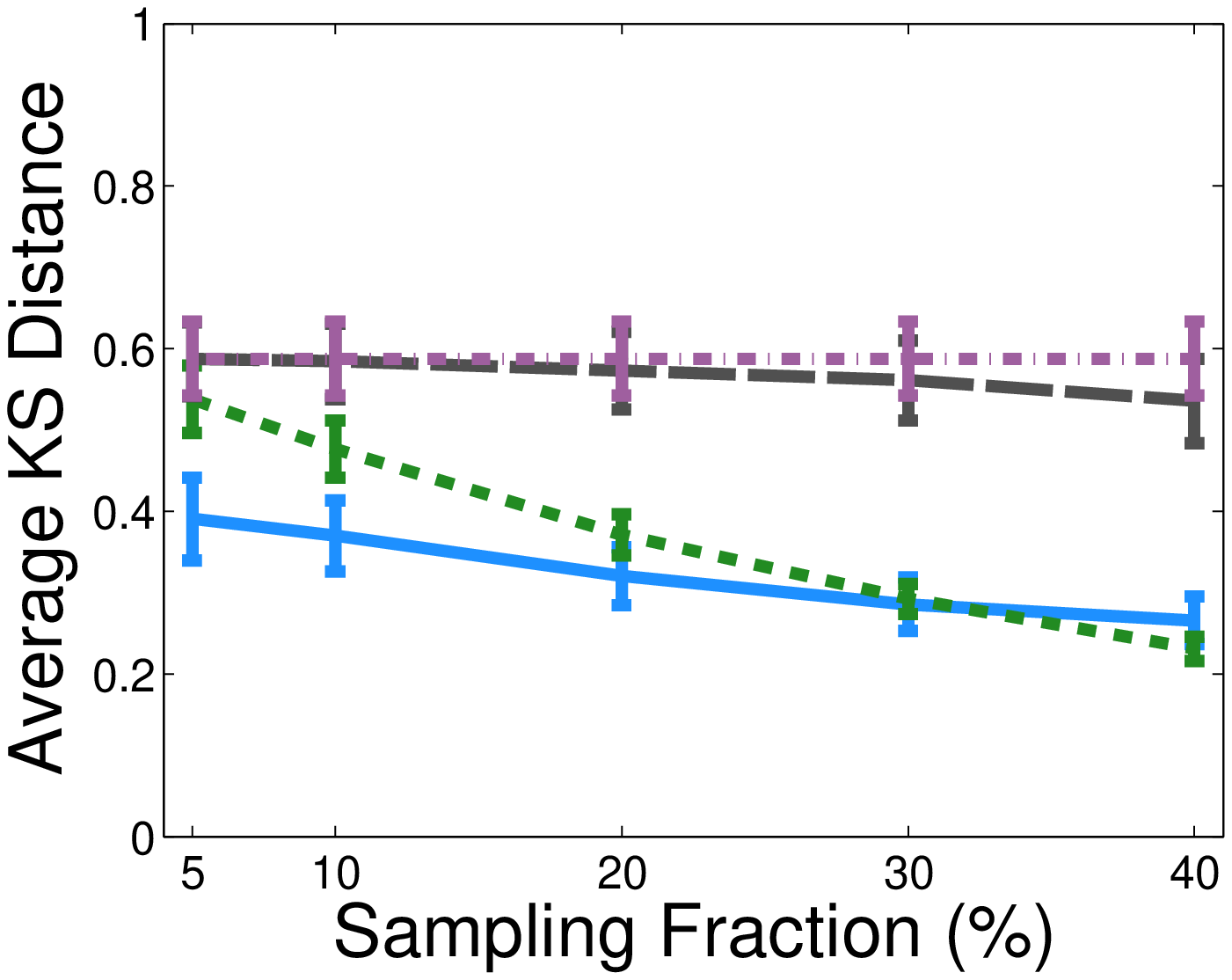}}
\hspace{-5.mm}
\subfigure[Kcore decomposition]{\label{fig:str_avg ks core}\includegraphics[width=0.268\textwidth]{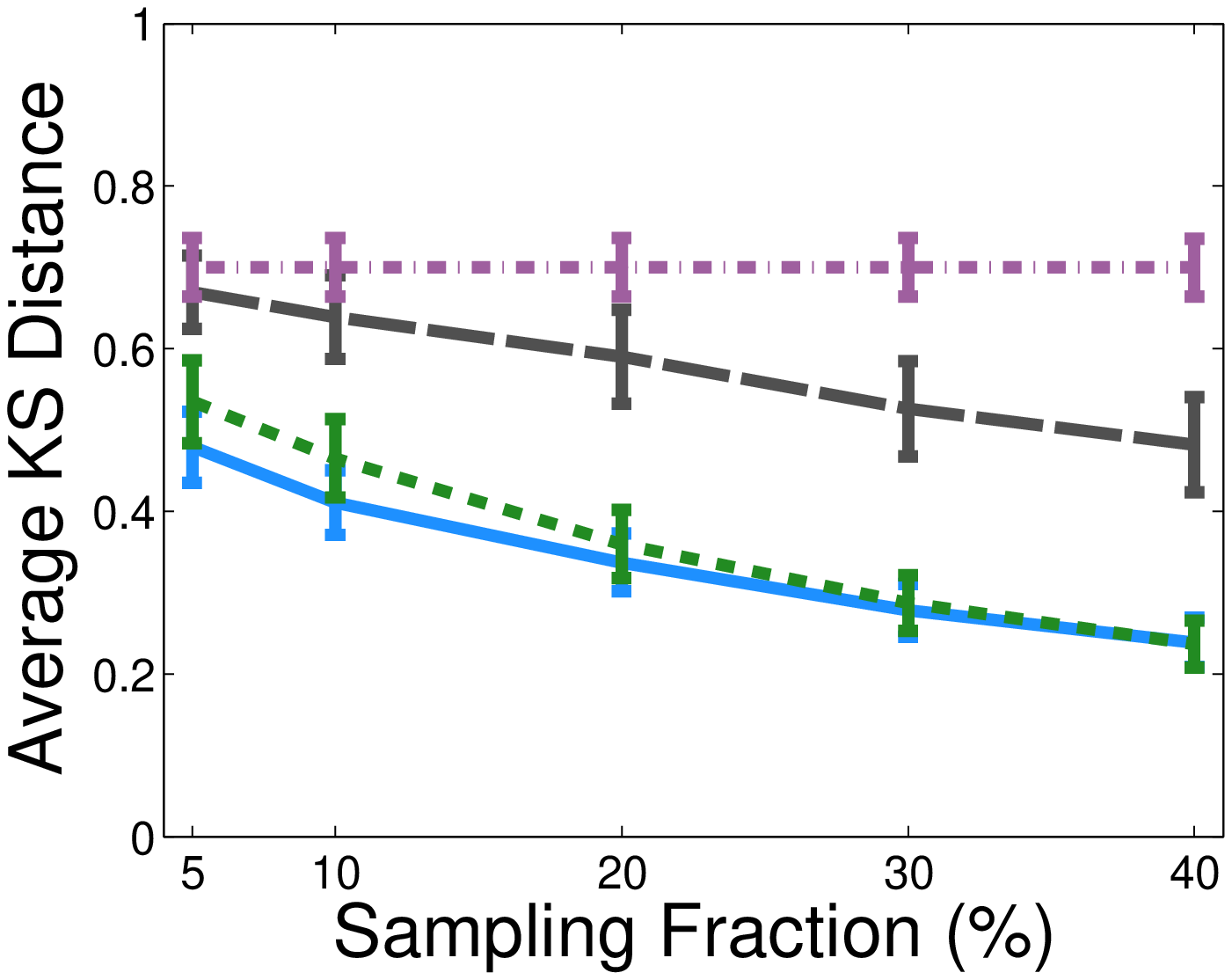}}\\
\vspace{-2mm}
\hspace{-2.mm}
\subfigure[Degree]{\label{fig:str_avg skl deg}\includegraphics[width=0.268\textwidth]{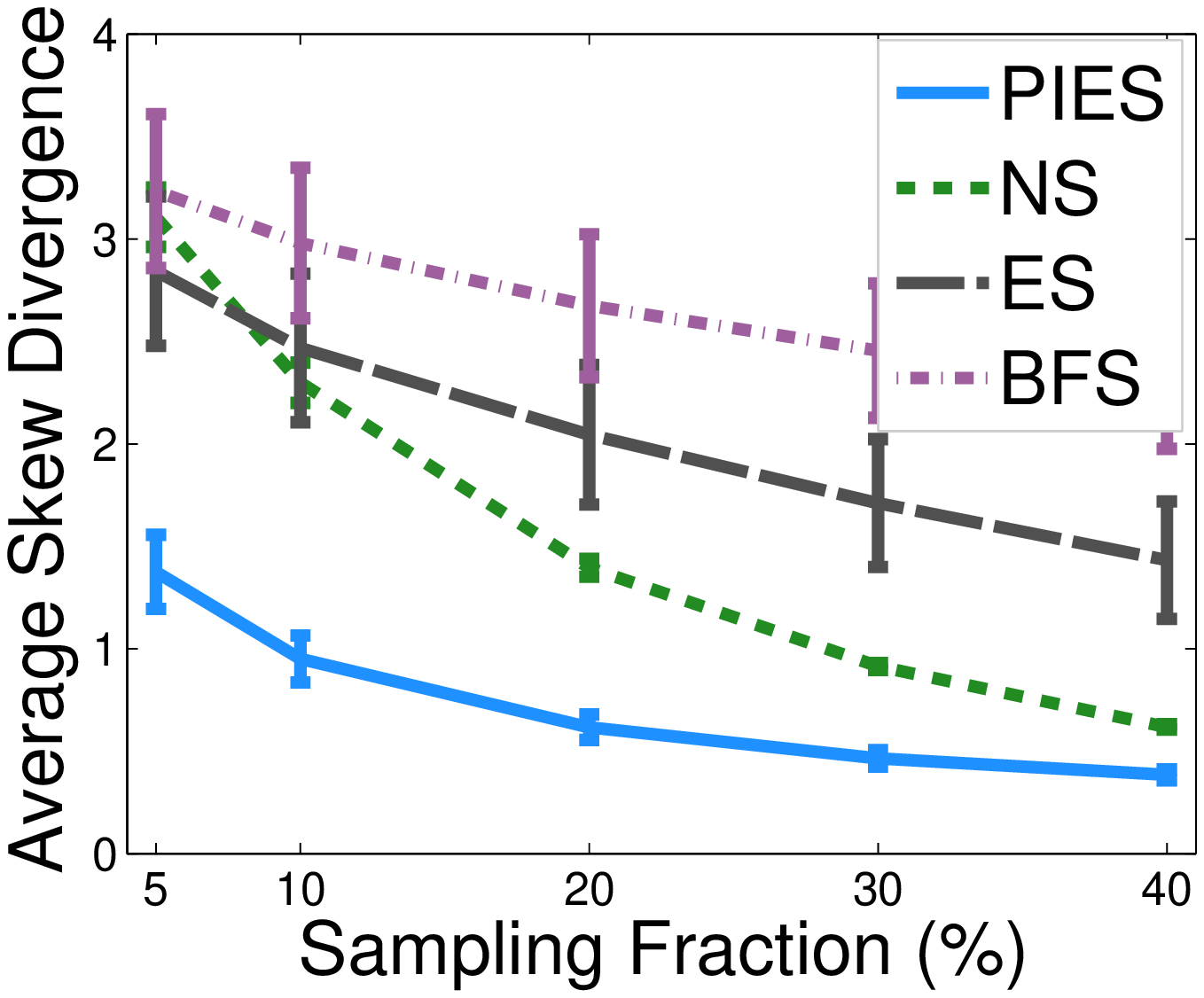}}
\hspace{-5.mm}
\subfigure[Path length]{\label{fig:str_avg skl plen}\includegraphics[width=0.268\textwidth]{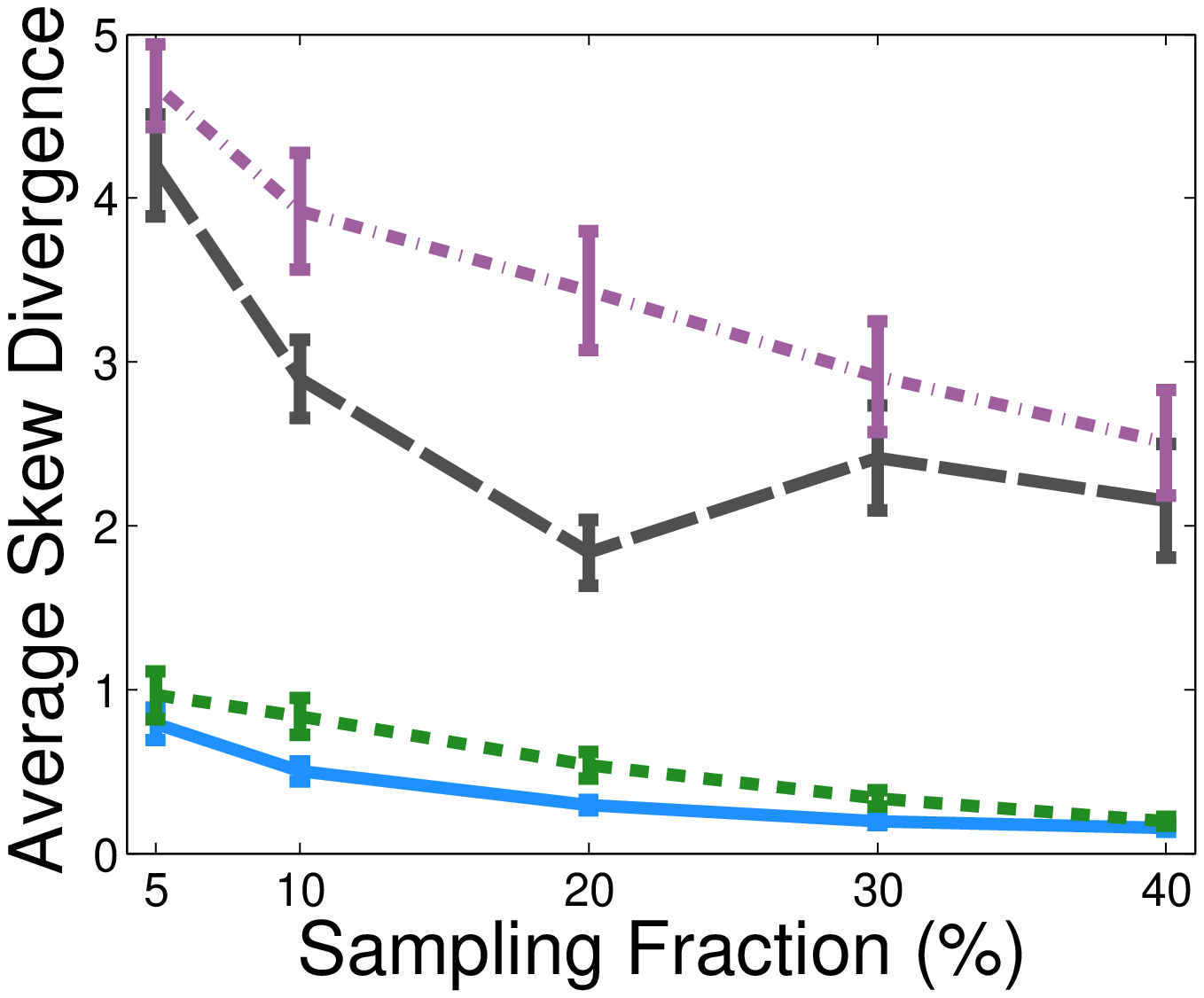}}
\hspace{-5.mm}
\subfigure[Clustering Coefficient]{\label{fig:str_avg skl cc}\includegraphics[width=0.268\textwidth]{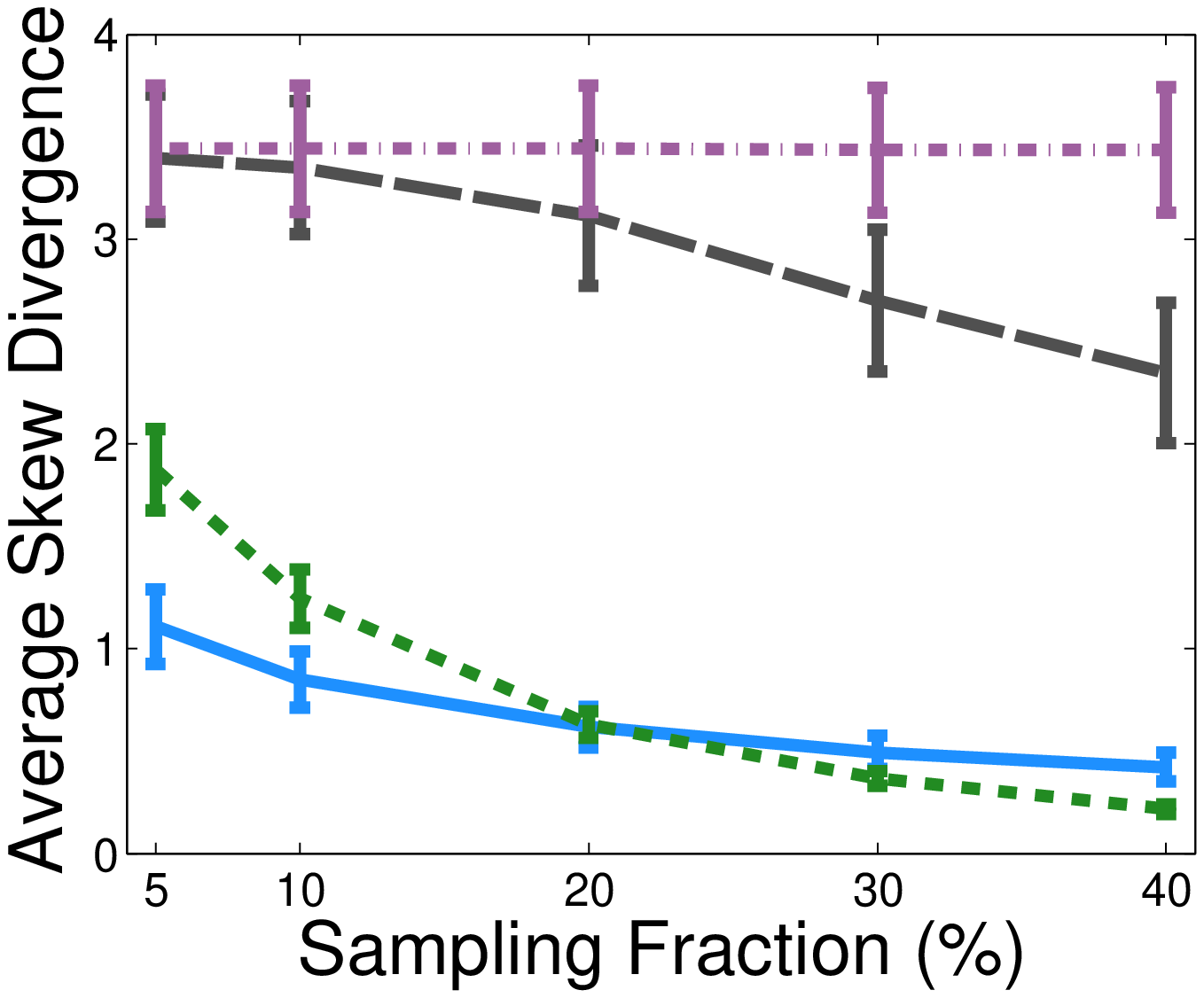}}
\hspace{-5.mm}
\subfigure[Kcore decomposition]{\label{fig:str_avg skl core}\includegraphics[width=0.268\textwidth]{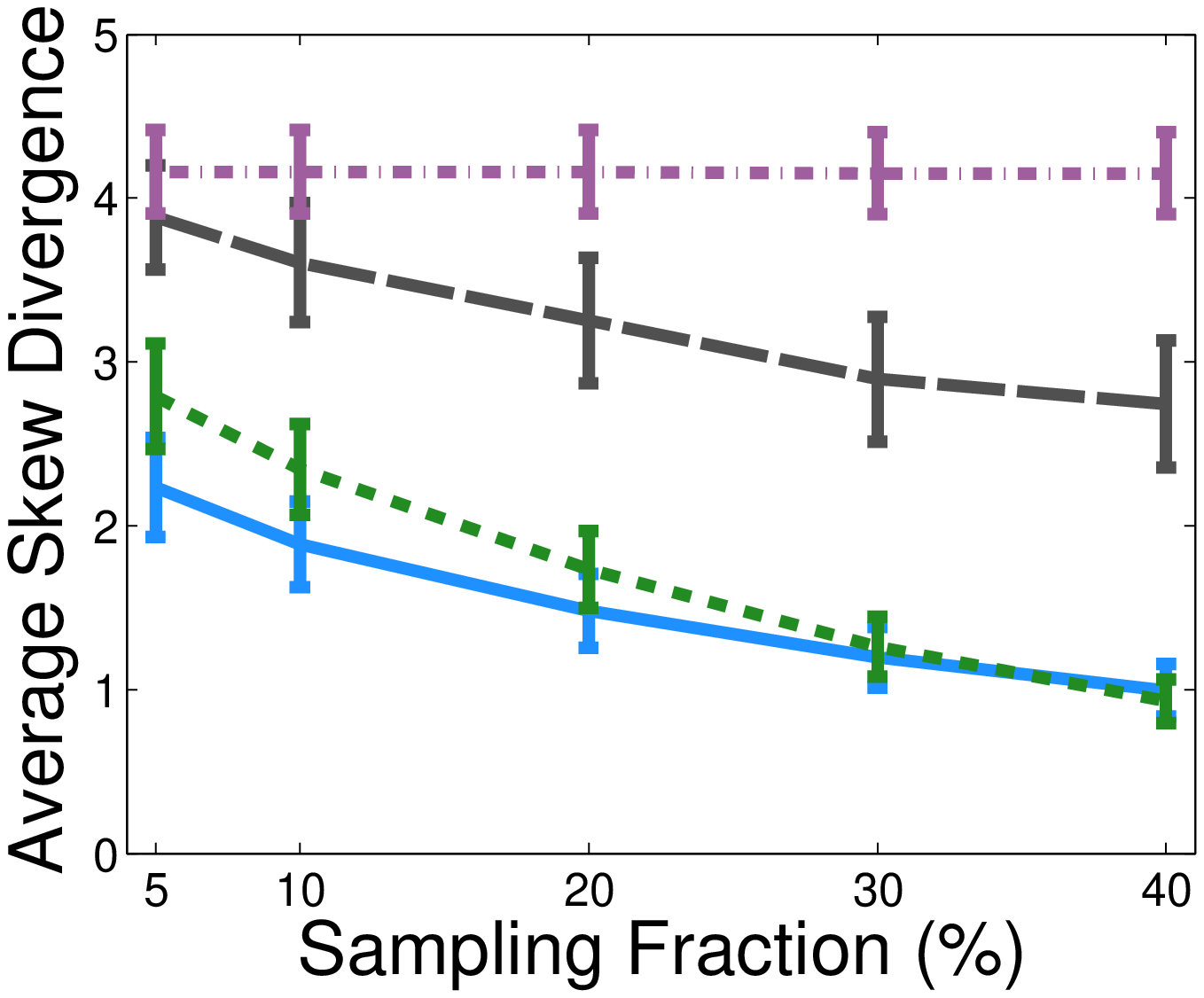}}
\vspace{-2mm}
\hspace{-2.mm}
\subfigure[Eigen Values]{\label{fig:str_avg evals}\includegraphics[width=0.268\textwidth]{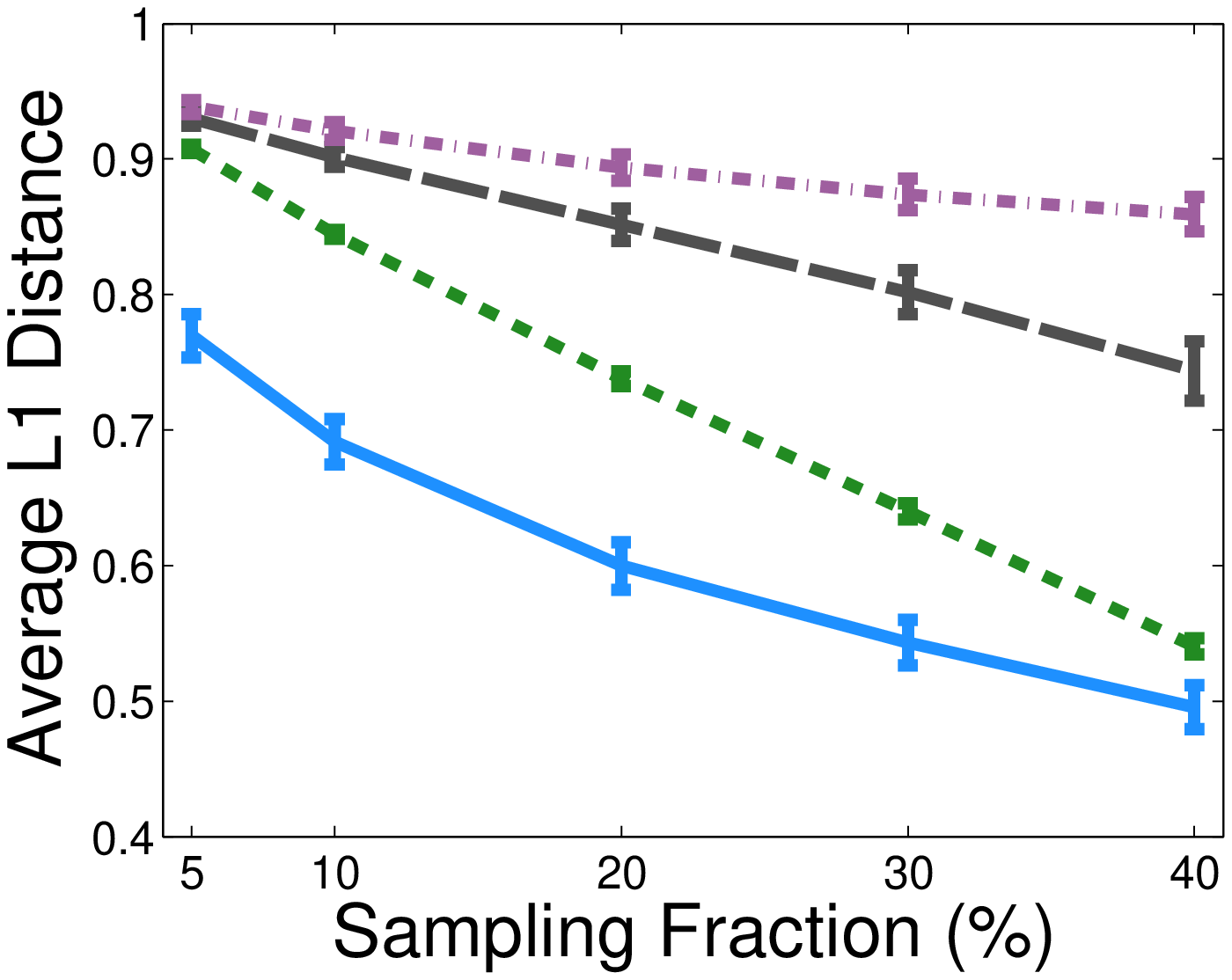}}
\hspace{-5.mm}
\subfigure[Network Value]{\label{fig:str_avg netvals}\includegraphics[width=0.268\textwidth]{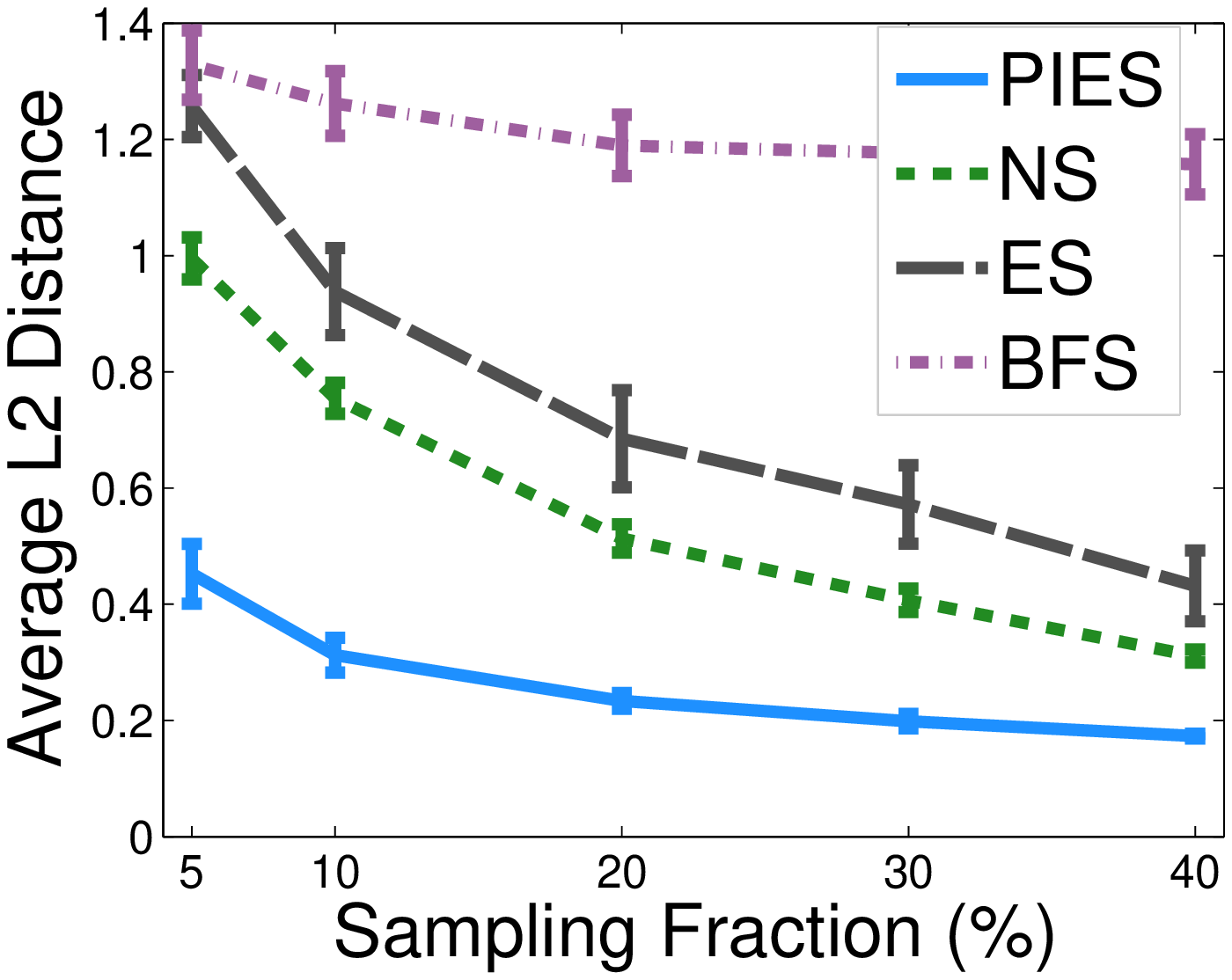}}
\vspace{-2mm}
\caption{(a-d) Average KS distance, (e-h) average skew divergence, and (i-j) average L1 and L2 distance respectively, across 6 datasets.
}\label{fig:all_distances2}
\vspace{-2mm}
\end{figure*}

\paragraph{Distance metrics} Figures~\ref{fig:str_avg ks deg}--\ref{fig:str_avg ks core} show the average KS statistic for degree, path length, clustering coefficient, and k-core distributions as an average over six datasets (as in section~\ref{sec:evaluation-stat}). PIES outperforms all other methods for the degree distribution statistic. NS performs almost as good as PIES for path length, clustering coefficient, and k-core distributions. As we explained in section~\ref{sec:sampling-stream}, PIES is biased to high degree nodes (due to its first phase) compared to NS. Both BFS and ES performs the worst among the four methods. This shows that the limited observability of the graph structure using a window of $100$ edges  does not facilitate effective breadth-first search. While increasing the window size may help improve the performance of BFS, we did not explore this as our focus was primarily on space-efficient sampling.  
Similar to the results of KS statistic, Figures~\ref{fig:str_avg skl deg}--\ref{fig:str_avg skl core} show the skew divergence statistic. 

Finally, Figures~\ref{fig:str_avg evals}--\ref{fig:str_avg netvals} show the L1 and L2 distance for eigenvalues and network values respectively. PIES outperforms all other methods. However, even though PIES performs the best, the distance is almost $50\%$ for the eigenvalues. This implies PIES is not suitable for capturing the eigenvalues of the graph.

\paragraph{Distributions} We plot the distributions of the six network statistics at the $20\%$ sample size. Figures~\ref{fig:stream_dist_comp_fbor}, ~\ref{fig:stream_dist_comp_condmat}, ~\ref{fig:stream_dist_comp_arxiv}, ~\ref{fig:stream_dist_comp_twcop}, ~\ref{fig:stream_dist_comp_email}, ~\ref{fig:stream_dist_comp_flickr}, and~\ref{fig:stream_dist_comp_socjor} (Appendix~\ref{appendix}) show the plots for all the distributions across the seven datasets. 
\begin{description}
\item[\textbf{Degree Distribution}] We observe across the six datasets, PIES outperforms the other methods for \textsc{Facebook}, \textsc{Twitter}, \textsc{Email-Univ}, \textsc{Flickr}, and \textsc{LiveJournal}. However, PIES only performs slightly better than NS for \textsc{HepPH} and \textsc{CondMAT}. This behavior appears to be related to the specific properties of the network datasets themselves. \textsc{HepPH} and \textsc{CondMAT} are more clustered and dense compared to other graphs used in the evaluation. We will discuss the behavior of the sampling methods for dense versus sparse graphs later in this section.
\vspace{2mm}
\item[\textbf{Path length Distribution}] PIES preserves the path length distribution of \textsc{Facebook}, \textsc{Twitter}, \textsc{Email-Univ}, \textsc{Flickr}, and \textsc{LiveJournal}, however, it overestimates the shortest path for \textsc{HepPH} and \textsc{CondMAT}.
\vspace{2mm}
\item[\textbf{Clustering  Distribution}] PIES generally underestimates the clustering coefficient in the graph by missing some of the clustering surrounding the sampled nodes. This behavior is more clear in \textsc{HepPH} and \textsc{CondMAT} since they are more clustered initially. 
\vspace{2mm}
\item[\textbf{K-Core Distribution}] Similar to the previous statistics, PIES outperforms the other methods for \textsc{Facebook}, \textsc{Twitter}, and \textsc{LiveJournal}. For \textsc{HepPH} and \textsc{CondMAT}, PIES performs almost as good as NS. In addition to the distribution of the core sizes, we compared the {\em max-core number} in the sampled subgraphs to their real counterparts for the $20\%$ sample size (Table~\ref{table:kcore_stream}). 
\vspace{2mm}
\item[\textbf{Eigenvalues}] While PIES captures the eigenvalues better than ES and BFS, its eigenvalues are orders of magnitude smaller than the real graph's eigenvalues. This implies that none of the streaming algorithms captures the eigenvalues (compared to \algo in section~\ref{sec:sampling-stat}).
\vspace{2mm}
\item[\textbf{Network values}] PIES accurately estimates the network values of most of the graphs compared to other methods.
\end{description}

\vspace{-2mm}
\begin{table*}[h!]
\centering\small
\begin{minipage}[c]{0.65\textwidth}
\caption{Comparison of {\em max-core-number} for the $20\%$ sample size for PIES, NS, ES, BFS versus Real value of $G$}
\begin{tabularx}{1.0\linewidth}{ l ccccc }
\toprule
\textbf{Graph} & \textbf{Real max core no.} & \textbf{PIES} & \textbf{NS} & \textbf{ES} & \textbf{BFS} \\
\midrule
\small
\textsc{HepPH}
&30&8\fivepc &8\fivepc&2&1\\
\textsc{CondMAT}
&25&7\fivepc &7\fivepc&2&1\\
\midrule
\textsc{Twitter}
&18&7\fivepc &4&3&1\\
\textsc{Facebook}
&16&6\fivepc &4&2&1\\
\textsc{Flickr}
&406&166\fivepc &81&19&1\\
\textsc{LiveJournal}
&372&117\fivepc &82&5&1\\
\midrule
\textsc{Email-UNIV}
&47&22\fivepc &15&3&1\\
\bottomrule
\end{tabularx}
\vspace{-1mm}
\label{table:kcore_stream}
\end{minipage}
\end{table*}

\begin{figure*}[h!t!]
\centering\small
\begin{minipage}[c]{0.8\textwidth}
\centering\small
\includegraphics[width=1.0\textwidth]{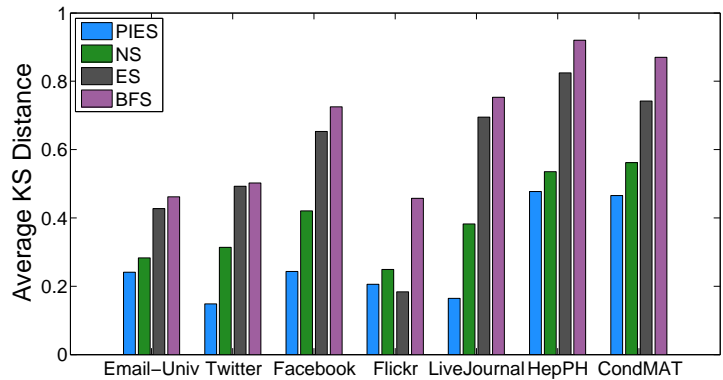}
\caption{Average KS Statistics for different networks (sorted in increasing order of clustering/density from left to right).}
\label{fig:all_datasets}
\vspace{-2mm}
\end{minipage}
\end{figure*}

\paragraph{Analysis of dense versus sparse graphs} Further to the discussion of the distributional results, we note that PIES is more accurate for sparse, less clustered graphs. To illustrate this, we report the performance of the stream sampling methods for each network in Figure~\ref{fig:all_datasets}, sorted from left to right in ascending order by clustering coefficient and density. Note that the bars represent the KS statistic (averaged over degree, path length, clustering, and k-core) for the $20\%$ sample size. Clearly, the KS statistic for all methods increases as the graph becomes more dense and clustered. PIES maintains a KS distance of approximately $\leq 24\%$ for five out of seven networks. These results indicate that PIES performs better in networks that are generally sparse and less clustered. This interesting result shows that PIES will be more suitable to sample rapidly changing graph streams that have lower density and less clustering---which is likely to be the case for many large-scale dynamic communication and activity networks.  

Moreover, we analyzed the number of isolated nodes for both NS and PIES in Table~\ref{table:isolated}. Since both NS and PIES sample nodes independently, it is expected that their sampled subgraph contains some nodes with zero degree (\ie, isolated nodes). This implies that PIES carries isolated nodes in the reservoir as the graph streams by. From the process of PIES, we know each time a new edge is sampled from the stream, its incident nodes replace randomly selected nodes from the reservoir. This random replacement policy could replace high degree nodes while other isolated nodes remain in the reservoir. With this observation in mind, we propose a modification for PIES such that a newly added node replaces the node with minimum degree which has stayed in the reservoir the longest amount of time without acquiring more edges. This strategy favors retaining high degree nodes over isolated and/or low degree nodes in the sample. We show the results of this modification in Tables~\ref{table:stream_ks} and ~\ref{table:stream_Ldist} which compare the KS distance, and L1/L2 distances respectively for each dataset (average over the two reasonable sample sizes $20\%$ and $30\%$). Note that we refer to the modification of PIES as "PIES (MIN)''. Clearly, modifying PIES in this manner (\ie, PIES (MIN)) achieved better results for dense graphs such as \textsc{HepPH} and \textsc{CondMAT}. 

\begin{table*}[h!t!]
\centering
\begin{minipage}[c]{0.45\textwidth}
\centering\small
\caption{Average probability of isolated nodes at the 20\% sample size for NS and PIES sampling methods.}
\begin{tabularx}{1.0\linewidth}{ l ccccc }
\toprule
\textbf{Graph} &  & &\textbf{PIES} &  & \textbf{NS}  \\
\midrule
\small
\textsc{HepPH}
& & &0.046& &0.15 \\
\textsc{CondMAT}
& & &0.14& &0.36 \\
\midrule
\textsc{Twitter}
& & &0.15& &0.51\\
\textsc{Facebook}
& & &0.13& &0.43\\
\textsc{Flickr}
& & &0.14& &0.56 \\
\textsc{LiveJournal}
& & &0.06& &0.36 \\
\midrule
\textsc{Email-UNIV}
& & &0.07& &0.51 \\
\bottomrule
\end{tabularx}
\label{table:isolated}
\vspace{-2mm}
\end{minipage}
\end{table*}

\noindent
\textit{\textbf{Summary}---We summarize the main empirical conclusions in this section:}

\textit{
\begin{enumerate}[(1)]
\vspace{-4mm}
\item Sampled subgraphs collected and constructed by PIES accurately preserve many network statistics (\eg, degree, path length, and k-core).
\item PIES produces better samples when the graph is sparse/less clustered (\eg, \textsc{Twitter} and \textsc{LiveJournal}).
\item We showed how PIES can be adapted to reduce the number of isolated nodes in the sample "PIES(MIN)".
\item PIES(MIN) can preserve the properties of dense graphs as well as certain statistics (\eg, eigenvalues) that are hard to be preserved by PIES.
\item The results show that the structure of the sampled subgraph $G_s$ depends on the manner in which the topology of the graph $G$, the
characteristics of $\eta$ (\eg, degree distribution), and the nature of the sampling method interact. In future work, we aim to study how to adapt the sample given prior knowledge of the graph properties.
\end{enumerate}
}

\begin{table*}[h!]
\centering\small
\caption{Average KS Distance for PIES, PIES (MIN), NS, ES, and BFS stream sampling methods.}
\begin{tabularx}{0.95\linewidth}{ l ccccccc }
\toprule
\textbf{Data} & \textbf{} & \textbf{}& \textbf{PIES} &  \textbf{PIES (MIN)} & \textbf{NS} &
\textbf{ES} & \textbf{BFS} \\
\midrule
\small
\multirow{3}{*}{\rotatebox{0}{\textsc{Email-UNIV}}}
& & \textsc{$Deg$}&0.2348&0.5803&0.4547&0.2186&0.3724\\
& & \textsc{$PL$}&0.204&0.5621&0.19&0.6989&0.5114\\
& & \textsc{$Clust$}&0.1108&0.4728&0.188&0.3302&0.3473\\
& & \textsc{$KCore$}&0.2819&0.5821&0.1985&0.3219&0.5759\\
\midrule
& &&\textbf{0.2079}\fivepc &0.5493&0.2578&0.3924&0.4518\\
\midrule
\multirow{3}{*}{\rotatebox{0}{\textsc{Twitter}}}
& & \textsc{$Deg$}&0.1521&0.2598&0.4667&0.3052&0.4194\\
& & \textsc{$PL$}&0.0528&0.3941&0.1243&0.617&0.4811\\
& & \textsc{$Clust$}&0.2462&0.2269&0.346&0.4673&0.482\\
& & \textsc{$KCore$}&0.1001&0.2886&0.2271&0.4393&0.5929\\
\midrule
& &&\textbf{0.1378}\fivepc    &0.2923&0.291&0.4572&0.4938\\
\midrule
\multirow{3}{*}{\rotatebox{0}{\textsc{Facebook}}}
& & \textsc{$Deg$}&0.1848&0.2357&0.3804&0.4912&0.6917\\
& & \textsc{$PL$}&0.2121&0.3171&0.4337&0.8762&0.9557\\
& & \textsc{$Clust$}&0.2594&0.2314&0.3496&0.4975&0.5017\\
& & \textsc{$KCore$}&0.2375&0.2447&0.3569&0.661&0.7275\\
\midrule
& &&\textbf{0.2234}\fivepc &0.2572&0.3802&0.6315&0.7192\\
\midrule
\multirow{3}{*}{\rotatebox{0}{\textsc{Flickr}}}
& & \textsc{$Deg$}&0.1503&0.399&0.514&0.0924&0.2706\\
& & \textsc{$PL$}&0.2845&0.4936&0.0789&0.1487&0.6763\\
& & \textsc{$Clust$}&0.1426&0.3754&0.2404&0.3156&0.3931\\
& & \textsc{$KCore$}&0.1654&0.4289&0.0595&0.1295&0.4541\\
\midrule
& &&0.1857&0.4242&0.2232&\textbf{0.1716}\fivepc &0.4485\\
\midrule
\multirow{3}{*}{\rotatebox{0}{\textsc{HepPH}}}
& & \textsc{$Deg$}&0.4103&0.1304&0.483&0.8585&0.8923\\
& & \textsc{$PL$}&0.306&0.1959&0.431&0.749&0.8676\\
& & \textsc{$Clust$}&0.4636&0.0393&0.3441&0.9156&0.9171\\
& & \textsc{$KCore$}&0.592&0.1674&0.6233&0.9402&0.9592\\
\midrule
& &&0.443&\textbf{0.1332}\fivepc &0.4704&0.8658&0.909\\
\midrule
\multirow{3}{*}{\rotatebox{0}{\textsc{CondMAT}}}
& & \textsc{$Deg$}&0.4042&0.1259&0.5006&0.6787&0.7471\\
& & \textsc{$PL$}&0.2944&0.2758&0.5211&0.6981&0.9205\\
& & \textsc{$Clust$}&0.5927&0.3285&0.5341&0.878&0.8853\\
& & \textsc{$KCore$}&0.4692&0.1512&0.4955&0.858&0.8909\\
\midrule
& &&0.4401&\textbf{0.2203}\fivepc &0.5128&0.7782&0.8609\\
\midrule
\midrule
\multicolumn{2}{c}{Average for all Datasets}&  & \textbf{0.2730}\fivepc &    0.3128  & 0.3559   & 0.5494   & 0.6472 \\
\bottomrule
\end{tabularx}
\vspace{-1mm}
\label{table:stream_ks}
\end{table*}

\begin{table*}[h!]
\centering\small
\caption{Average L1/L2 Distance for PIES, PIES (MIN), NS, ES, and BFS stream sampling.}
\begin{tabularx}{0.95\linewidth}{ l ccccccc }
\toprule
\textbf{Data} & \textbf{} & \textbf{}& \textbf{PIES} &  \textbf{PIES (MIN)} & \textbf{NS} &
\textbf{ES} & \textbf{BFS} \\
\midrule
\small
\multirow{2}{*}{\rotatebox{0}{\textsc{Email-UNIV}}}
& & \textsc{$Eigen Val$}&0.4487&\textbf{0.074}\fivepc&0.7018&0.7588&0.838\\
& & \textsc{$Net Val$}&0.199&\textbf{0.0201}\fivepc&0.5785&0.3799&1.007\\
\midrule
\multirow{2}{*}{\rotatebox{0}{\textsc{Twitter}}}
& & \textsc{$Eigen Val$}&0.4981&\textbf{0.1851}\fivepc&0.6411&0.7217&0.7964\\
& & \textsc{$Net Val$}&0.1431&\textbf{0.043}\fivepc&0.3108&0.4271&0.8385\\
\midrule
\multirow{2}{*}{\rotatebox{0}{\textsc{Facebook (NO)}}}
& & \textsc{$Eigen Val$}&0.591&\textbf{0.1143}\fivepc&0.6771&0.8417&0.9018\\
& & \textsc{$Net Val$}&0.306&\textbf{0.0617}\fivepc&0.5383&1.0984&1.5027\\
\midrule
\multirow{2}{*}{\rotatebox{0}{\textsc{Flickr}}}
& & \textsc{$EigenVal$}&0.5503&\textbf{0.0049}\fivepc &0.7227&0.8491&0.9298\\
& & \textsc{$NetVal$}&0.1657&\textbf{0.0005}\fivepc &0.5626&0.0574&1.5193\\
\midrule
\multirow{2}{*}{\rotatebox{0}{\textsc{HepPH}}}
& & \textsc{$EigenVal$}&0.7083&\textbf{0.2825}\fivepc&0.7232&0.9373&0.95\\
& & \textsc{$NetVal$}&0.3&\textbf{0.1817}\fivepc&0.3198&1.0821&1.2477\\
\midrule
\multirow{2}{*}{\rotatebox{0}{\textsc{CondMAT}}}
& & \textsc{$Eigen Val$}&0.6278&\textbf{0.1475}\fivepc&0.6843&0.8507&0.8875\\
& & \textsc{$Net Val$}&0.2254&\textbf{0.06}\fivepc&0.3235&0.7514&0.9853\\
\bottomrule
\end{tabularx}
\vspace{-1mm}
\label{table:stream_Ldist}
\end{table*}

%% file: applications.tex
\section{Practical Applications of Network Sampling}
\label{sec:applications}
In Section~\ref{sec:foundations}, we discussed how network sampling arises in many different applications (\eg, social science, data mining). Most research in network sampling has focused on how to collect a sample that closely match {\em topological} properties of the network~\cite{leskovec2006slg,Maiya2011kdd}. However, since the topological properties are never entirely preserved, it is also important to study how the sampling process impacts the investigation of applications overlaid on the networks. 
One such study recently investigated the impact of sampling methods on the discovery of the information diffusion process~\cite{de2010does}. The study shows that sampling methods which considers both topology and user context improves on other naive methods. In this section, we consider the impact of sampling on relational learning. Network sampling is a core part of relational learning and it comes in many different problems. For example, learning models, evaluation of learning algorithms, learning ensembles of models, and active learning.

However, network sampling can produce samples with imbalance in class membership and bias in topological features (\eg, path length, clustering) due to missing nodes/edges---thus the sampling process can significantly impact the accuracy of relational classification. This bias may result from the size of the sample, the sampling method, or both. While, most previous work in relational learning has focused on analyzing a single {\em input network} and research has considered how to further split the input network into training and testing networks for evaluation \cite{Korner:2006,macskassy:07,neville:icdm09}, the fact that the input network is often itself {\em sampled} from an unknown target network has largely been ignored. There has been little focus on {\em how} the construction of the input networks may impact the evaluation of relational algorithms. 

In this section, we study the question of how the choice of the sampling method can impact {\em parameter estimation} and {\em performance evaluation} of relational classification algorithms. We aim to evaluate the impact of network sampling on relational classification using two different goals:
\begin{enumerate}
\item \textit{Parameter estimation:} we study the impact of network sampling on the estimation of class priors, \ie, probability of class labels, goal $3$.
\item \textit{Performance evaluation:} we study the impact of network sampling on the estimation of classification accuracy of relational learners, \ie, goal $2$.
\end{enumerate}

\subsection*{Case Study: Relational Classification}

Conventional classification algorithms focus on the problem of identifying the unknown class (\eg, group) to which an entity (\eg., person) belongs. Classification models are learned from a training set of (disjoint) entities, which are assumed to be independent and identically distributed (i.i.d.) and drawn from the underlying population of instances. However, relational classification problems differs from this conventional view in that entities violate the i.i.d. assumption. In relational data, entities (\eg, users in social networks) can exhibit complex dependencies. For example, friends often share similar interests (\eg, political views). 

Recently, there have been a great deal of research in relational learning and classification. For example,~\cite{Friedman99learningprobabilistic}
and~\cite{Taskar01probabilisticclassification} outline probabilistic relational learning algorithms that search the space for relational attributes and structures of neighbors to improve the classification accuracy. Further, Macskassy proposed a simple relational neighbor classifier (weighted-vote relational neighbor wvRN) that requires no learning and iteratively classifying the entities of a relational network based only on the relational structure \cite{macskassy:07}. Macskassy showed that wvRN often performs competitively to other relational learning algorithms. 

\paragraph{Impact of sampling on parameter estimation}
Let $a$ be the node attribute representing the class label of any node $v_i \in V$ in graph $G$. We denote $\mathcal{C}=\{c_1,c_2,...\}$ as the set of possible class labels, where $c_l$ is the class label of node $v_i$ (\ie, $a(v_i)=c_l$).

We study the impact of network sampling on the estimation of of class priors in $G$ (\ie, the distribution of class labels), using the following procedure: 
\begin{enumerate}
\item Choose a set of nodes $S$ from $V$ using a sampling algorithm $\sigma$.
\item For each node $v_i \in S$, observe $v_i$'s class label. 
\item Estimate the class label distribution$\hat{p}_{c_l}$ from $S$, using the following equation, 

\begin{eqnarray*} 
\hat{p}_{c_l}={1 \over {|S|}} {\sum_{v_i \in S}{1_{(a(v_i)=c_l)}}}
\end{eqnarray*}
\end{enumerate}

\begin{figure*}[!h]
\centering
\vspace{-2.mm}
\subfigure[Political Blogs]{\label{fig:polblogs_cdist}\includegraphics[width=0.40\textwidth]{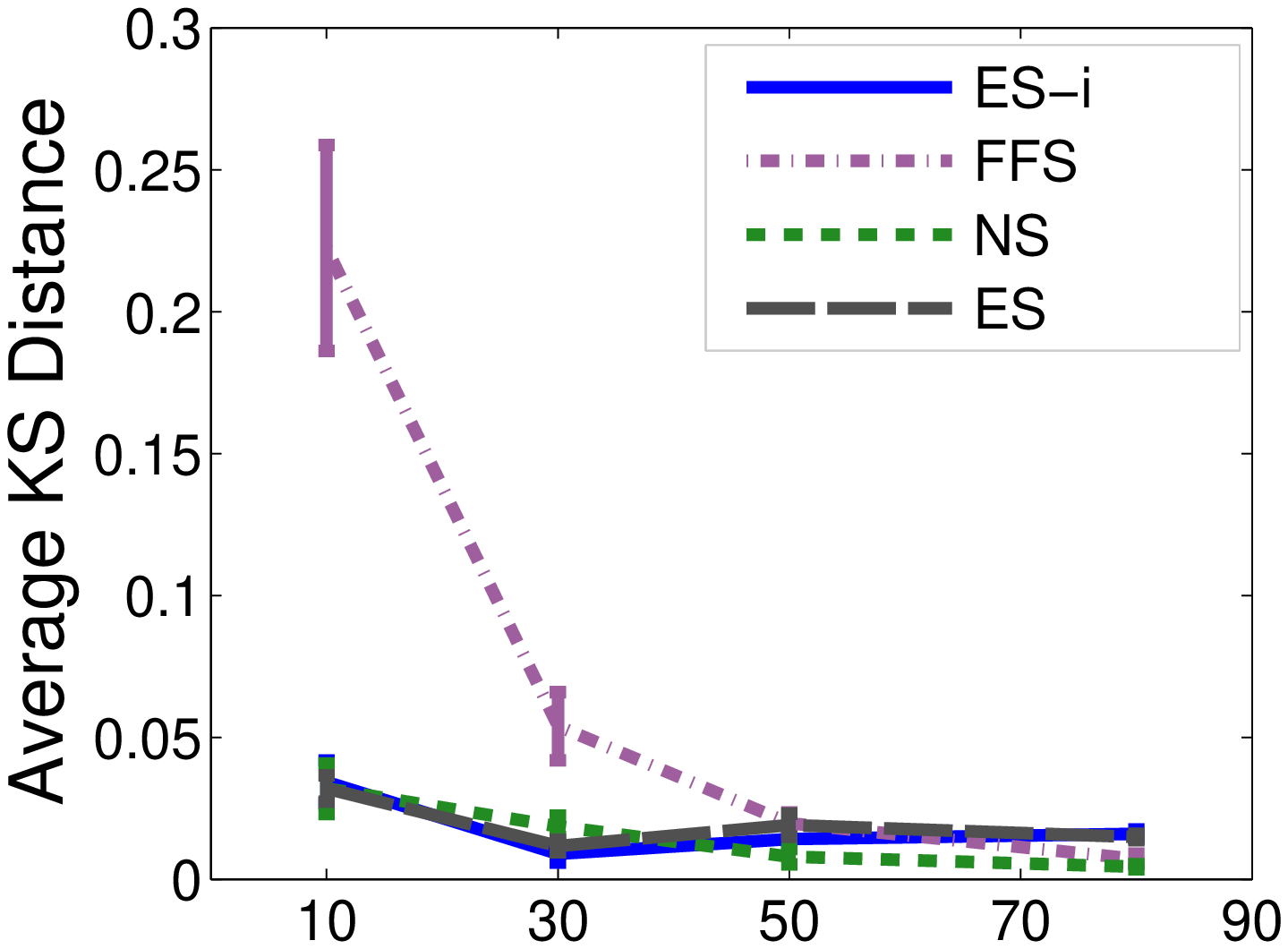}}
\hspace{4.mm}
\subfigure[Facebook]{\label{fig:FBPU_cdist}\includegraphics[width=0.40\textwidth]{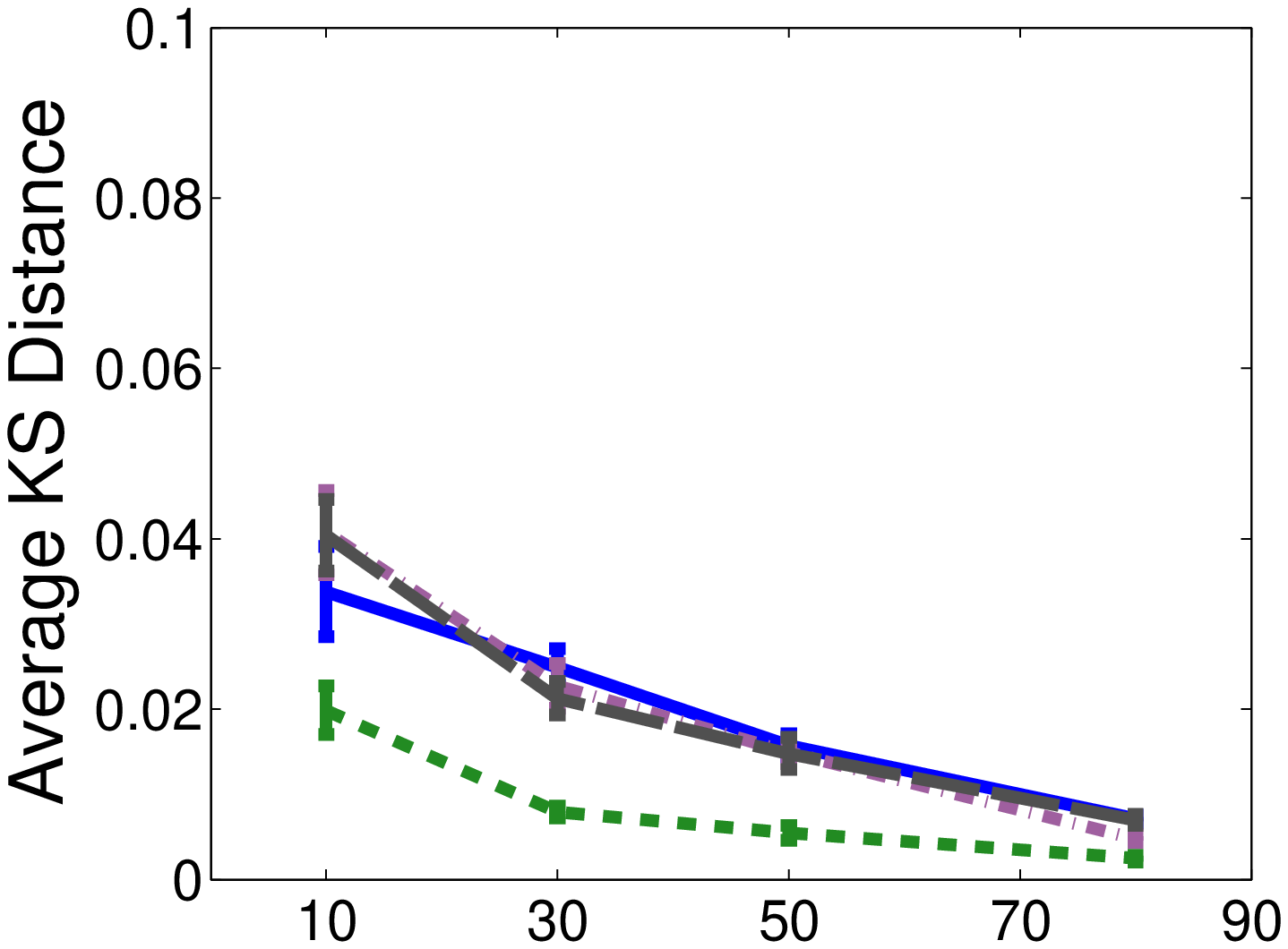}}
\vspace{-2.mm}
\hspace{-2.mm}
\subfigure[CoRA]{\label{fig:CorA_cdist}\includegraphics[width=0.40\textwidth]{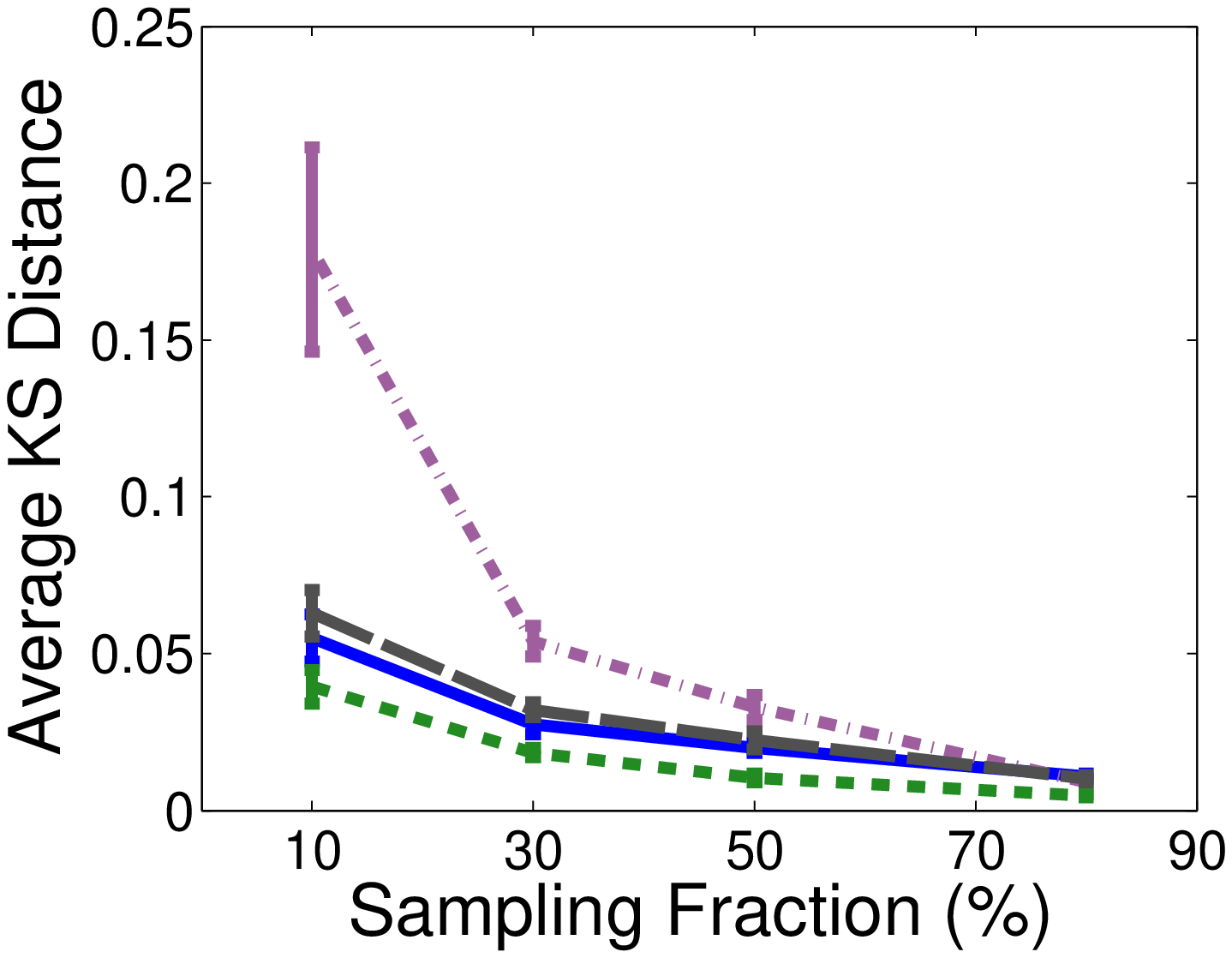}}
\hspace{4.mm}
\subfigure[Citeseer]{\label{fig:cite_cdist}\includegraphics[width=0.40\textwidth]{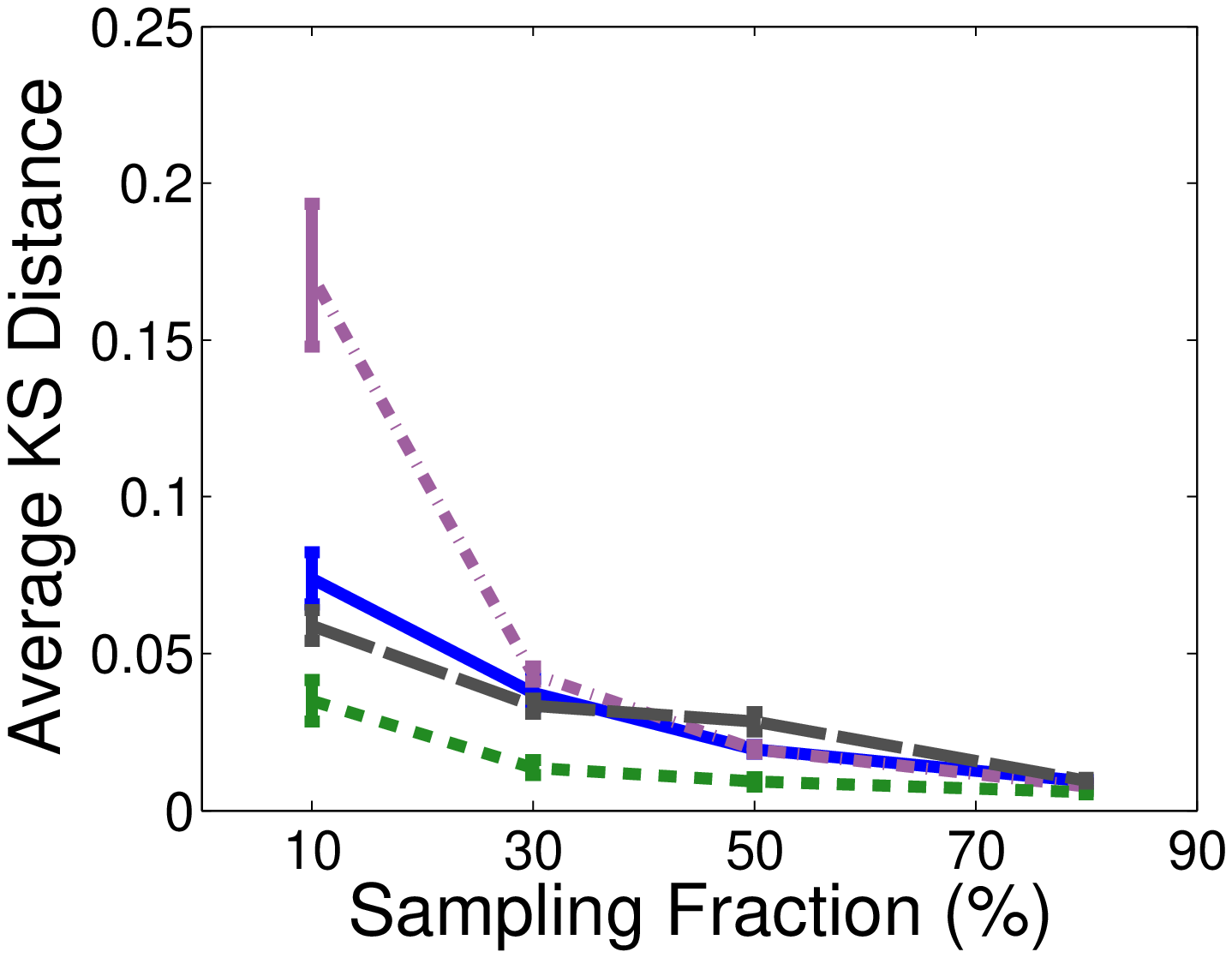}}
\hspace{-2.mm}
\caption{Average KS distance of class priors for NS, ES, FFS, and \algo.}
\label{fig:cdist}
\vspace{-1mm}
\end{figure*}

In our experiments, we consider four real networks: two citation networks \textsc{CoRA} with 2708 nodes and \textsc{Citeseer} with 3312 nodes~\cite{sen:aimag08}, \textsc{Facebook} collected from Facebook Purdue network with 7315 users with their political views~\cite{xiang10}, and a single day snapshot of 1490 political blogs that shows the interactions between liberal and conservative blogs~\cite{adamic2005political}.
We sample a subset of the nodes using NS, ES, \algo, and FFS, where the sample size is between $10\%-80\%$ of the graph $G$. For each sample size, we take the average of ten different runs. Then, we compare the estimated class prior to the actual class prior in the full graph $G$ using the average KS distance measure. As plotted in Figures~\ref{fig:polblogs_cdist}--\ref{fig:cite_cdist}, node sampling (NS) estimates the class priors more accurately than other methods, however, we note that FFS produces a large bias in most of the graphs (at $10\%$ sample size).

\paragraph{Impact of sampling on classification accuracy}
Let $\mathcal{R}$ be a relational classifier which takes a graph $G$ as input. The goal is to predict the class labels of nodes in $G$. Therefore, $\mathcal{R}$ uses a proportion of nodes in graph $G$ with known class labels as a {\em training set} to learn a model. Afterwards, $\mathcal{R}$ is used to predict the label of the remaining (unlabeled) nodes in $G$ (\ie, {\em test set}). Generally, the performance of $\mathcal{R}$ can be evaluated based on the accuracy of the predicted class labels---we calculate the accuracy using area under the ROC curve (AUC) measure.    

We study the impact of network sampling on the accuracy of relational classification using the following procedure:

\begin{enumerate}
\item Sample a subgraph $G_s$ from $G$ using a sampling algorithm $\sigma$.
\item Estimate the classification accuracy of a classifier $\mathcal{R}$ on $G_s$: $\hat{auc}=\mathcal{R}(G_s)$
\end{enumerate}

We compare the actual classification accuracy on $G$ to the estimated classification accuracy on $G_s$. Formally, we compare $auc=\mathcal{R}(G)$ to $\hat{auc}=\mathcal{R}(G_s)$. So then, $G_s$ is said to be representative to $G$, if $\hat{auc} \approx {auc}$.
 
In our experiments, we use the weighted-vote relational neighbor classifier (wvRN) as our base classifier $\mathcal{R}$ \cite{macskassy:07}. In wvRN, the class membership probability of a node $v_i$ belonging to class $c_l$ is defined as:
\begin{eqnarray*} 
P(c_l|v_i) = {1 \over Z} \sum_{v_j \in \mathcal{N}(v_i)} {w(v_i,v_j) * P(c_l|v_j)}
\end{eqnarray*}

\noindent
where $\mathcal{N}(v_i)$ is the set of neighbors of node $v_i$,  ${w(v_i,v_j)}$ is the weight of the edge $e_{ij}=(v_i,v_j)$, and $Z=\sum_{v_j \in \mathcal{N}(v_i)} {w(v_i,v_j)}$ is the normalization term. 

\begin{figure*}[!h]
\centering
\vspace{-2.mm}
\subfigure[Political Blogs]{\label{fig:polblogs_auc10}\includegraphics[width=0.40\textwidth]{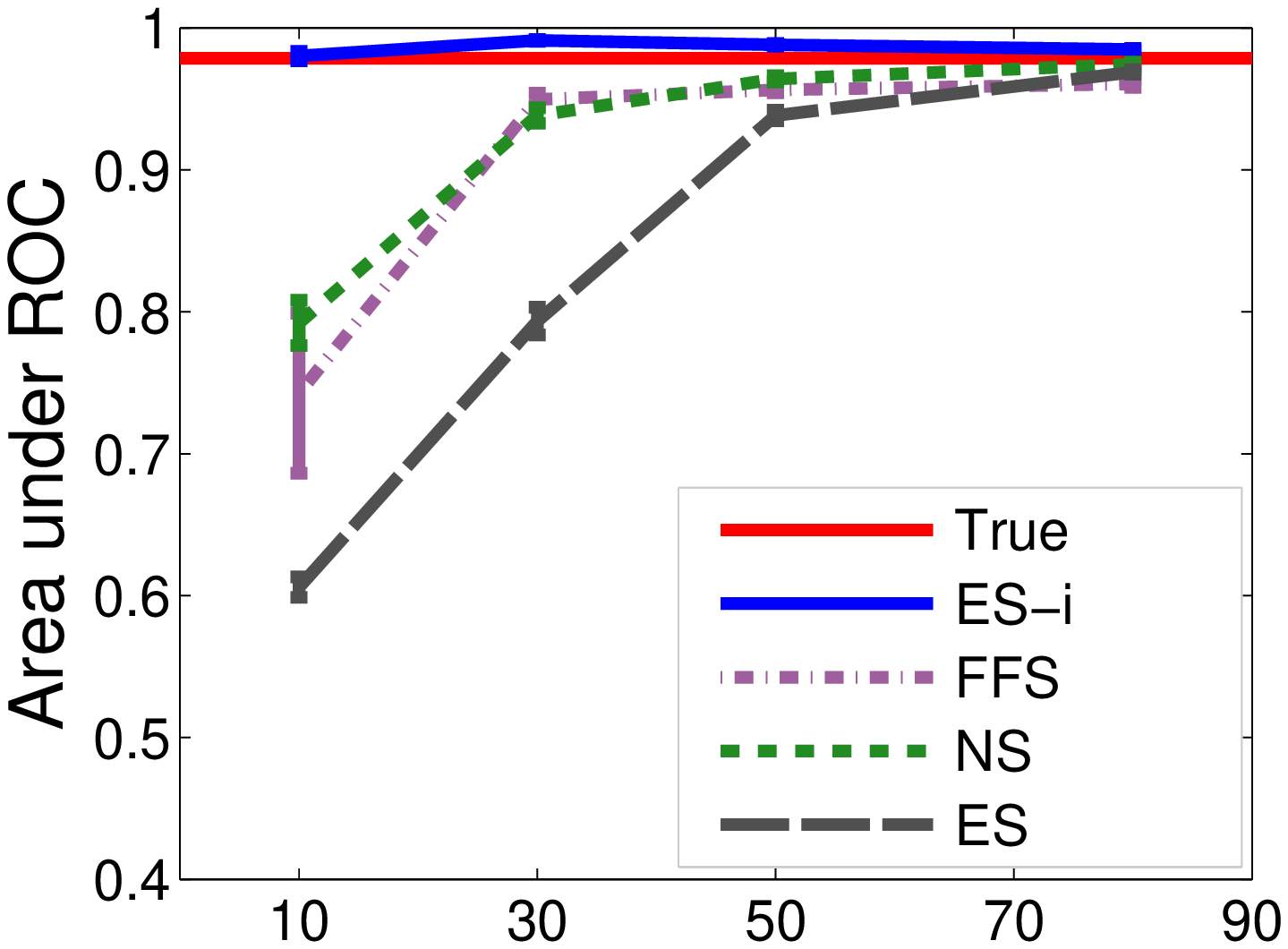}}
\hspace{4.mm}
\subfigure[Facebook]{\label{fig:FBPU_auc10}\includegraphics[width=0.40\textwidth]{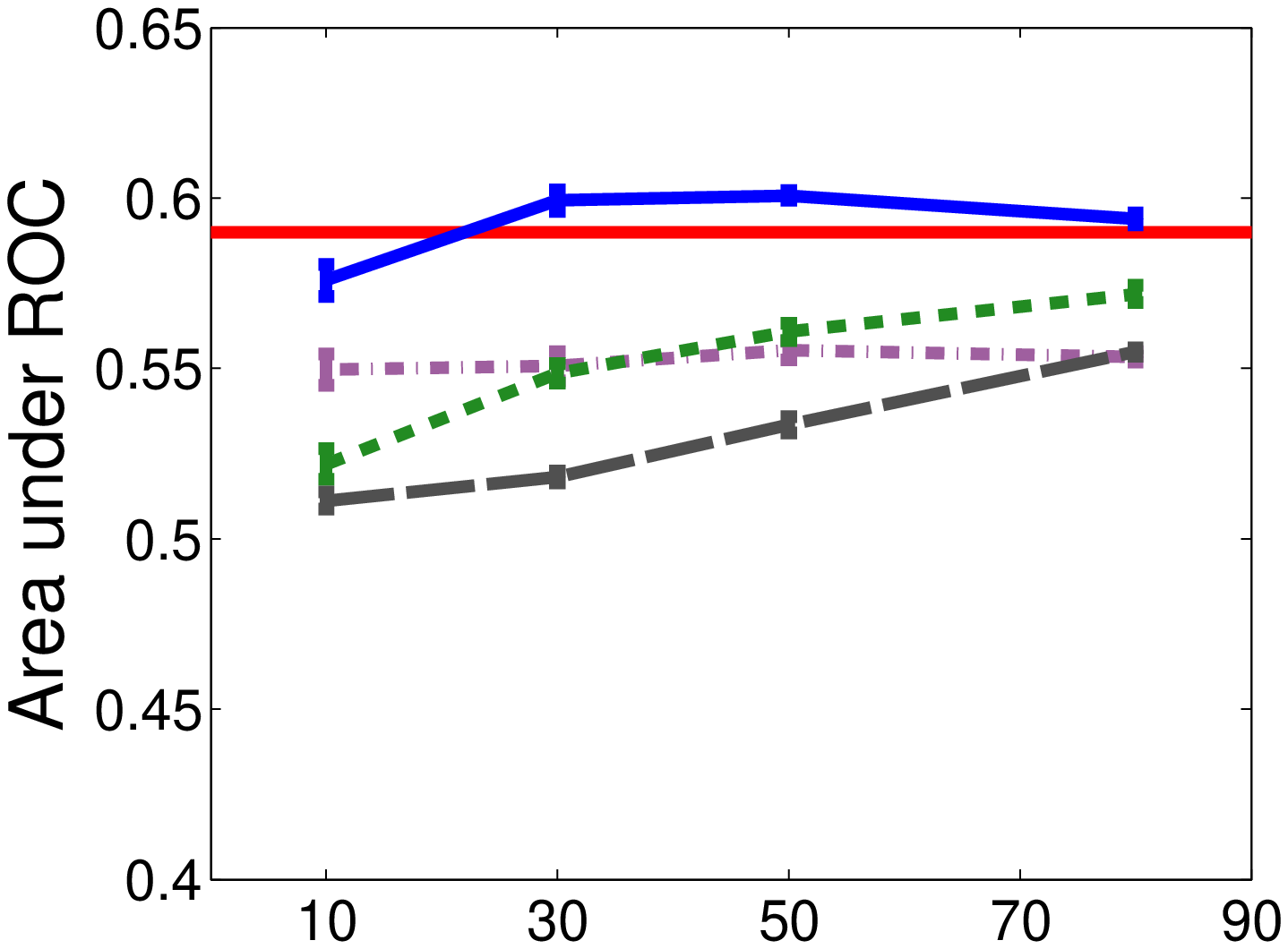}}
\vspace{-2.mm}
\hspace{-2.mm}
\subfigure[CoRA]{\label{fig:CorA_auc10}\includegraphics[width=0.40\textwidth]{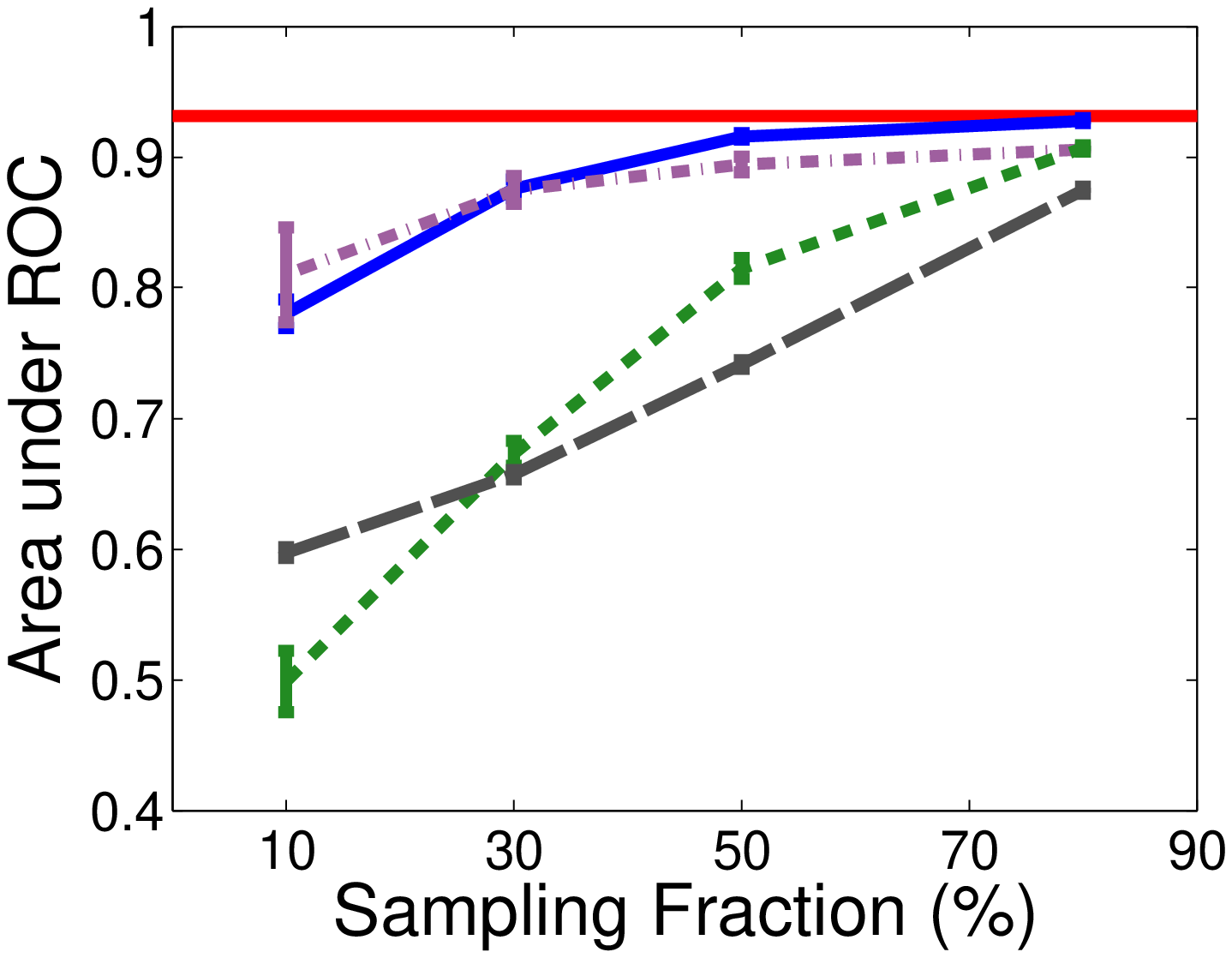}}
\hspace{4.mm}
\subfigure[Citeseer]{\label{fig:cite_auc10}\includegraphics[width=0.40\textwidth]{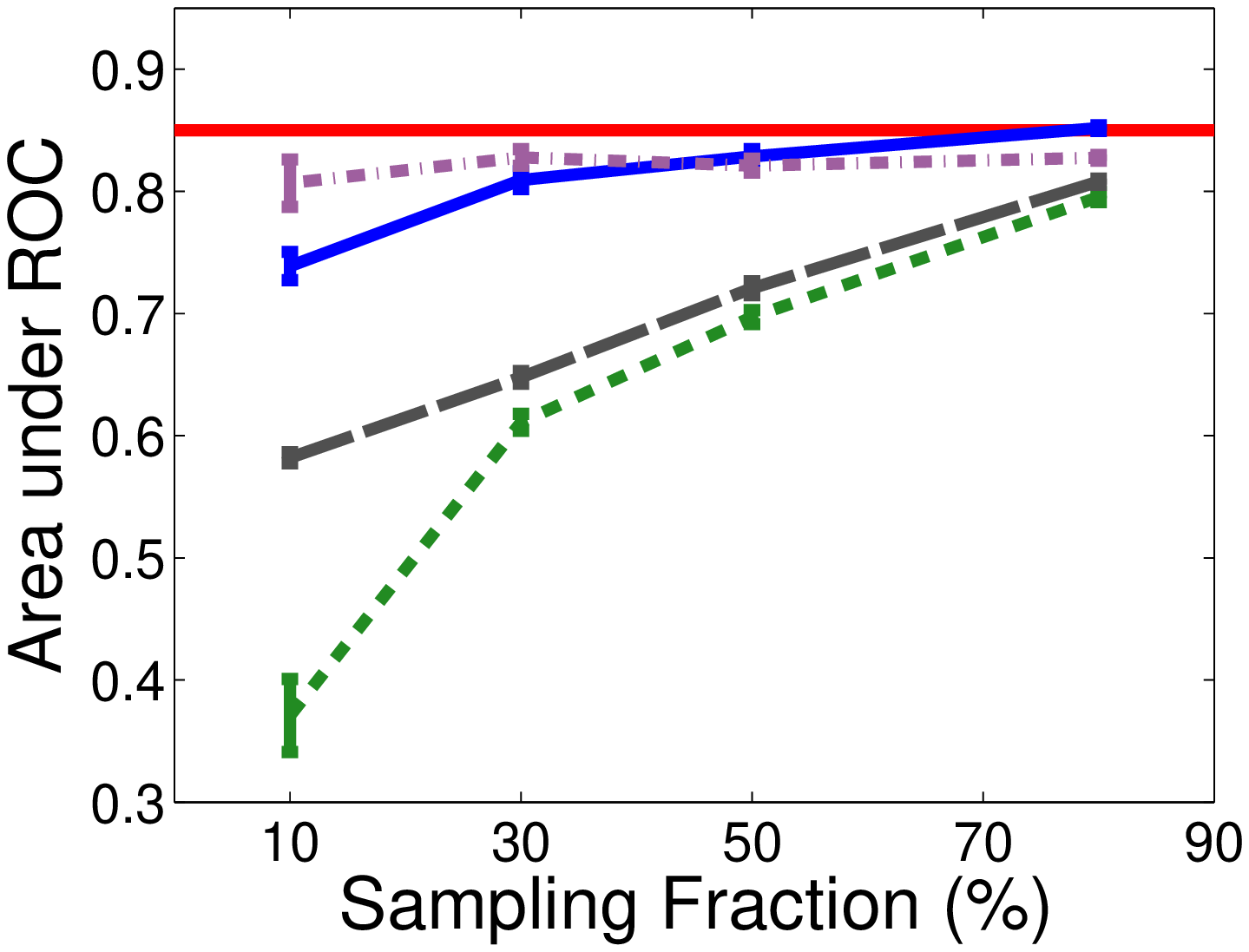}}
\vspace{-2.mm}
\caption{Classification accuracy versus sampling fraction with 10\% initially labeled nodes. }
\label{fig:auc_10}
\vspace{-4mm}
\end{figure*}

\begin{figure*}[!h]
\centering
\vspace{-2.mm}
\subfigure[Political Blogs]{\label{fig:polblogs_diffL}\includegraphics[width=0.40\textwidth]{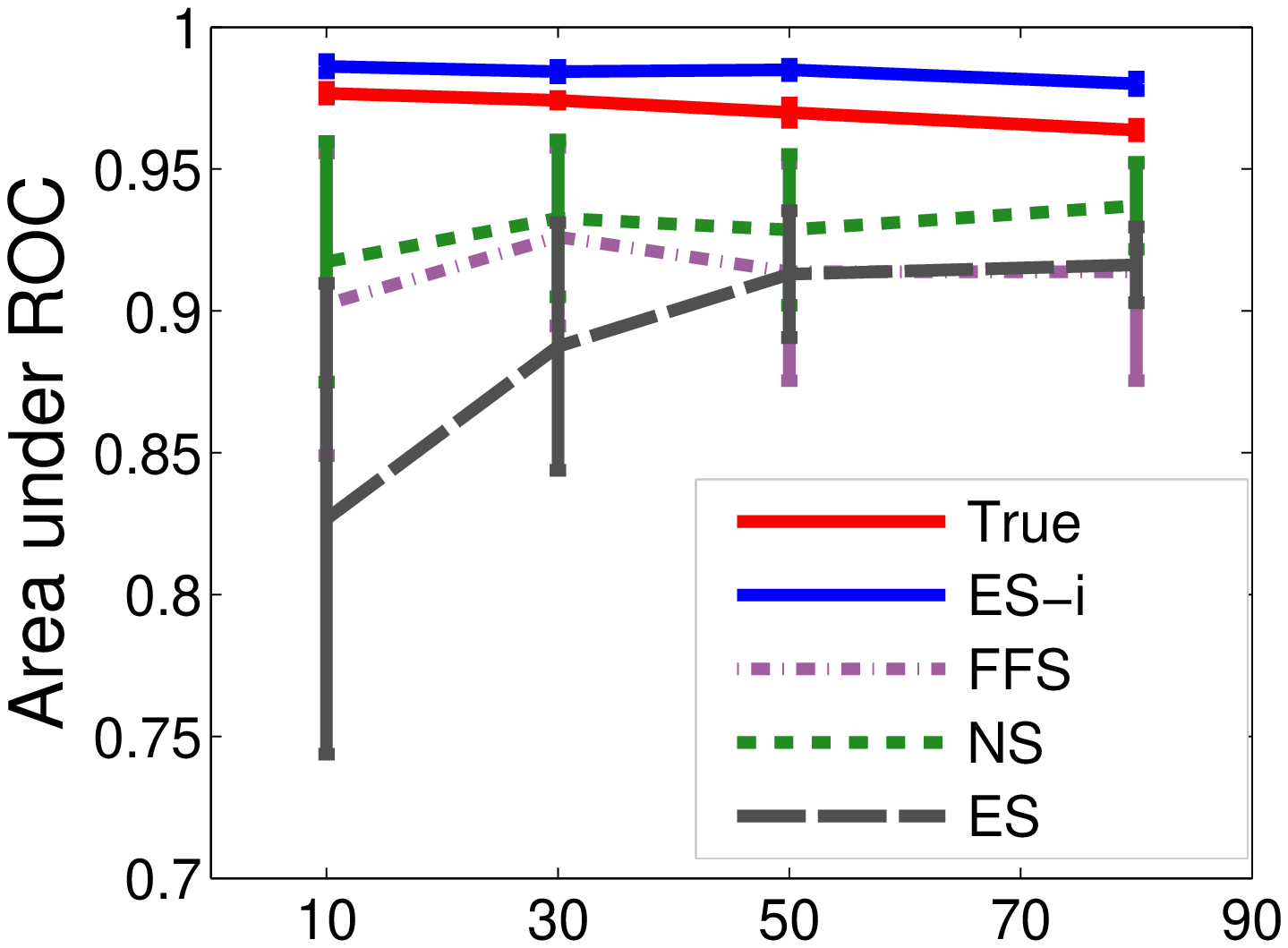}}
\hspace{4.mm}
\subfigure[Facebook]{\label{fig:FBPU_diffL}\includegraphics[width=0.40\textwidth]{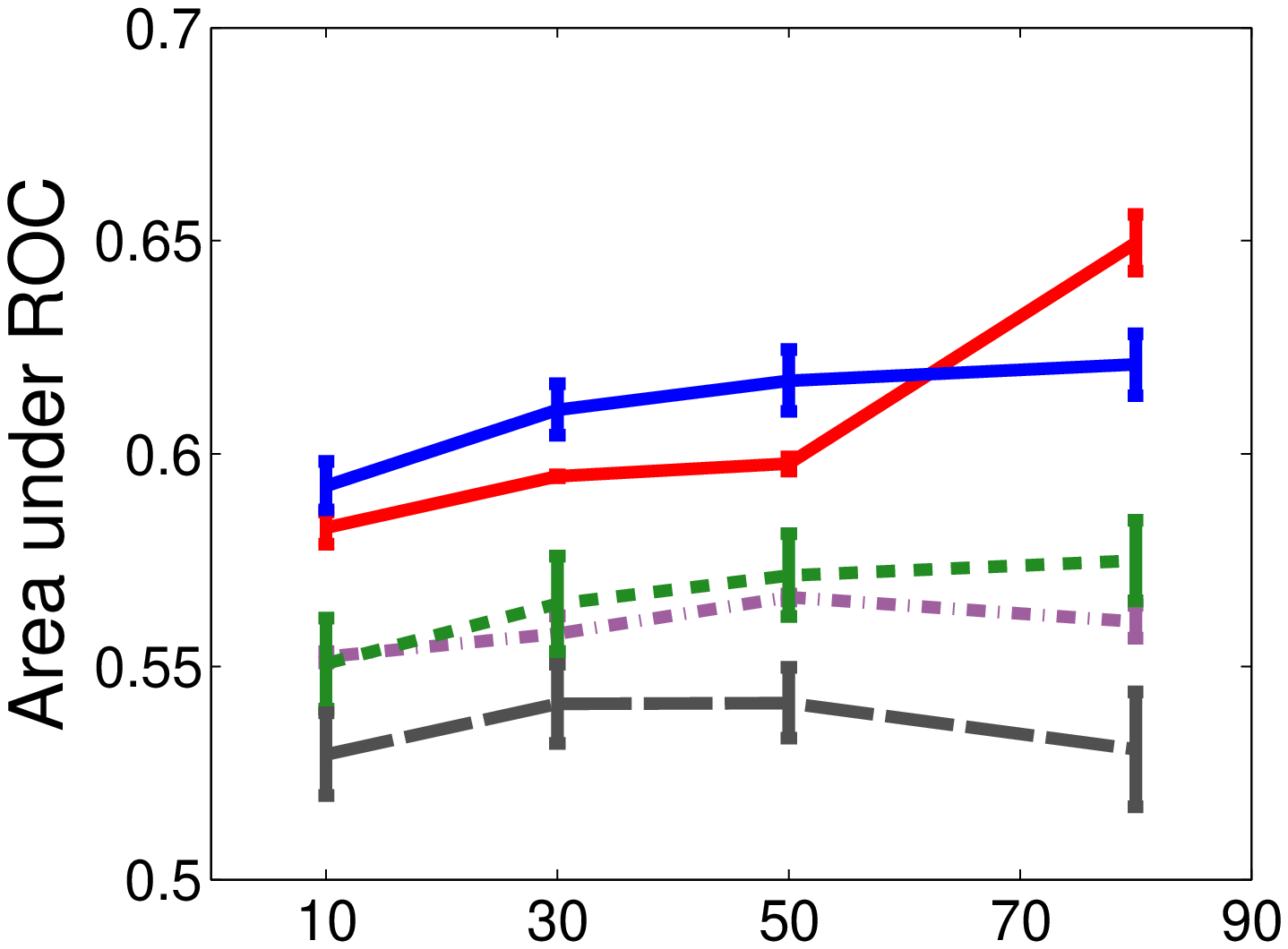}}
\vspace{-2.mm}
\hspace{-2.mm}
\subfigure[CoRA]{\label{fig:CorA_diffL}\includegraphics[width=0.40\textwidth]{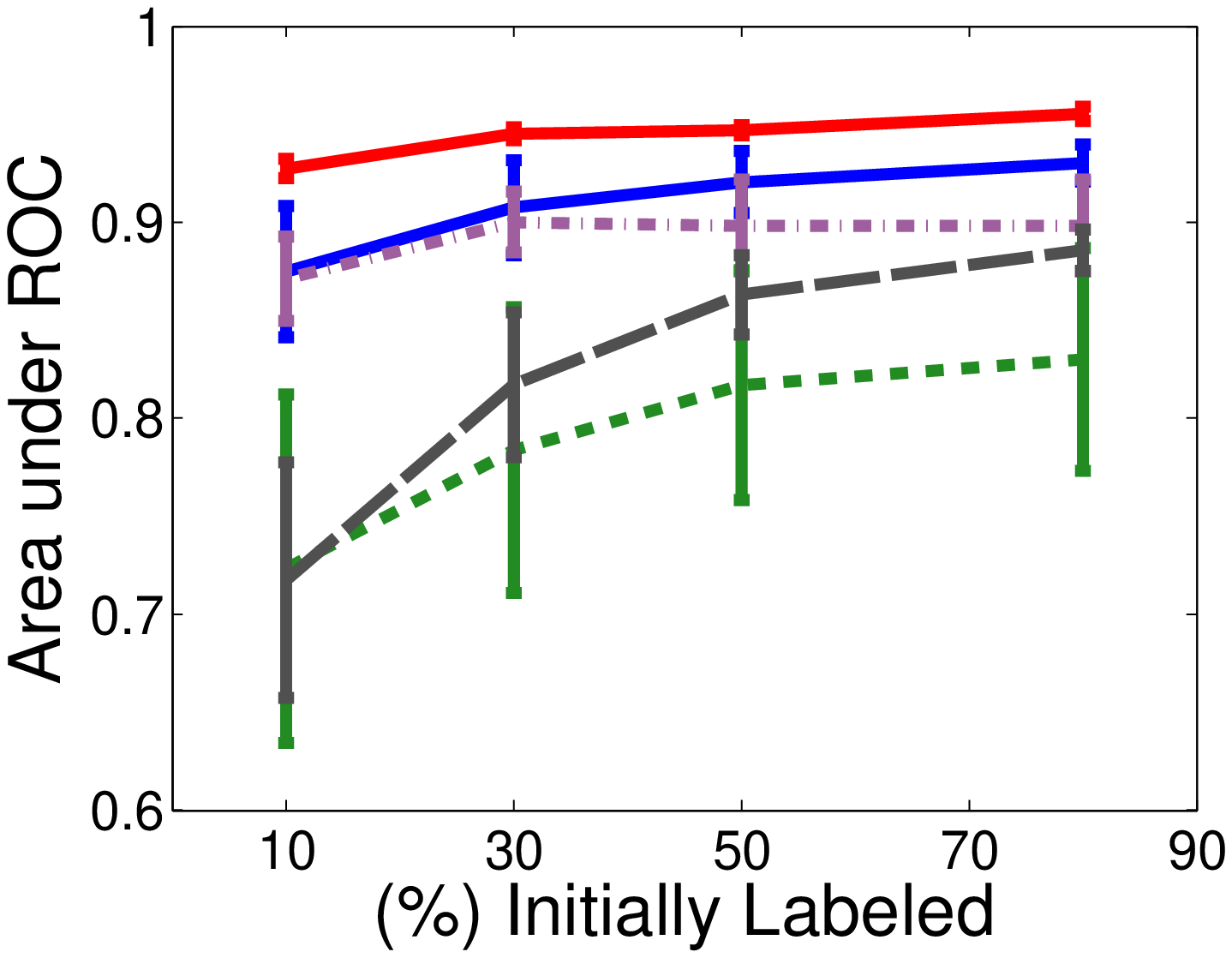}}
\hspace{4.mm}
\subfigure[Citeseer]{\label{fig:cite_diffL}\includegraphics[width=0.40\textwidth]{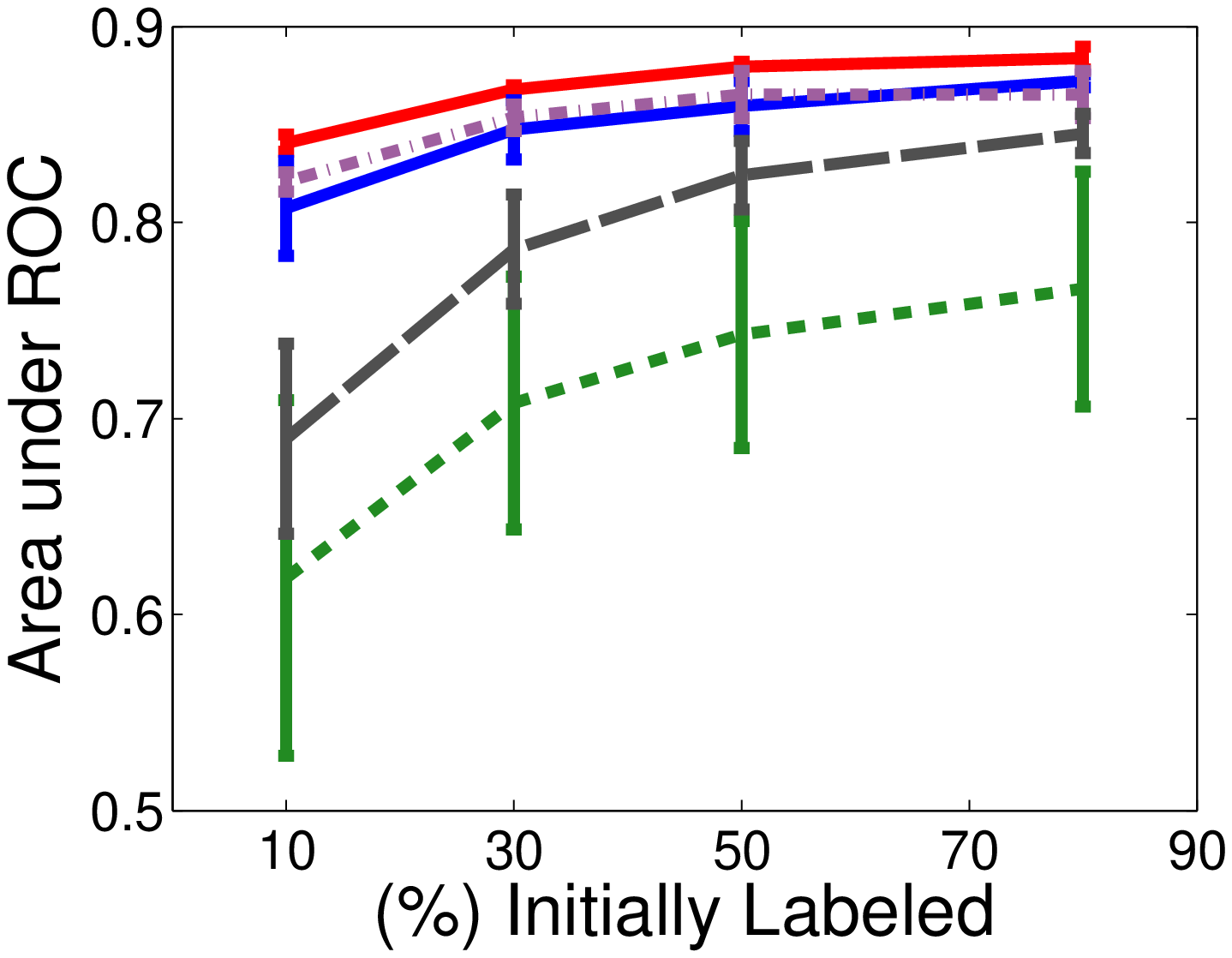}}
\vspace{-2.mm}
\caption{Classification accuracy versus proportion of labeled nodes.}
\label{fig:auc_10}
\vspace{-2mm}
\end{figure*}

We follow the common methodology used in \cite{macskassy:07} to compute the classification accuracy. First, we vary the proportion of randomly selected labeled nodes from $10\%-80\%$; and we use 5-fold cross validation to compute the average AUC. Then, we repeat this procedure for both the graph $G$ and the sample subgraph $G_s$. Note that AUC is calculated for the most prevalent class. Figures ~\ref{fig:polblogs_auc10} -- ~\ref{fig:cite_auc10} show the plots of AUC versus the sample size ($\phi=10\%-80\%$) with $10\%$ labeled nodes. Similarly, Figures~\ref{fig:polblogs_diffL} -- ~\ref{fig:cite_diffL} show the plots of AUC versus the the proportion of labeled nodes, such that the AUC is an average of all sample sizes ($10\%-80\%$). We observe that AUC of $G$ is generally underestimated for sample sizes $< 30\%$ in the case of NS, ES, and FFS. However, generally \algo performs better than other sampling methods and converges to the "True'' AUC on $G$. 

\paragraph{Summary} We conclude that many sampling methods fail to satisfy the different two goals (\ie, parameter versus accuracy estimation). For example, while node sampling estimates the class prior better than other methods, it cannot estimate the classification accuracy. Edge sampling performs similar to node sampling. In addition, Forest fire sampling is generally non robust for estimating class priors (for $\phi \leq 30\%$). Generally, \algo provides a good balance for satisfying the two goals with a little bias at the smaller sample sizes. 

%% file: conclusions.tex
\section{Conclusions and Future Work}
\label{sec:conclusions}
In this paper, we outlined a framework for the general problem of network sampling, by highlighting the different goals of study (from different research fields), population and units of interest, and classes of network sampling methods. This framework should facilitate the comparison of different sampling algorithms (strengths and weaknesses) relative to the particular goal of study. In addition, we proposed and discussed a spectrum of computational models for network sampling methods, going from the simple and least restrictive model of sampling from static graphs to the more realistic and restrictive model of sampling from streaming graphs. Within the context of the proposed spectrum, we designed a family of sampling methods based on the concept of graph induction that generalize across the full spectrum of computational models (from static to streaming), while efficiently preserving many of the topological properties of static and streaming graphs. Our experimental results indicate that our family of sampling methods more accurately preserves the underlying properties of the graph for both static and streaming graphs. Finally, we studied the impact of network sampling algorithms on the parameter estimation and performance evaluation of relational classification algorithms. Our results indicate our sampling method produces accurate estimates of classification accuracy. Concerning future work, we aim to investigate the performance of sub-linear time stream sampling methods for sampling a representative subgraph as well as extending to other task-based evaluations.

%% file: appendix.tex
\appendix
\section{Appendix A}
\label{appendix}
\setcounter{section}{1}
\subsection{Distributions for Static Graphs (at 20\% sample size)}

\begin{figure}[!h]
\centering
\subfigure{\label{fig:fbor deg dist}\includegraphics[width=0.33\linewidth]{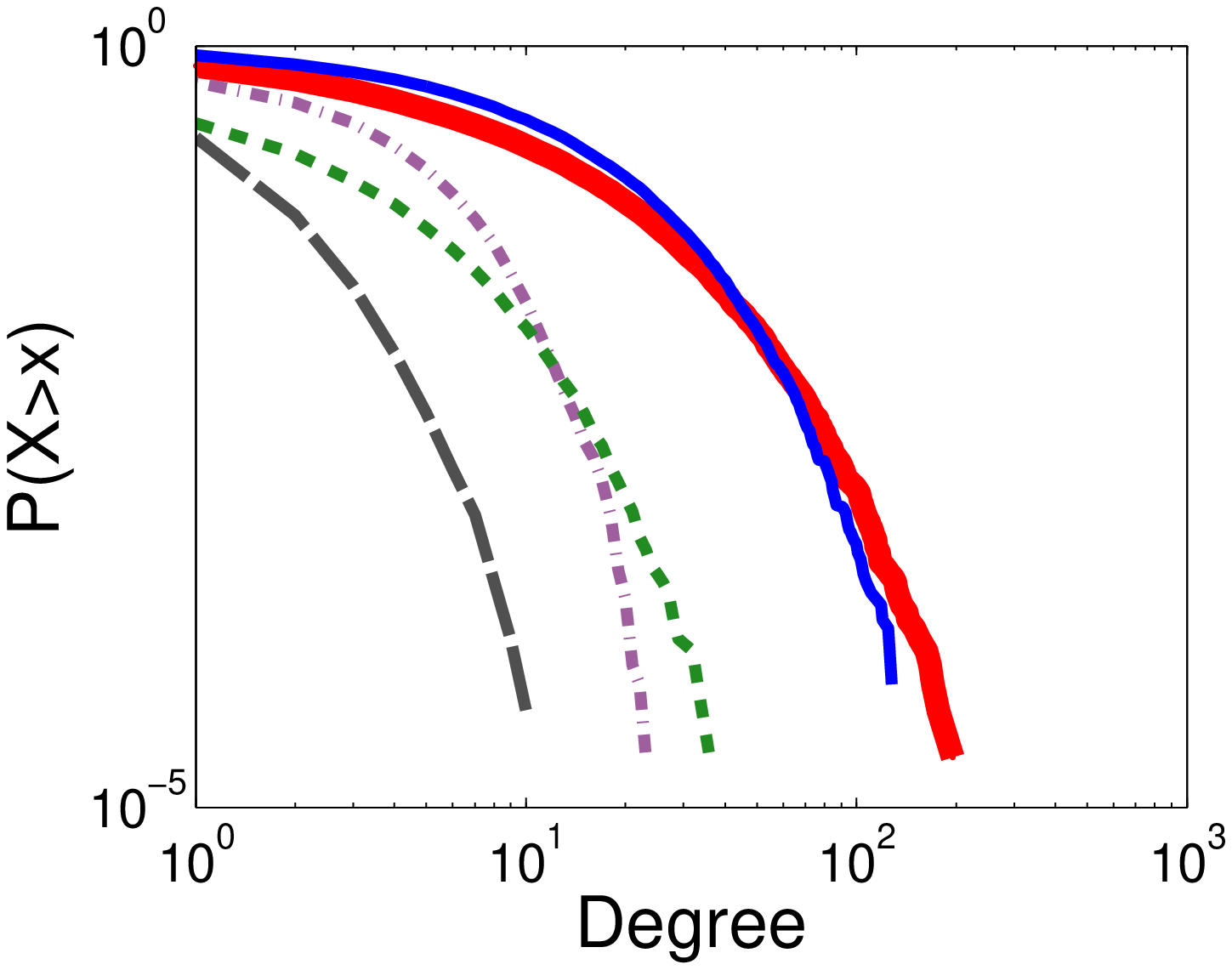}}
\hspace{-5.mm}
\subfigure{\label{fig:fbor pl dist}\includegraphics[width=0.33\linewidth]{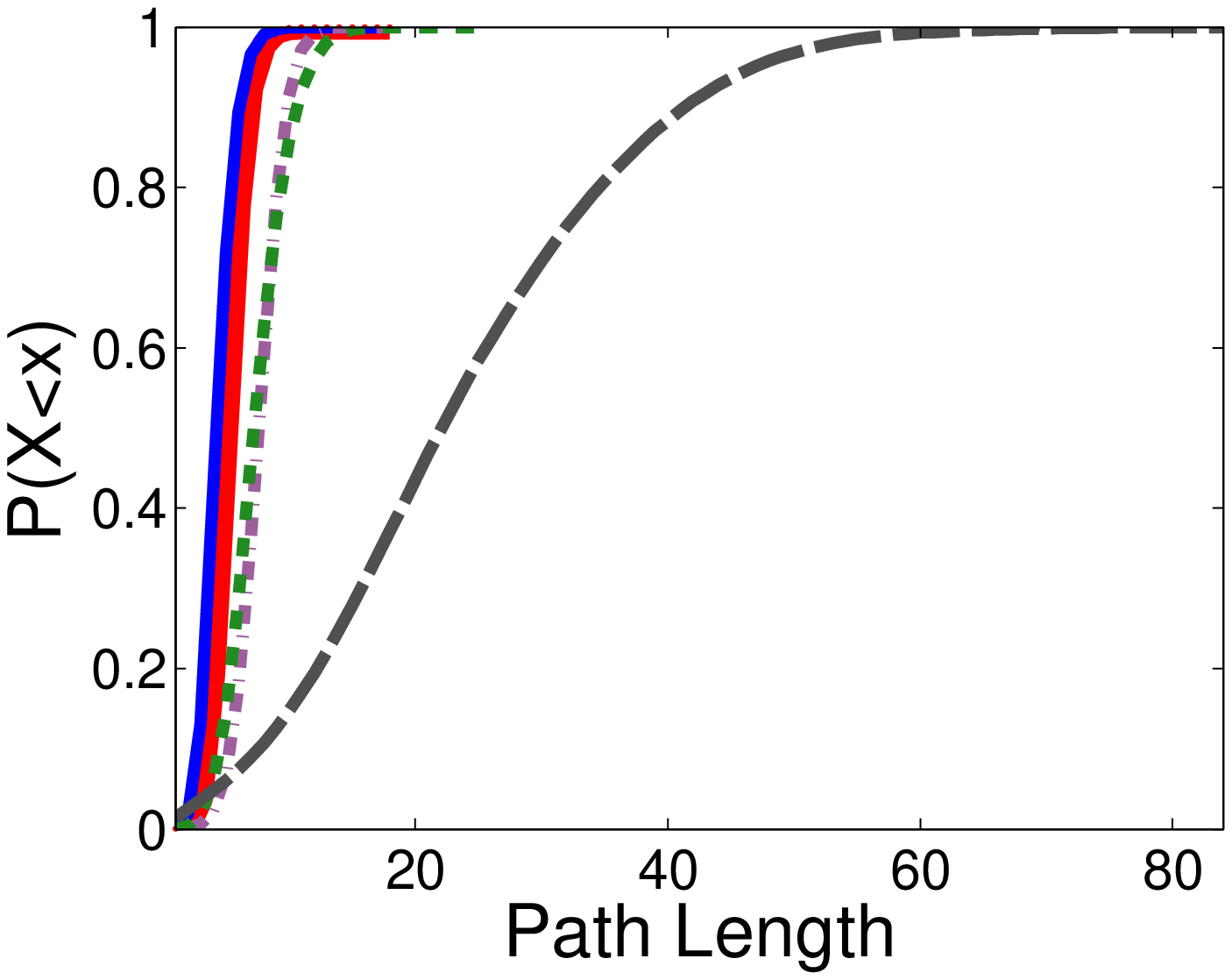}}
\hspace{-5.mm}
\subfigure{\label{fig:fbor cc dist}\includegraphics[width=0.33\linewidth]{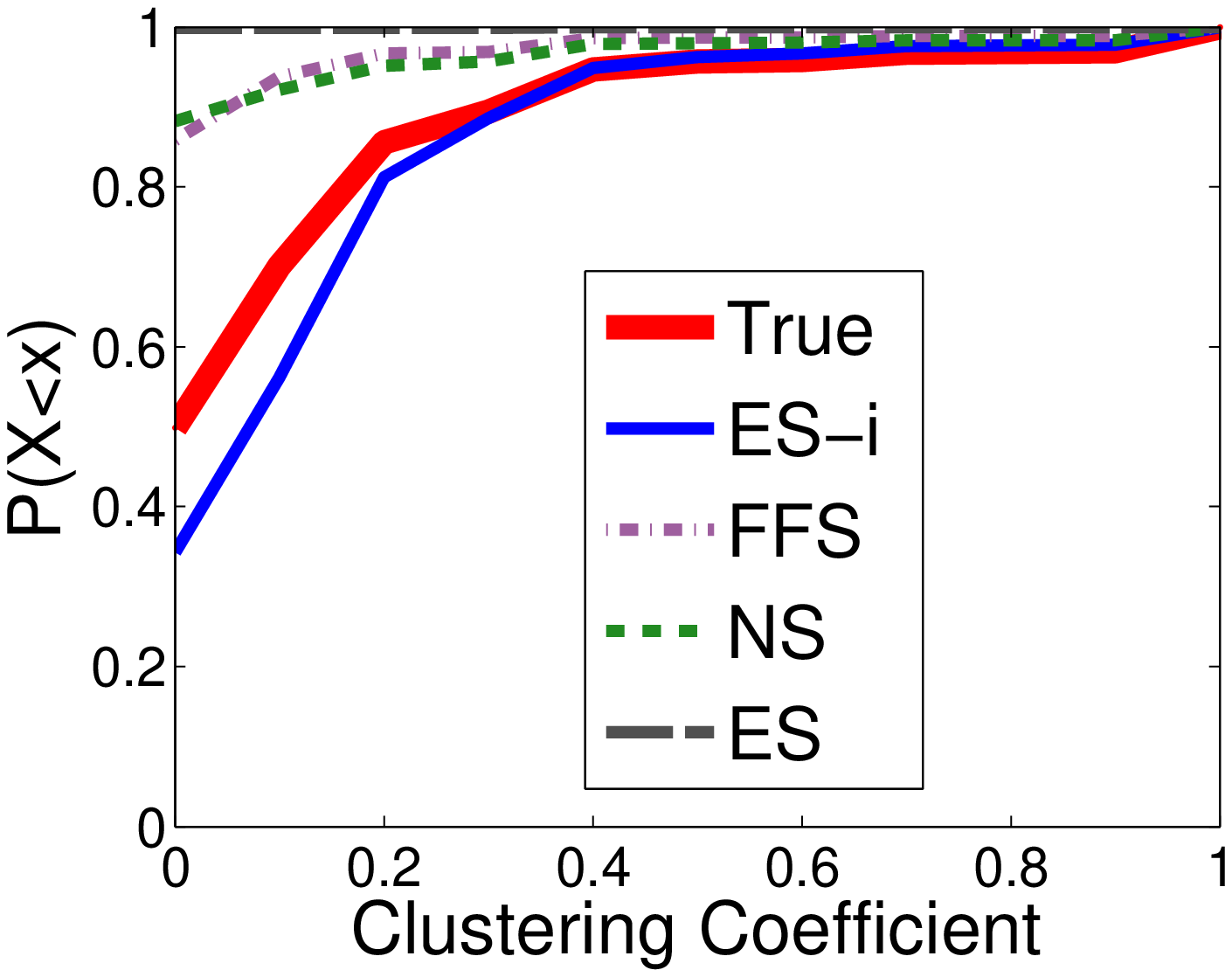}}
\vspace{-2mm}
\hspace{-4.mm}
\subfigure{\label{fig:fbor deg dist}\includegraphics[width=0.33\linewidth]{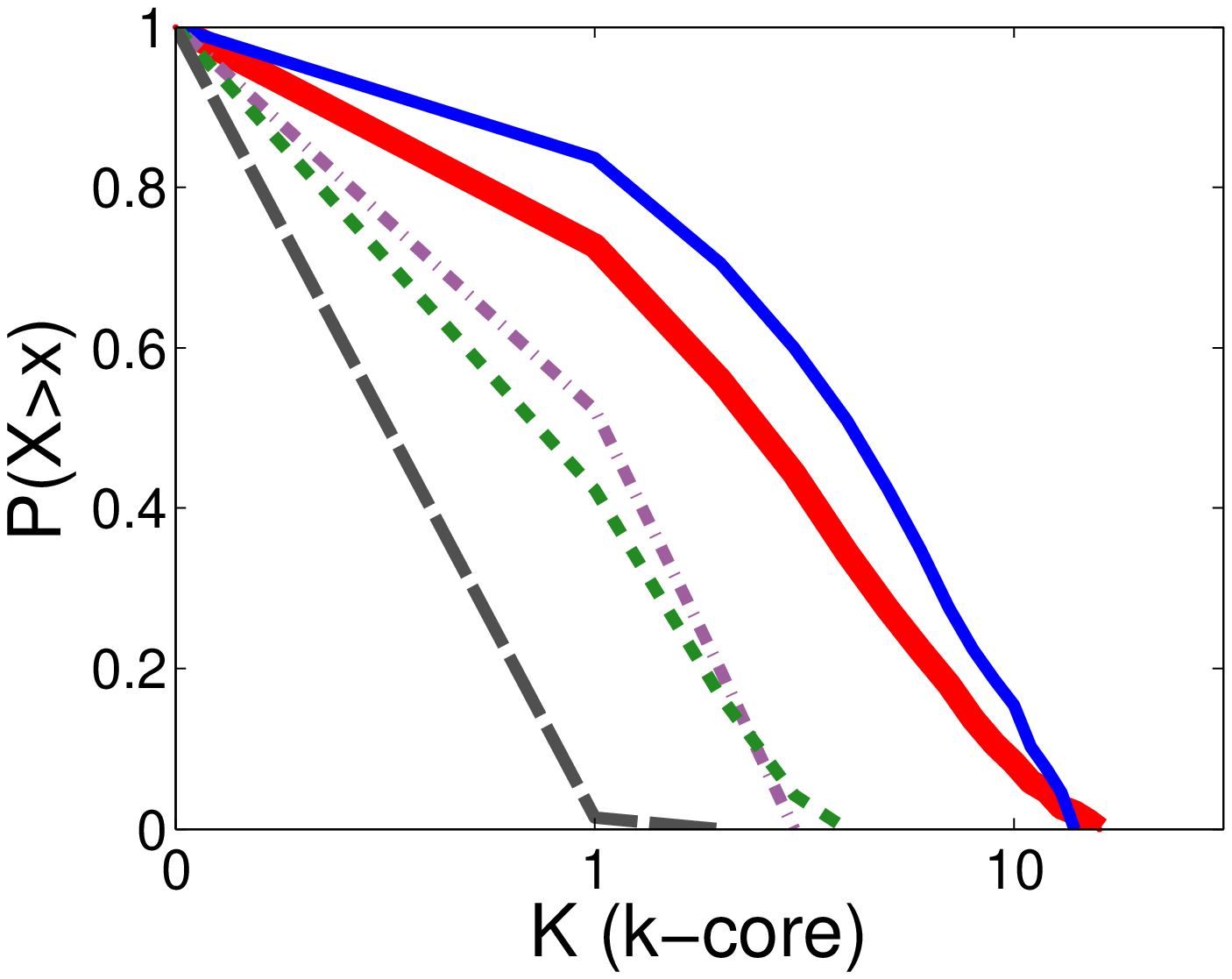}}
\hspace{-5.mm}
\subfigure{\label{fig:fbor pl dist}\includegraphics[width=0.33\linewidth]{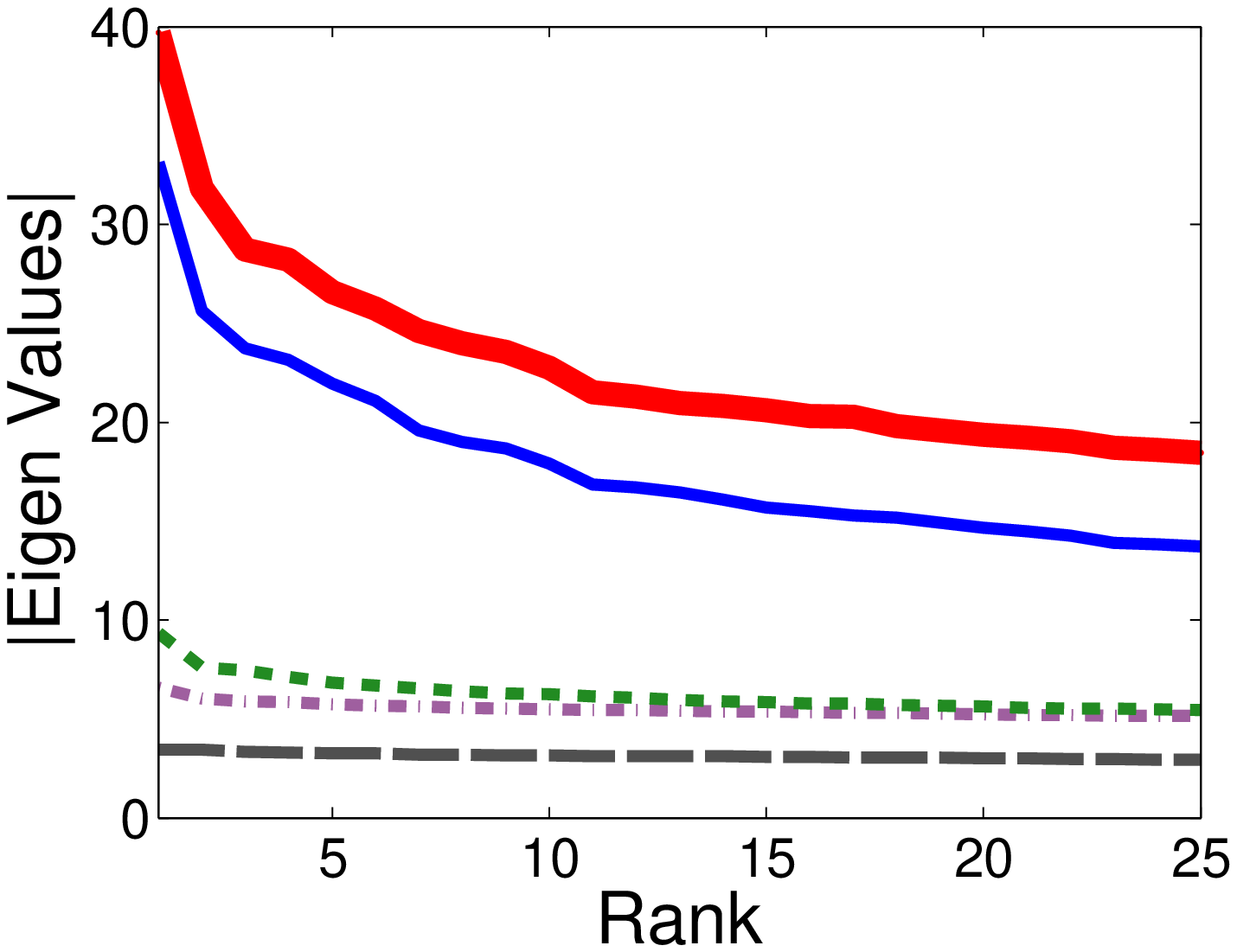}}
\hspace{-5.mm}
\subfigure{\label{fig:fbor cc dist}\includegraphics[width=0.33\linewidth]{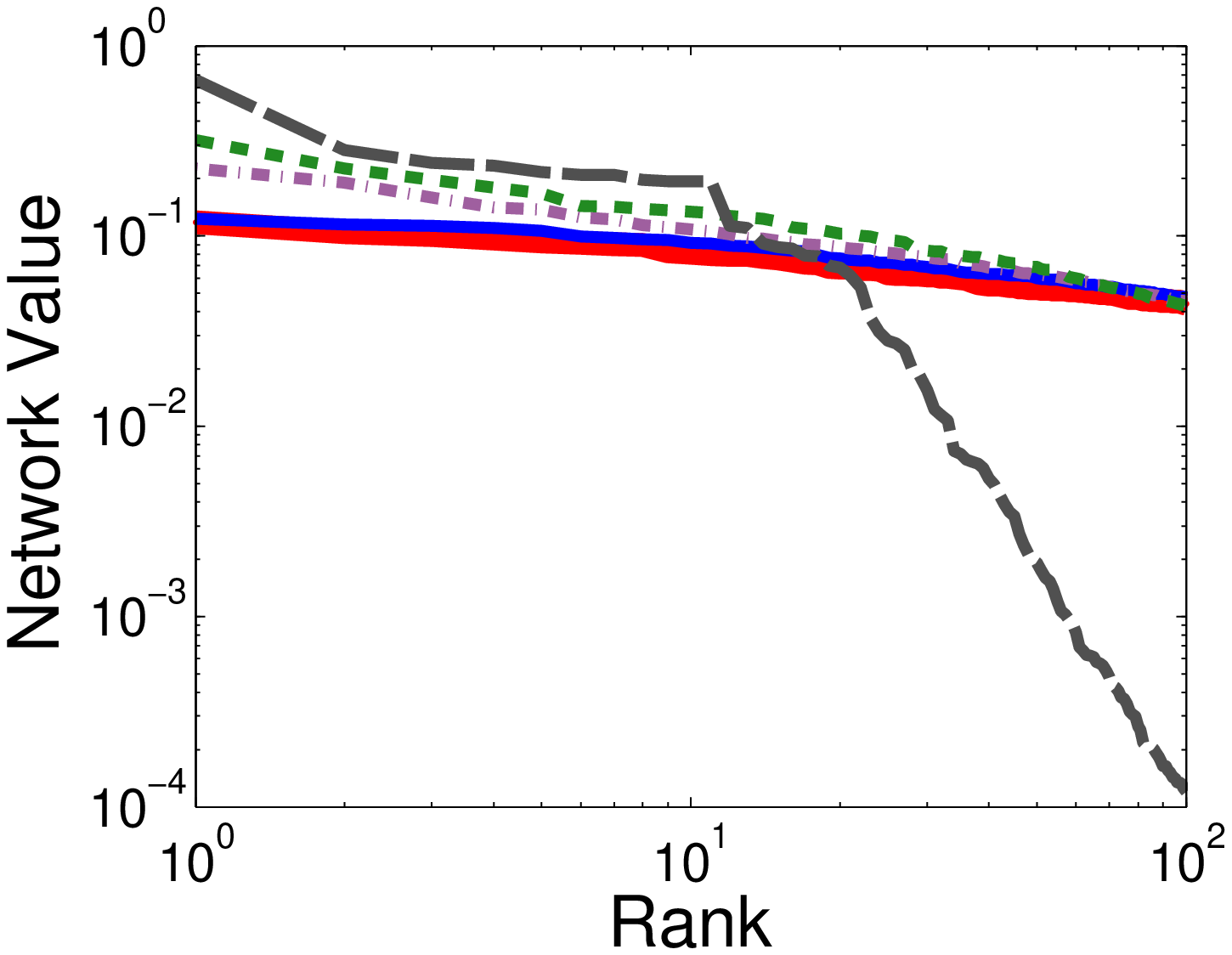}}
\caption{\textsc{Facebook} Graph}
\label{fig:dist_comp_fbor}
\vspace{-3.mm}
\end{figure}

\begin{figure}[!h]
\centering
\subfigure{\label{fig:arxiv deg dist}\includegraphics[width=0.33\linewidth]{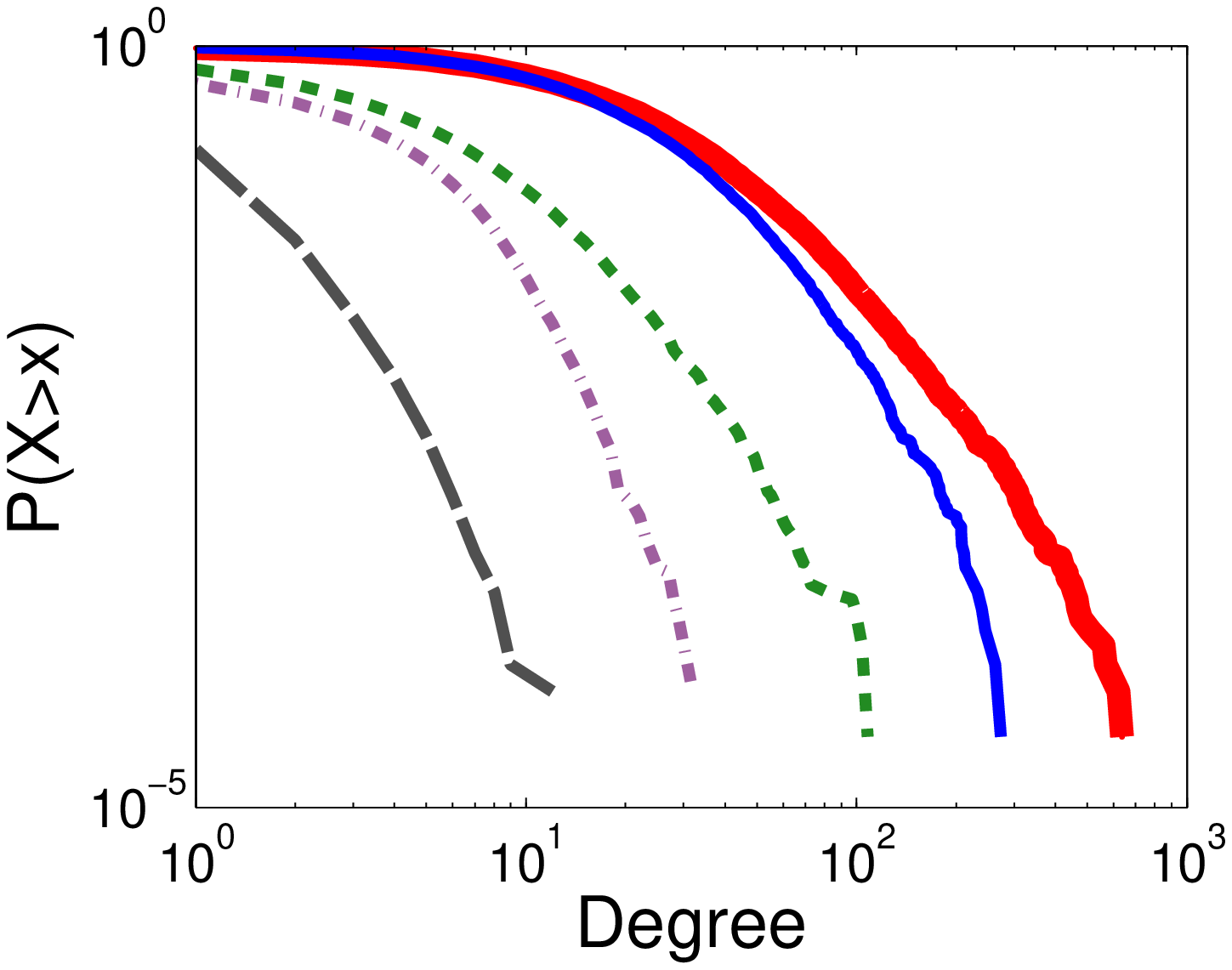}}
\hspace{-5.mm}
\subfigure{\label{fig:arxiv pl dist}\includegraphics[width=0.33\linewidth]{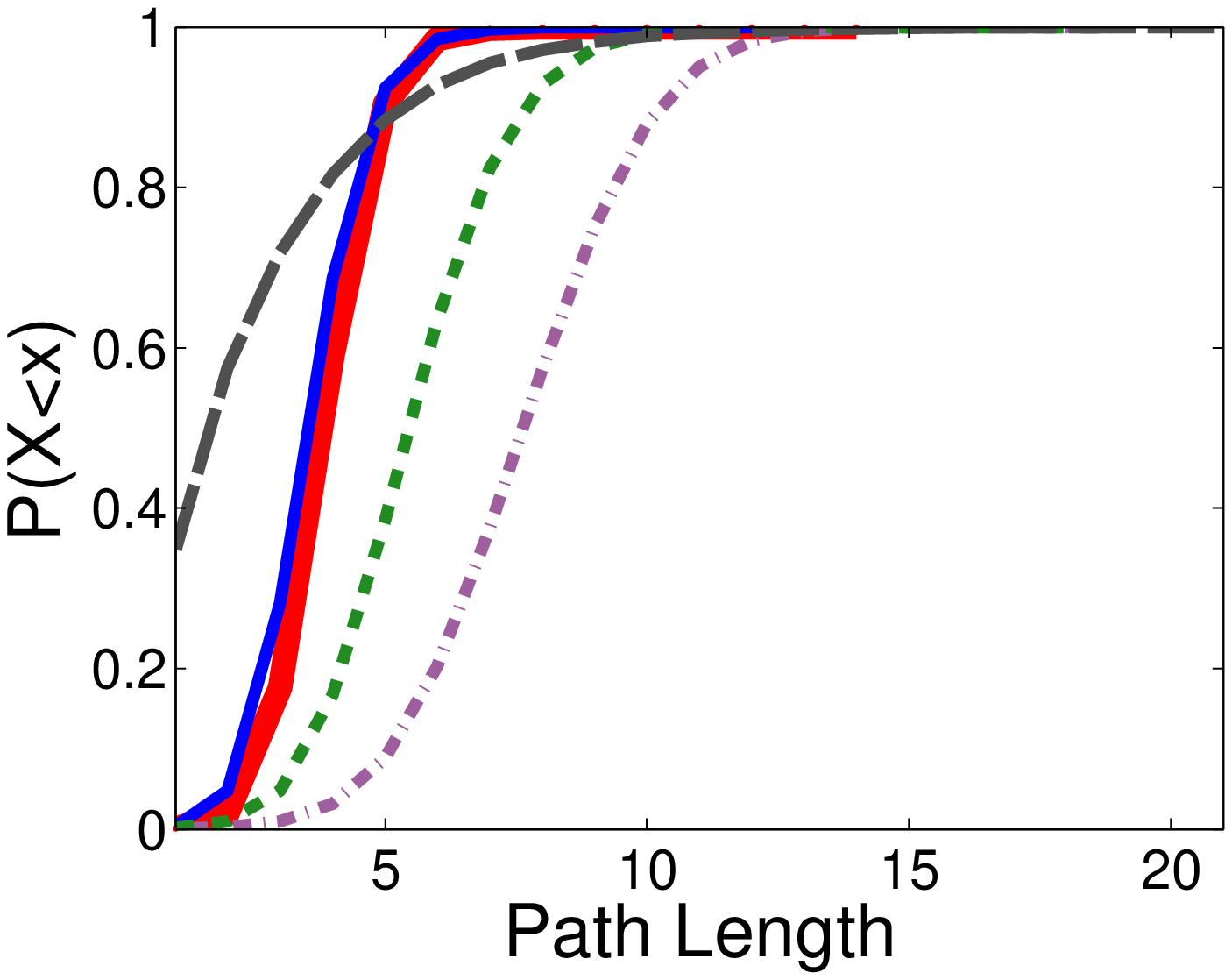}}
\hspace{-5.mm}
\subfigure{\label{fig:arxiv cc dist}\includegraphics[width=0.33\linewidth]{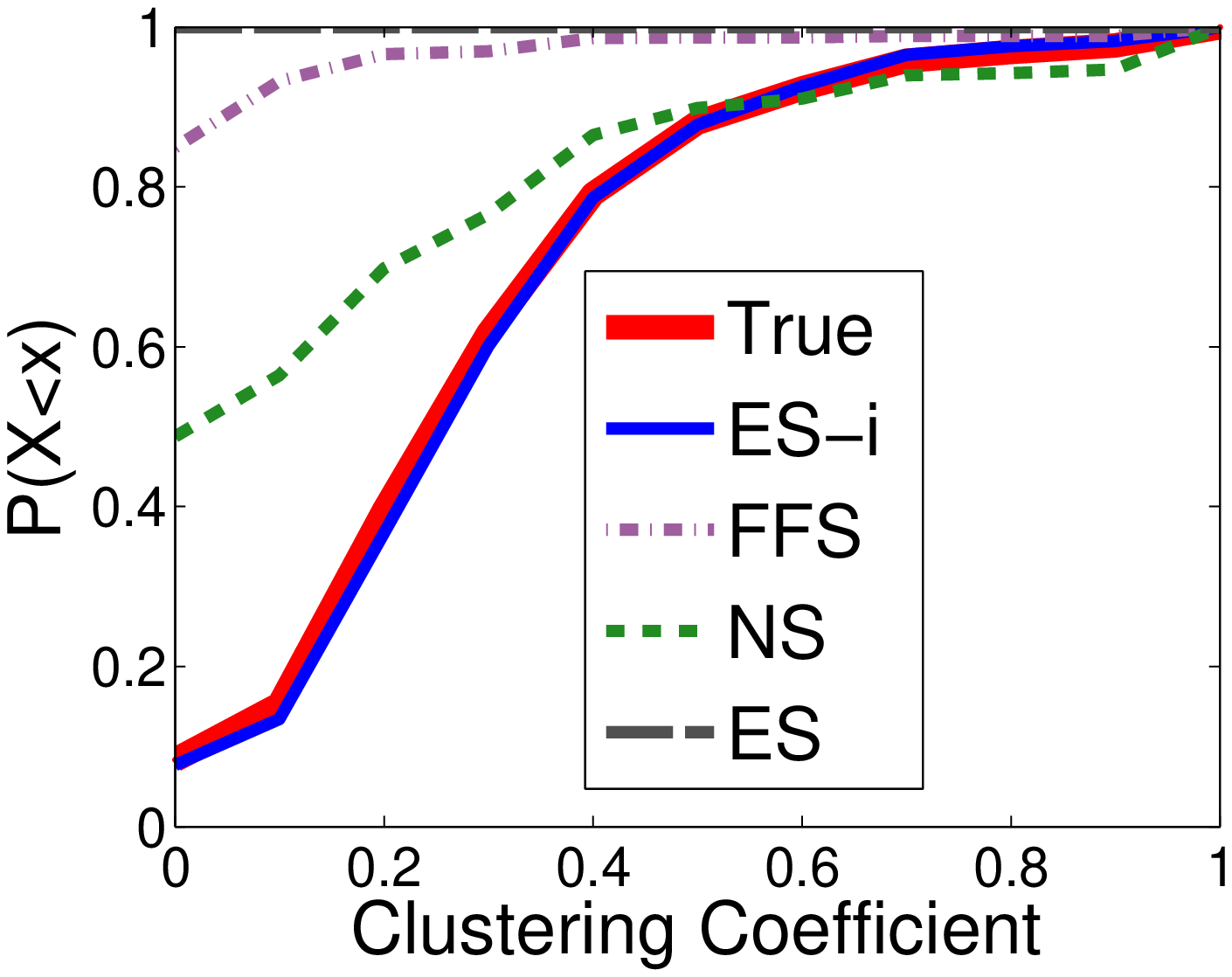}}
\vspace{-2mm}
\hspace{-4.mm}
\subfigure{\label{fig:arxiv deg dist}\includegraphics[width=0.33\linewidth]{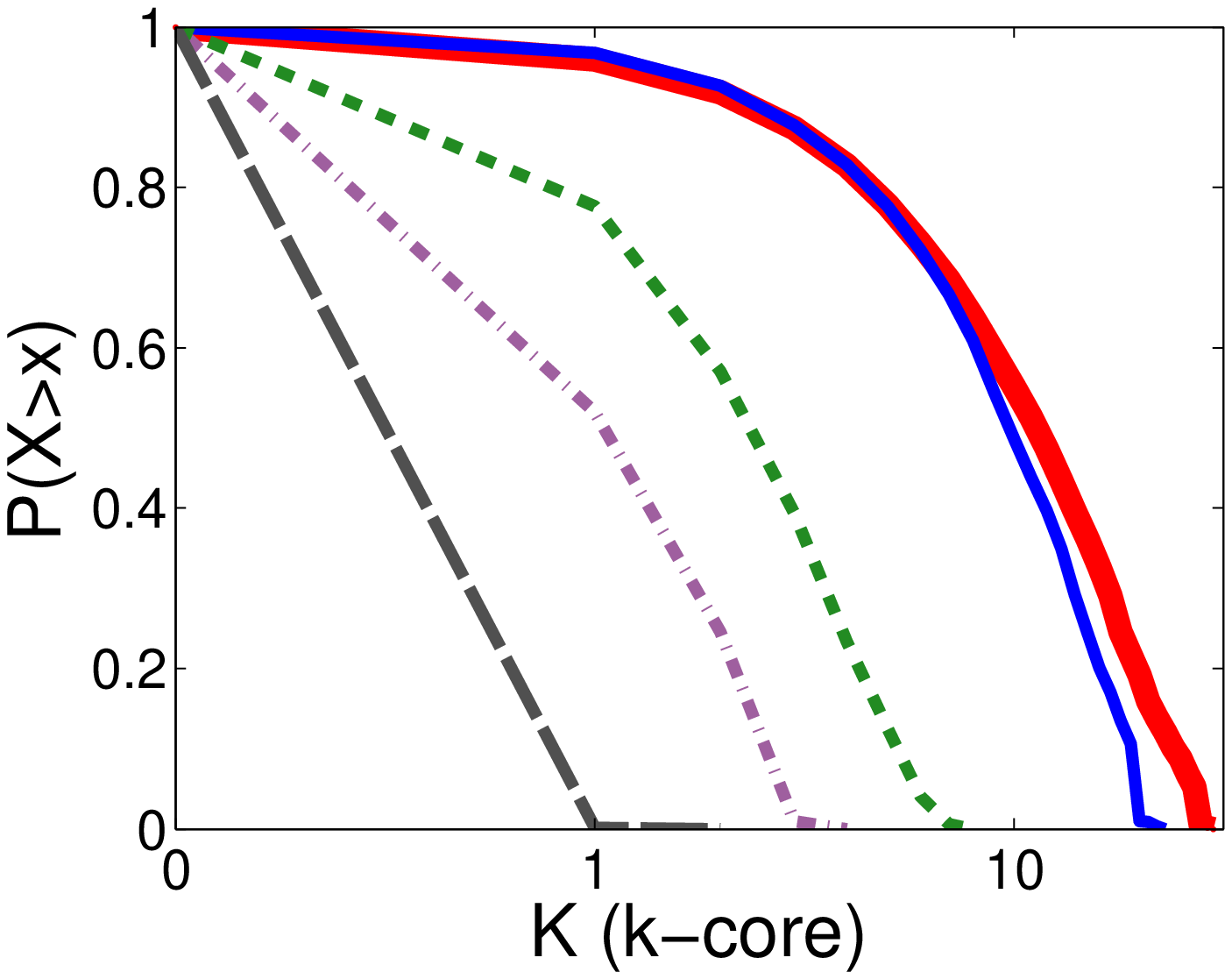}}
\hspace{-5.mm}
\subfigure{\label{fig:arxiv pl dist}\includegraphics[width=0.33\linewidth]{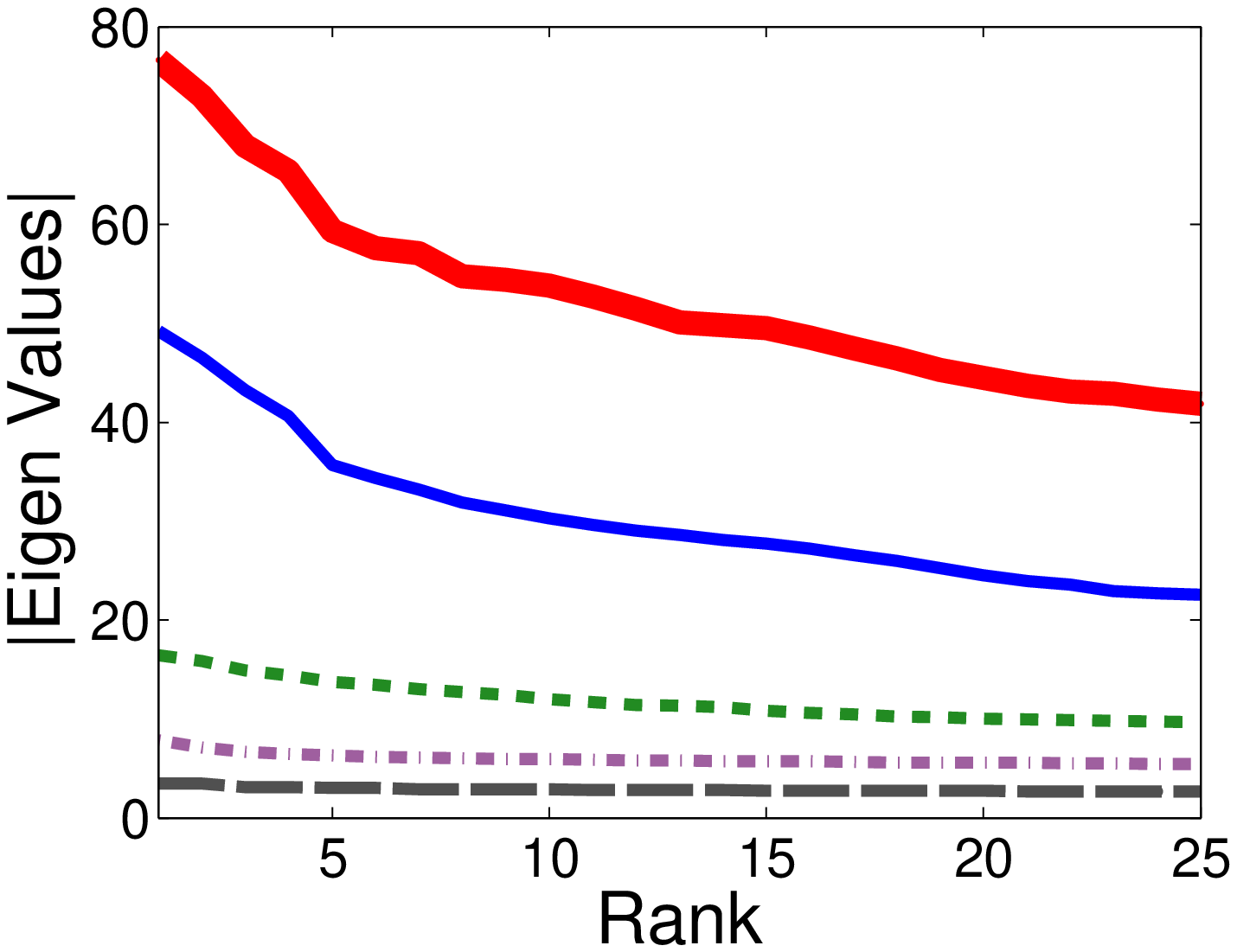}}
\hspace{-5.mm}
\subfigure{\label{fig:arxiv cc dist}\includegraphics[width=0.33\linewidth]{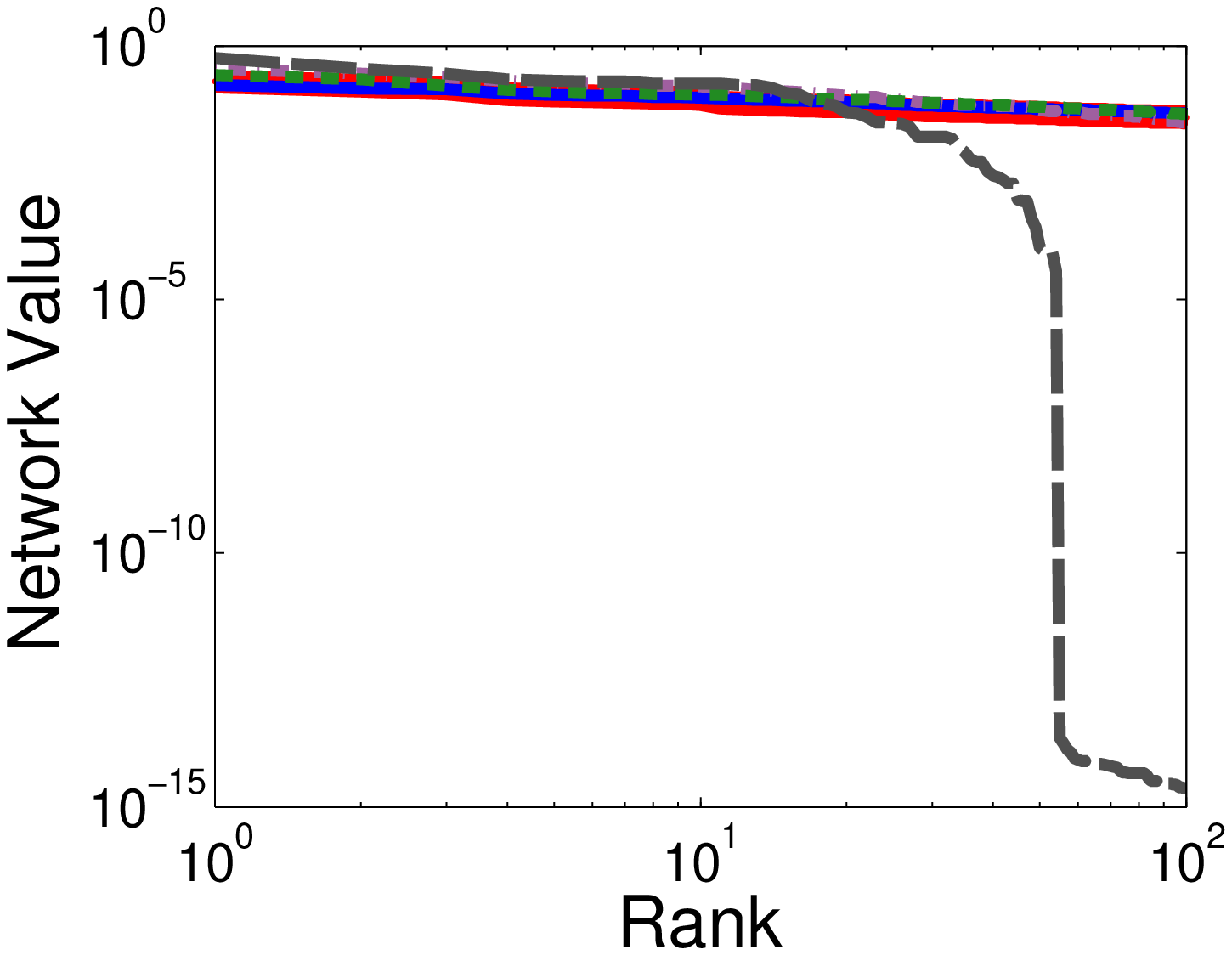}}
\caption{\textsc{HepPH} Graph}
\label{fig:dist_comp_arxiv}
\vspace{-3.mm}
\end{figure}

\newpage

\begin{figure}[!h]
\centering
\subfigure{\label{fig:condmat deg dist}\includegraphics[width=0.33\linewidth]{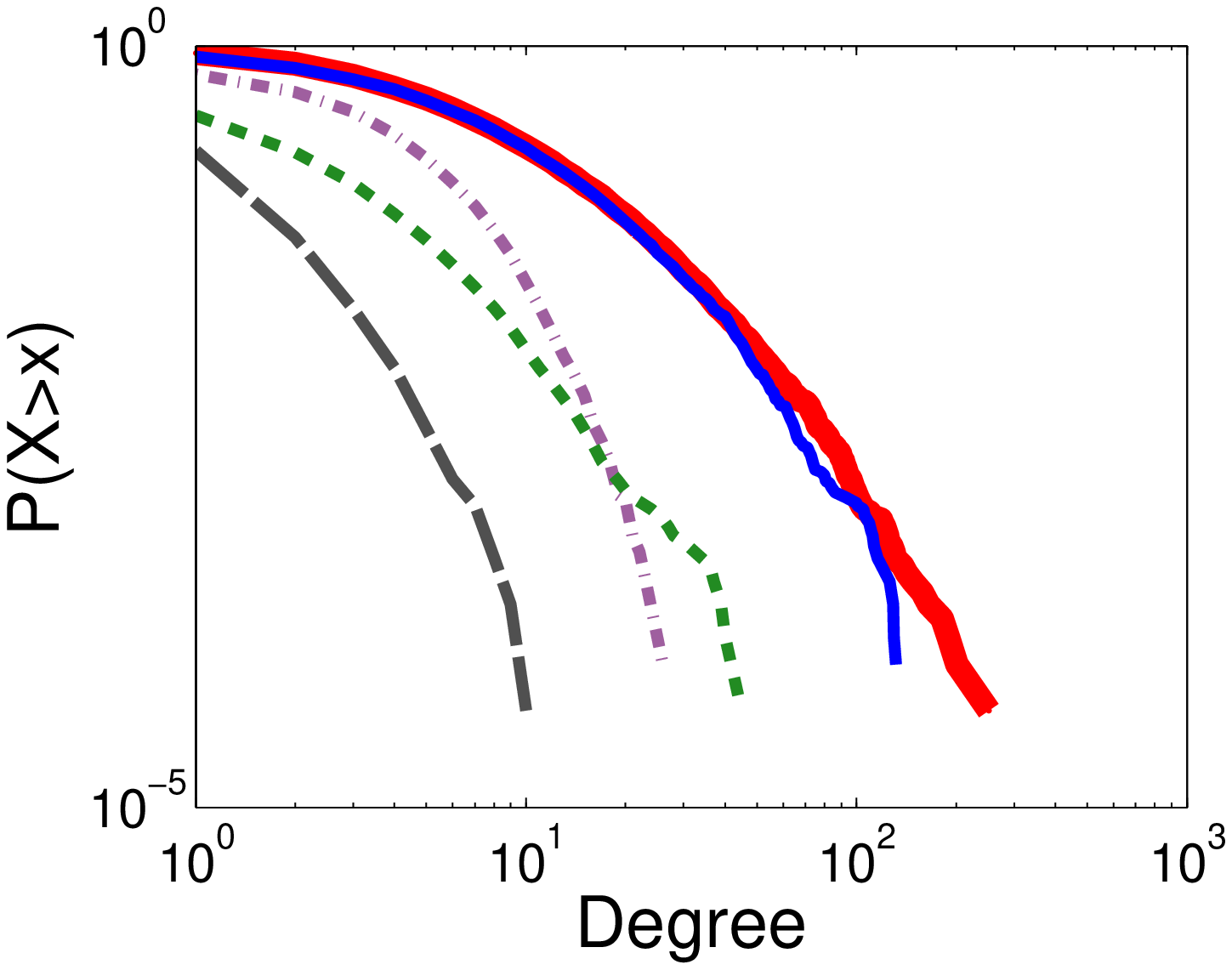}}
\hspace{-5.mm}
\subfigure{\label{fig:condmat pl dist}\includegraphics[width=0.33\linewidth]{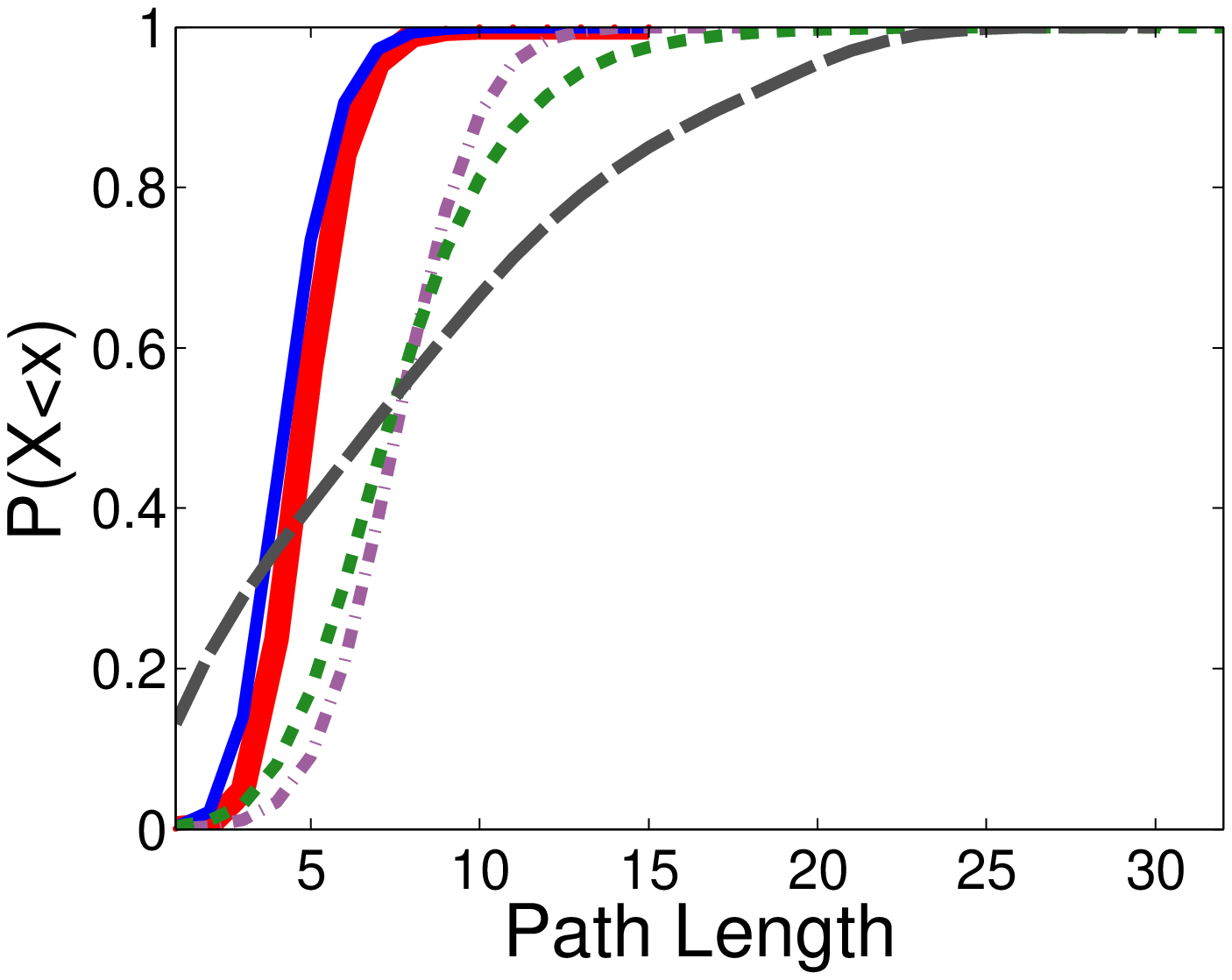}}
\hspace{-5.mm}
\subfigure{\label{fig:condmat cc dist}\includegraphics[width=0.33\linewidth]{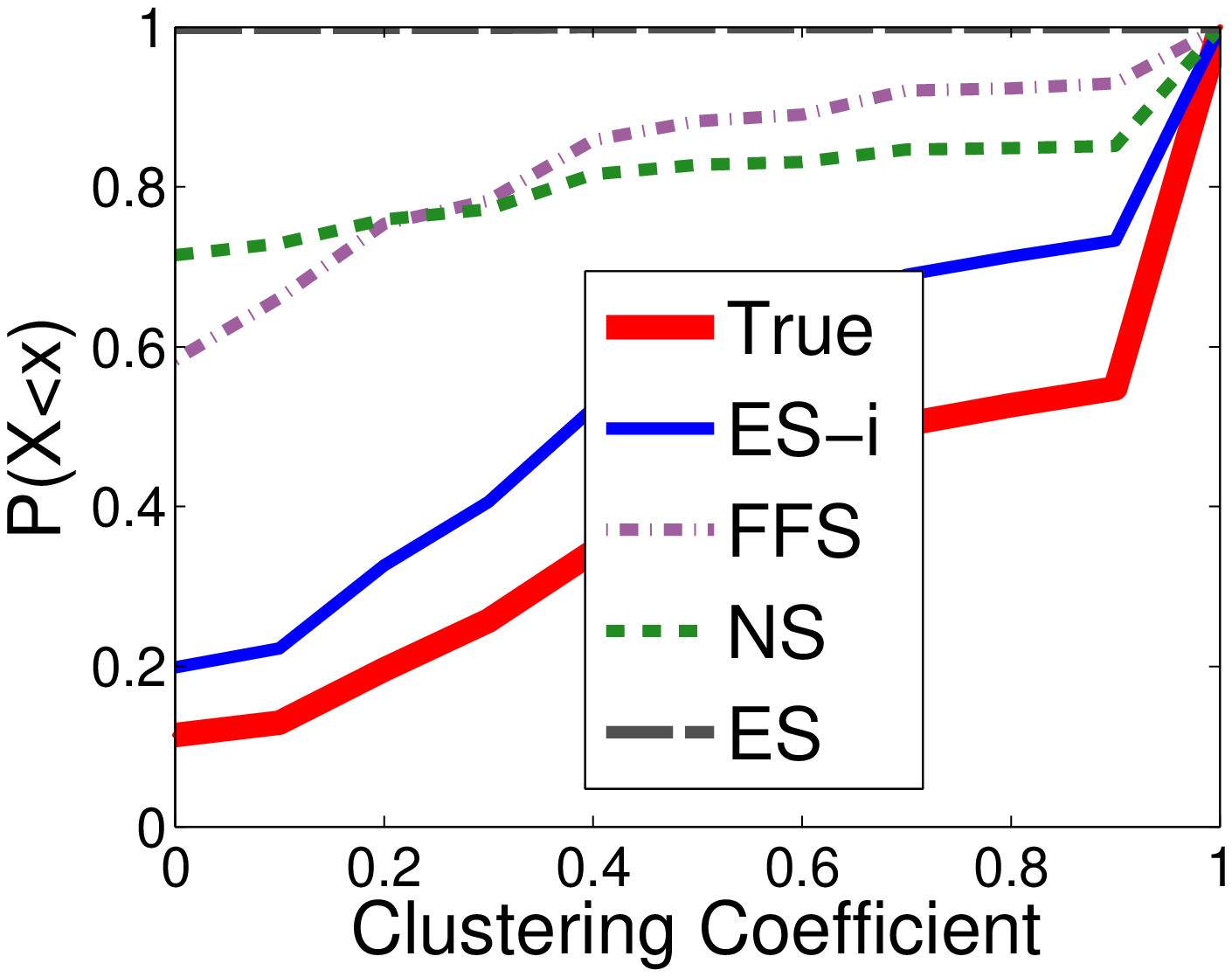}}
\vspace{-2mm}
\hspace{-2.mm}
\subfigure{\label{fig:condmat deg dist}\includegraphics[width=0.33\linewidth]{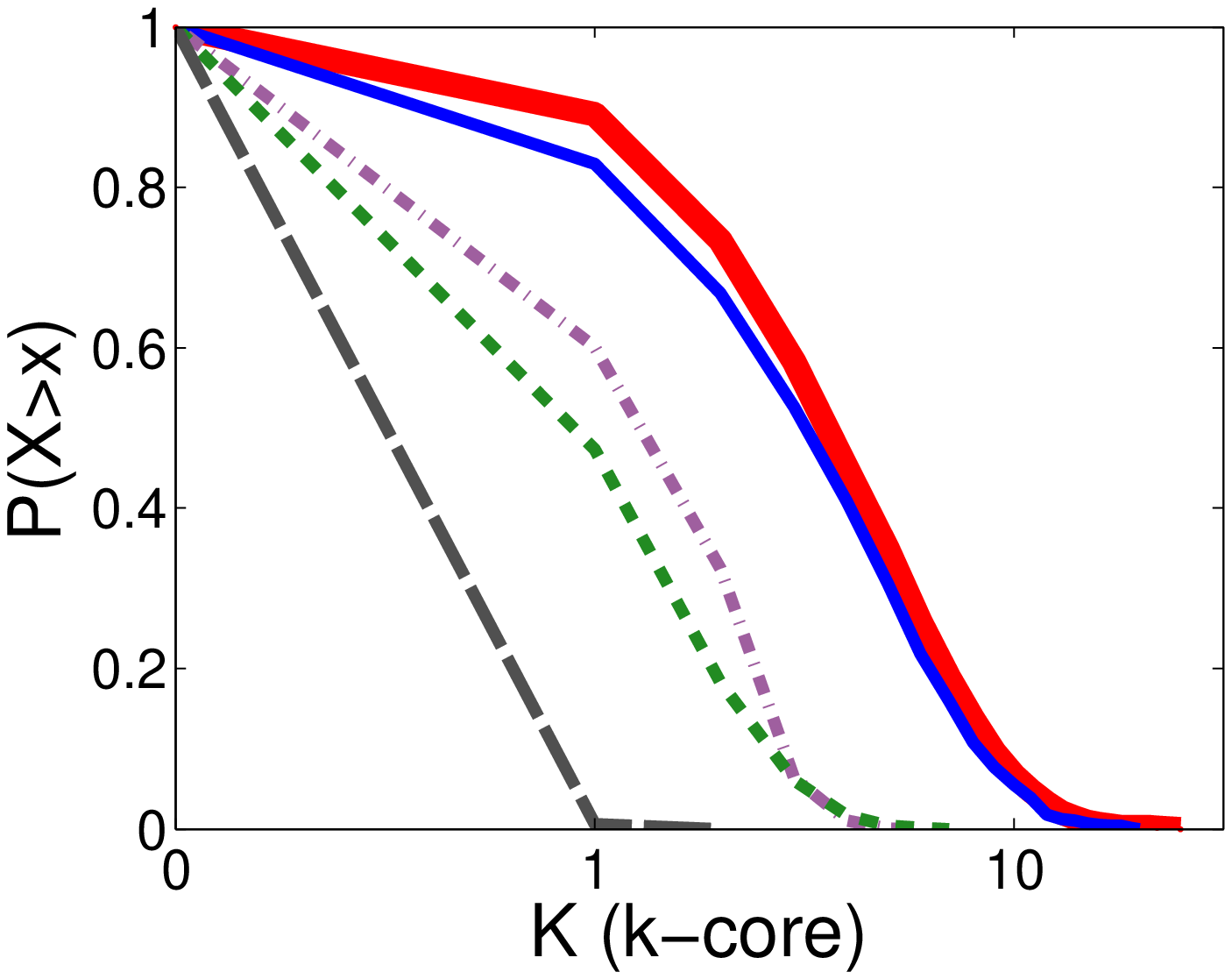}}
\hspace{-5.mm}
\subfigure{\label{fig:condmat pl dist}\includegraphics[width=0.33\linewidth]{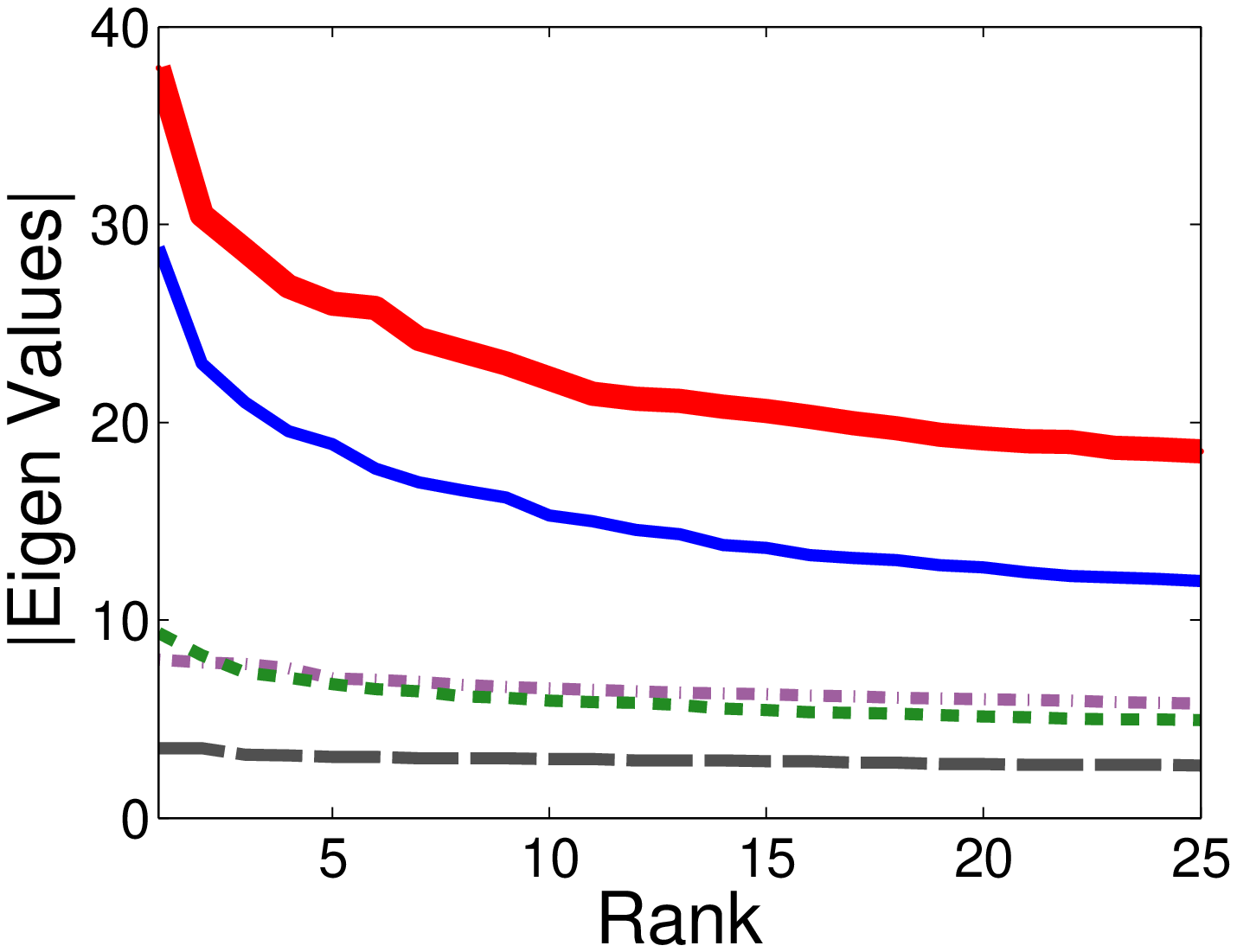}}
\hspace{-5.mm}
\subfigure{\label{fig:condmat cc dist}\includegraphics[width=0.33\linewidth]{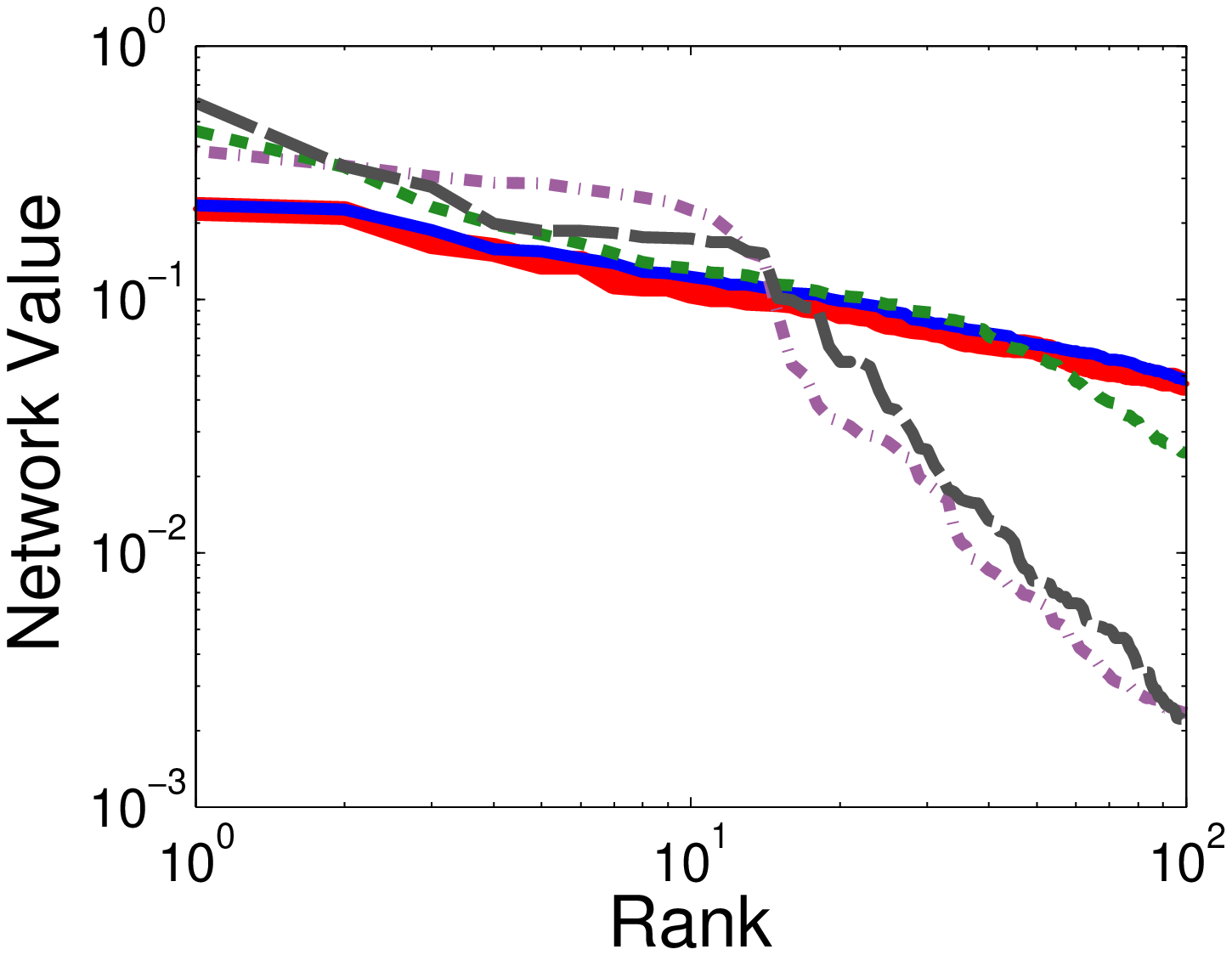}}
\caption{\textsc{CondMAT} Graph}
\label{fig:dist_comp_condmat}
\vspace{-3.mm}
\end{figure}

\begin{figure}[!h]
\centering
\vspace{-3.mm}
\subfigure{\label{fig:twcop deg dist}\includegraphics[width=0.33\linewidth]{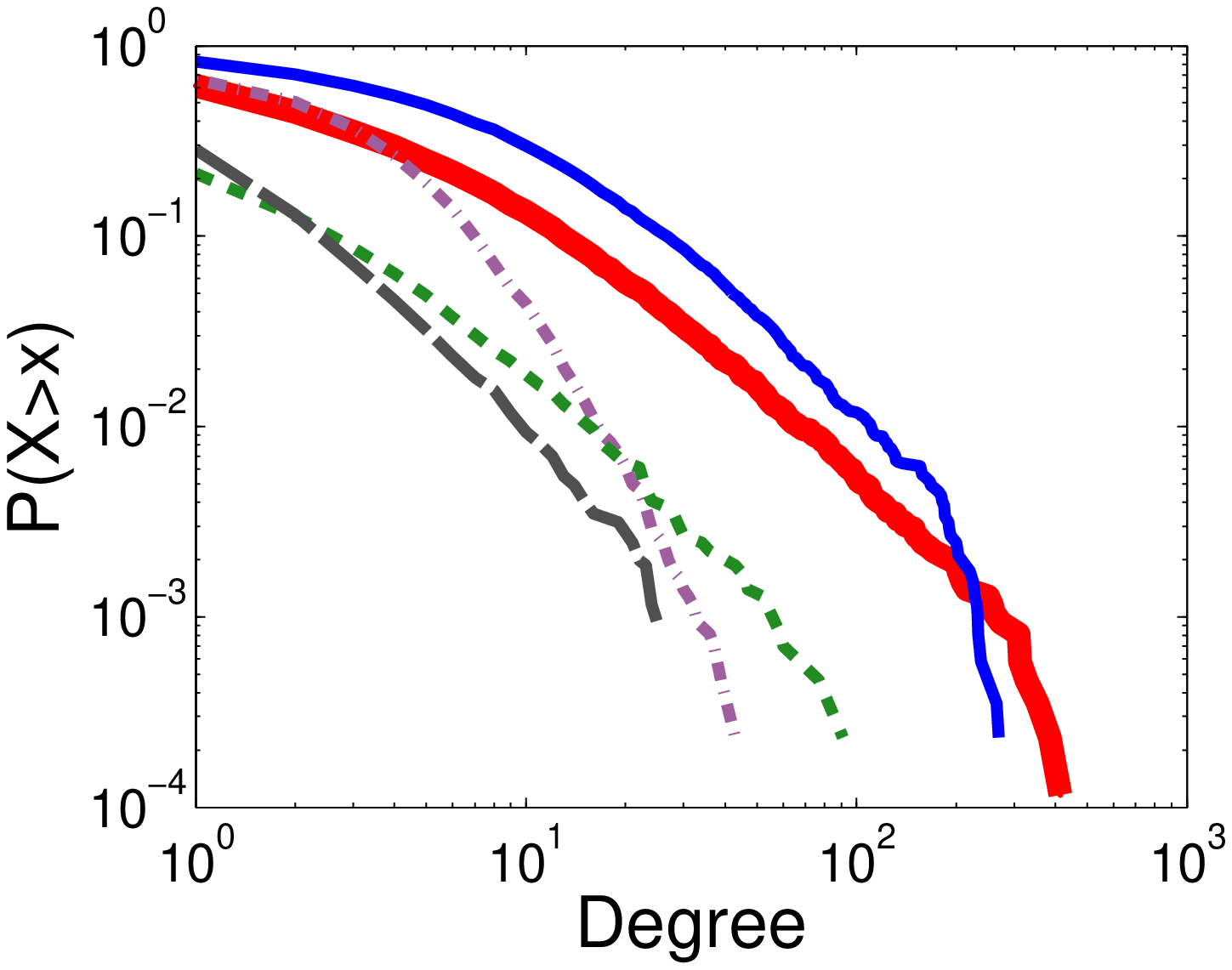}}
\hspace{-5.mm}
\subfigure{\label{fig:twcop pl dist}\includegraphics[width=0.33\linewidth]{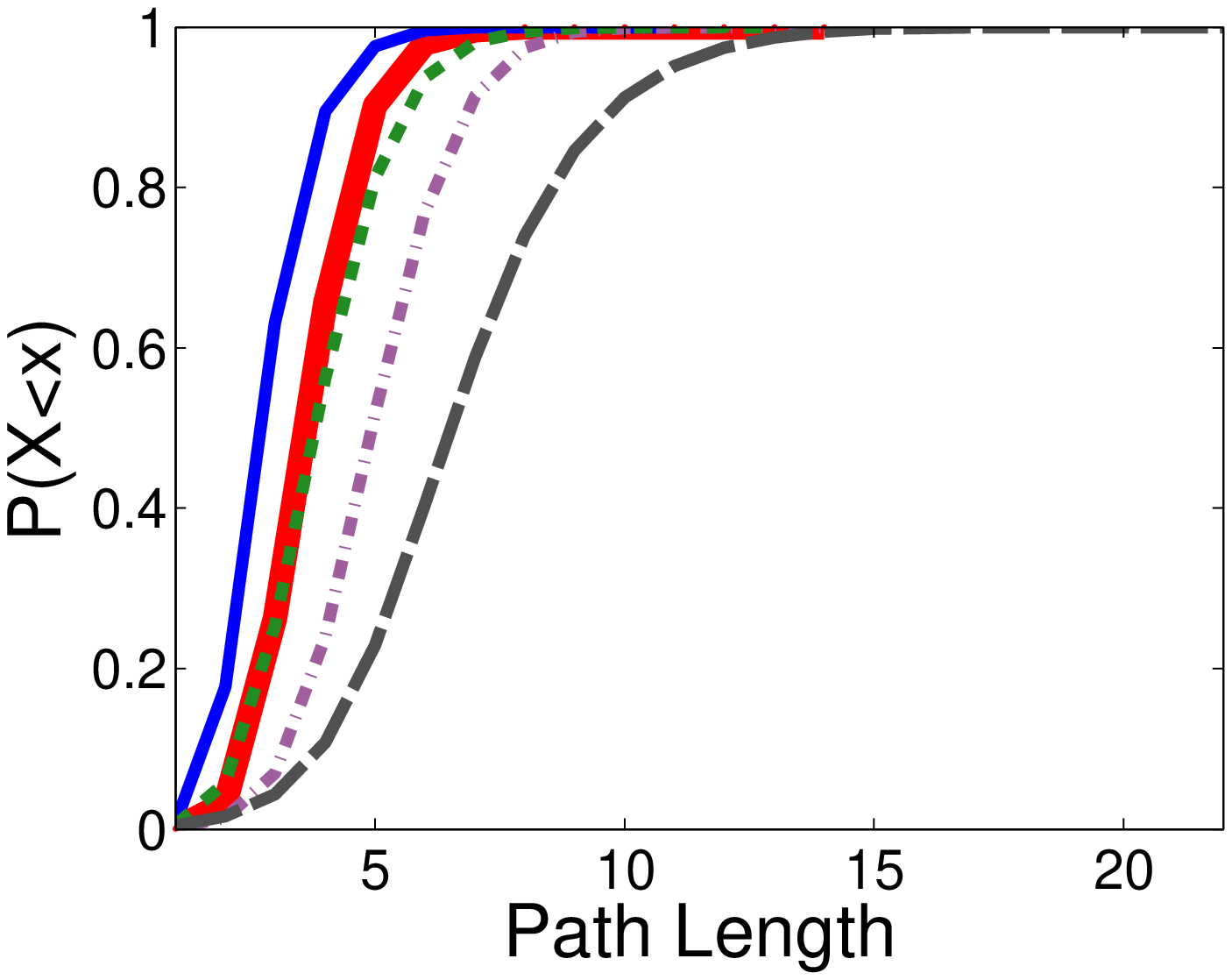}}
\hspace{-5.mm}
\subfigure{\label{fig:twcop cc dist}\includegraphics[width=0.33\linewidth]{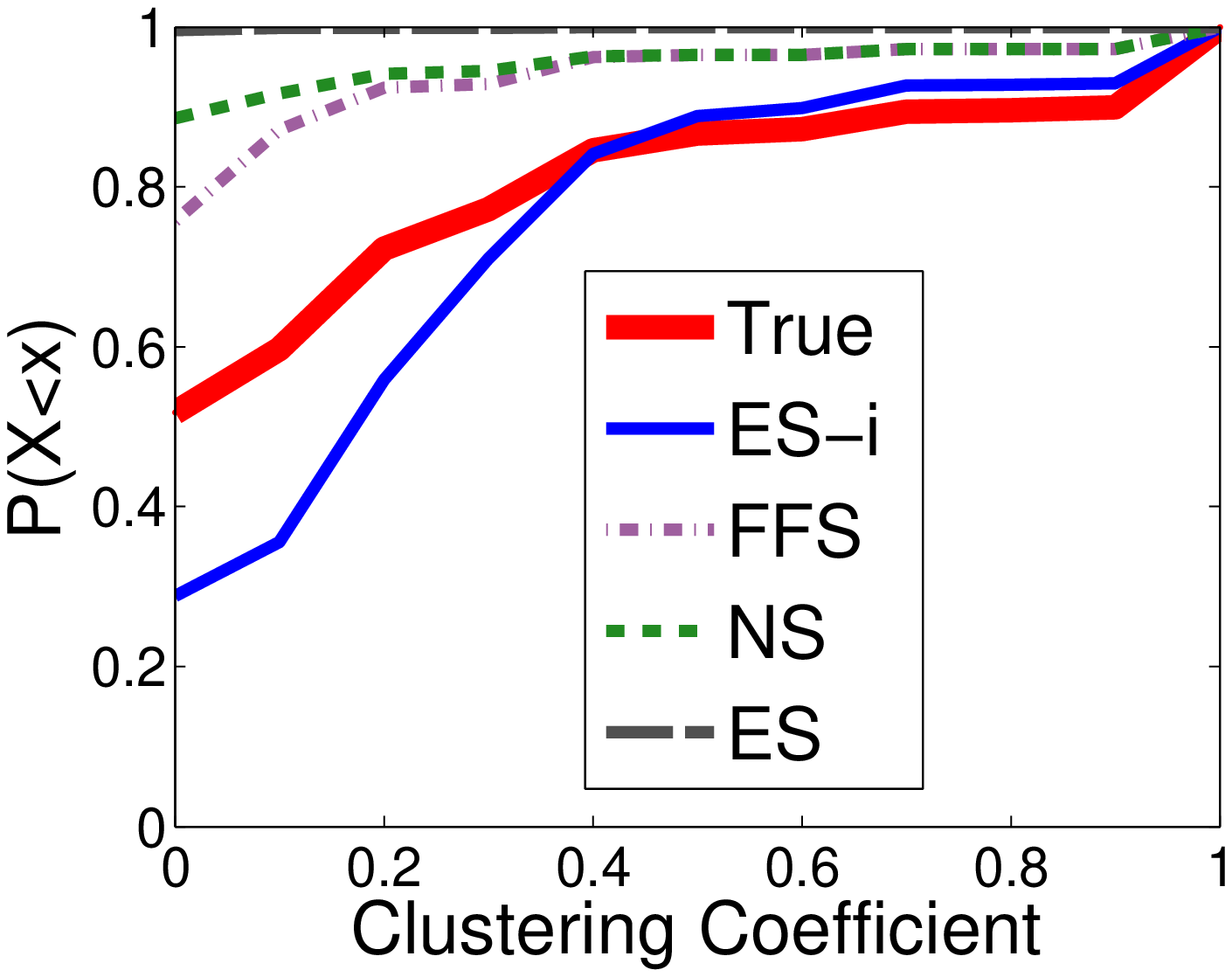}}
\vspace{-2mm}
\hspace{-2.mm}
\subfigure{\label{fig:twcop deg dist}\includegraphics[width=0.33\linewidth]{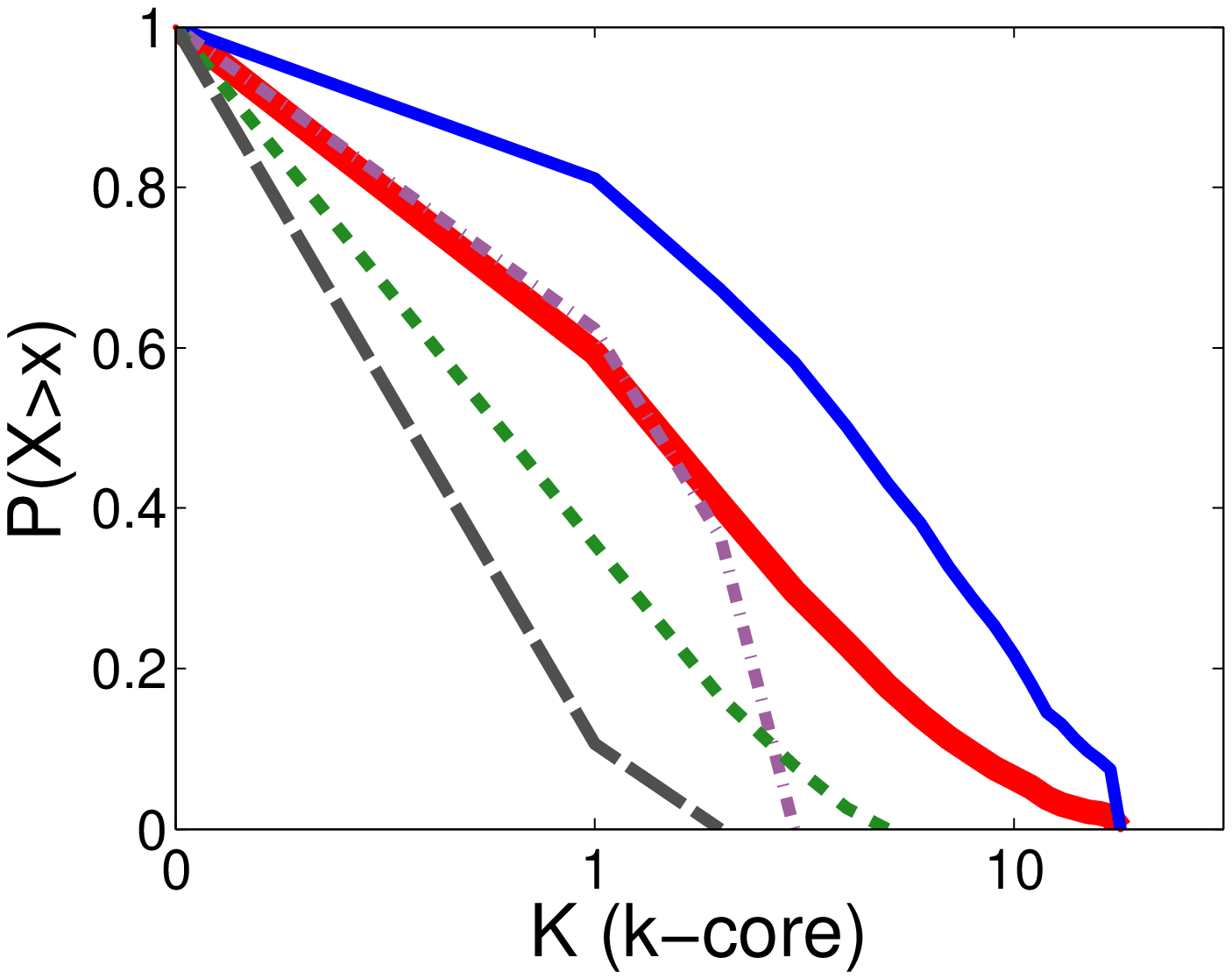}}
\hspace{-5.mm}
\subfigure{\label{fig:twcop pl dist}\includegraphics[width=0.33\linewidth]{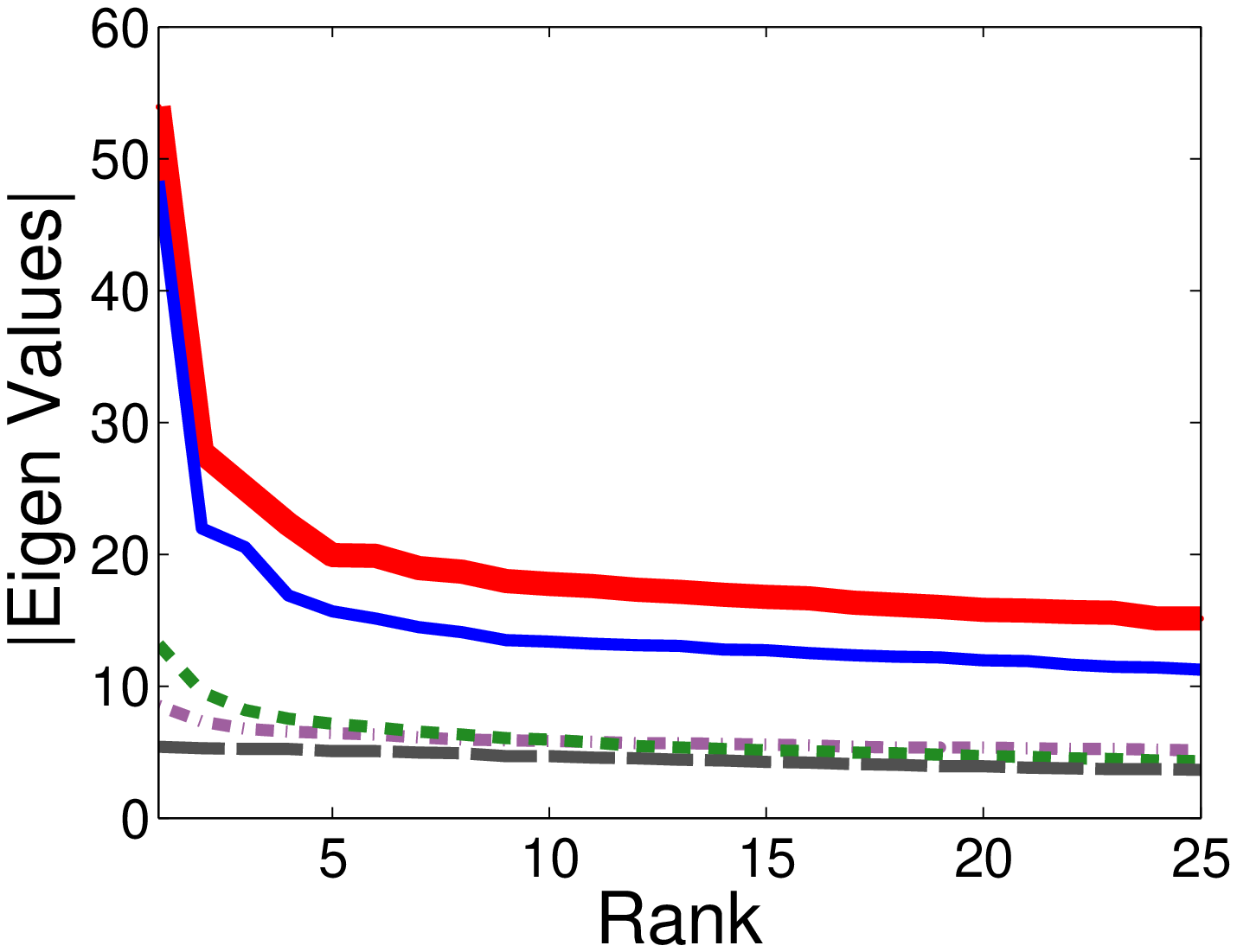}}
\hspace{-5.mm}
\subfigure{\label{fig:twcop cc dist}\includegraphics[width=0.33\linewidth]{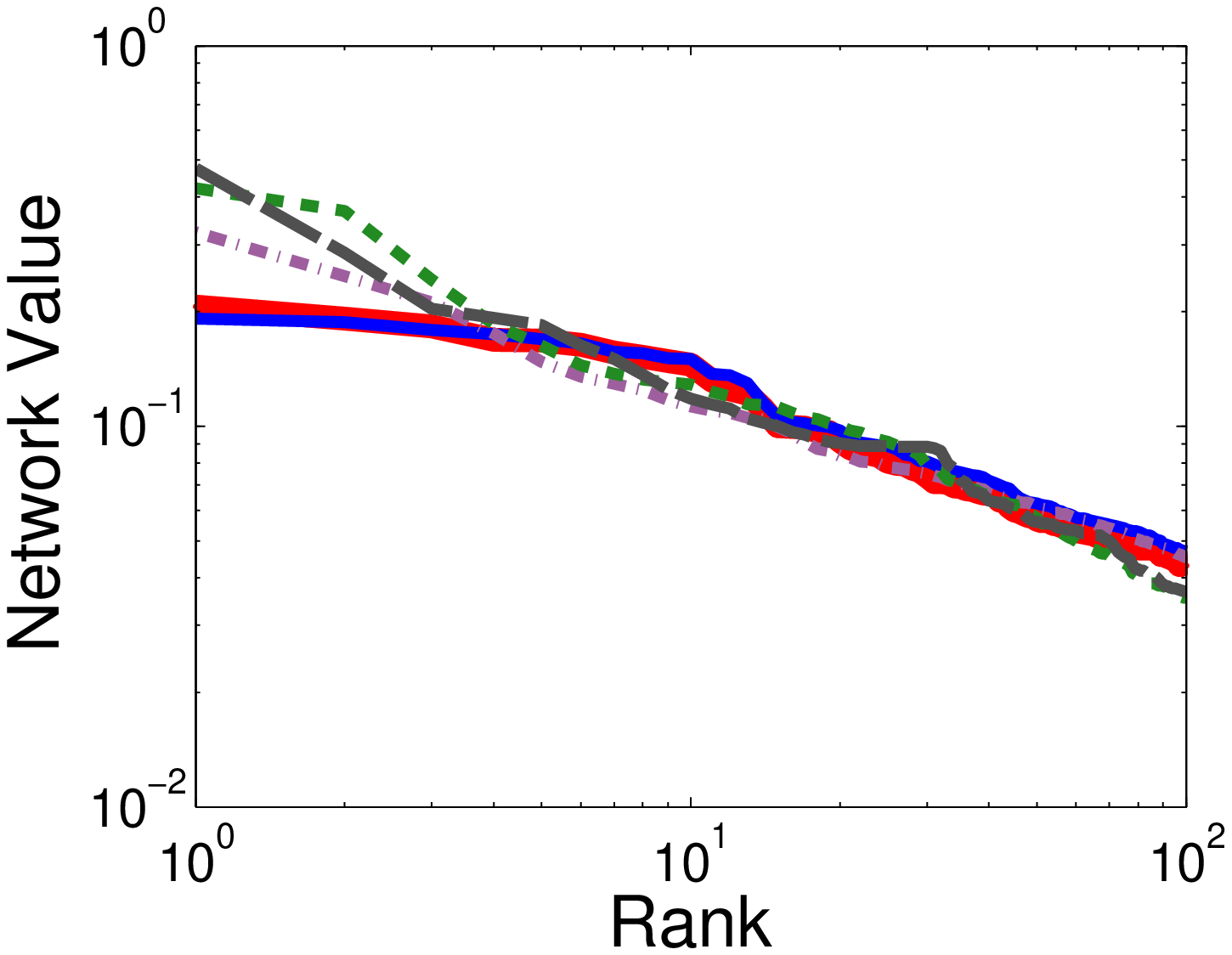}}
\caption{\textsc{Twitter} Graph}
\label{fig:dist_comp_twcop}
\vspace{-3.mm}
\end{figure}

\newpage

\begin{figure}[!h]
\centering
\subfigure{\label{fig:emailpu deg dist}\includegraphics[width=0.33\linewidth]{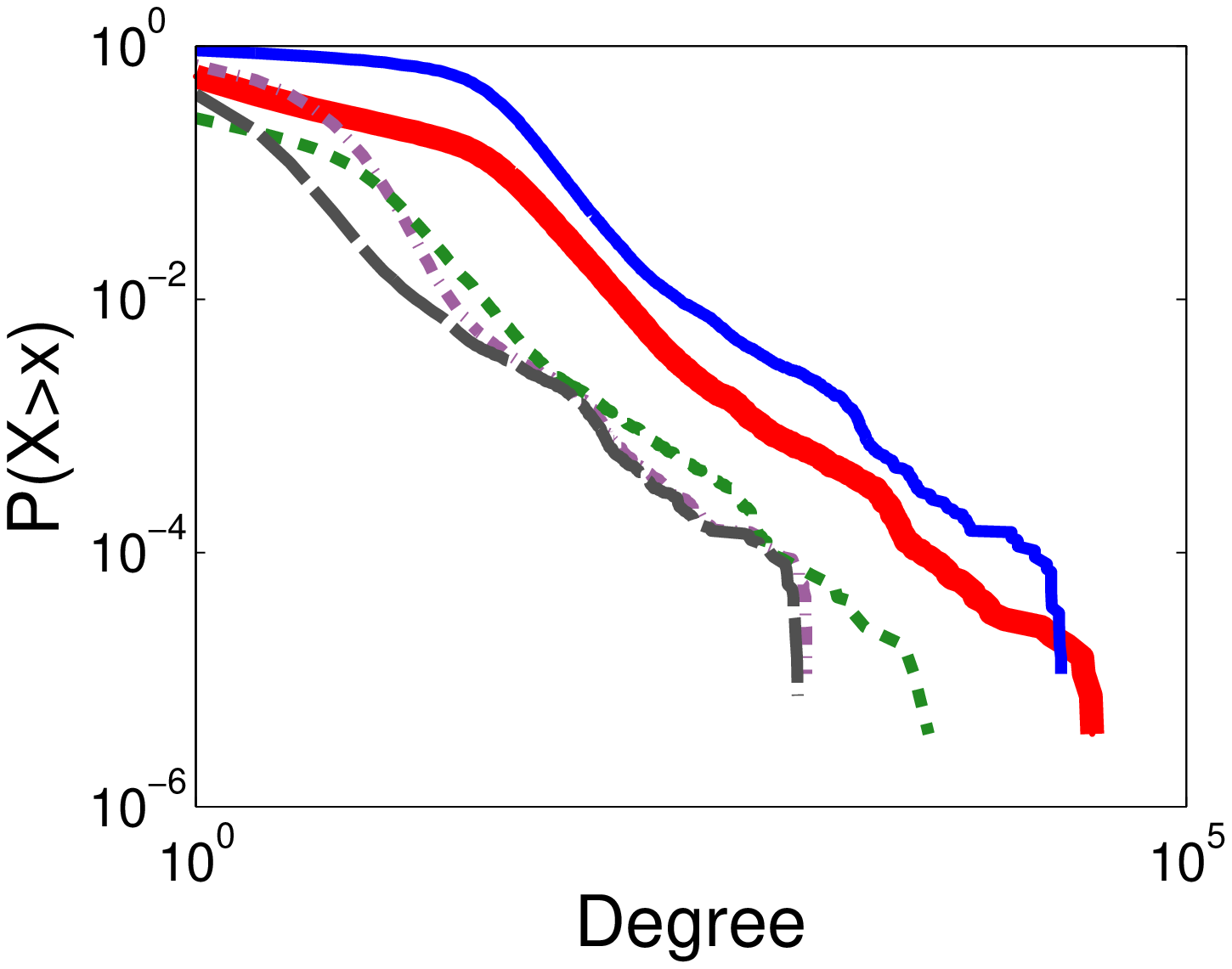}}
\hspace{-5.mm}
\subfigure{\label{fig:emailpu pl dist}\includegraphics[width=0.33\linewidth]{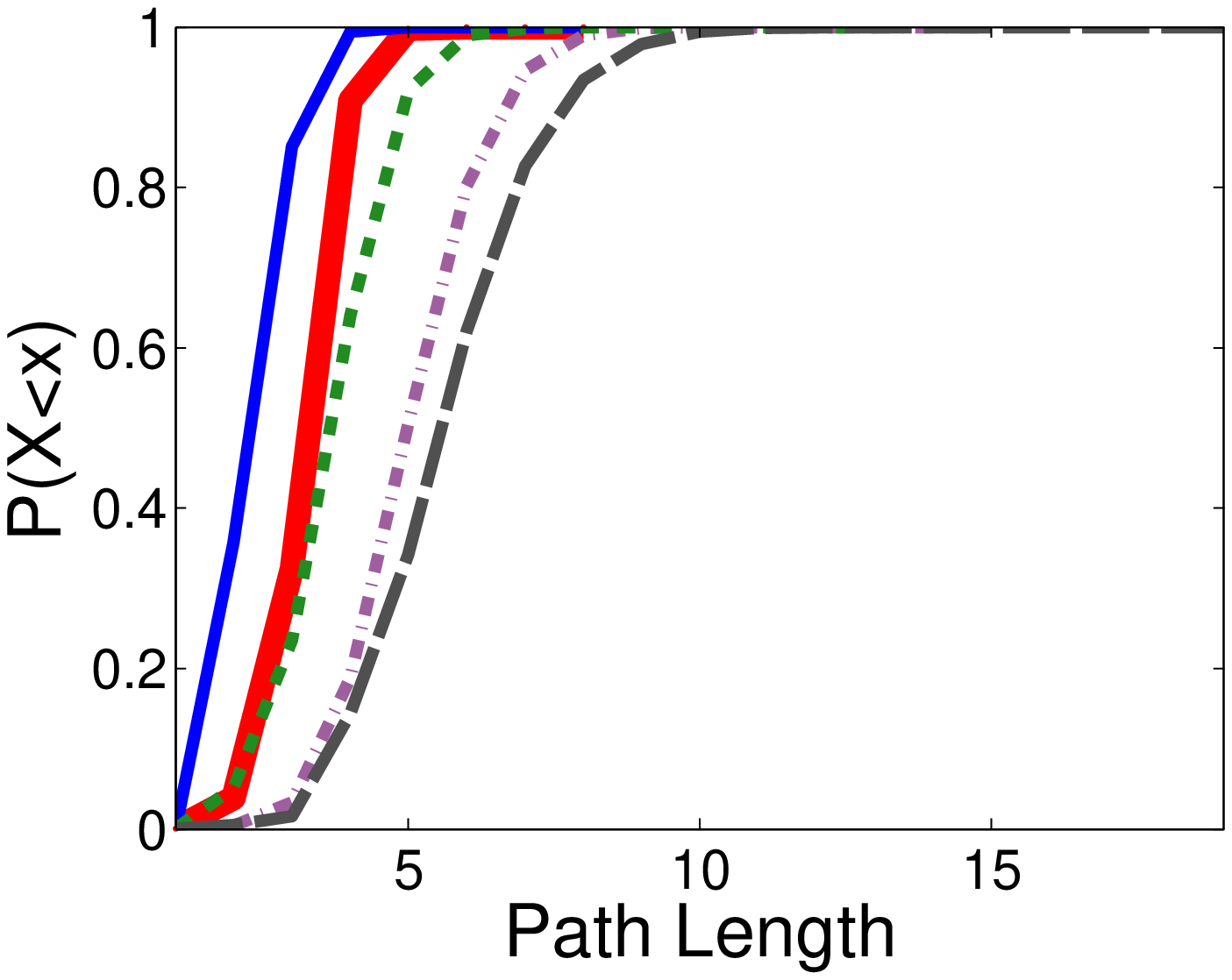}}
\hspace{-5.mm}
\subfigure{\label{fig:emailpu cc dist}\includegraphics[width=0.33\linewidth]{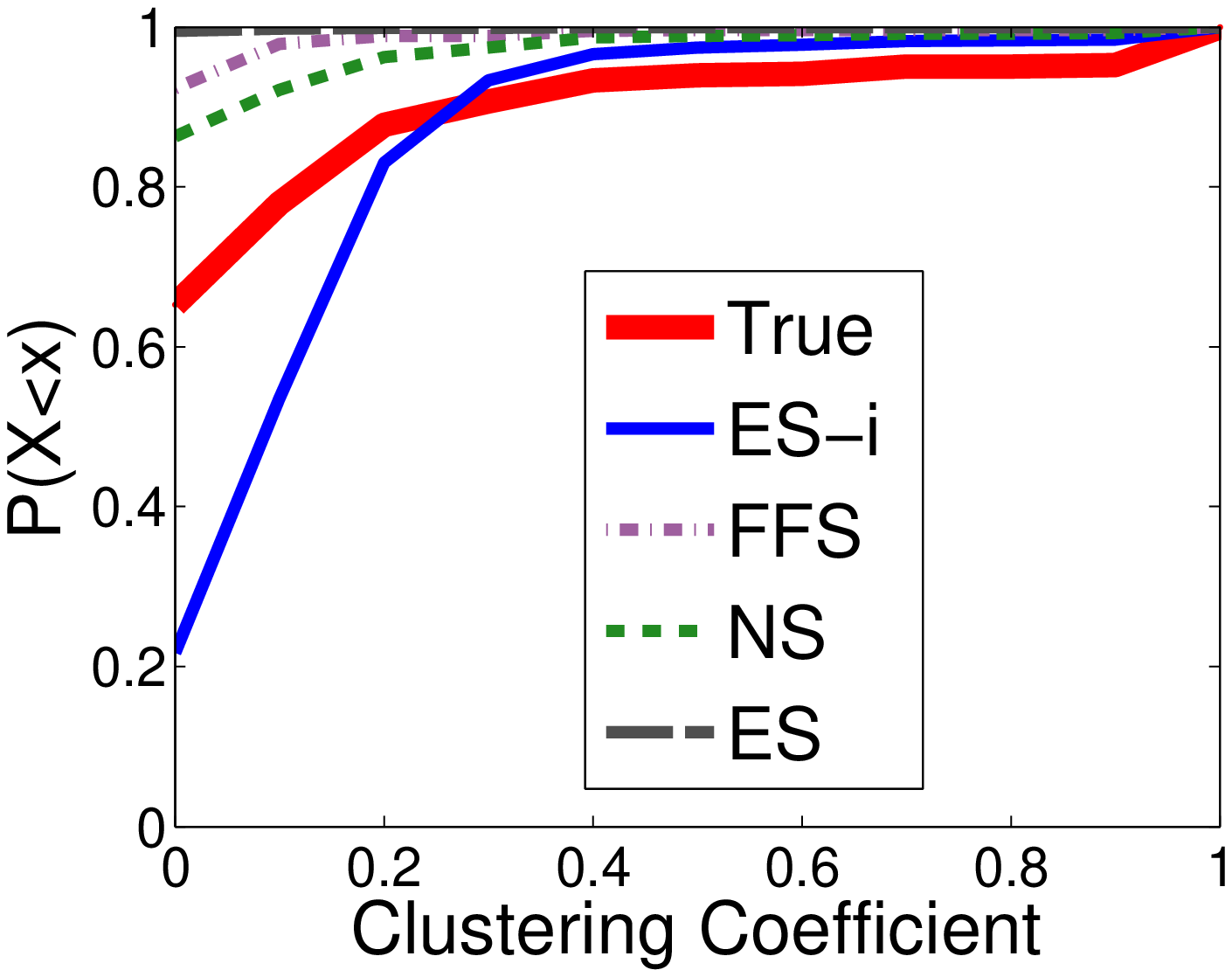}}
\vspace{-2mm}
\hspace{-2.mm}
\subfigure{\label{fig:emailpu deg dist}\includegraphics[width=0.33\linewidth]{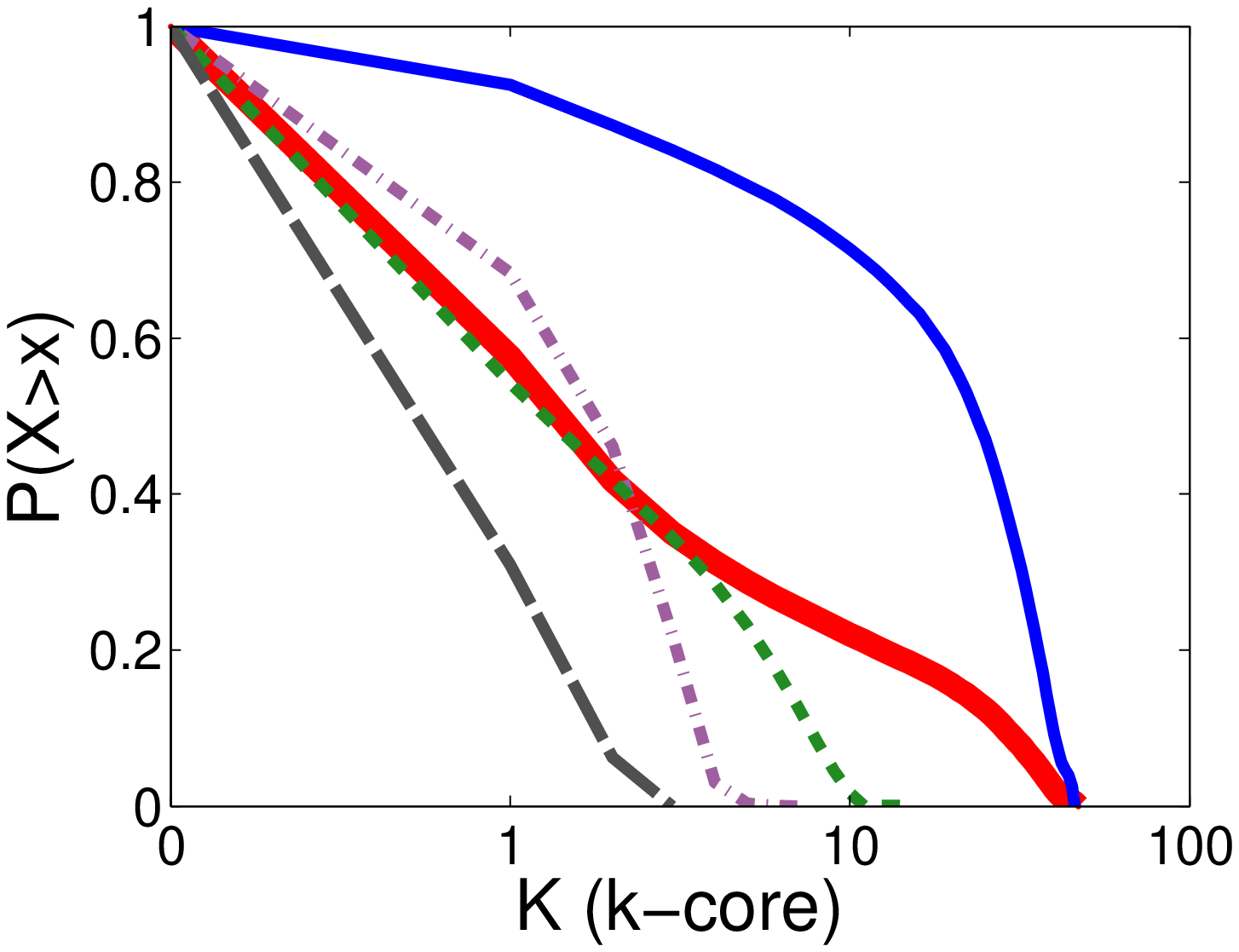}}
\hspace{-5.mm}
\subfigure{\label{fig:emailpu pl dist}\includegraphics[width=0.33\linewidth]{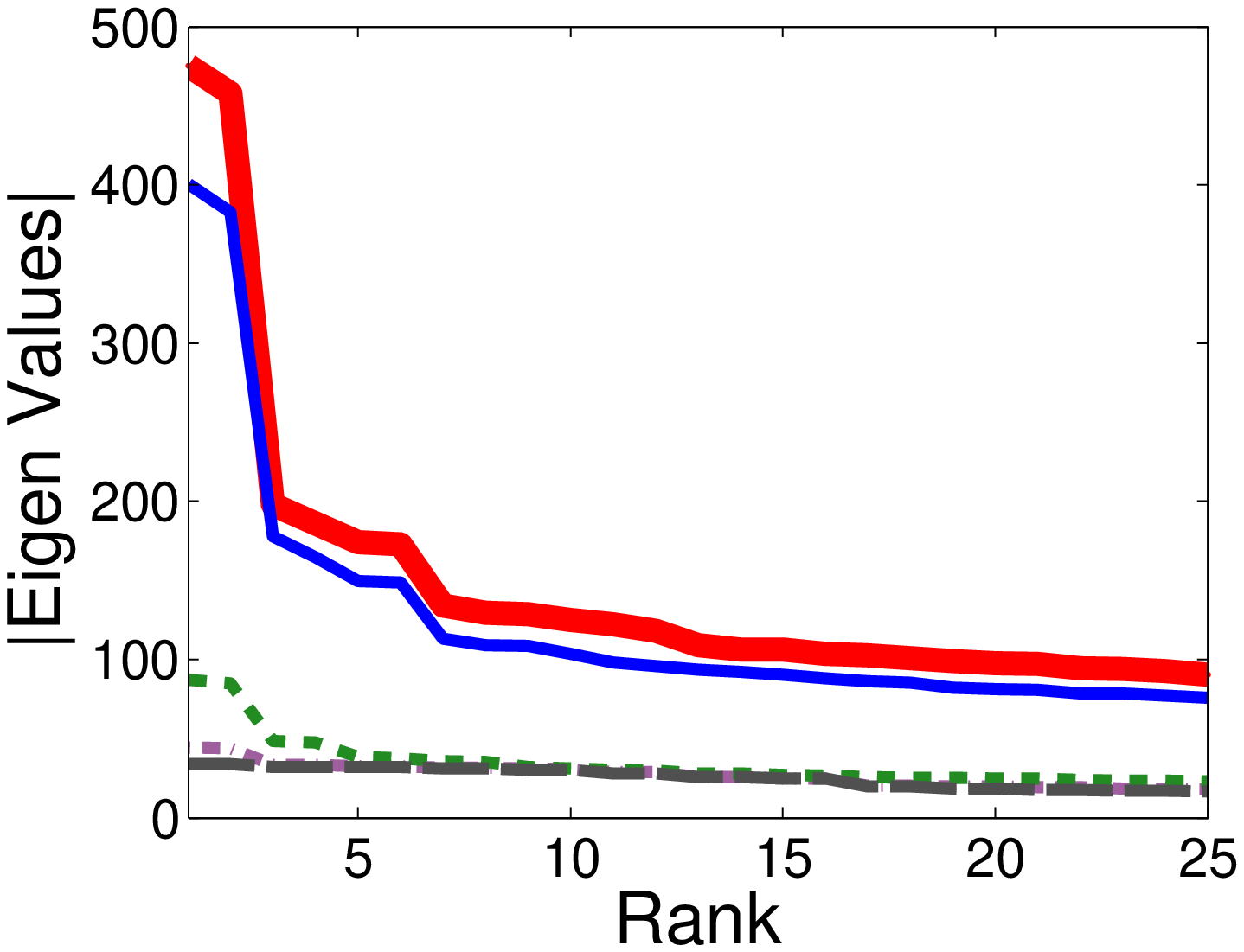}}
\hspace{-5.mm}
\subfigure{\label{fig:emailpu cc dist}\includegraphics[width=0.33\linewidth]{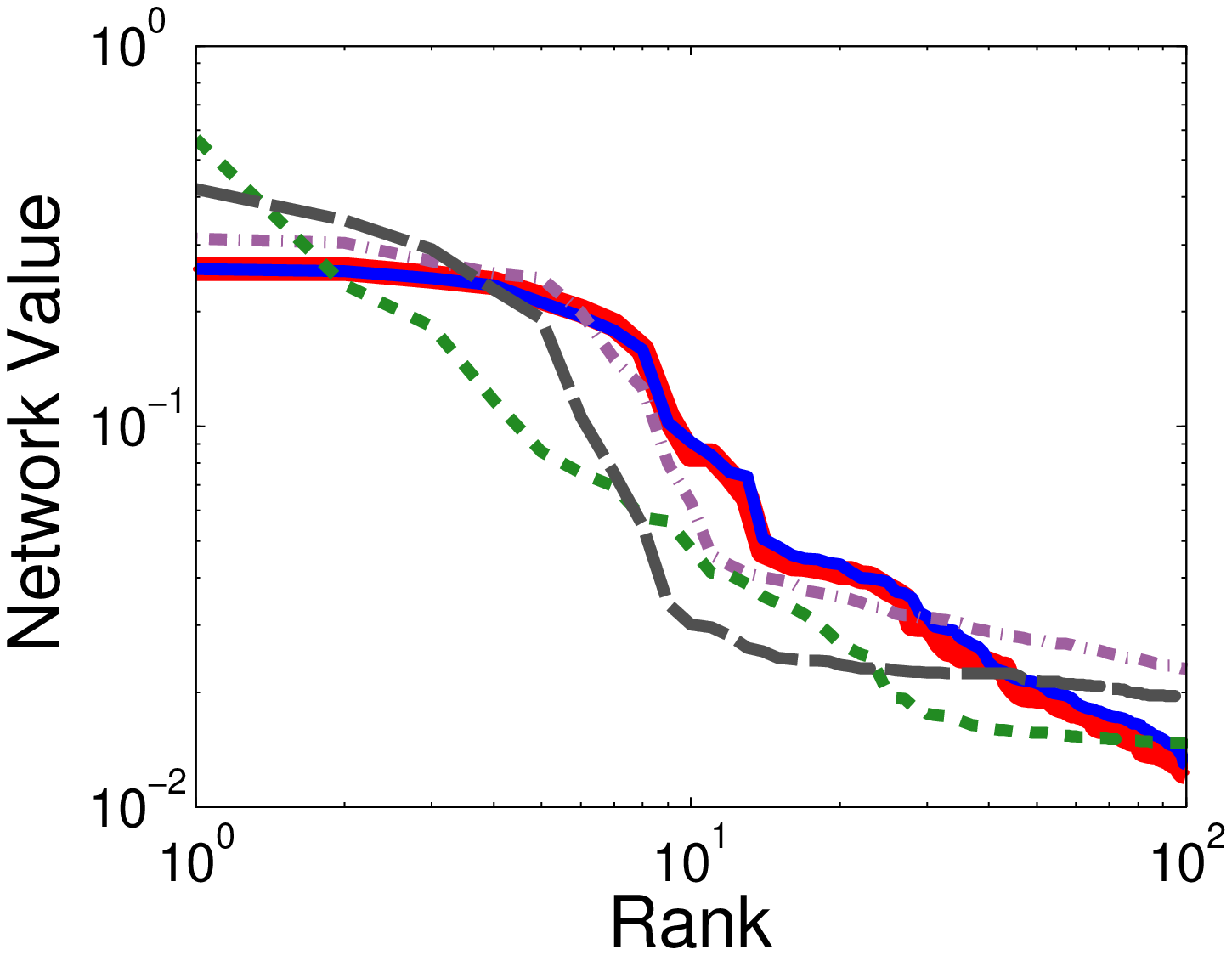}}
\caption{\textsc{Email-Univ} Graph}
\label{fig:dist_comp_email}
\vspace{-3.mm}
\end{figure}

\begin{figure}[!h]
\centering
\subfigure{\label{fig:socjor deg dist}\includegraphics[width=0.33\linewidth]{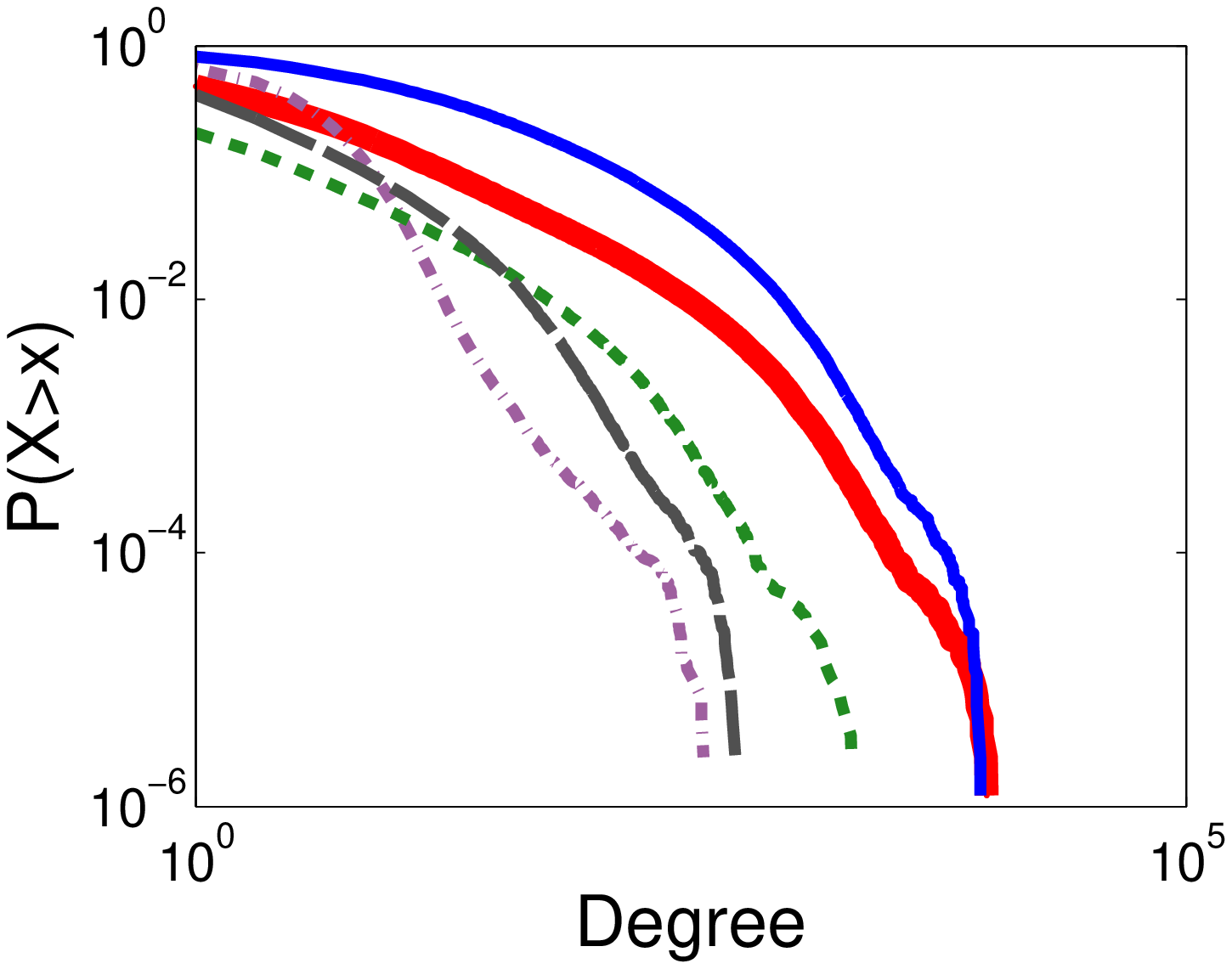}}
\hspace{-5.mm}
\subfigure{\label{fig:socjor pl dist}\includegraphics[width=0.33\linewidth]{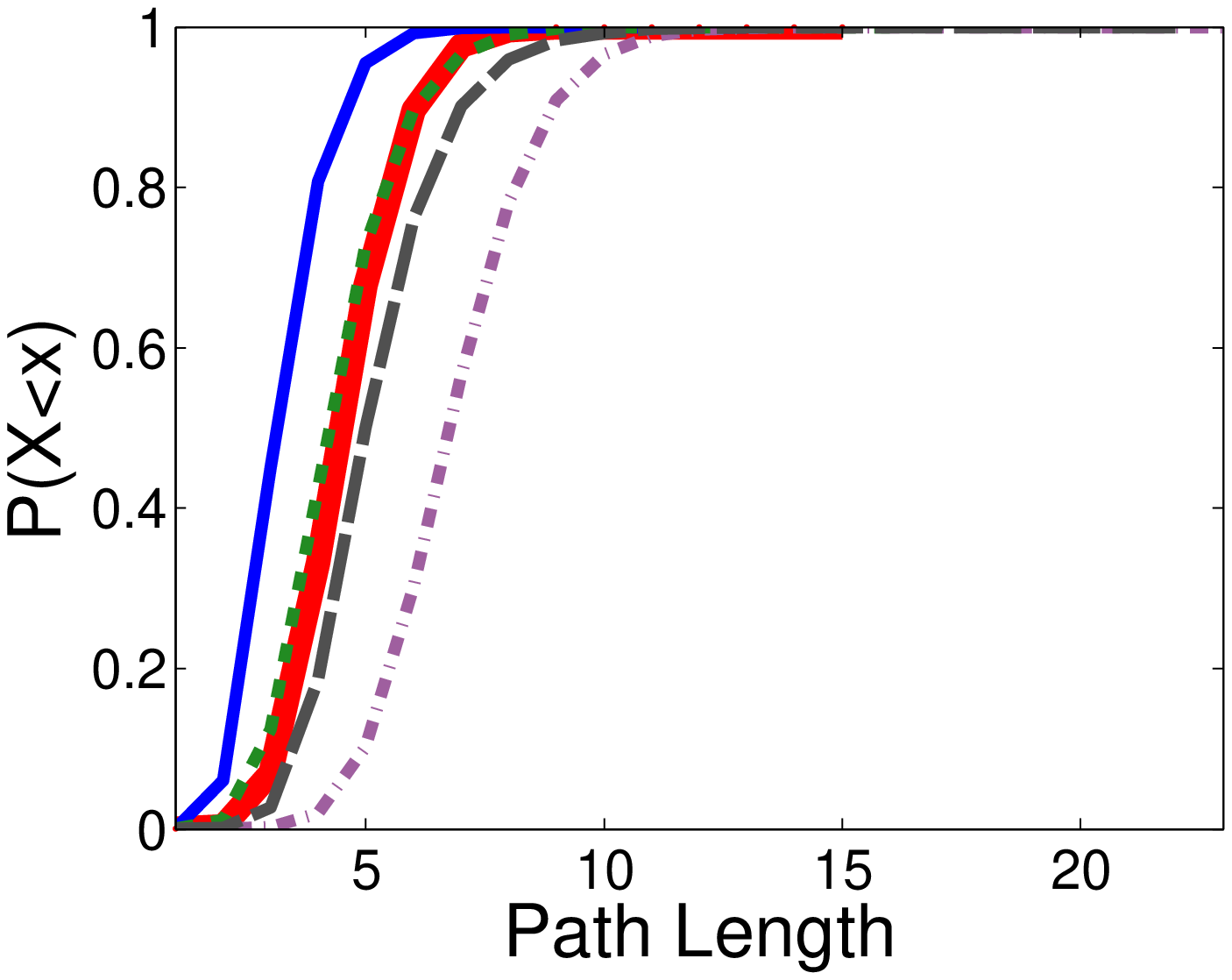}}
\hspace{-5.mm}
\subfigure{\label{fig:socjor cc dist}\includegraphics[width=0.33\linewidth]{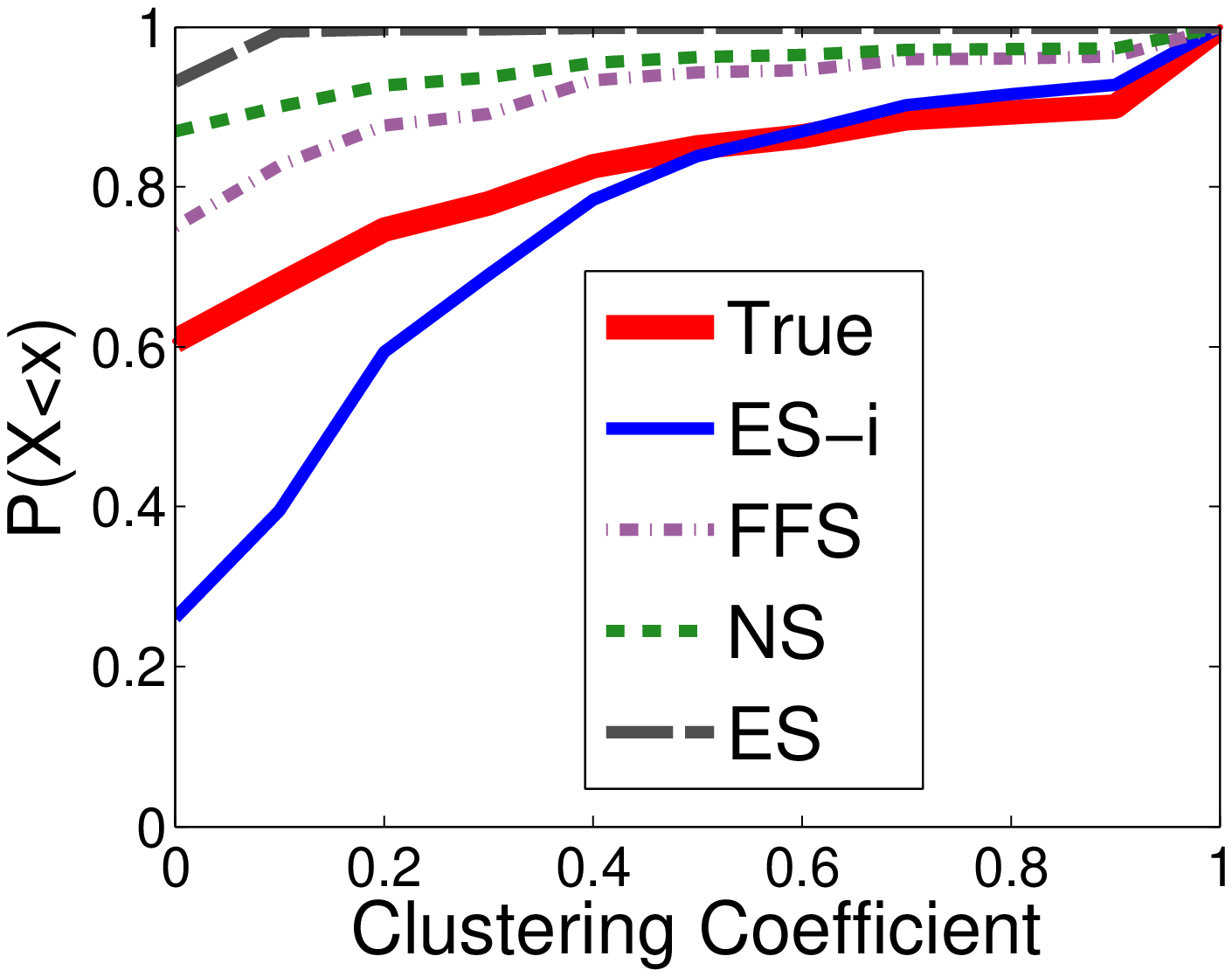}}
\vspace{-2mm}
\hspace{-2.mm}
\subfigure{\label{fig:socjor deg dist}\includegraphics[width=0.33\linewidth]{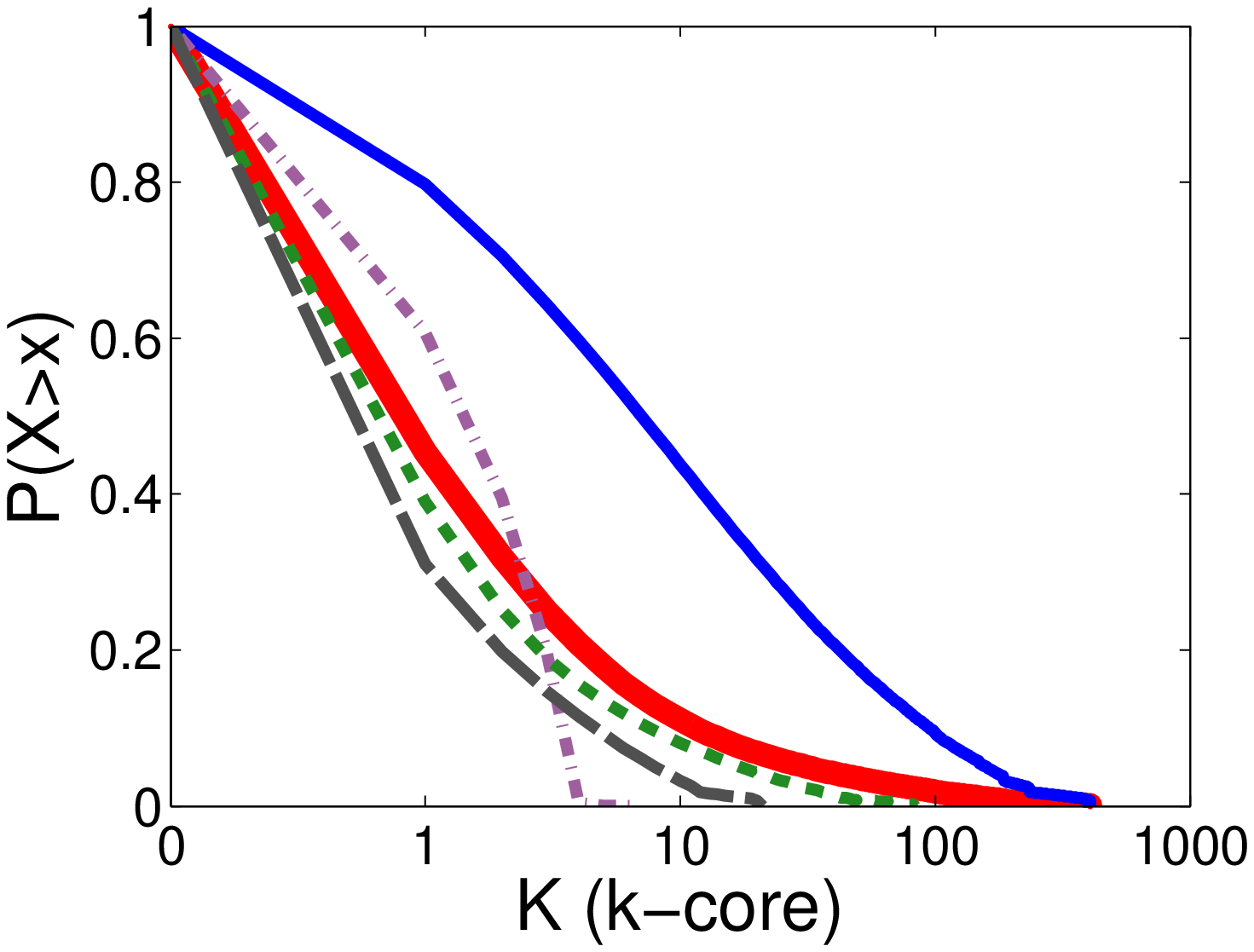}}
\hspace{-5.mm}
\subfigure{\label{fig:socjor pl dist}\includegraphics[width=0.33\linewidth]{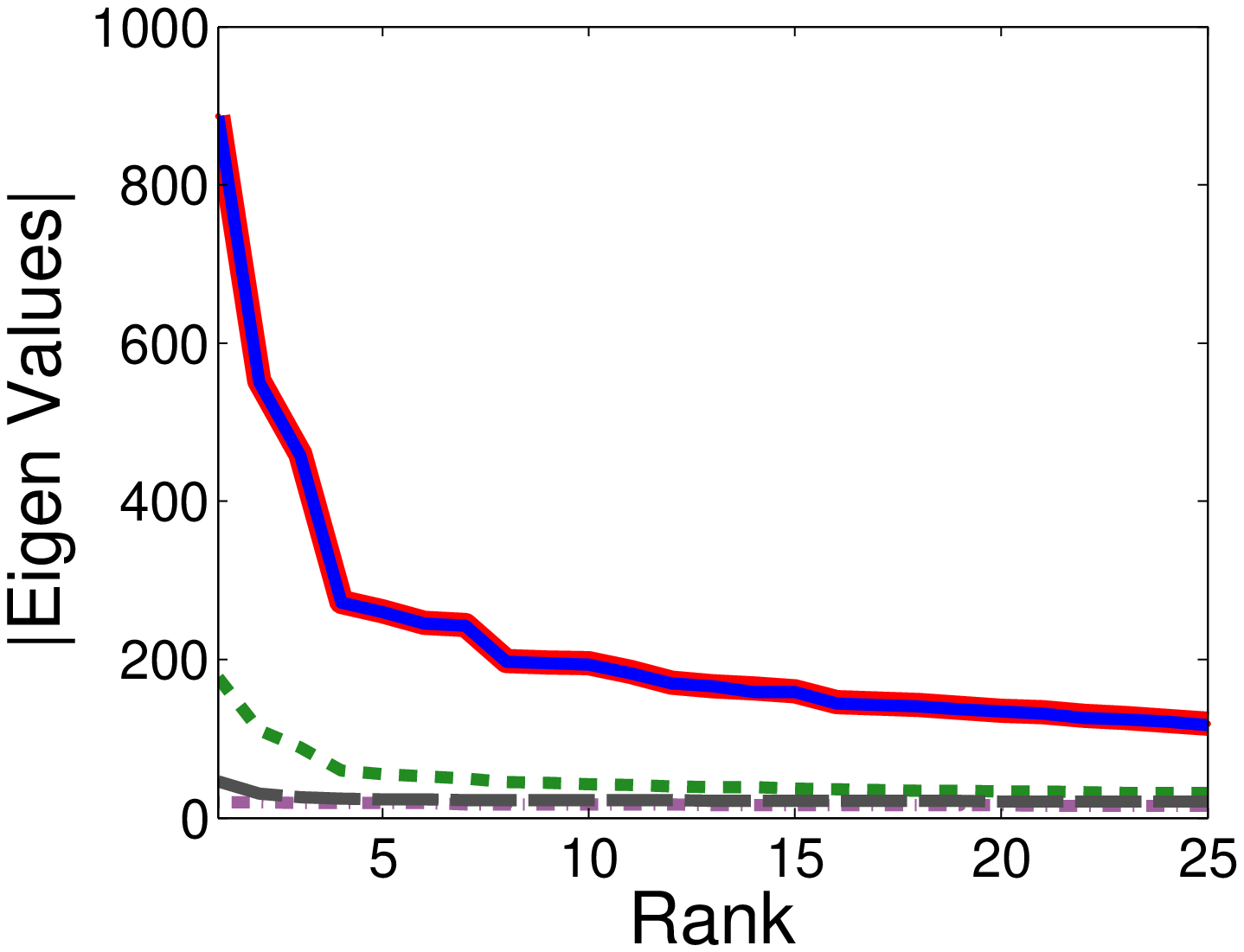}}
\hspace{-5.mm}
\subfigure{\label{fig:socjor cc dist}\includegraphics[width=0.33\linewidth]{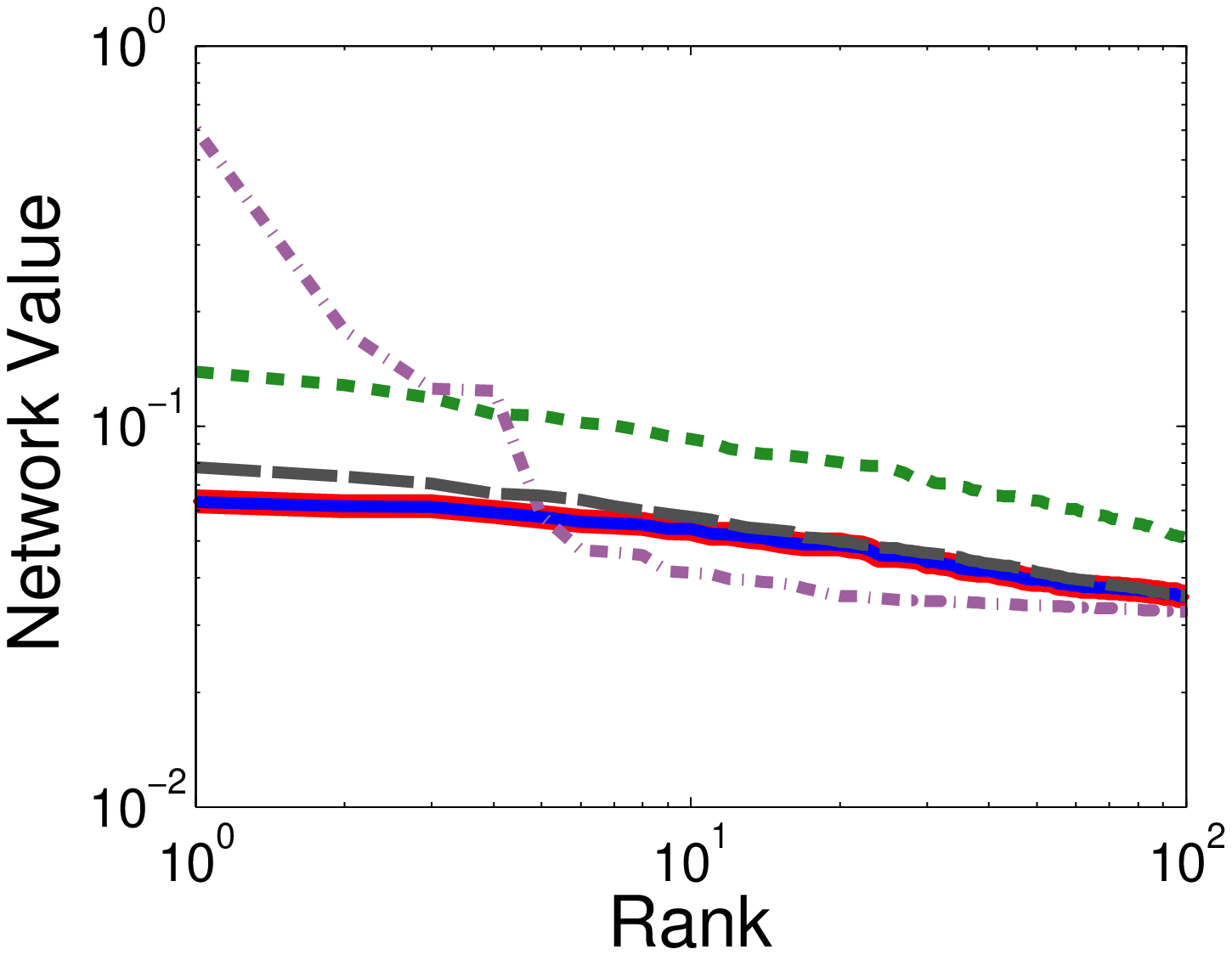}}
\caption{\textsc{Flickr} Graph}
\label{fig:dist_comp_flickr}
\vspace{-3.mm}
\end{figure}

\newpage

\begin{figure}[!h]
\centering
\subfigure{\label{fig:socjor deg dist}\includegraphics[width=0.33\linewidth]{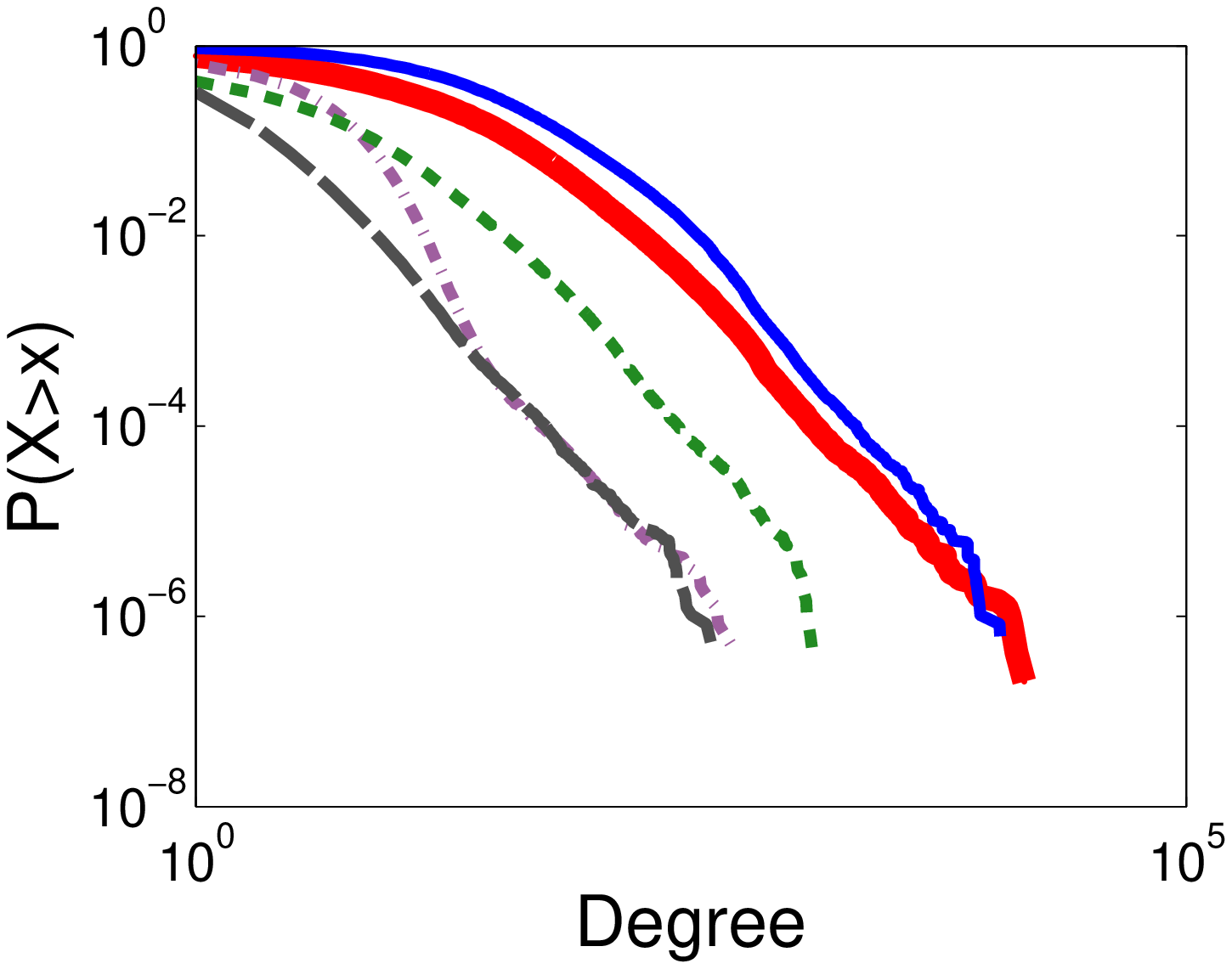}}
\hspace{-5.mm}
\subfigure{\label{fig:socjor pl dist}\includegraphics[width=0.33\linewidth]{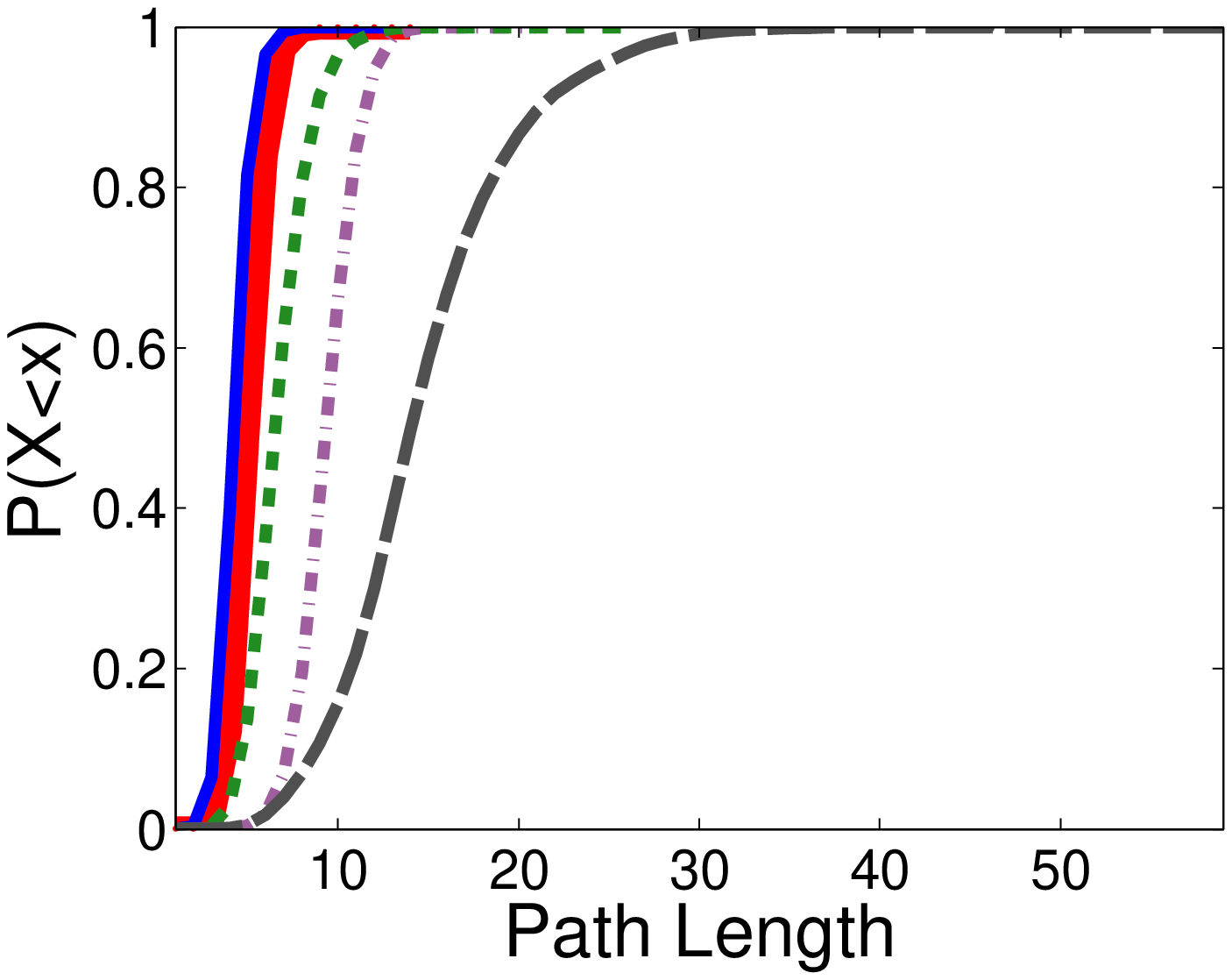}}
\hspace{-5.mm}
\subfigure{\label{fig:socjor cc dist}\includegraphics[width=0.33\linewidth]{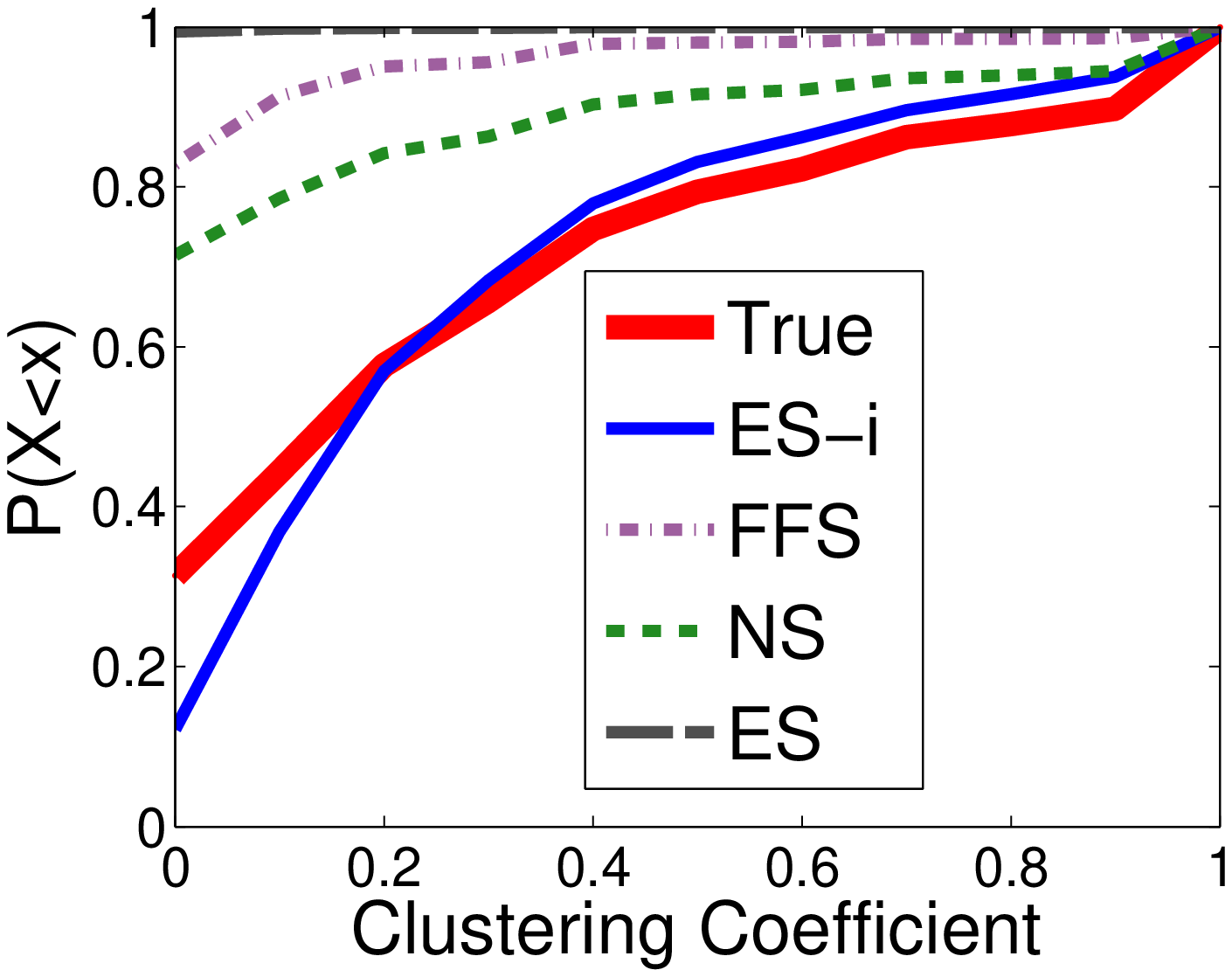}}
\vspace{-2mm}
\hspace{-2.mm}
\subfigure{\label{fig:socjor deg dist}\includegraphics[width=0.33\linewidth]{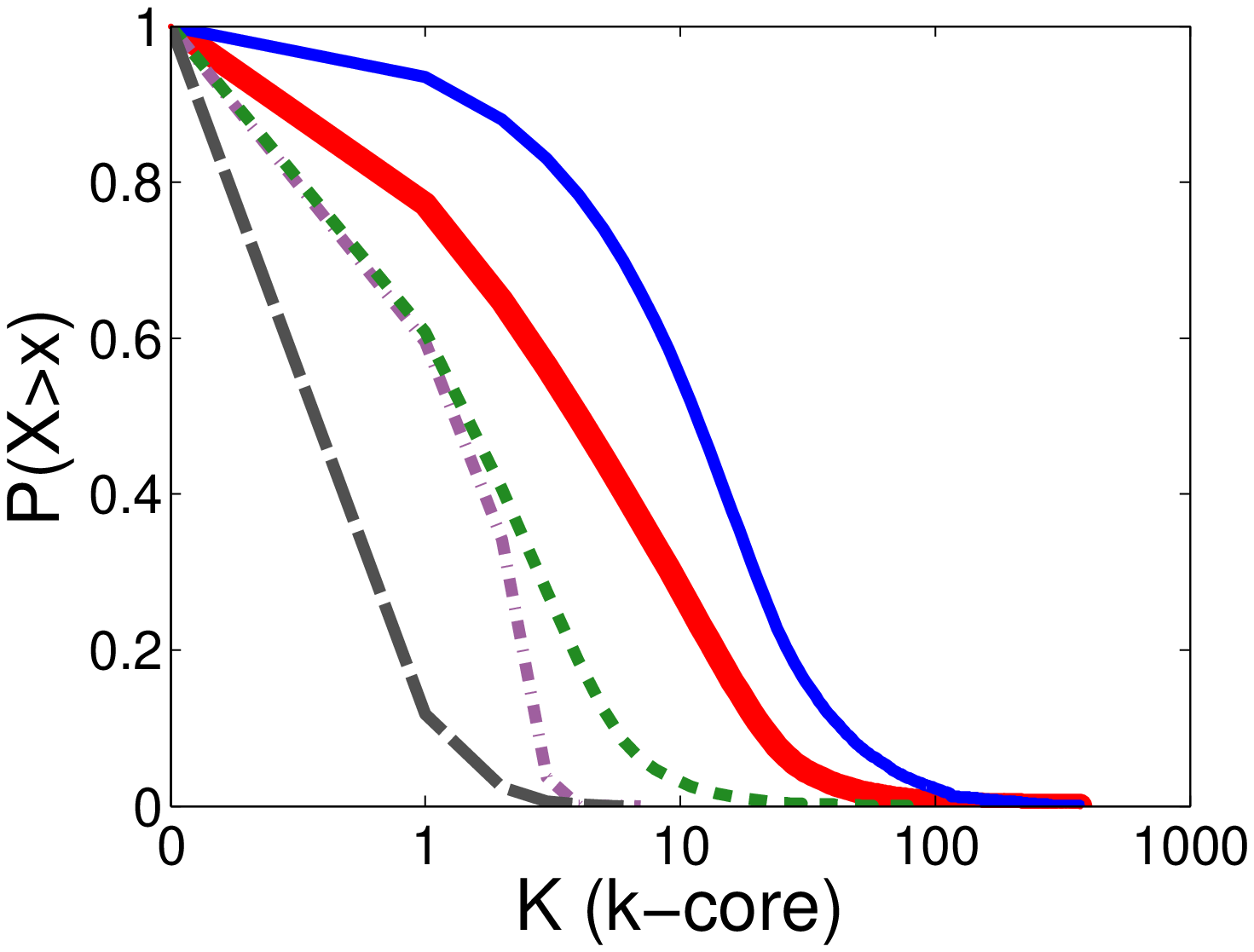}}
\hspace{-5.mm}
\subfigure{\label{fig:socjor pl dist}\includegraphics[width=0.33\linewidth]{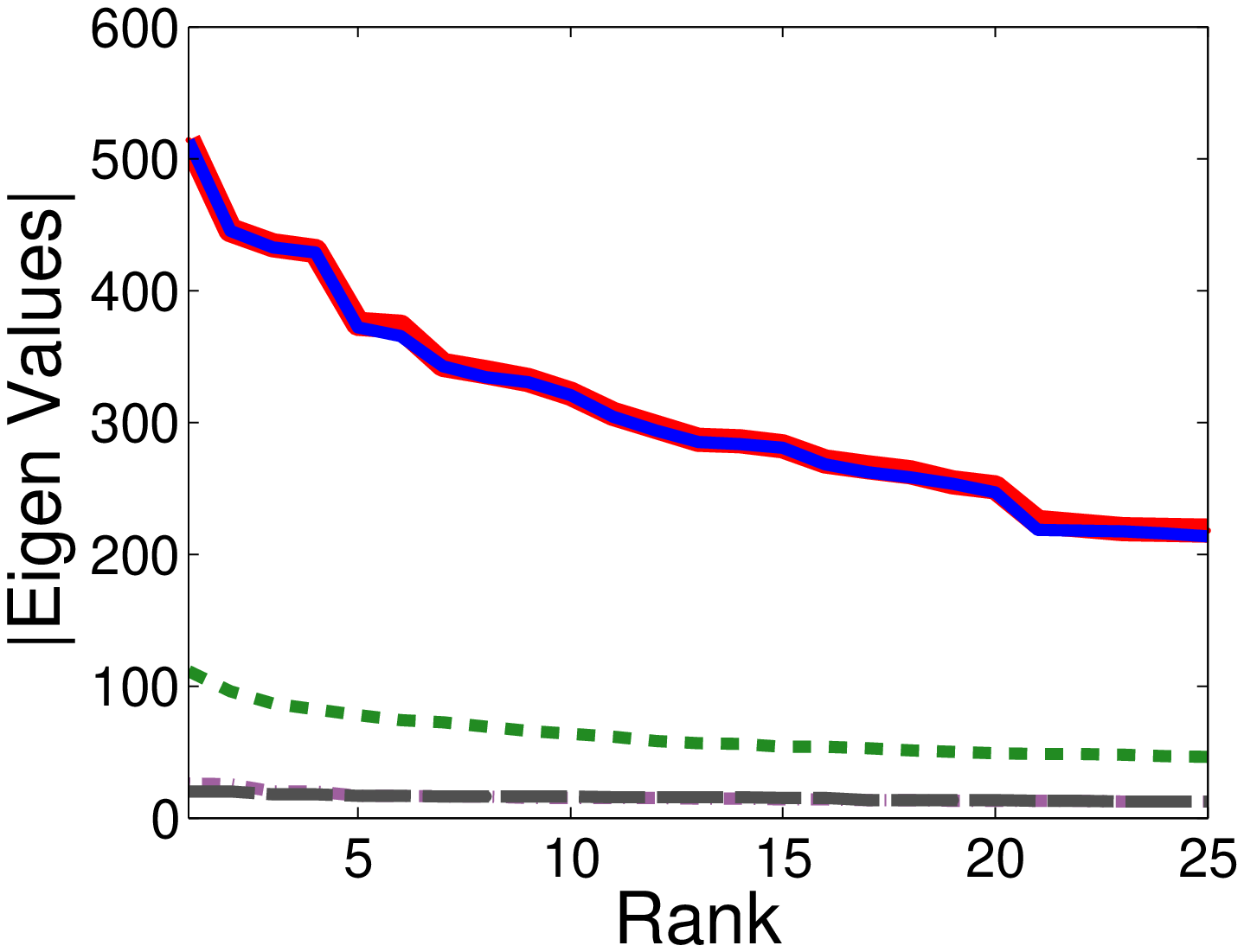}}
\hspace{-5.mm}
\subfigure{\label{fig:socjor cc dist}\includegraphics[width=0.33\linewidth]{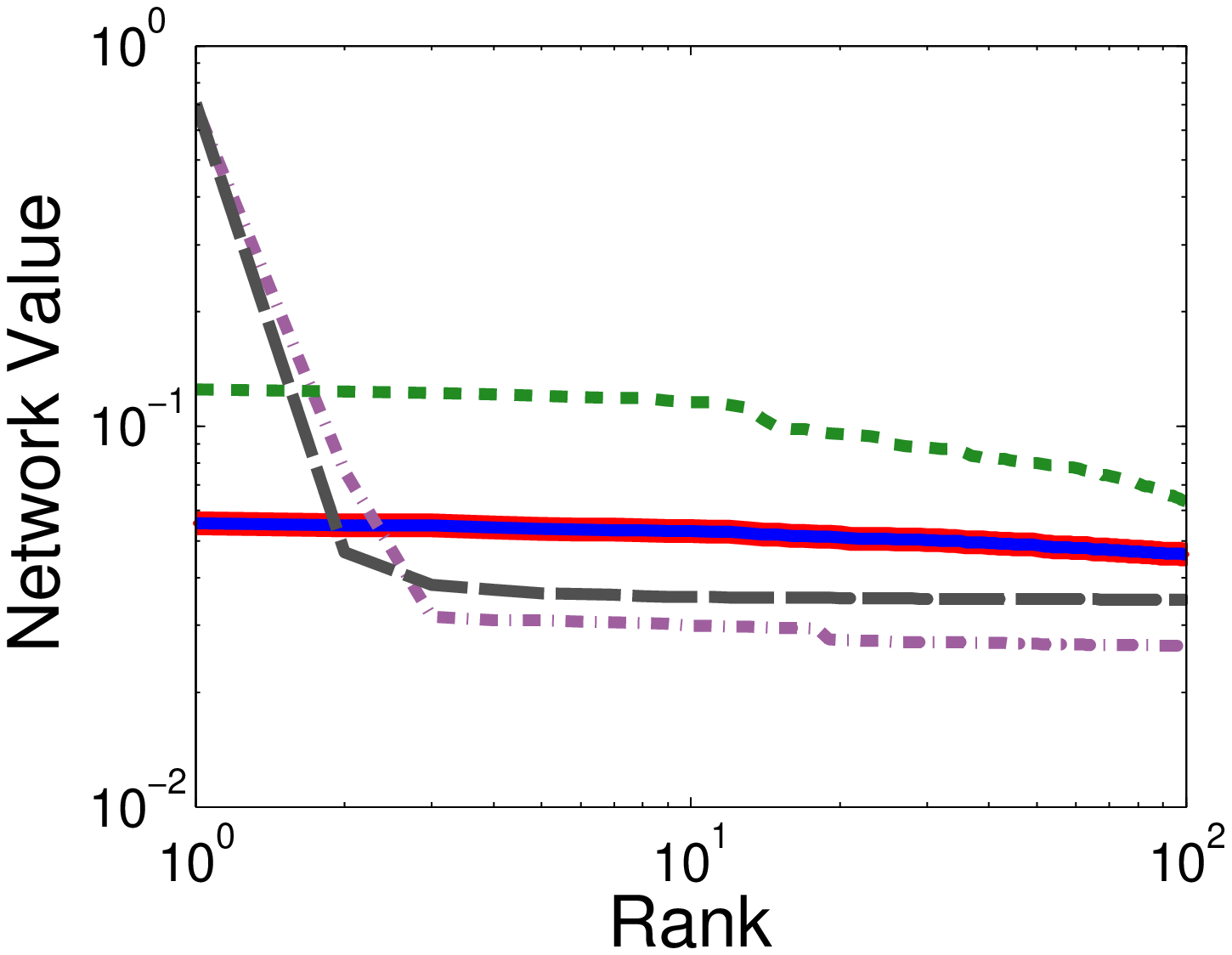}}
\caption{\textsc{LiveJournal} Graph}
\label{fig:dist_comp_socjor}
\vspace{-3.mm}
\end{figure}

\newpage
\subsection{Distributions for Streaming Graphs (at 20\% sample size)}
\begin{figure}[!h]
\centering
\subfigure{\label{fig:fbor deg dist}\includegraphics[width=0.33\linewidth]{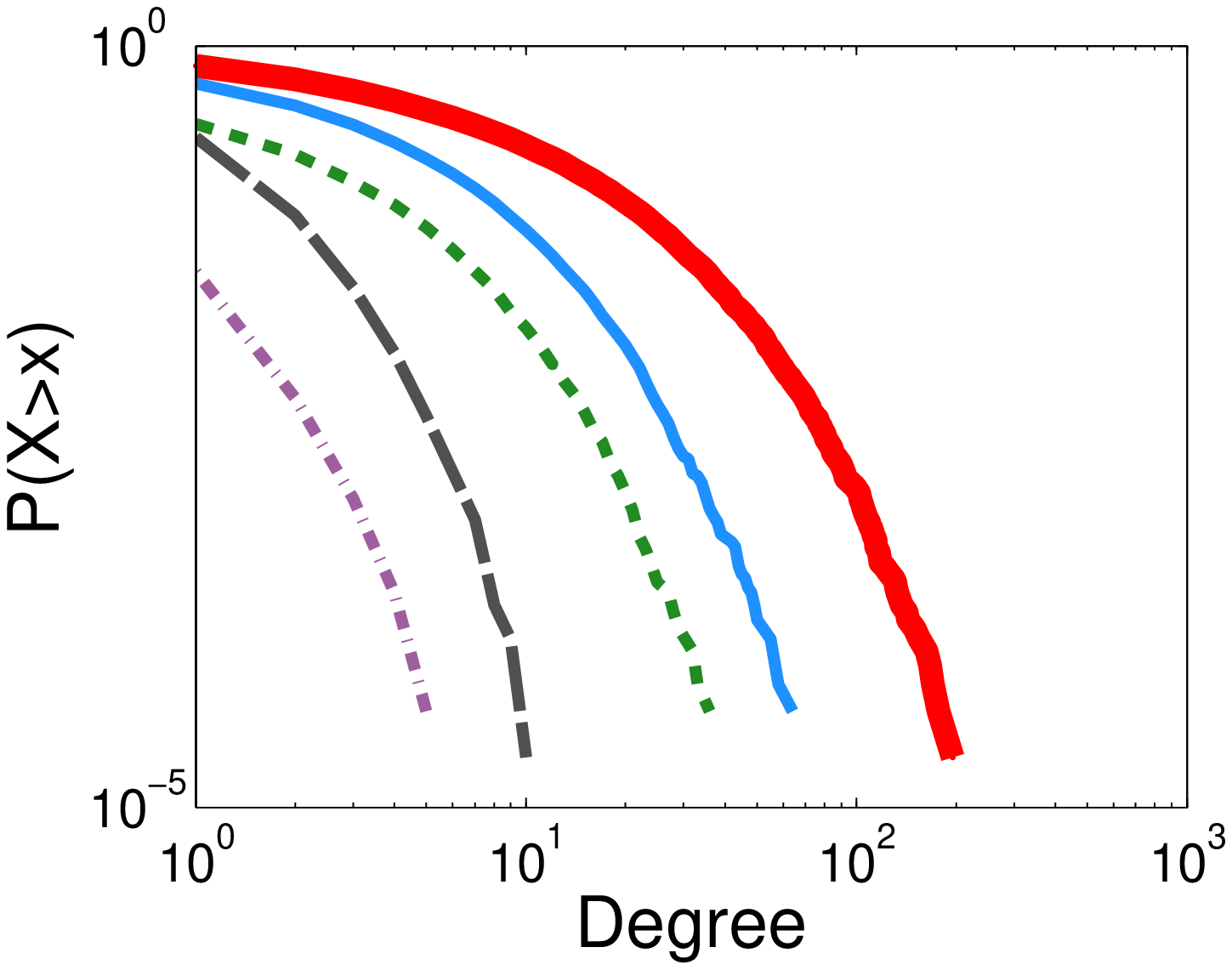}}
\hspace{-5.mm}
\subfigure{\label{fig:fbor pl dist}\includegraphics[width=0.33\linewidth]{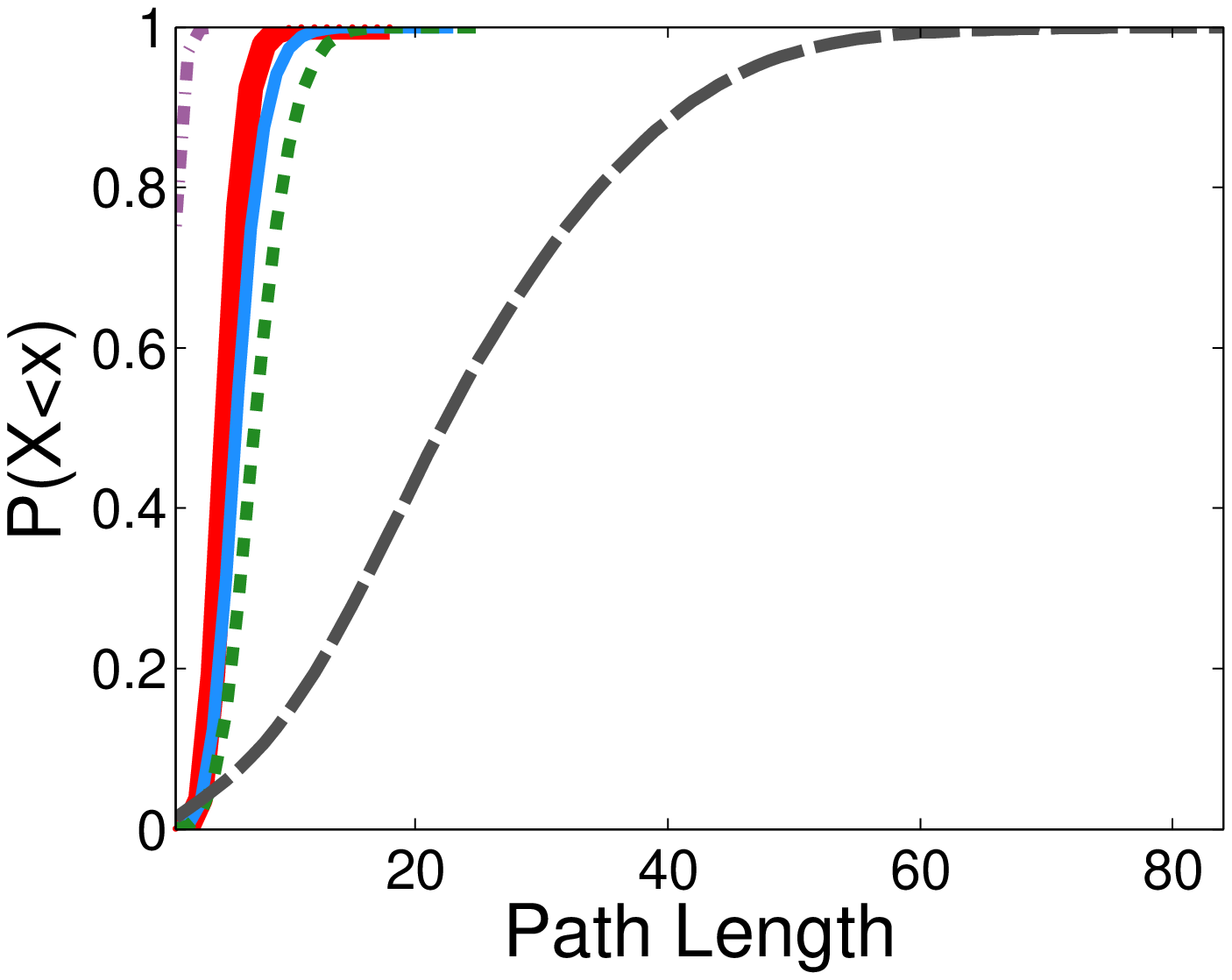}}
\hspace{-5.mm}
\subfigure{\label{fig:fbor cc dist}\includegraphics[width=0.33\linewidth]{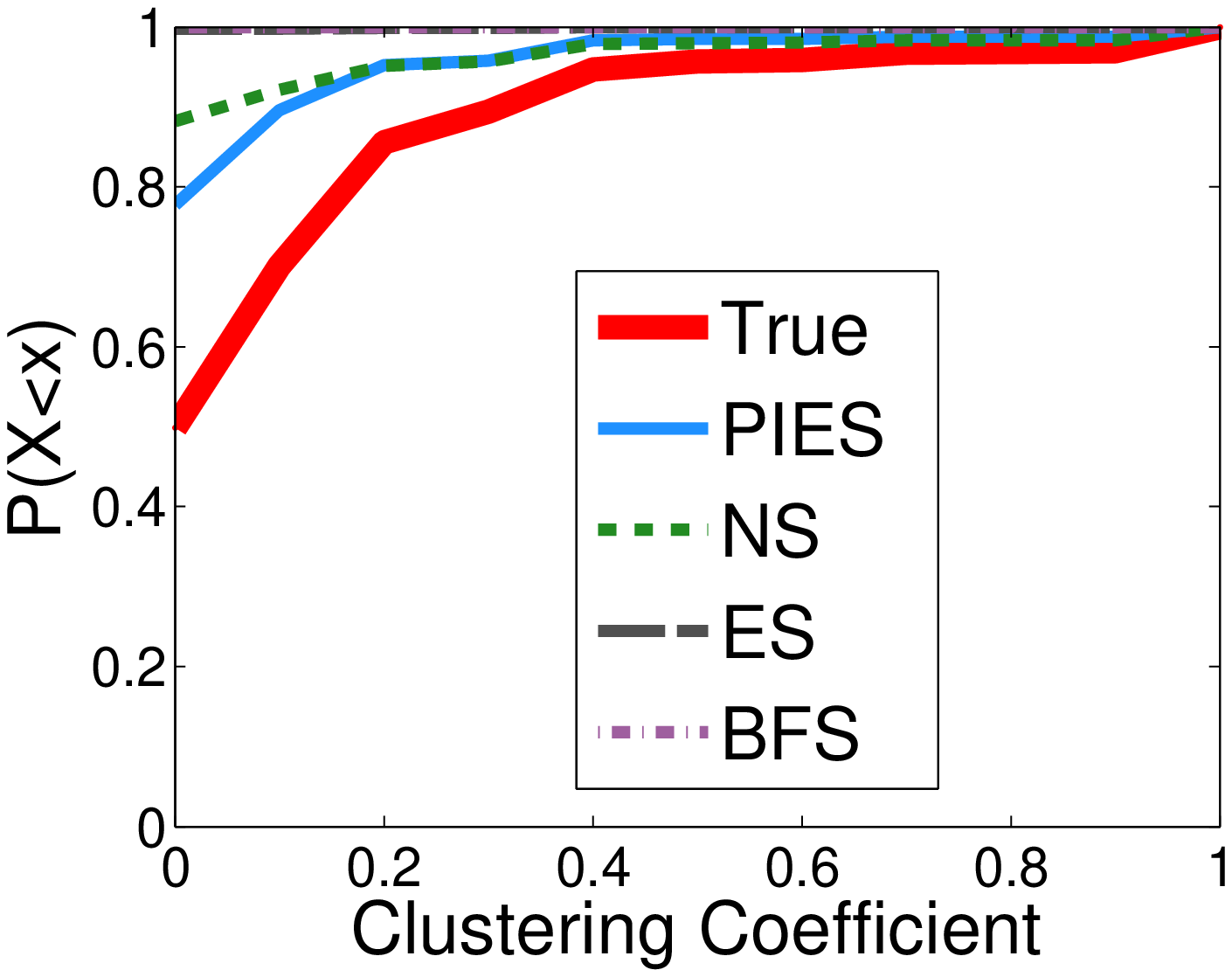}}
\vspace{-2mm}
\hspace{-2.mm}
\subfigure{\label{fig:fbor deg dist}\includegraphics[width=0.33\linewidth]{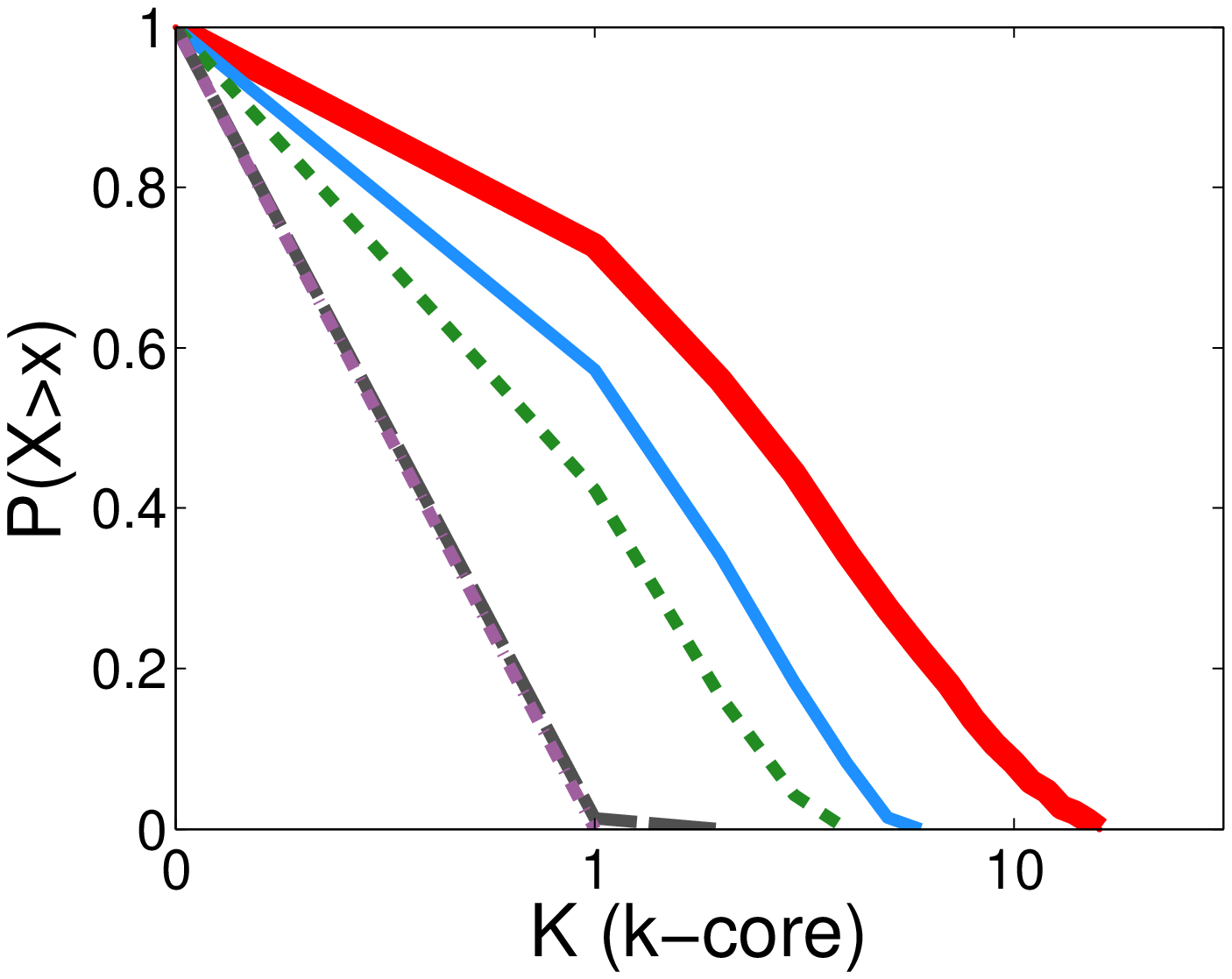}}
\hspace{-5.mm}
\subfigure{\label{fig:fbor pl dist}\includegraphics[width=0.33\linewidth]{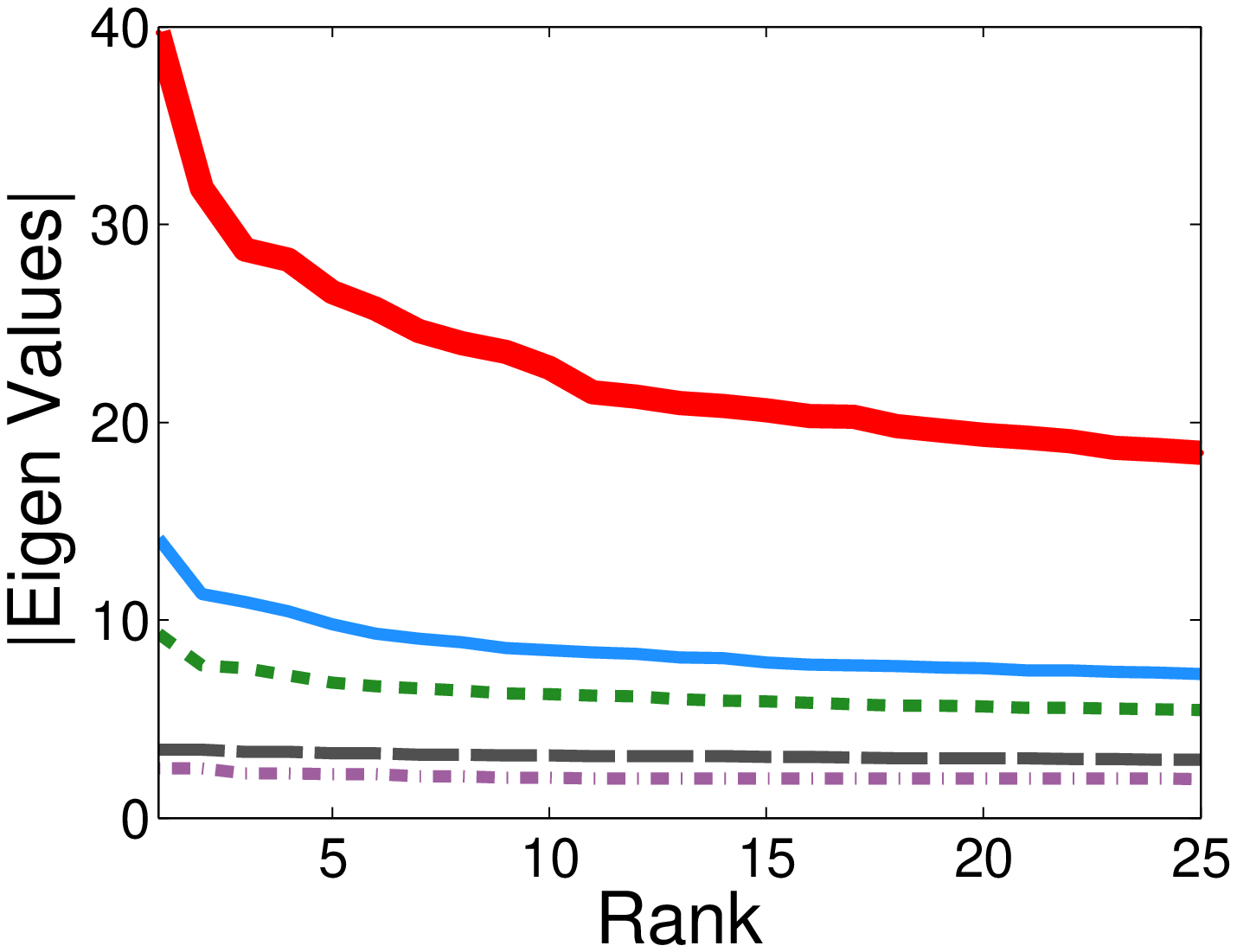}}
\hspace{-5.mm}
\subfigure{\label{fig:fbor cc dist}\includegraphics[width=0.33\linewidth]{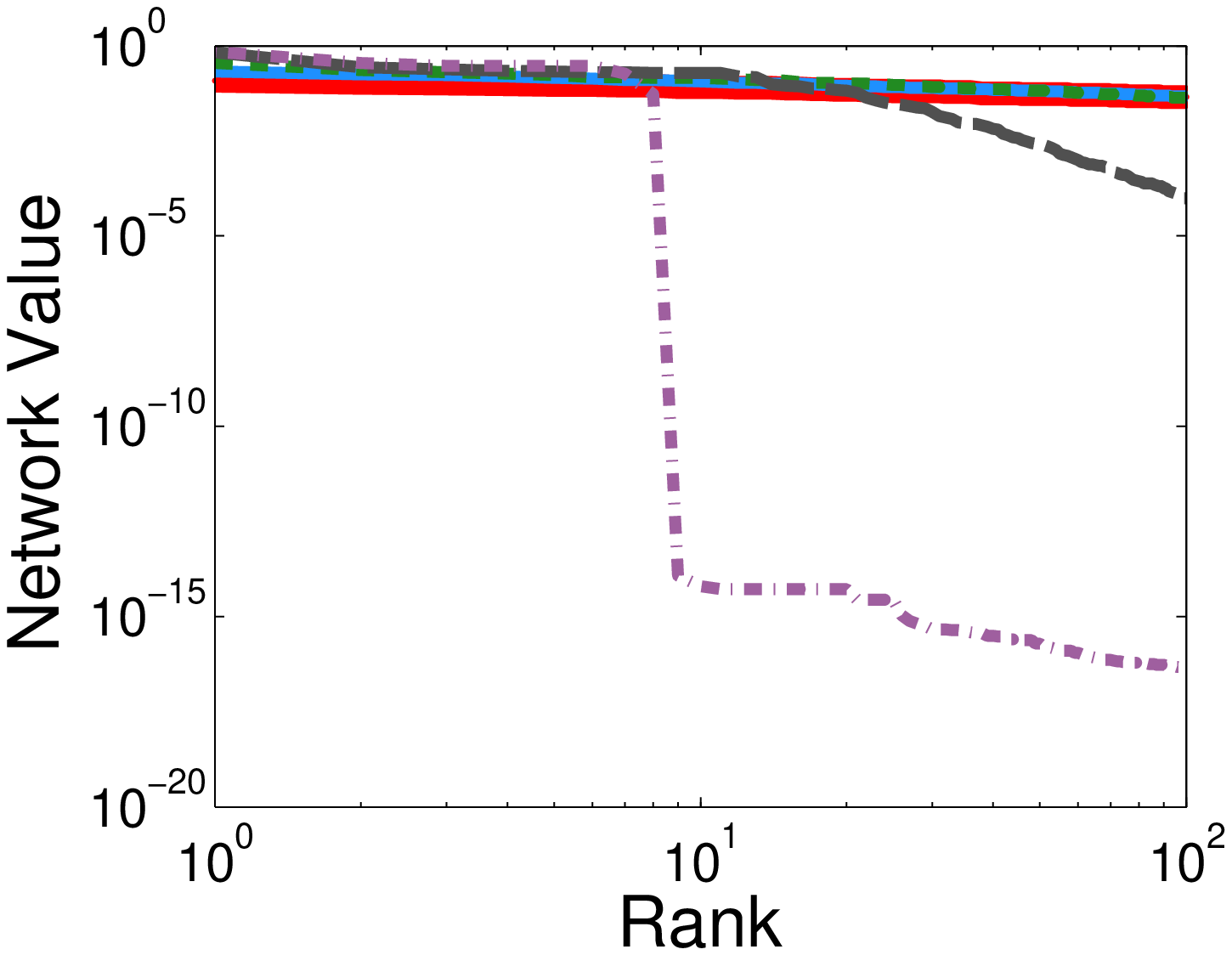}}
\caption{\textsc{Facebook} Graph}
\label{fig:stream_dist_comp_fbor}
\vspace{-3.mm}
\end{figure}

\begin{figure}[!h]
\centering
\subfigure{\label{fig:arxiv deg dist}\includegraphics[width=0.33\linewidth]{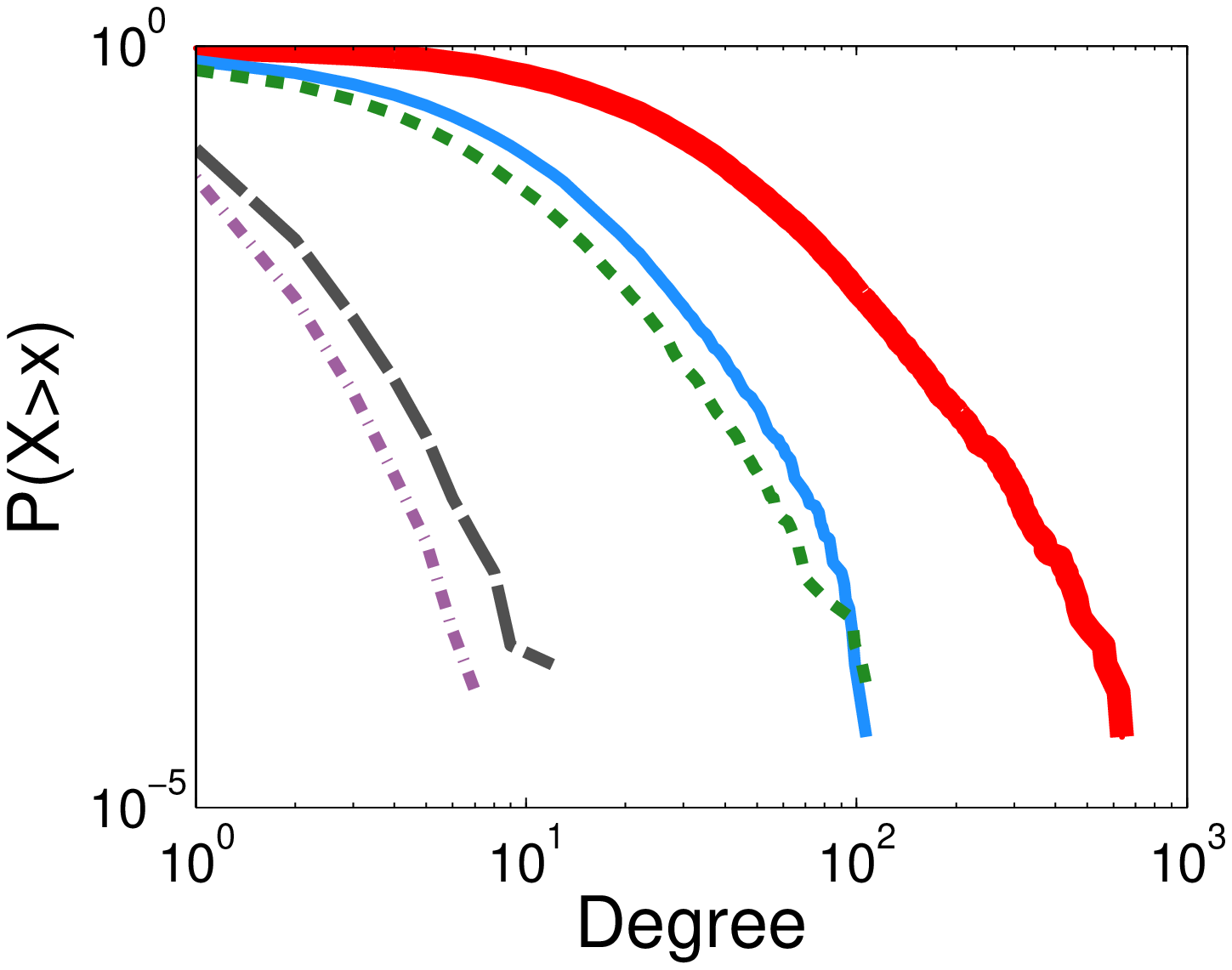}}
\hspace{-5.mm}
\subfigure{\label{fig:arxiv pl dist}\includegraphics[width=0.33\linewidth]{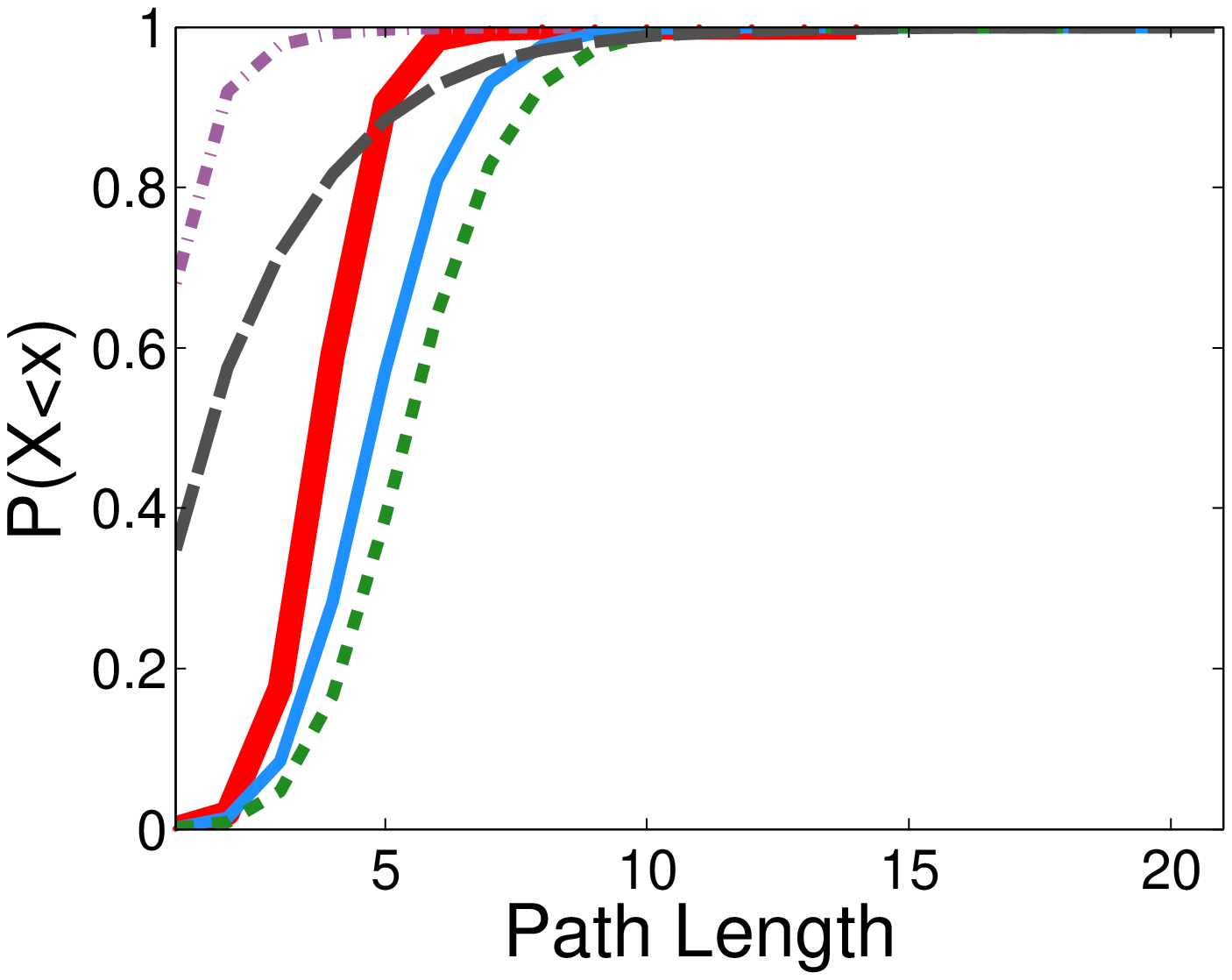}}
\hspace{-5.mm}
\subfigure{\label{fig:arxiv cc dist}\includegraphics[width=0.33\linewidth]{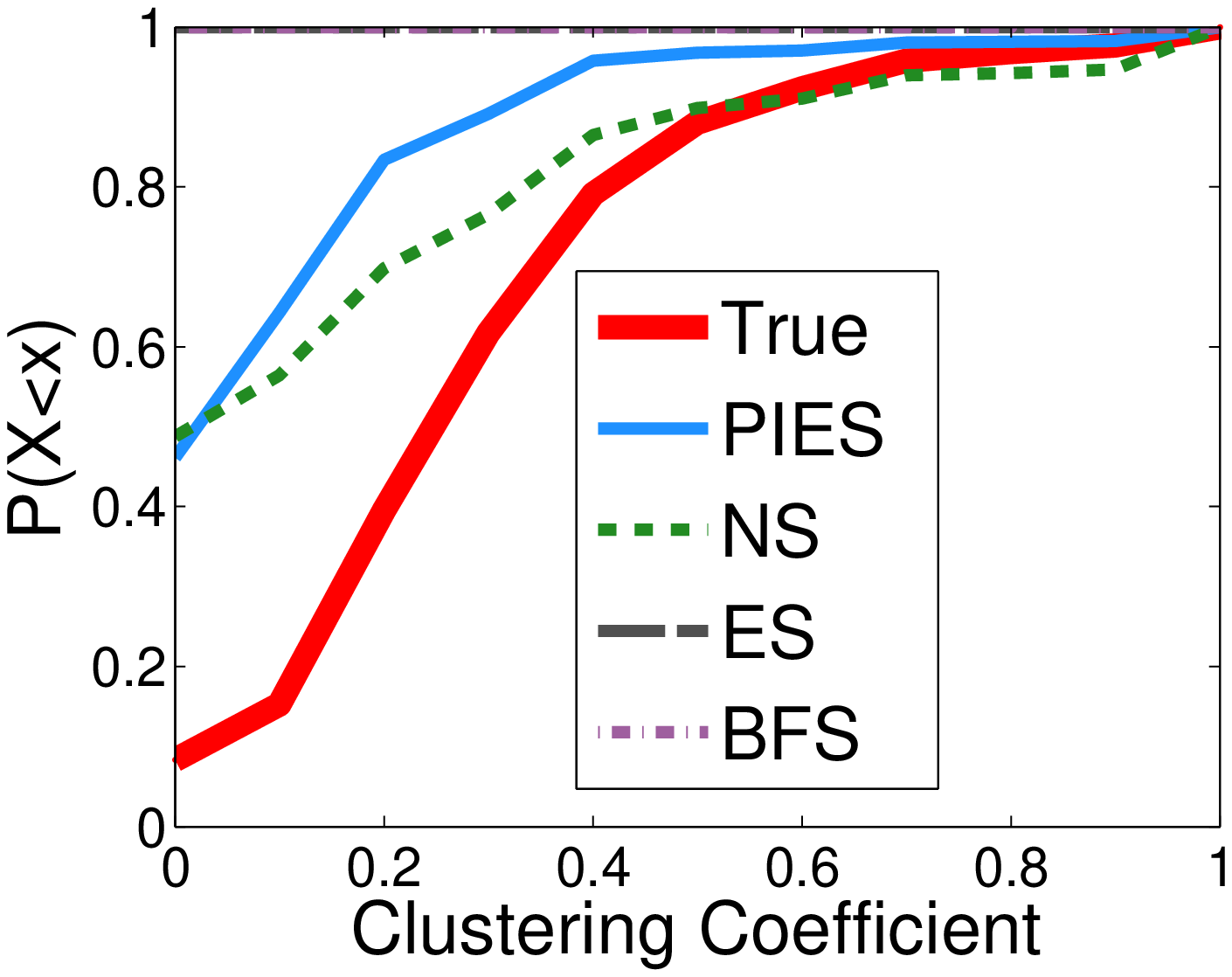}}
\vspace{-2mm}
\hspace{-2.mm}
\subfigure{\label{fig:arxiv deg dist}\includegraphics[width=0.33\linewidth]{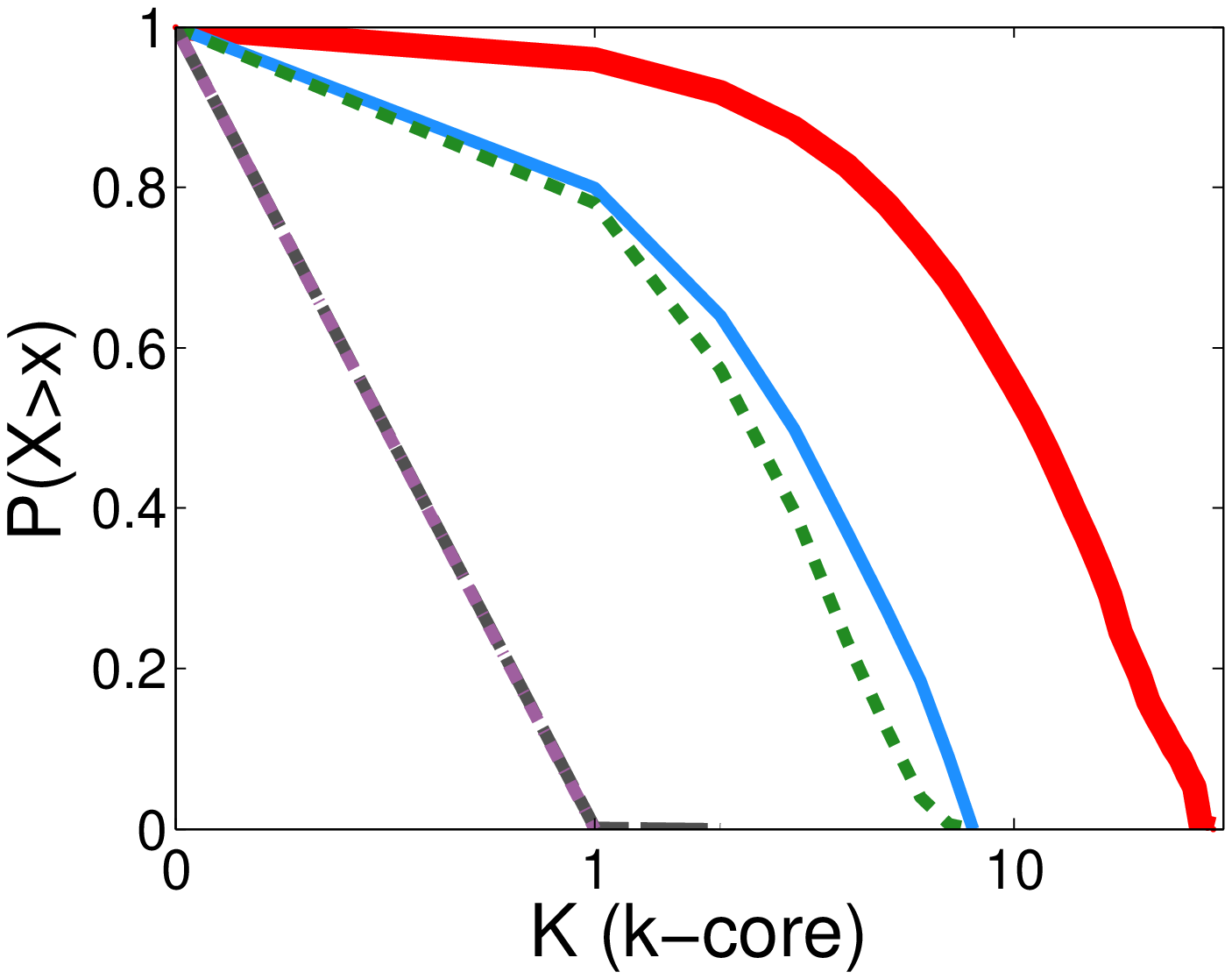}}
\hspace{-5.mm}
\subfigure{\label{fig:arxiv pl dist}\includegraphics[width=0.33\linewidth]{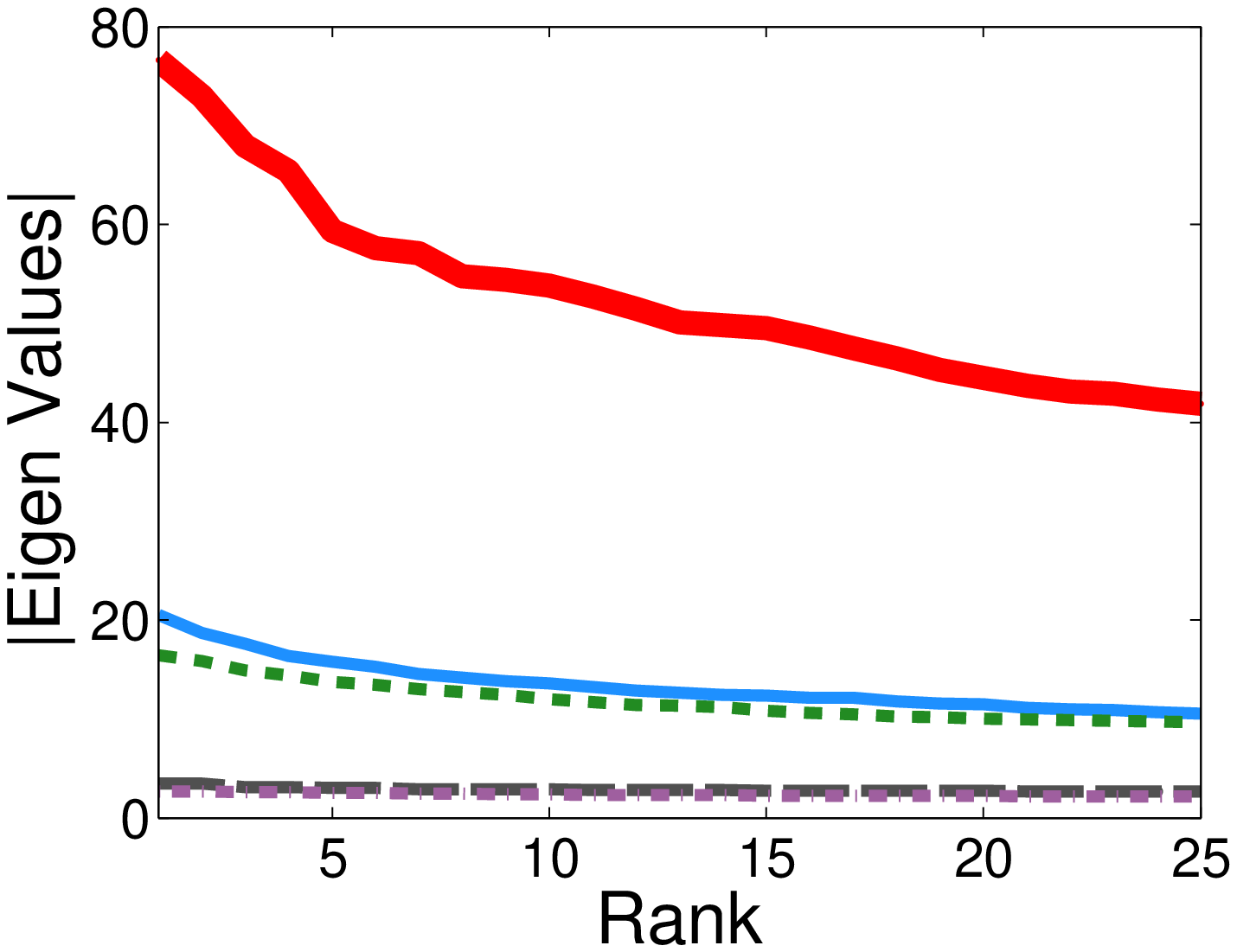}}
\hspace{-5.mm}
\subfigure{\label{fig:arxiv cc dist}\includegraphics[width=0.33\linewidth]{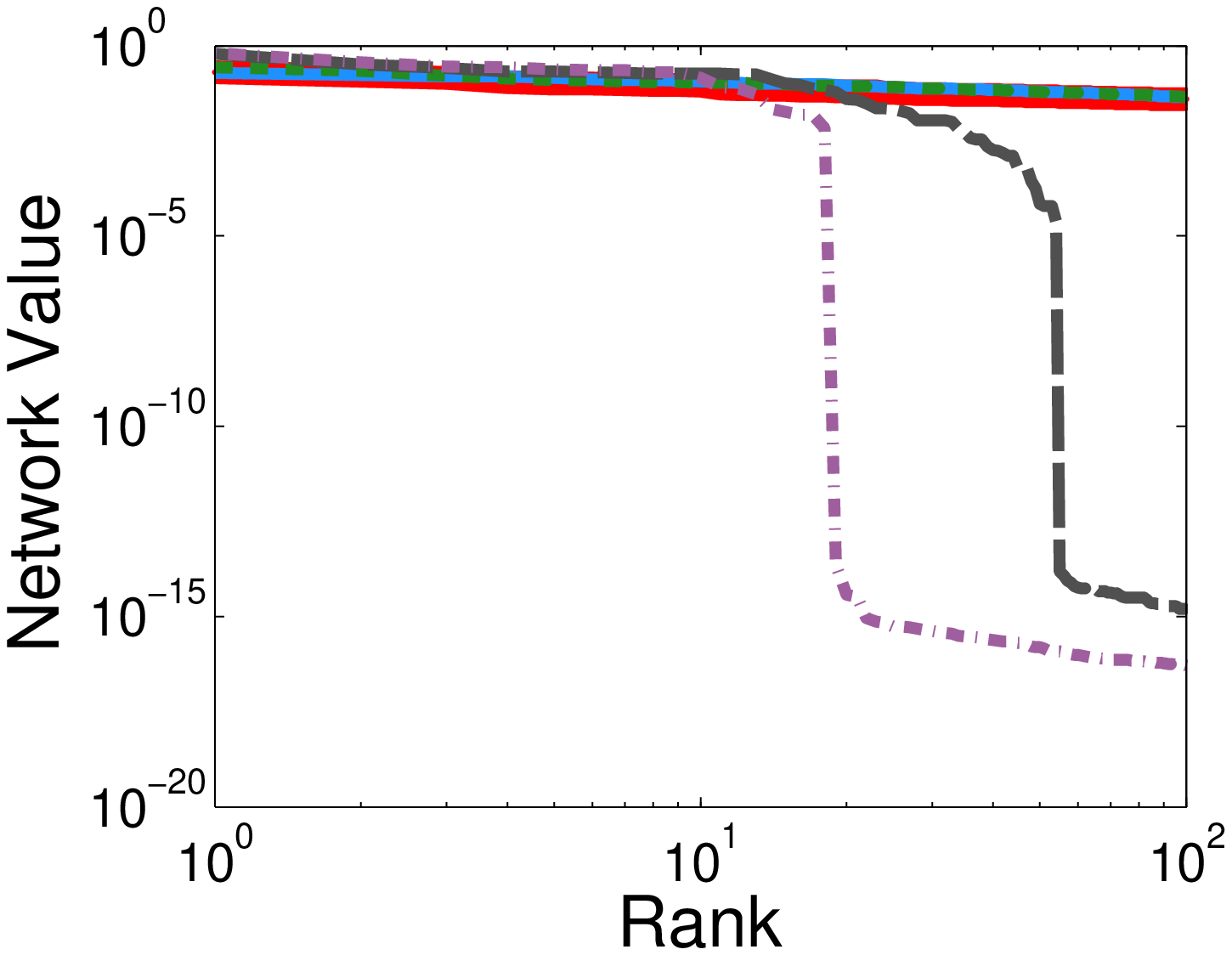}}
\caption{\textsc{HepPH} Graph}
\label{fig:stream_dist_comp_arxiv}
\vspace{-3.mm}
\end{figure}

\newpage

\begin{figure}[!h]
\centering
\subfigure{\label{fig:condmat deg dist}\includegraphics[width=0.33\linewidth]{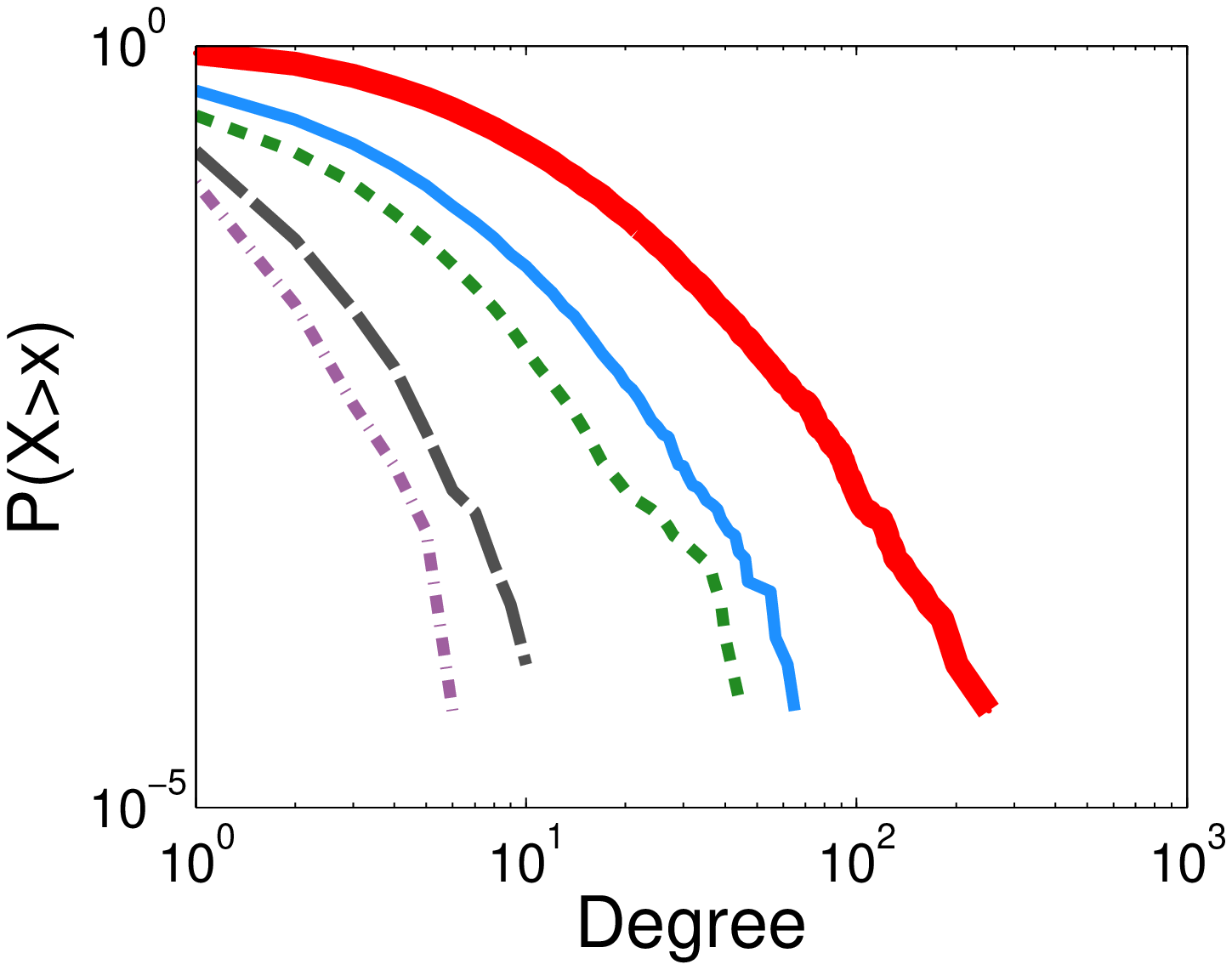}}
\hspace{-5.mm}
\subfigure{\label{fig:condmat pl dist}\includegraphics[width=0.33\linewidth]{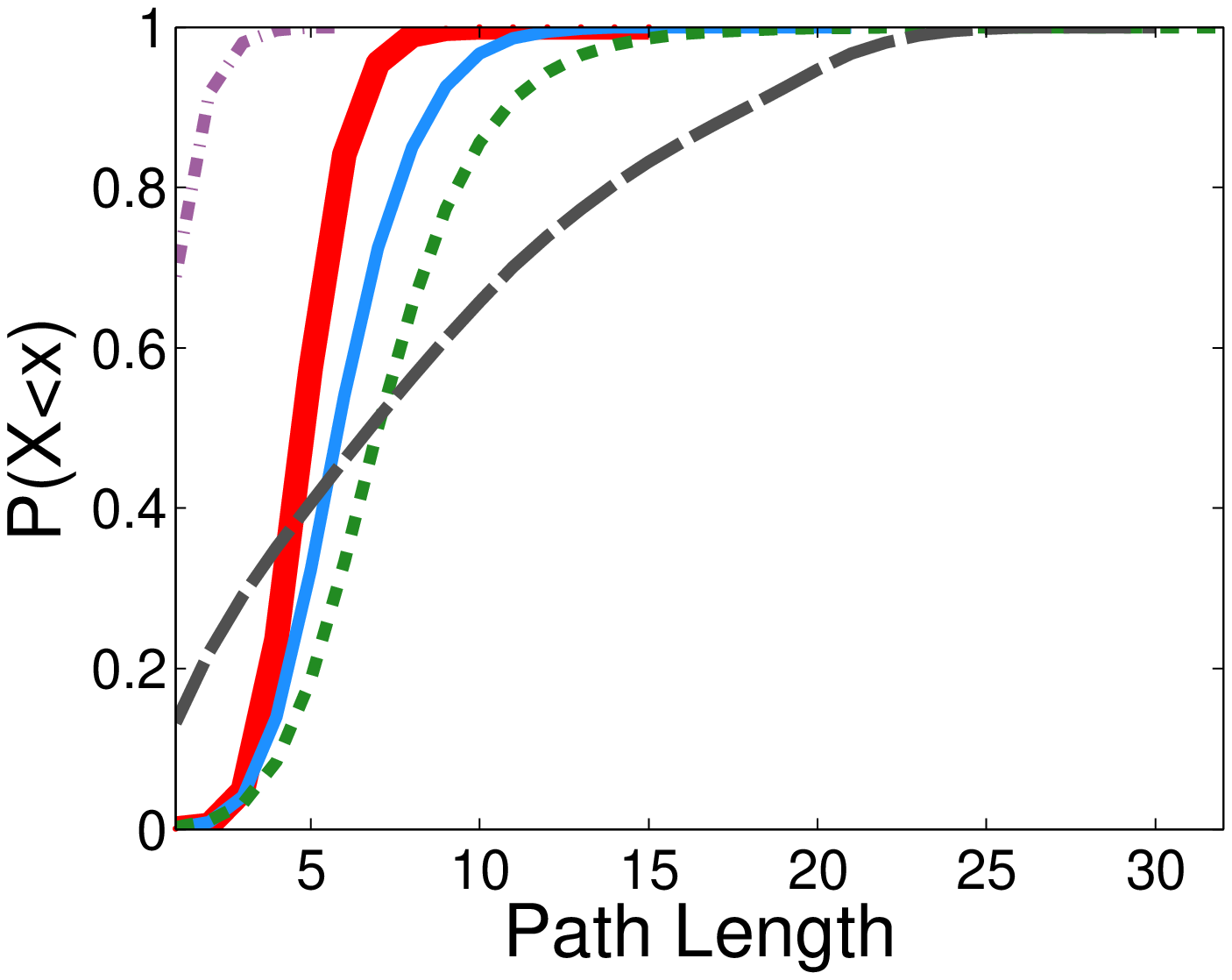}}
\hspace{-5.mm}
\subfigure{\label{fig:condmat cc dist}\includegraphics[width=0.33\linewidth]{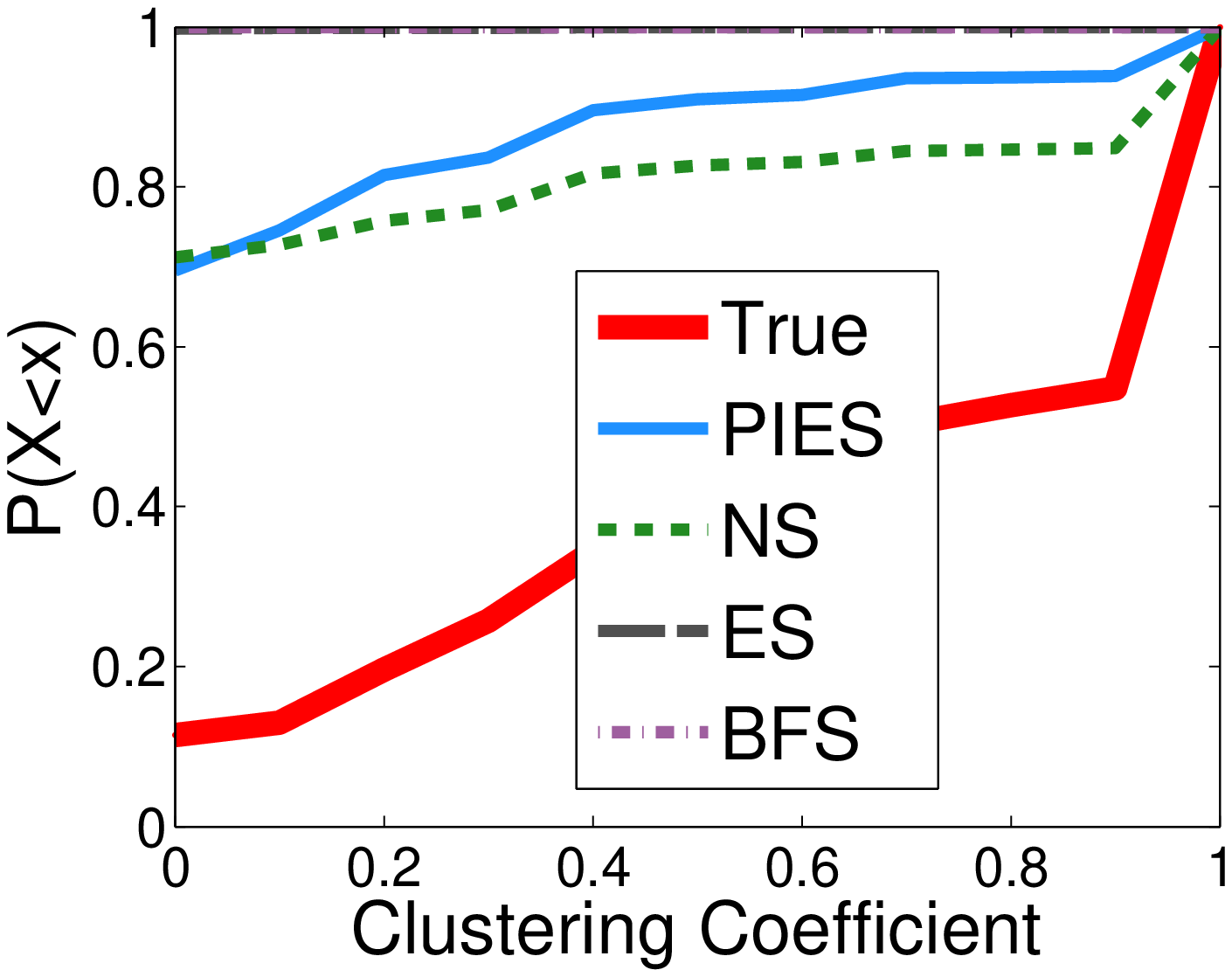}}
\vspace{-2mm}
\hspace{-2.mm}
\subfigure{\label{fig:condmat deg dist}\includegraphics[width=0.33\linewidth]{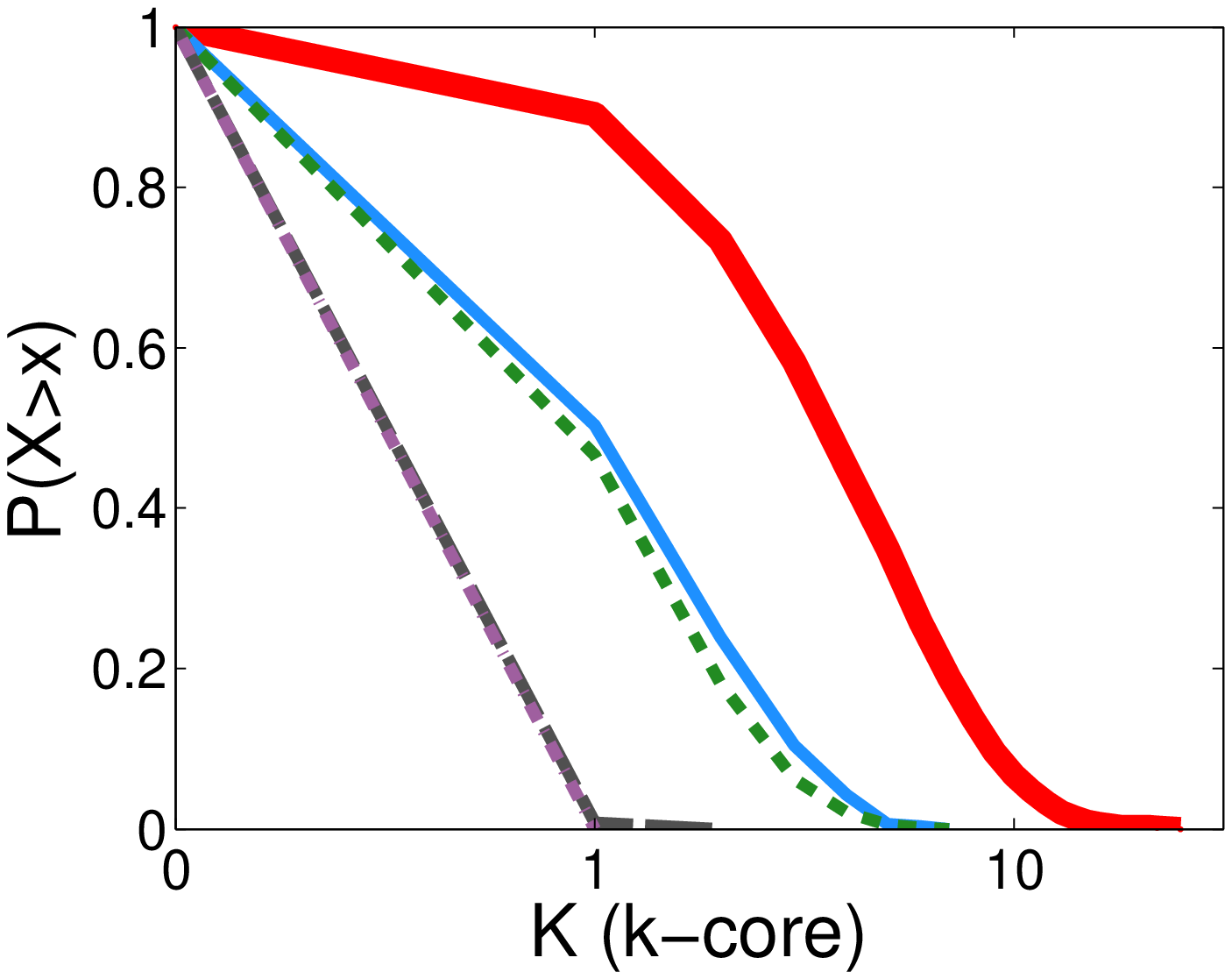}}
\hspace{-5.mm}
\subfigure{\label{fig:condmat pl dist}\includegraphics[width=0.33\linewidth]{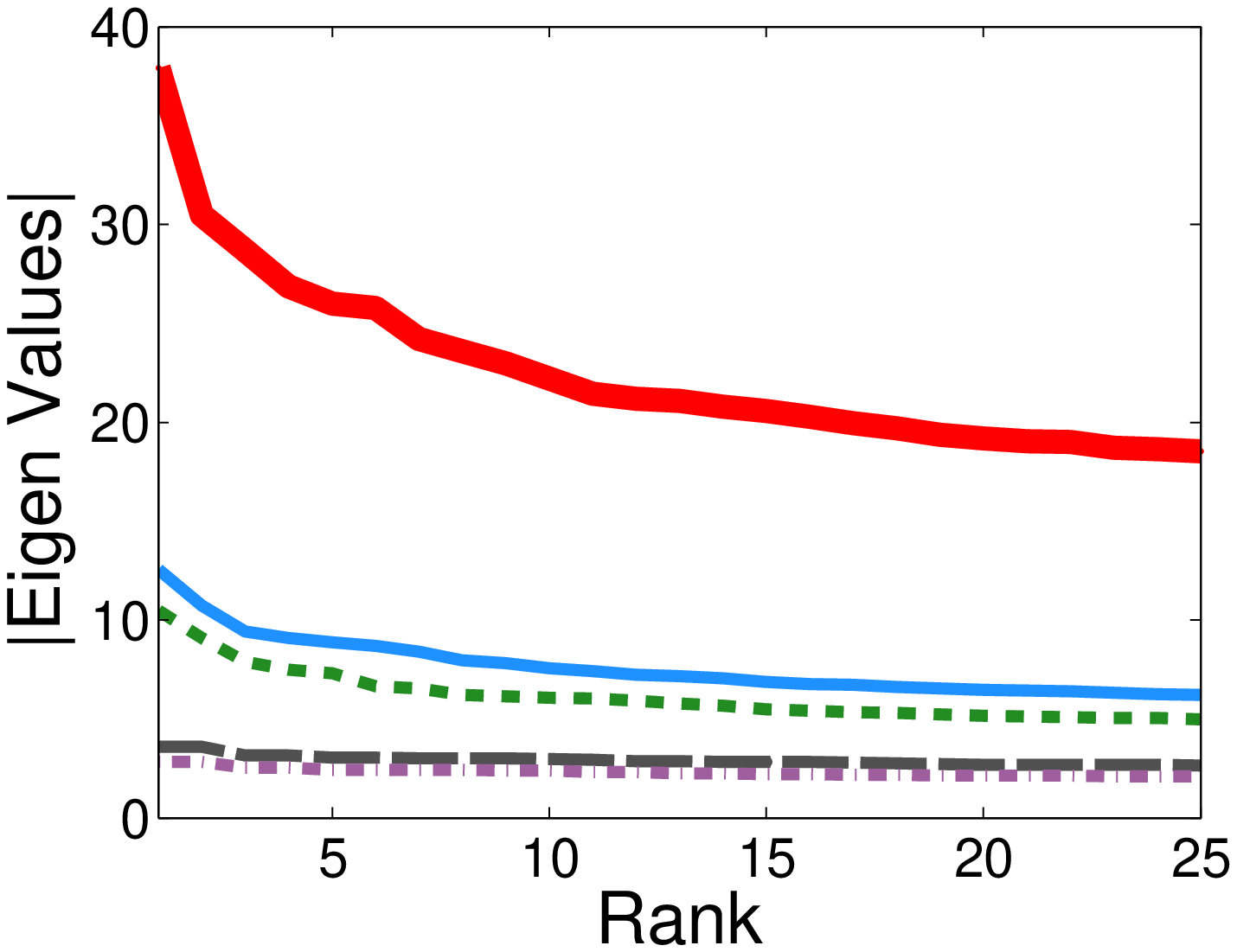}}
\hspace{-5.mm}
\subfigure{\label{fig:condmat cc dist}\includegraphics[width=0.33\linewidth]{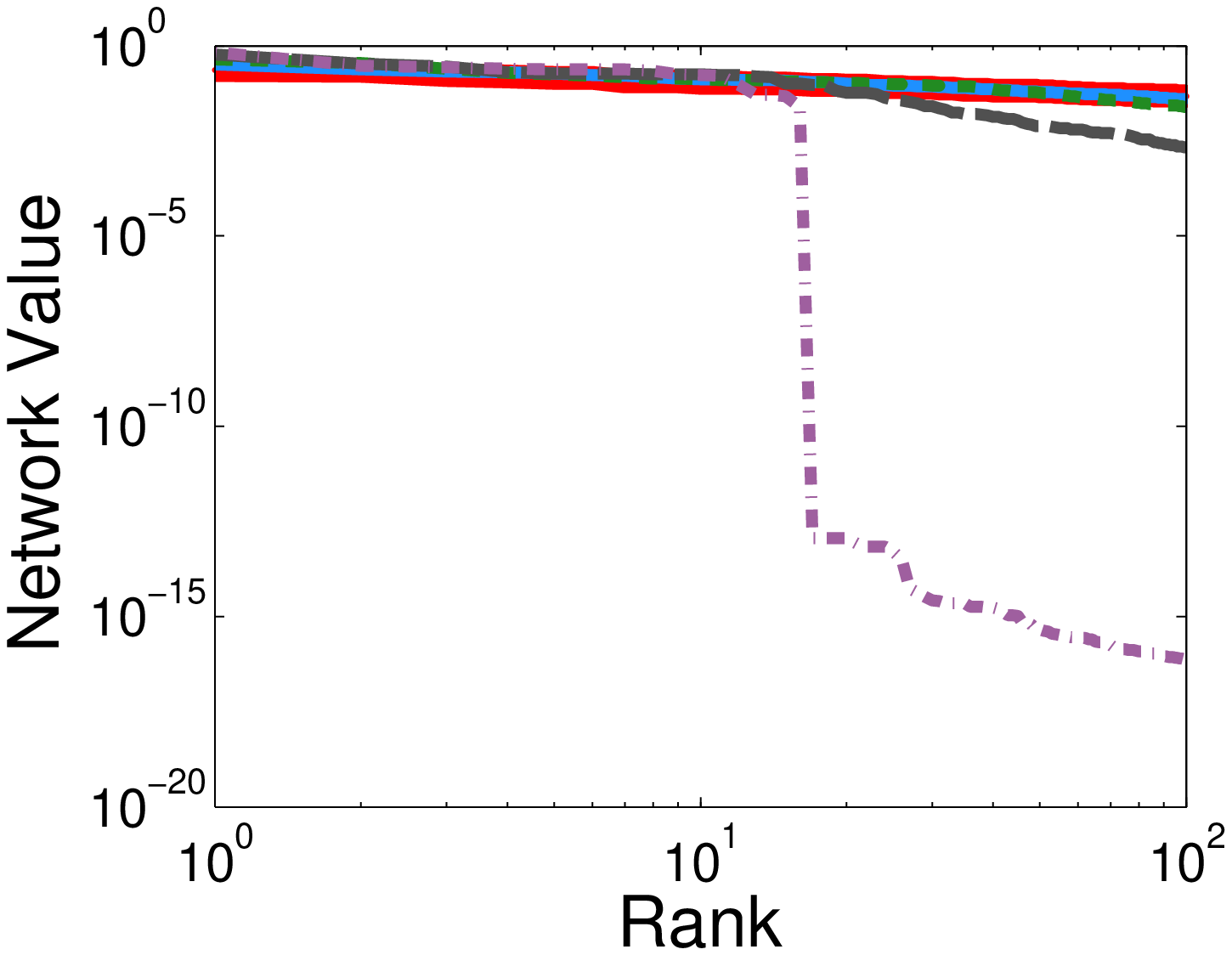}}
\caption{\textsc{CondMAT} Graph}
\label{fig:stream_dist_comp_condmat}
\vspace{-3.mm}
\end{figure}

\begin{figure}[!h]
\centering
\vspace{-3.mm}
\subfigure{\label{fig:twcop deg dist}\includegraphics[width=0.33\linewidth]{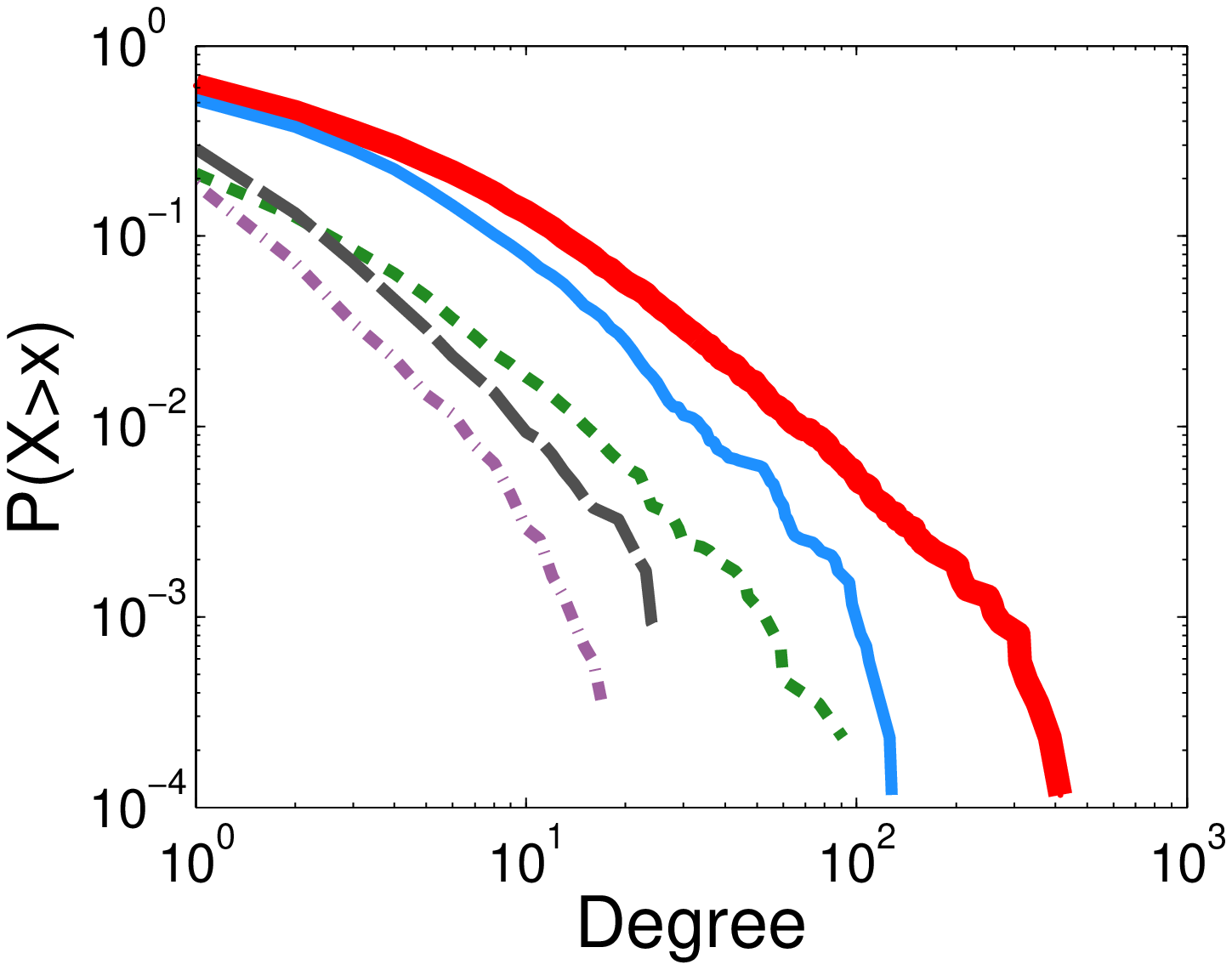}}
\hspace{-5.mm}
\subfigure{\label{fig:twcop pl dist}\includegraphics[width=0.33\linewidth]{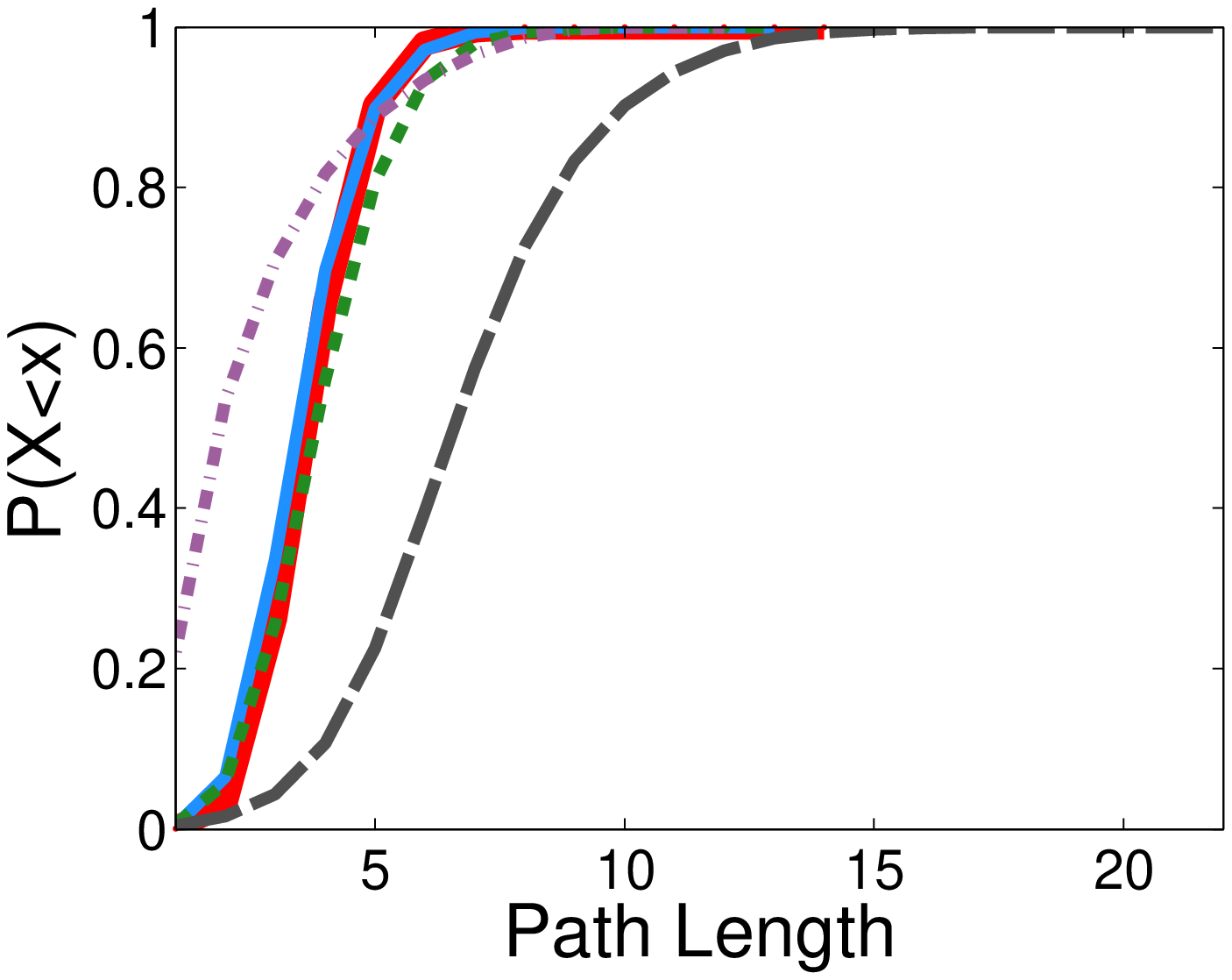}}
\hspace{-5.mm}
\subfigure{\label{fig:twcop cc dist}\includegraphics[width=0.33\linewidth]{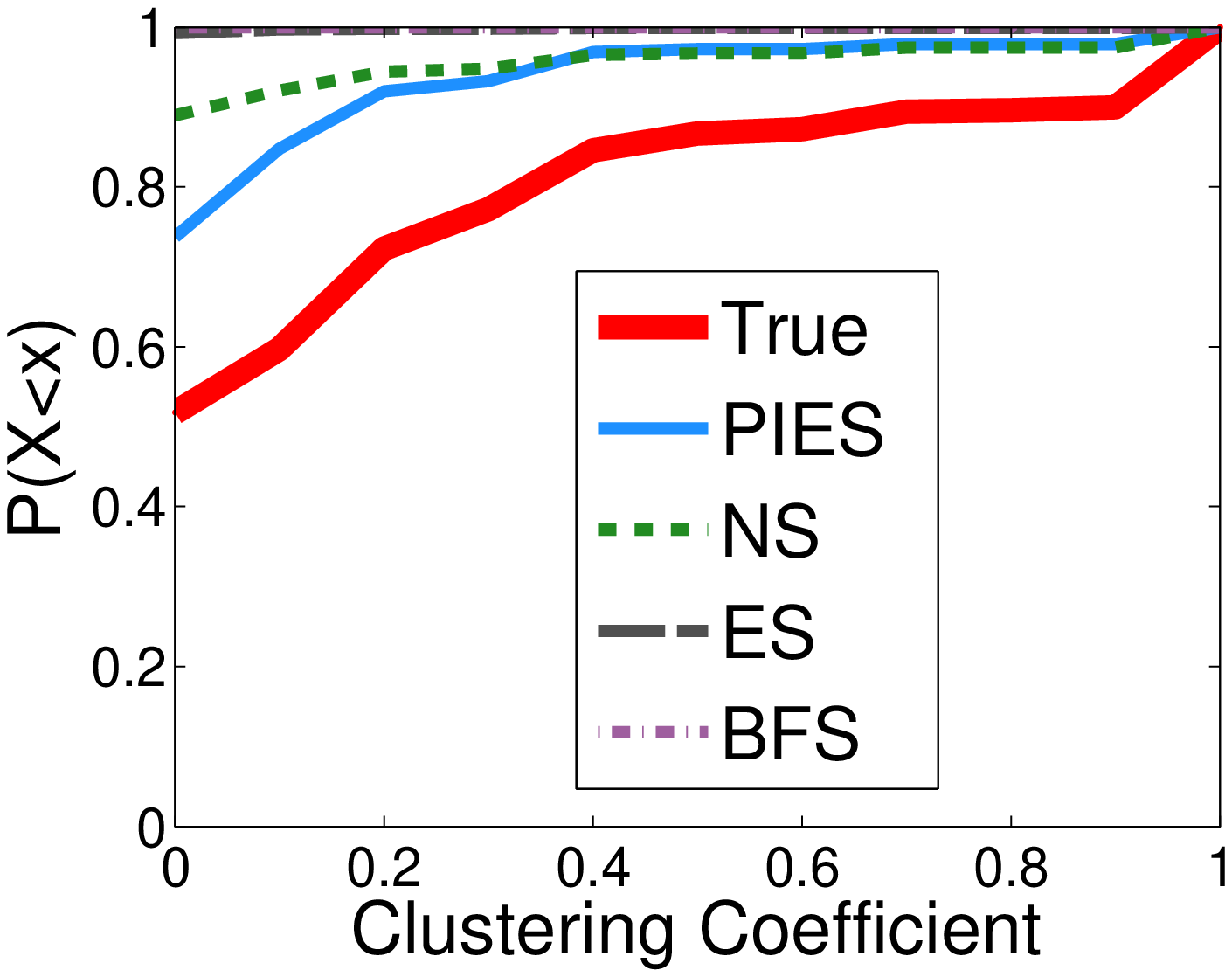}}
\vspace{-2mm}
\hspace{-2.mm}
\subfigure{\label{fig:twcop deg dist}\includegraphics[width=0.33\linewidth]{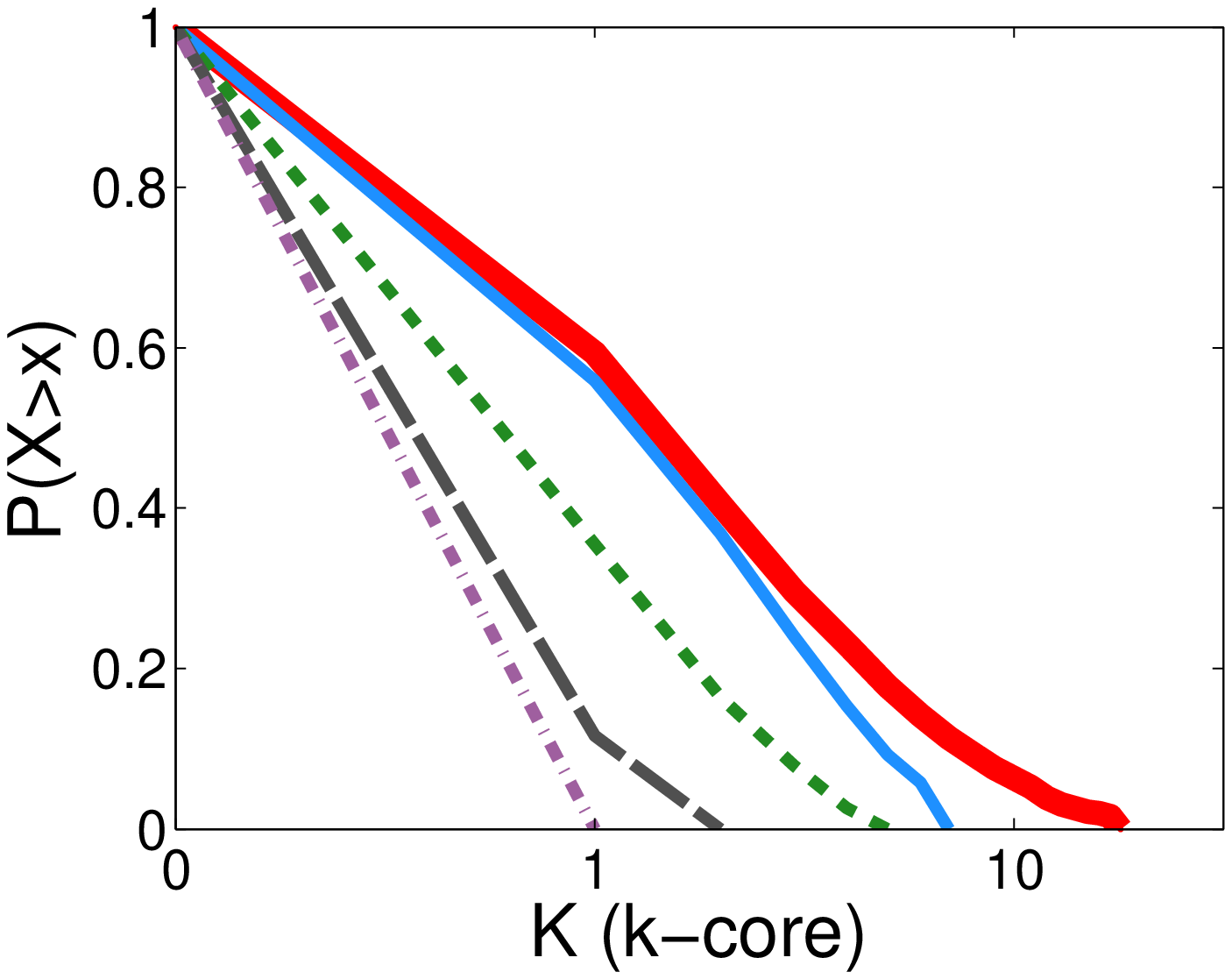}}
\hspace{-5.mm}
\subfigure{\label{fig:twcop pl dist}\includegraphics[width=0.33\linewidth]{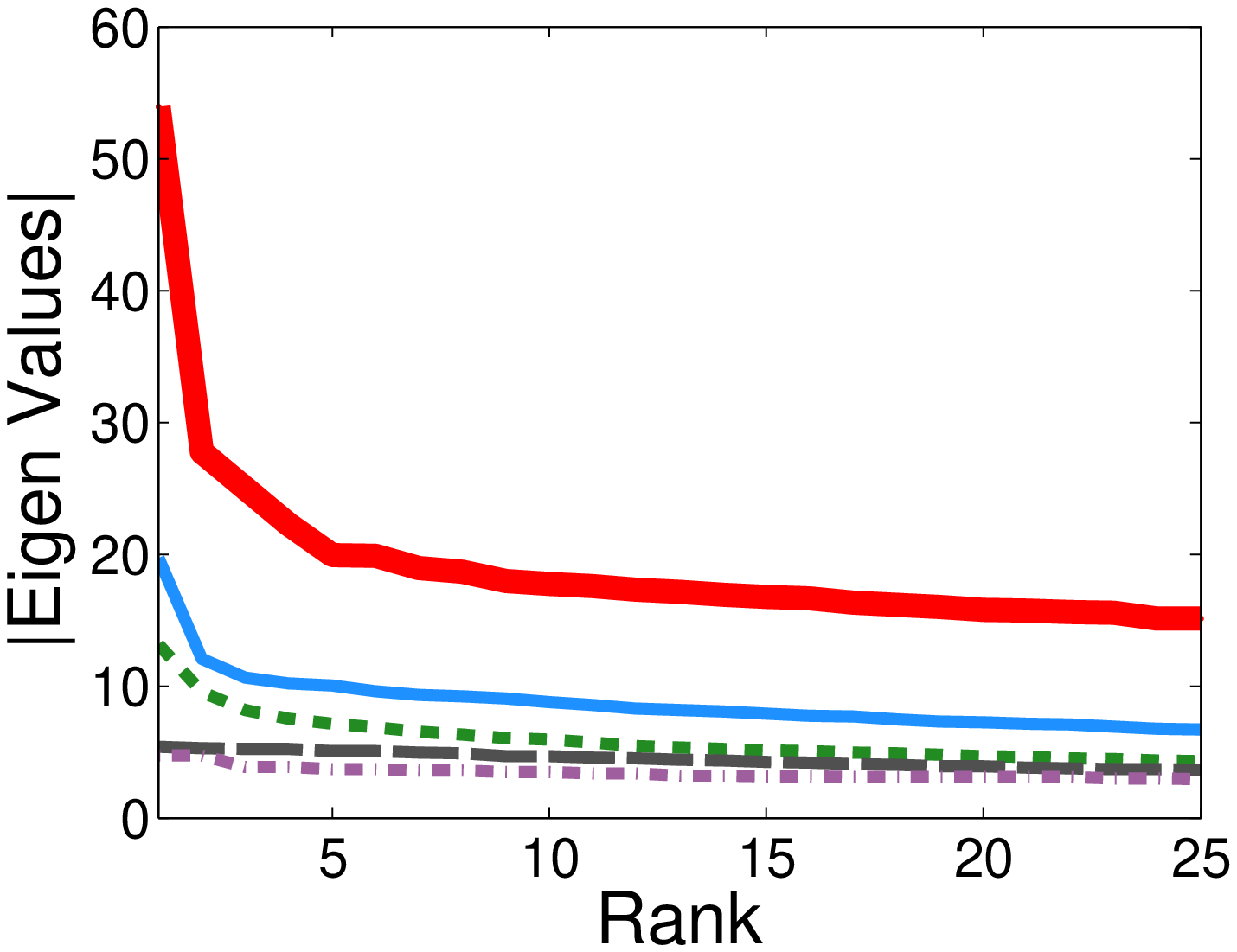}}
\hspace{-5.mm}
\subfigure{\label{fig:twcop cc dist}\includegraphics[width=0.33\linewidth]{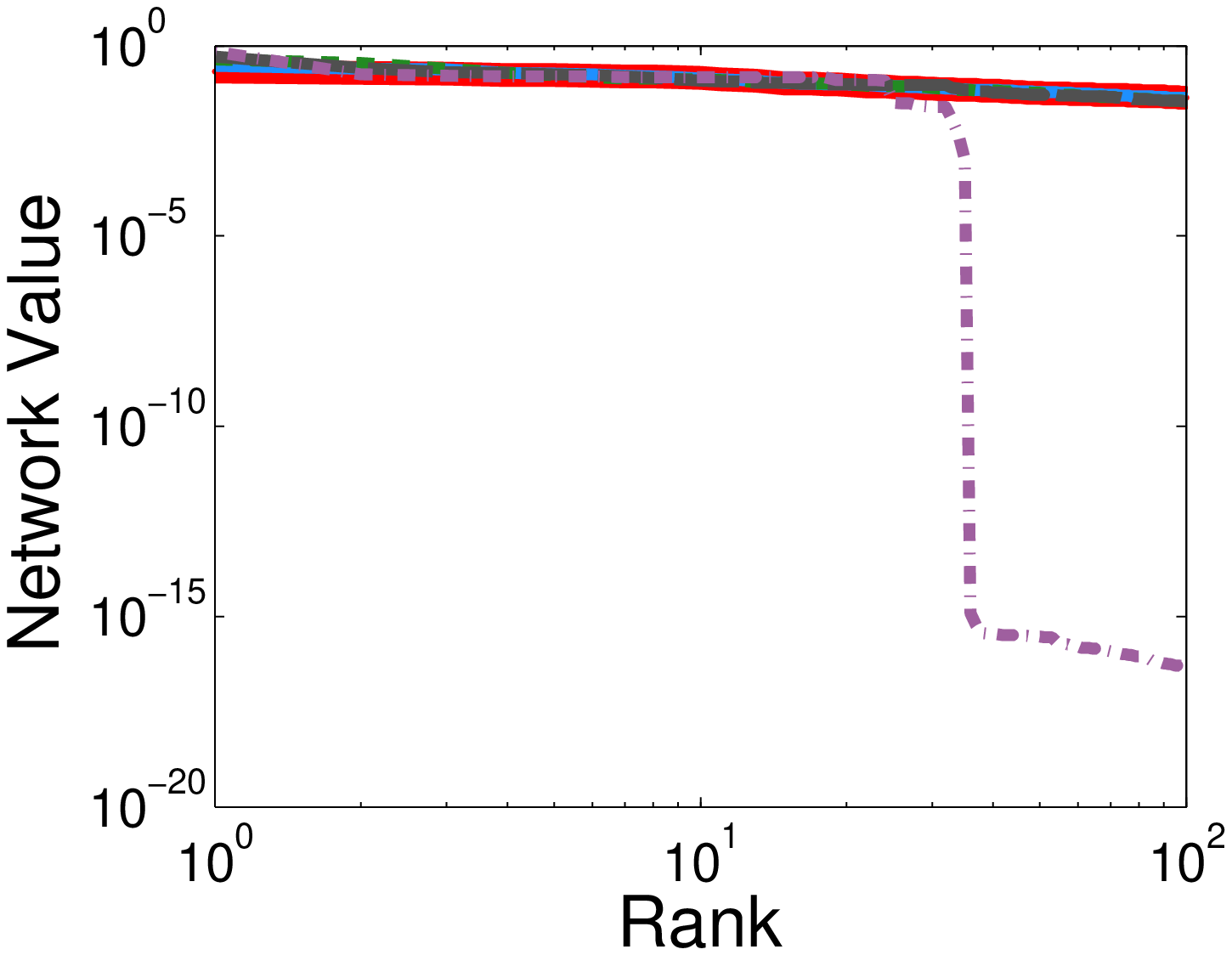}}
\caption{\textsc{Twitter} Graph}
\label{fig:stream_dist_comp_twcop}
\vspace{-3.mm}
\end{figure}

\newpage

\begin{figure}[!h]
\centering
\subfigure{\label{fig:emailpu deg dist}\includegraphics[width=0.33\linewidth]{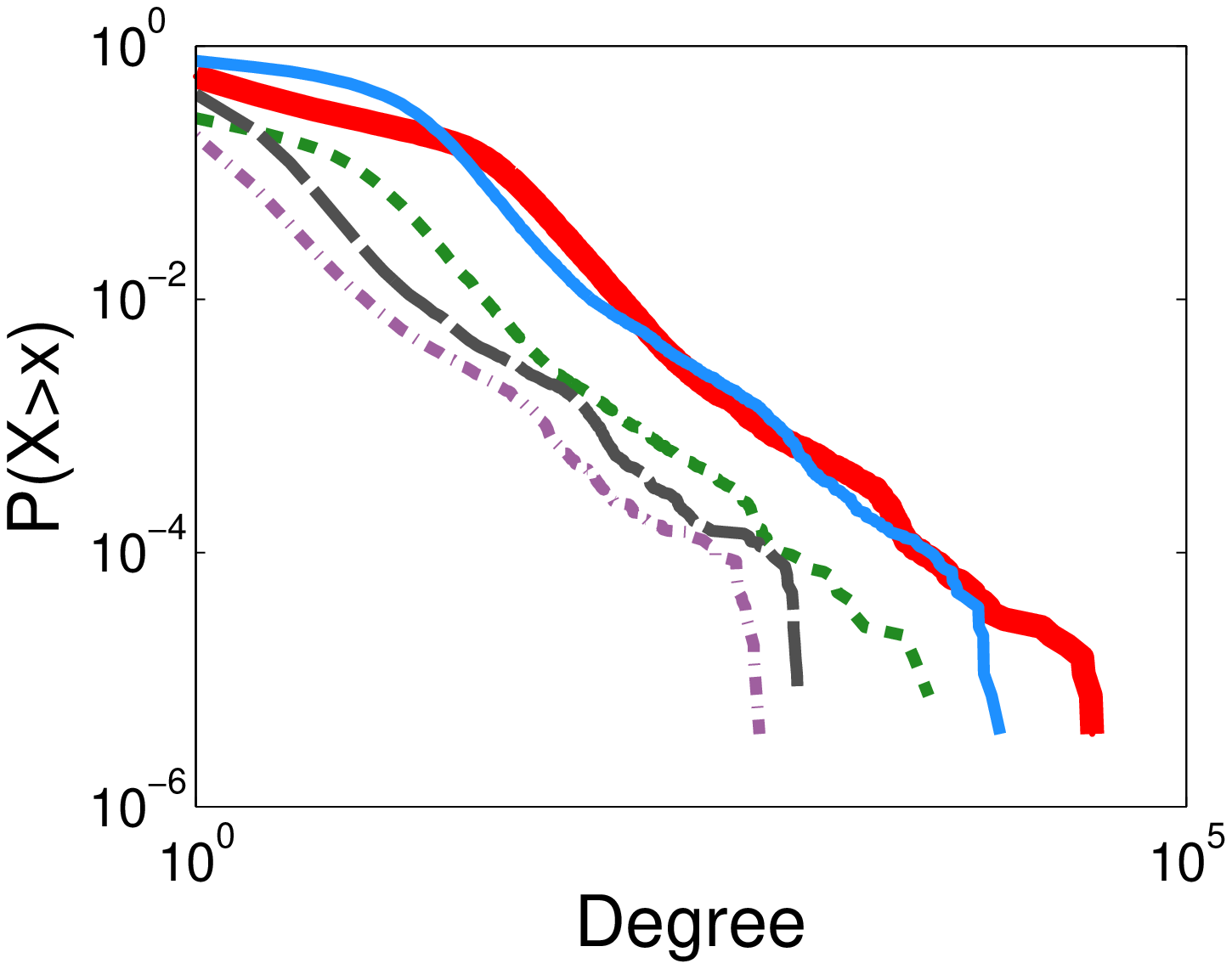}}
\hspace{-5.mm}
\subfigure{\label{fig:emailpu pl dist}\includegraphics[width=0.33\linewidth]{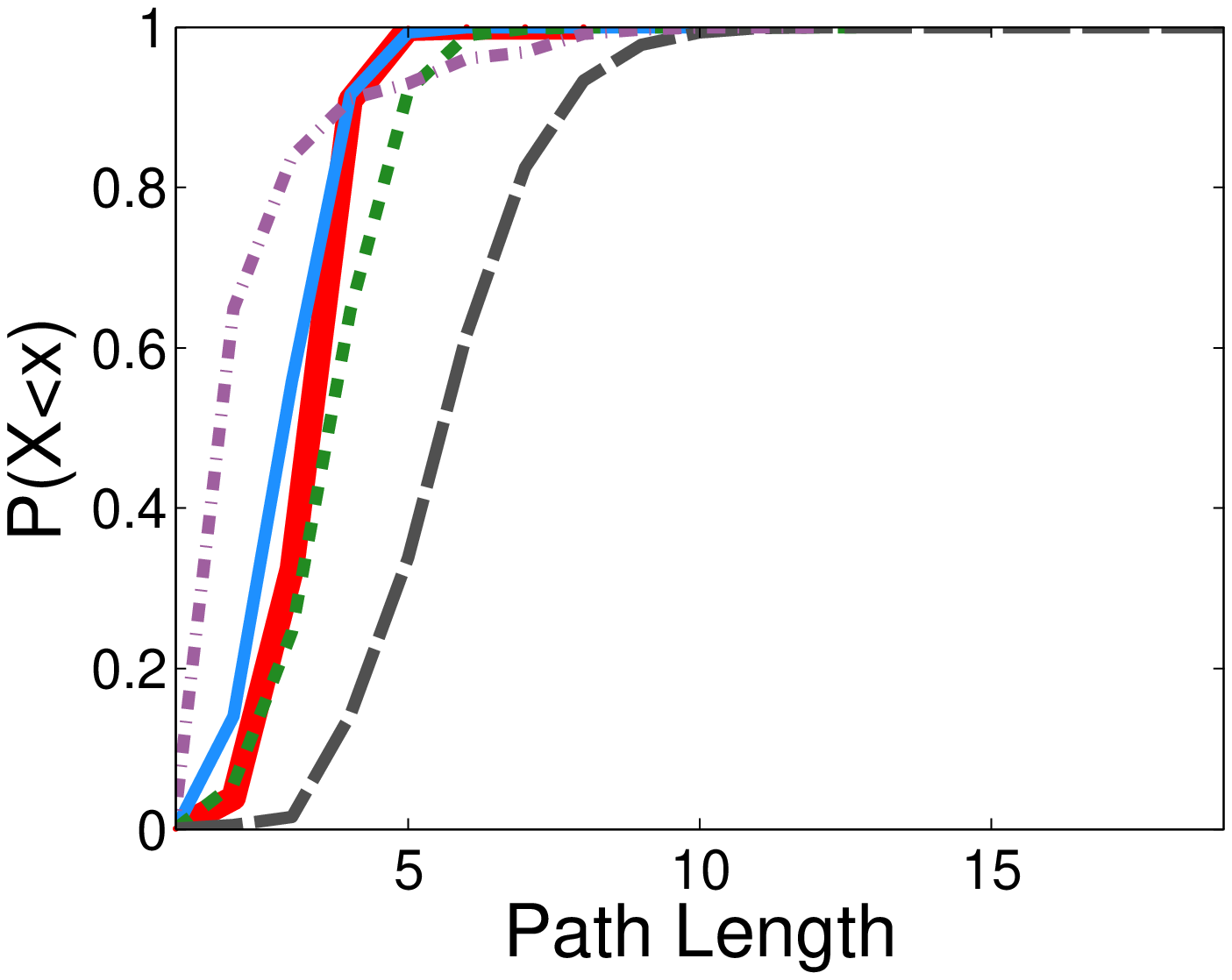}}
\hspace{-5.mm}
\subfigure{\label{fig:emailpu cc dist}\includegraphics[width=0.33\linewidth]{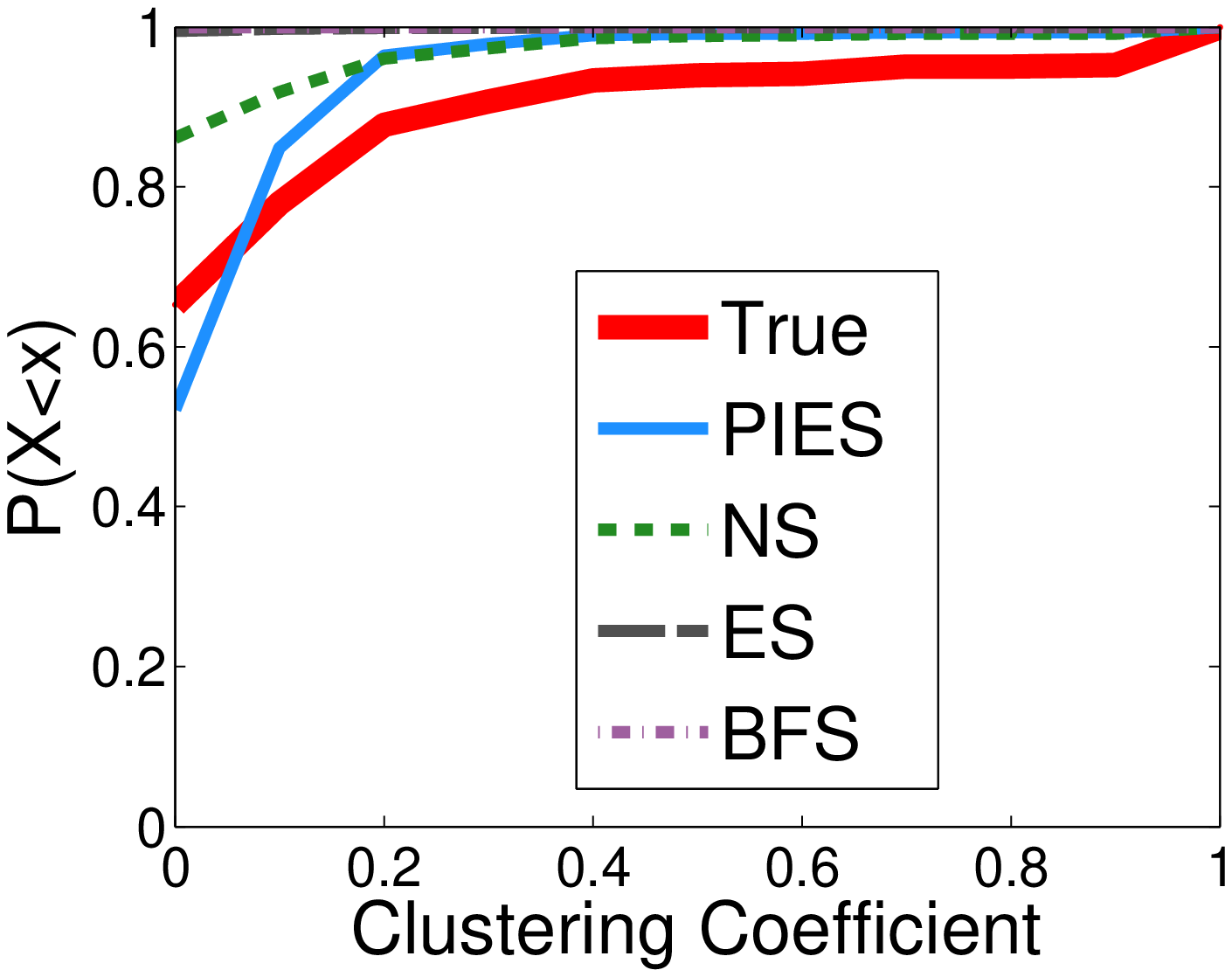}}
\vspace{-2mm}
\hspace{-2.mm}
\subfigure{\label{fig:emailpu deg dist}\includegraphics[width=0.33\linewidth]{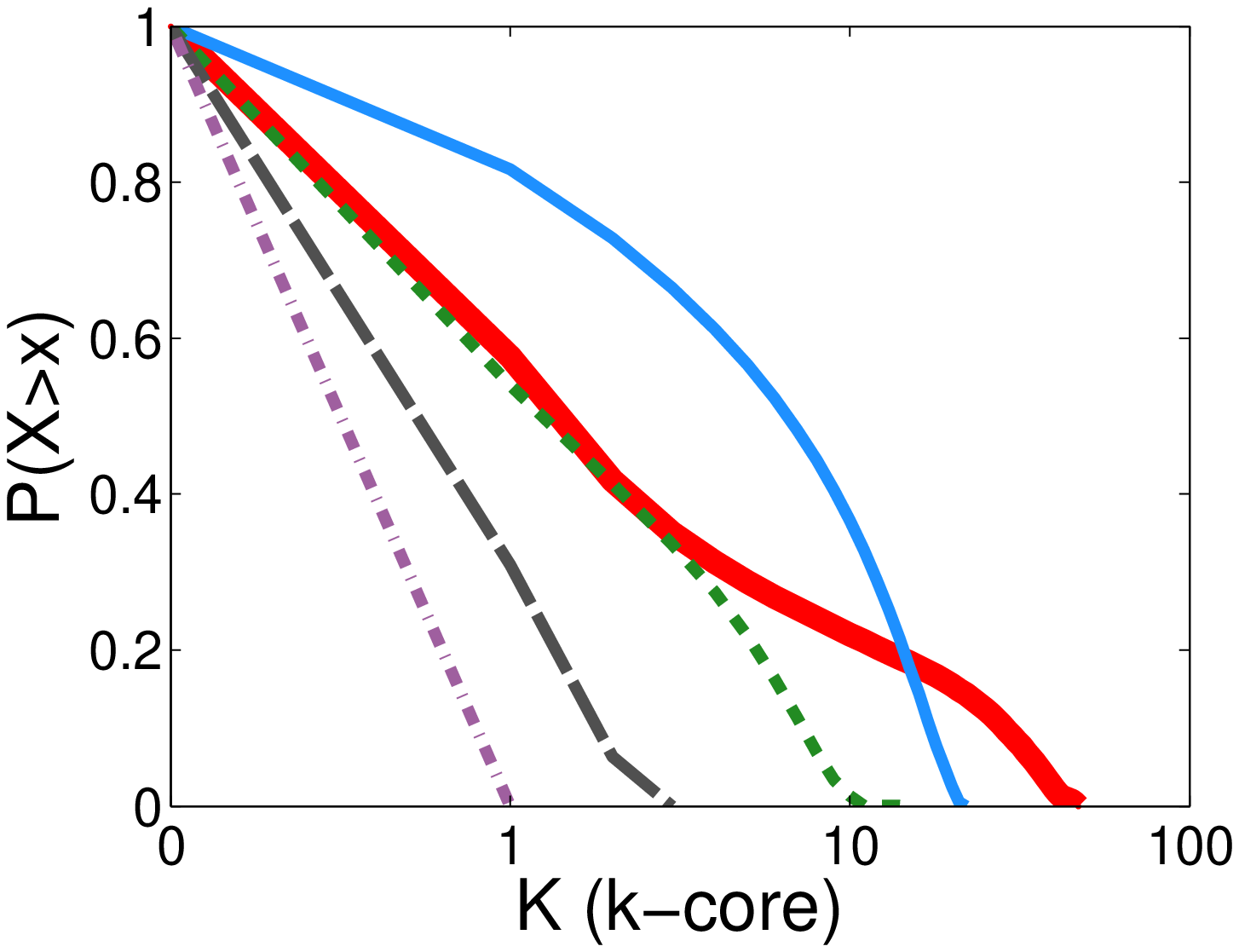}}
\hspace{-5.mm}
\subfigure{\label{fig:emailpu pl dist}\includegraphics[width=0.33\linewidth]{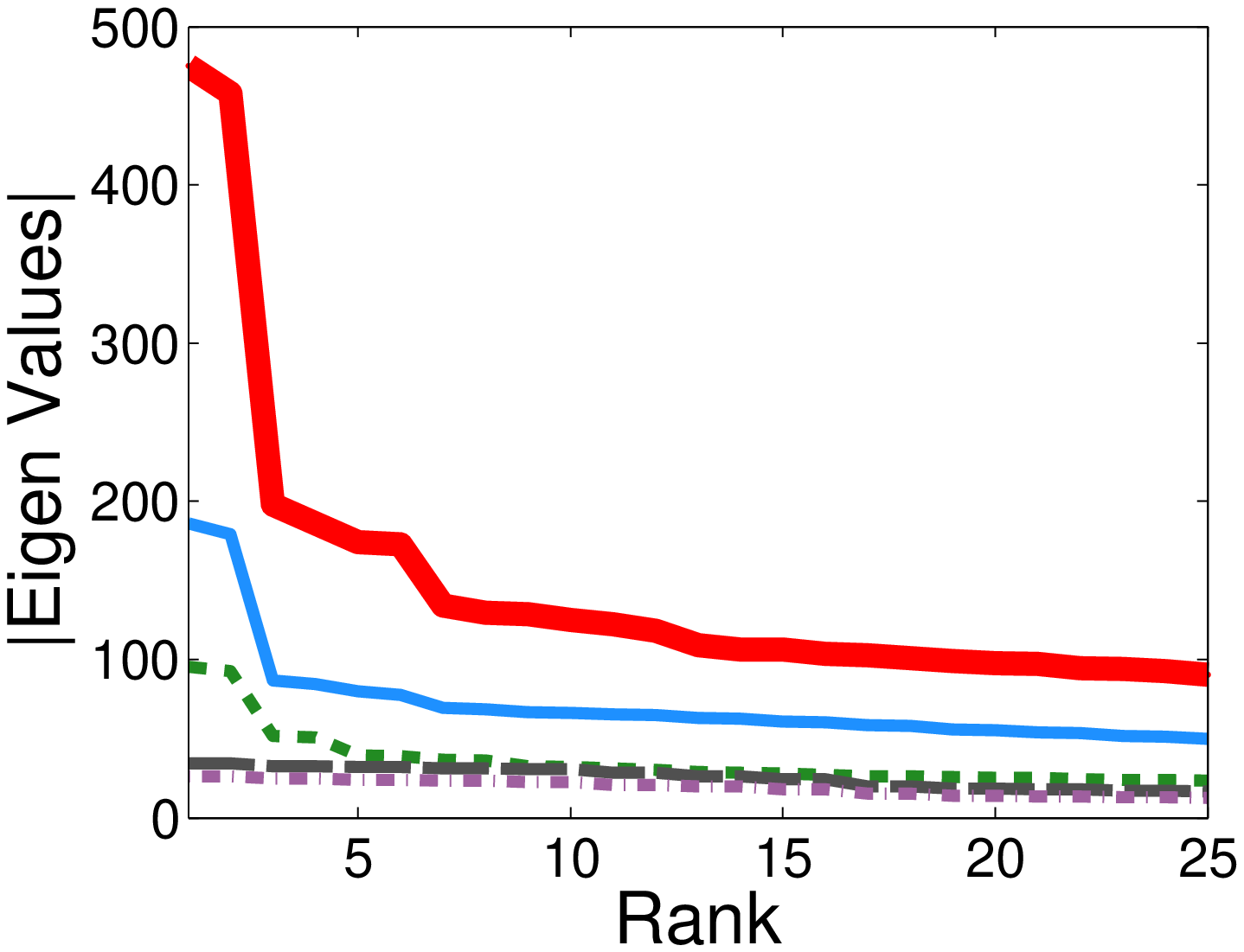}}
\hspace{-5.mm}
\subfigure{\label{fig:emailpu cc dist}\includegraphics[width=0.33\linewidth]{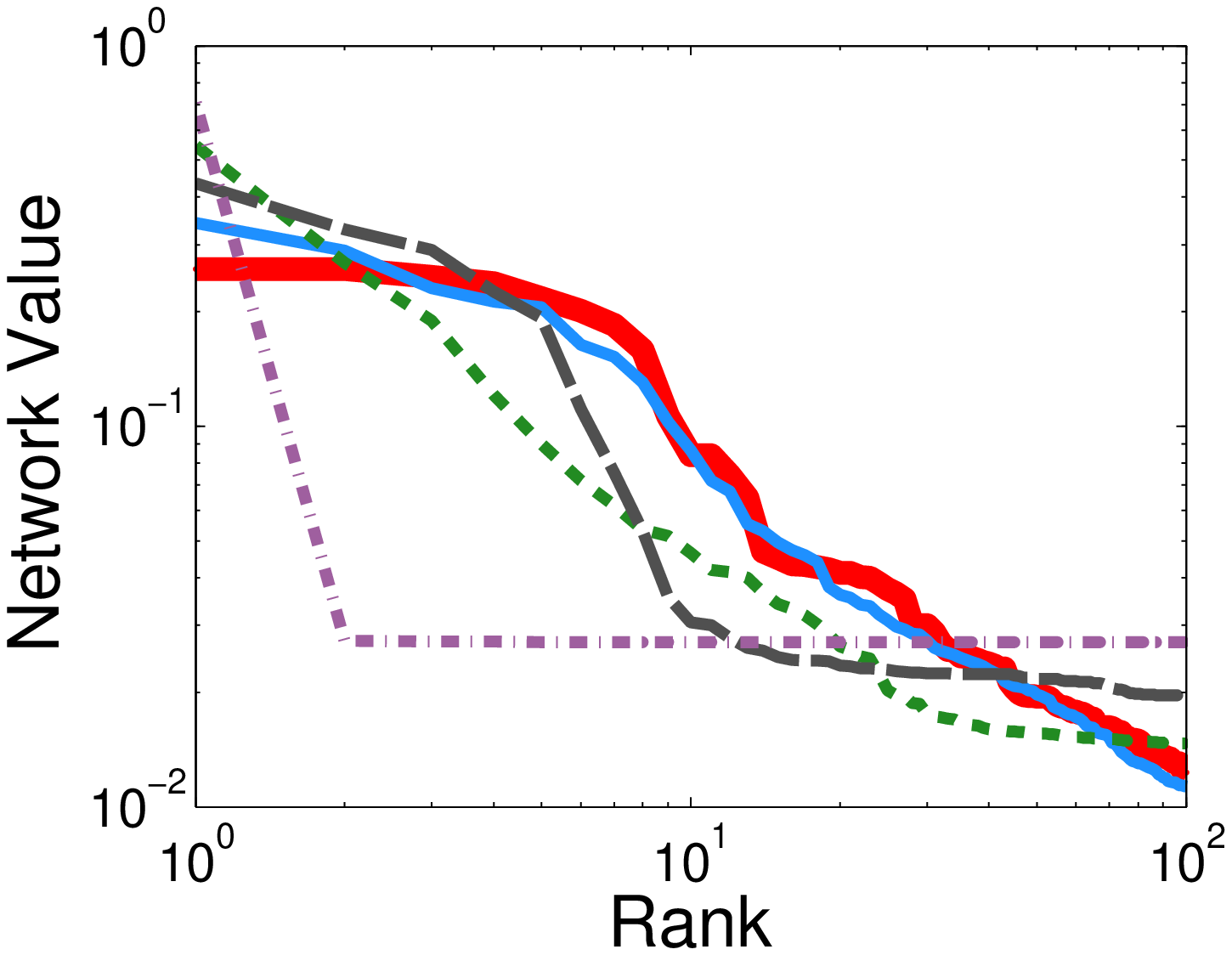}}
\caption{\textsc{Email-Univ} Graph}
\label{fig:stream_dist_comp_email}
\vspace{-3.mm}
\end{figure}

\begin{figure}[!h]
\centering
\subfigure{\label{fig:socjor deg dist}\includegraphics[width=0.33\linewidth]{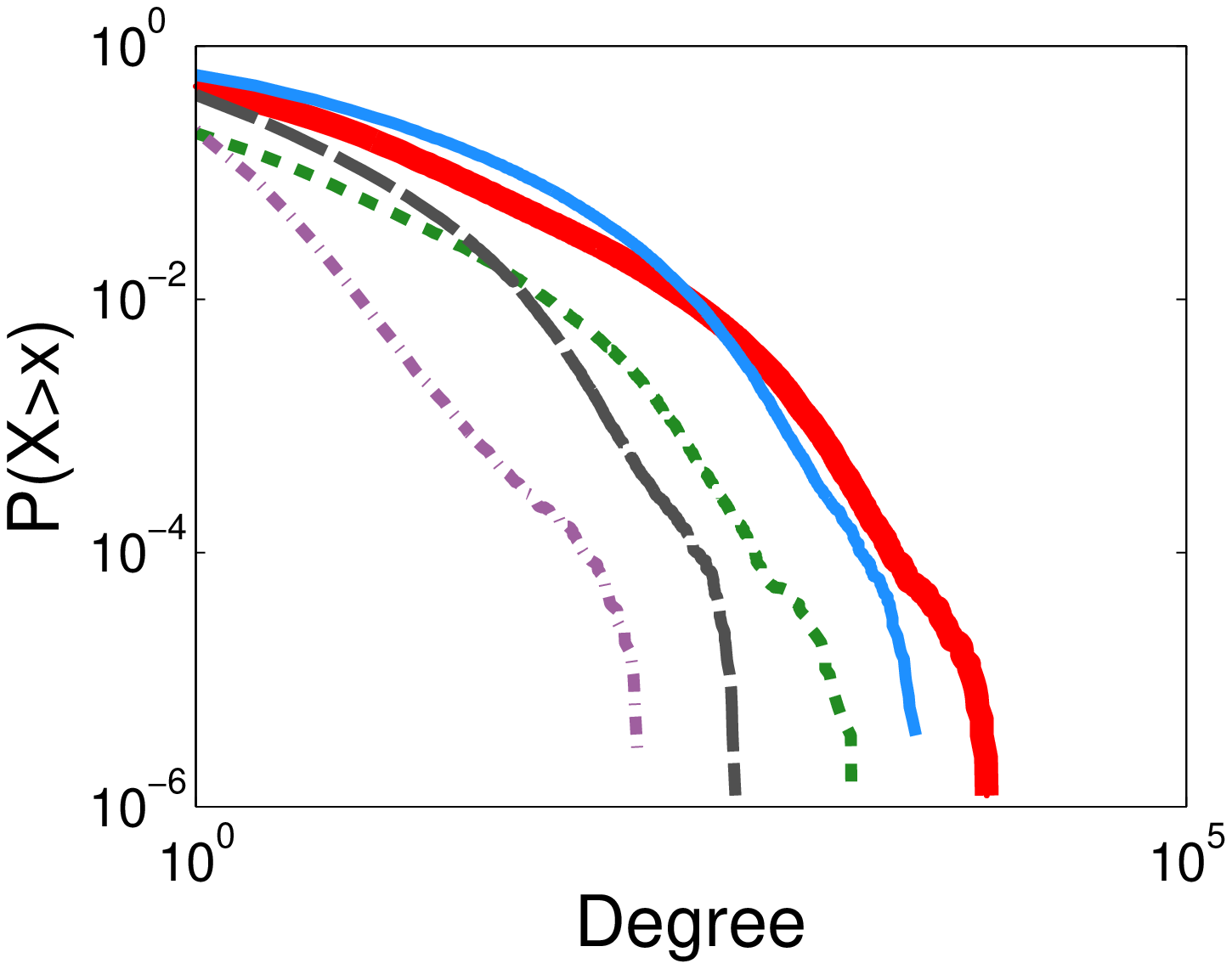}}
\hspace{-5.mm}
\subfigure{\label{fig:socjor pl dist}\includegraphics[width=0.33\linewidth]{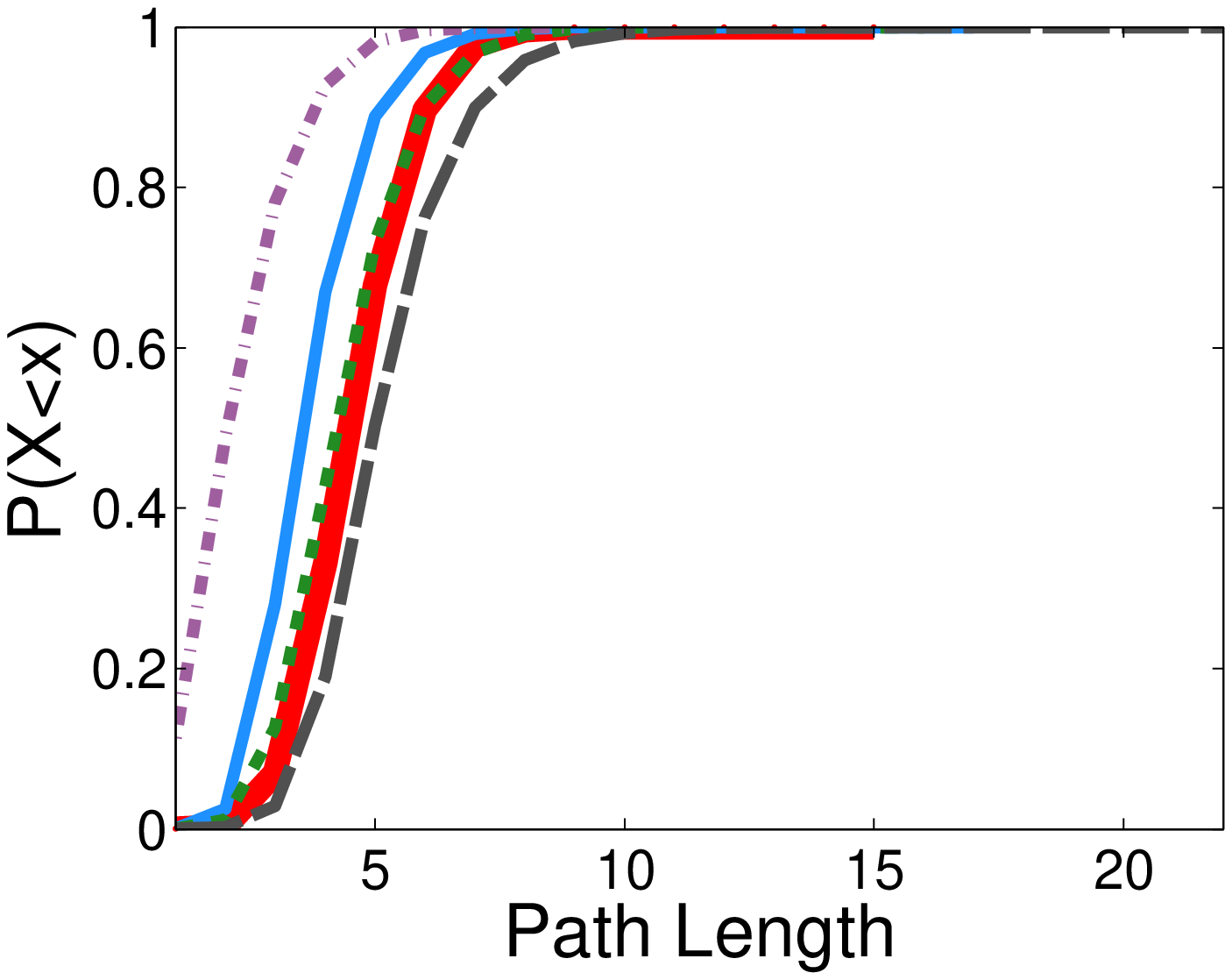}}
\hspace{-5.mm}
\subfigure{\label{fig:socjor cc dist}\includegraphics[width=0.33\linewidth]{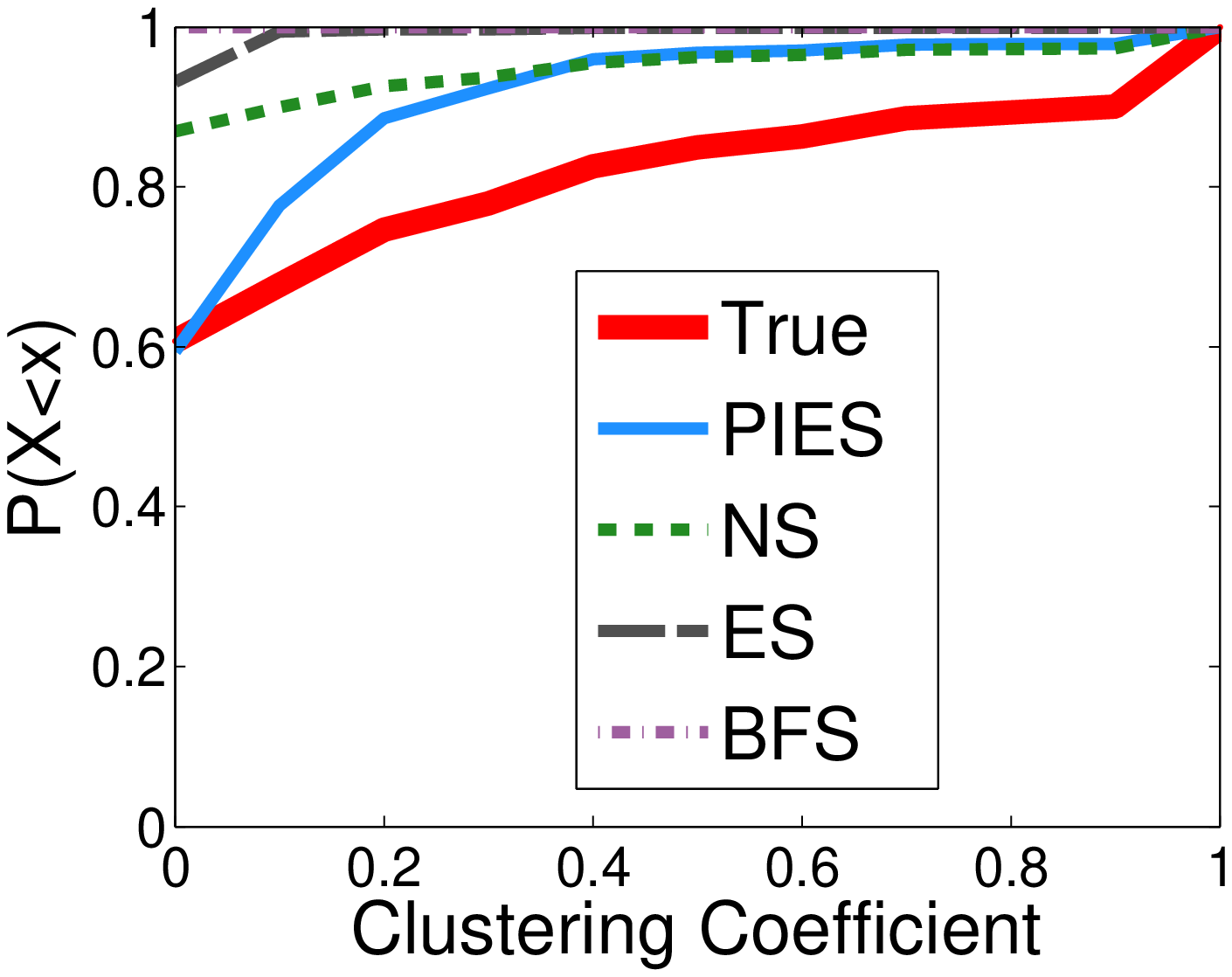}}
\vspace{-2mm}
\hspace{-2.mm}
\subfigure{\label{fig:socjor deg dist}\includegraphics[width=0.33\linewidth]{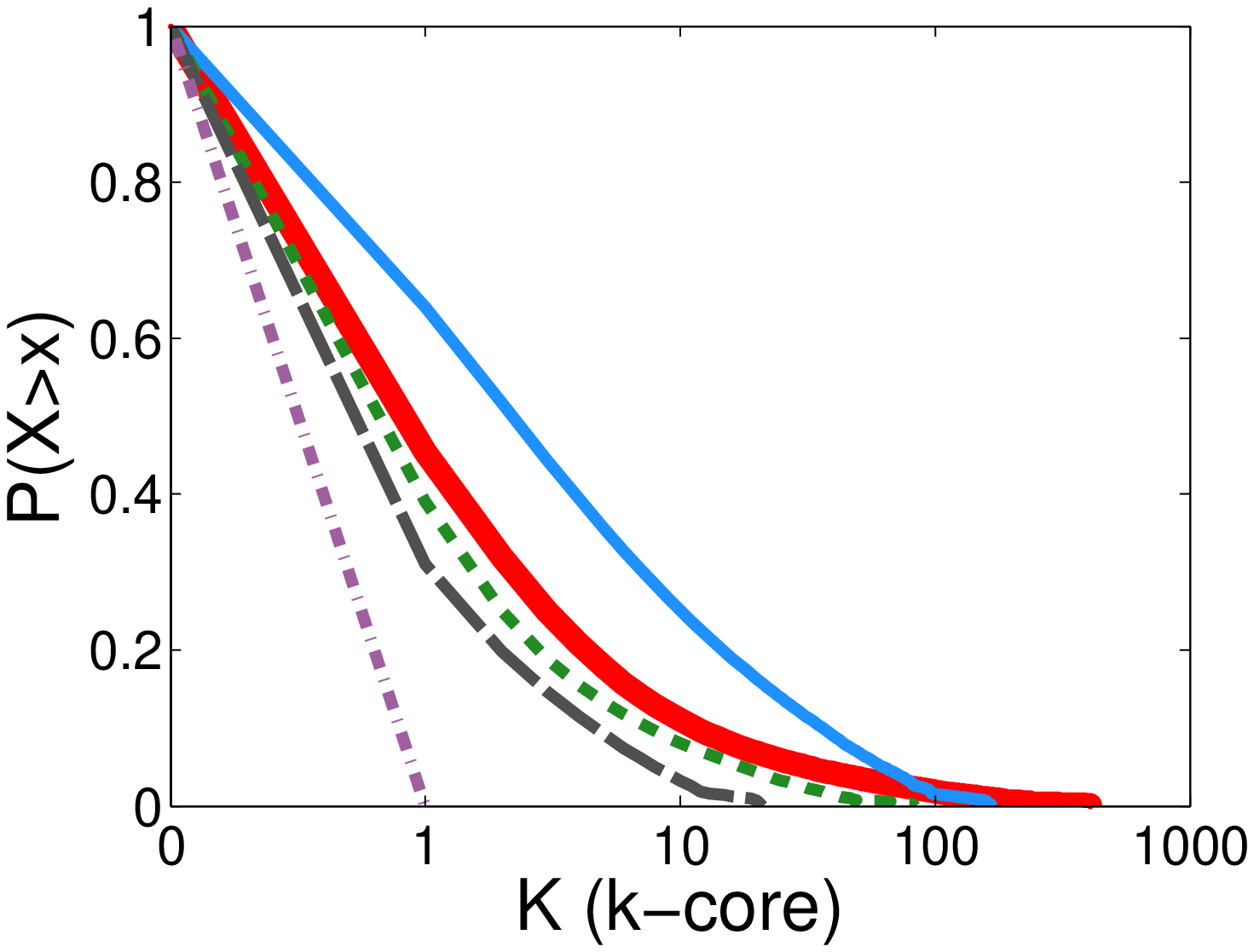}}
\hspace{-5.mm}
\subfigure{\label{fig:socjor pl dist}\includegraphics[width=0.33\linewidth]{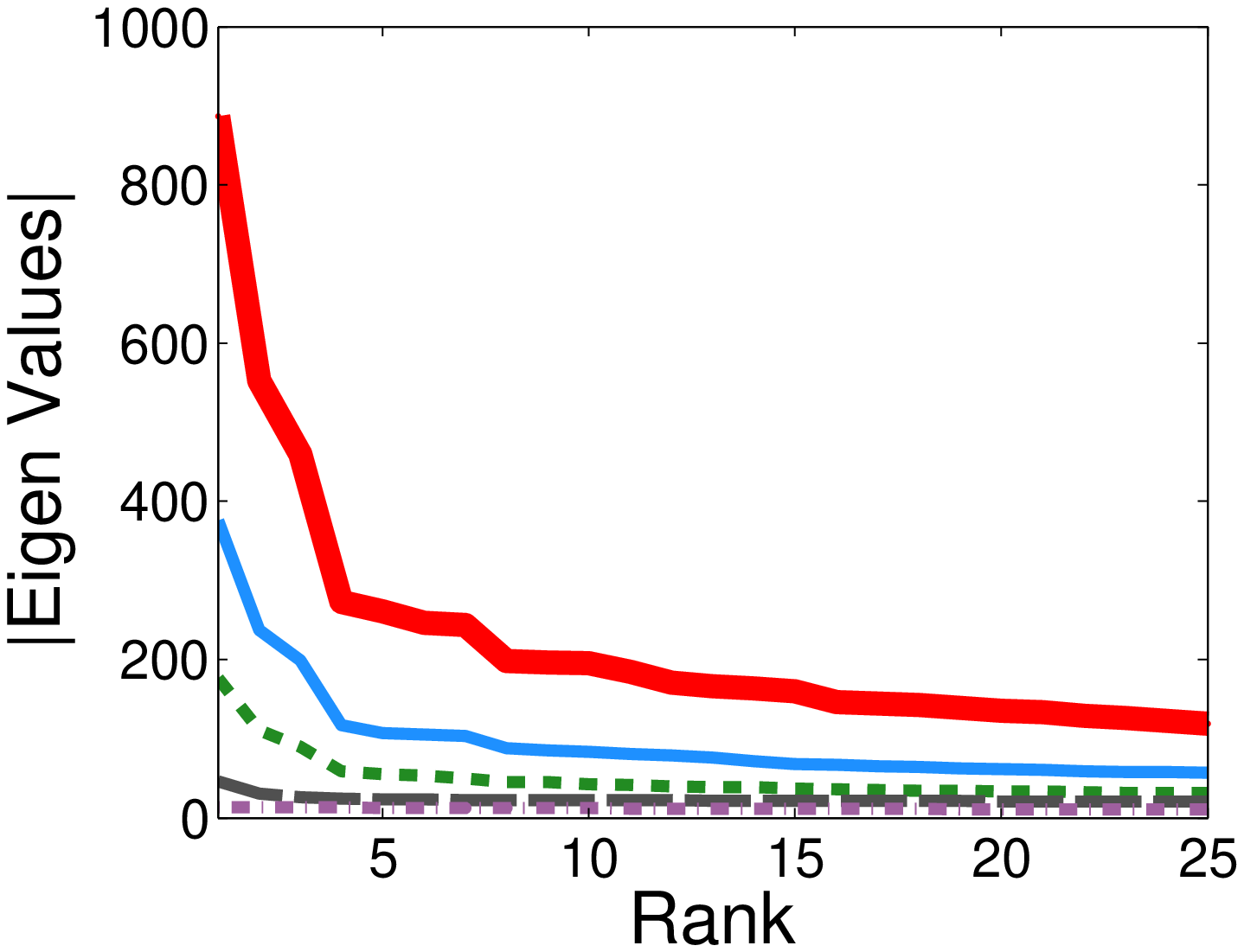}}
\hspace{-5.mm}
\subfigure{\label{fig:socjor cc dist}\includegraphics[width=0.33\linewidth]{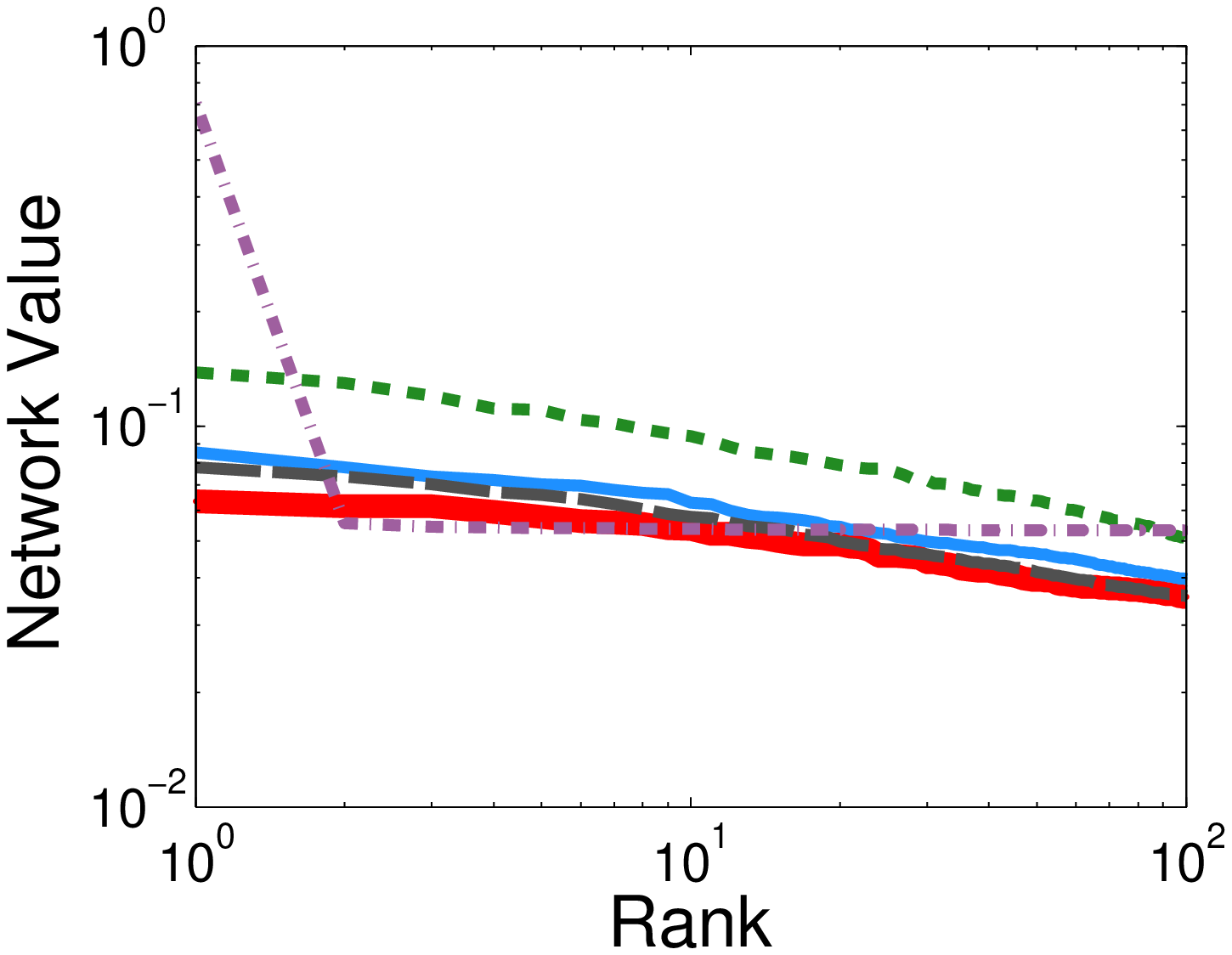}}
\caption{\textsc{Flickr} Graph}
\label{fig:stream_dist_comp_flickr}
\vspace{-3.mm}
\end{figure}

\newpage

\begin{figure}[!h]
\centering
\subfigure{\label{fig:socjor deg dist}\includegraphics[width=0.33\linewidth]{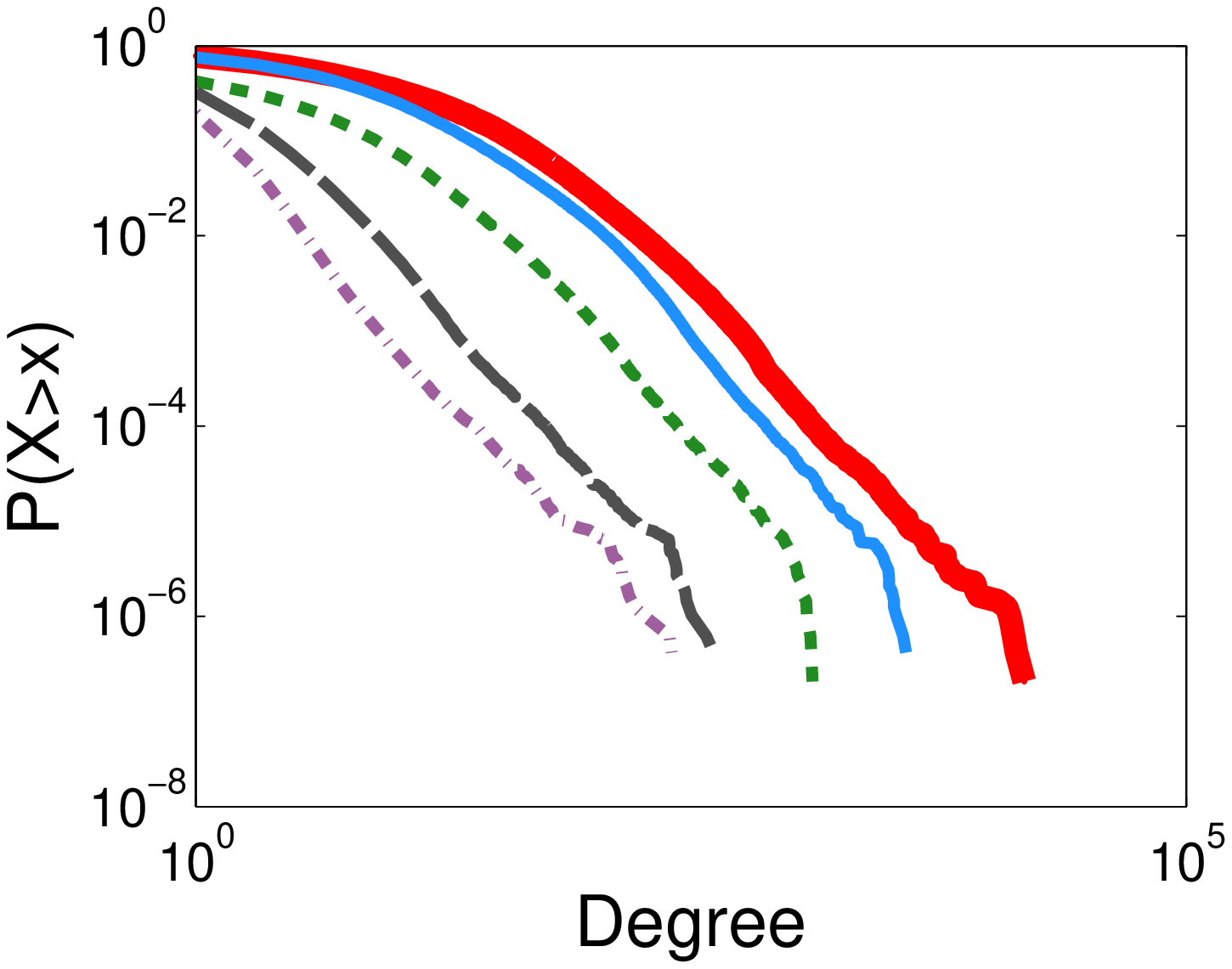}}
\hspace{-5.mm}
\subfigure{\label{fig:socjor pl dist}\includegraphics[width=0.33\linewidth]{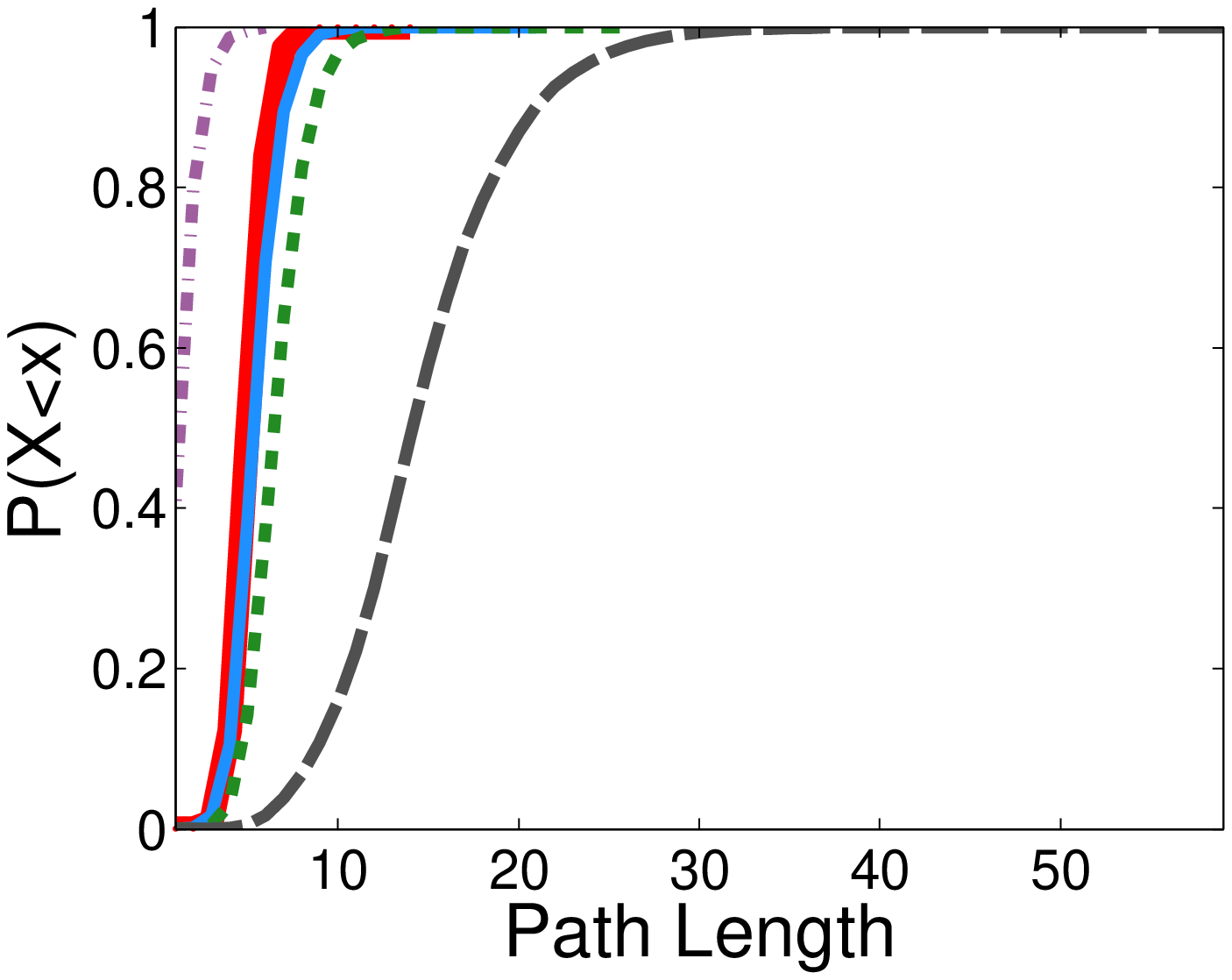}}
\hspace{-5.mm}
\subfigure{\label{fig:socjor cc dist}\includegraphics[width=0.33\linewidth]{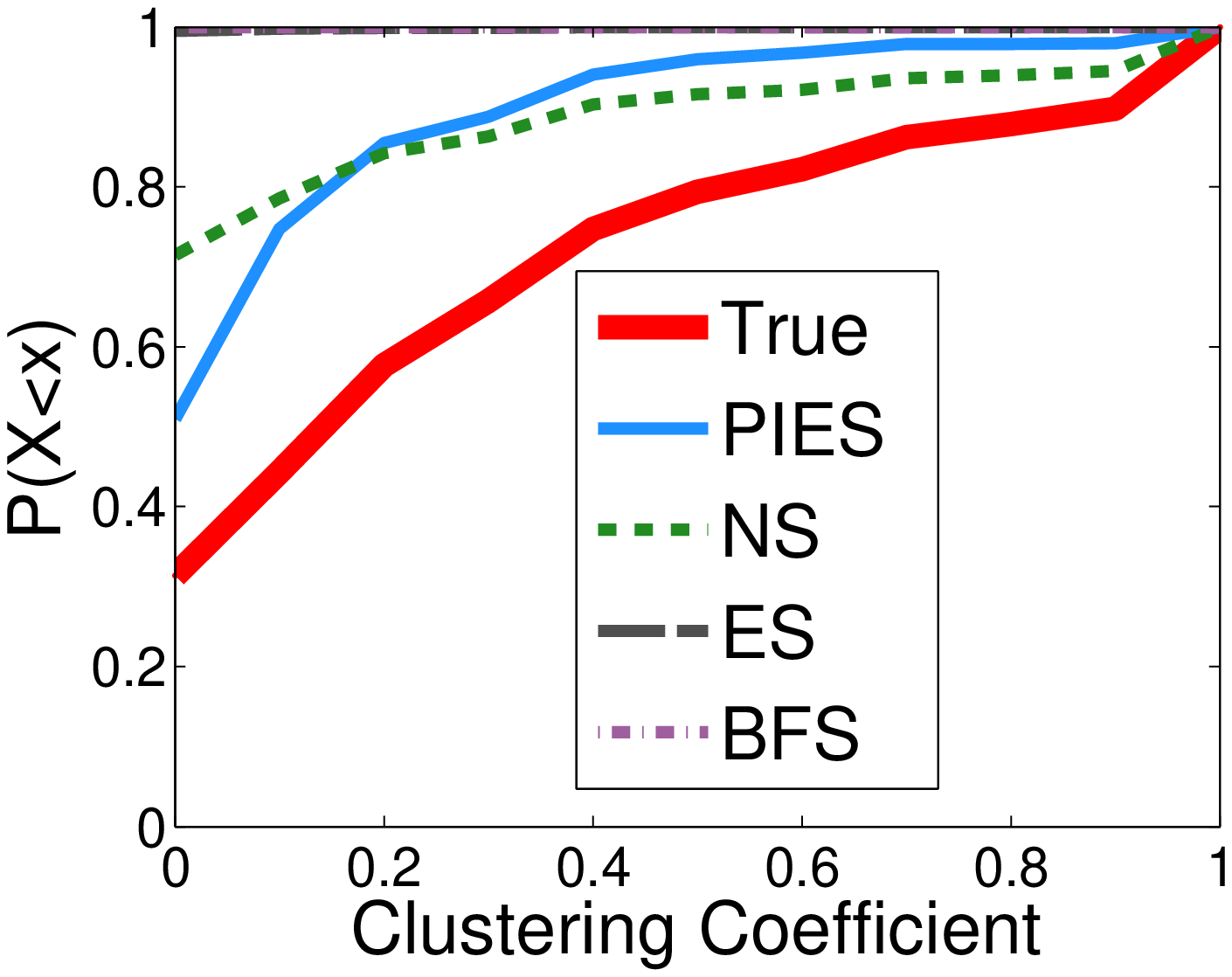}}
\vspace{-2mm}
\hspace{-2.mm}
\subfigure{\label{fig:socjor deg dist}\includegraphics[width=0.33\linewidth]{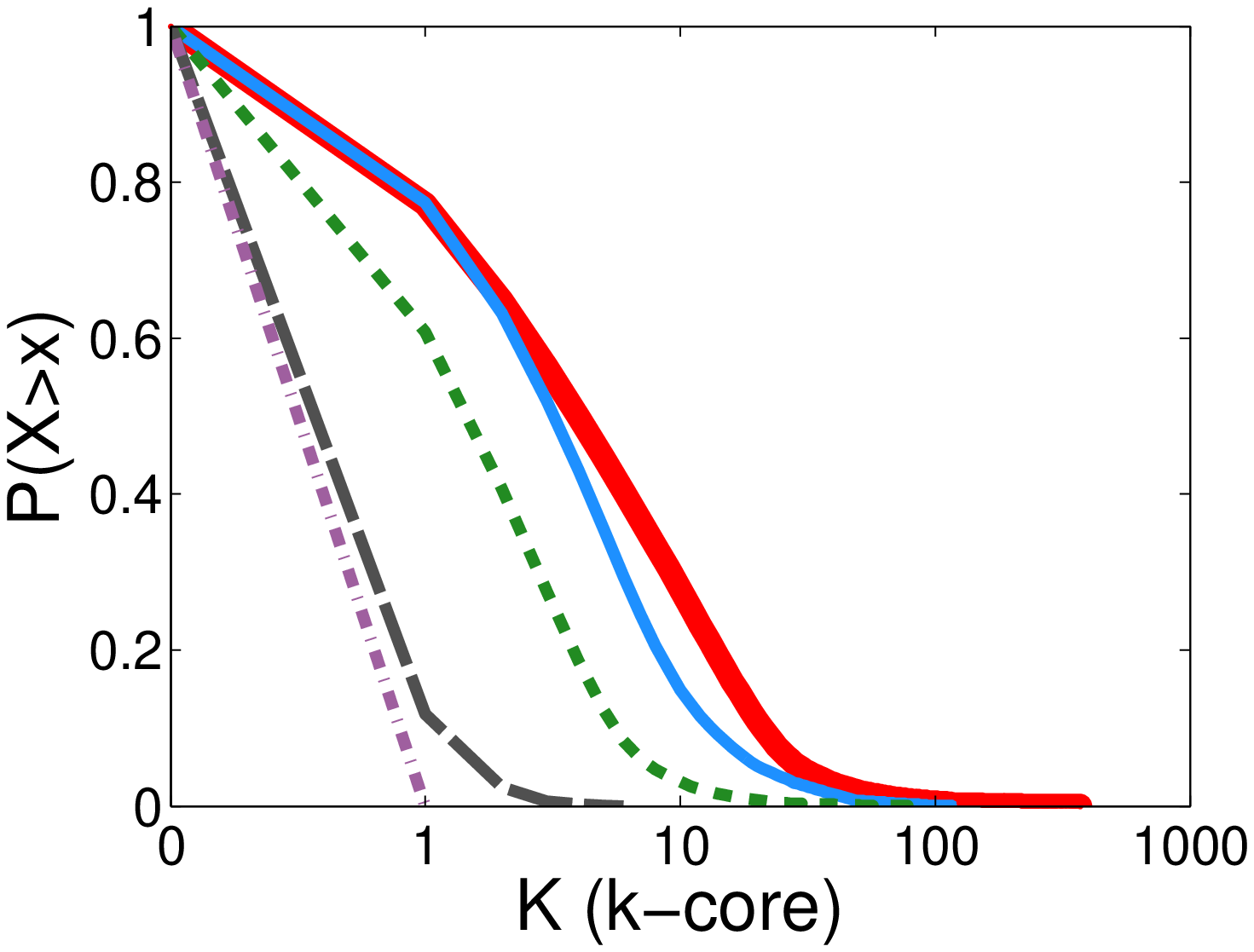}}
\hspace{-5.mm}
\subfigure{\label{fig:socjor pl dist}\includegraphics[width=0.33\linewidth]{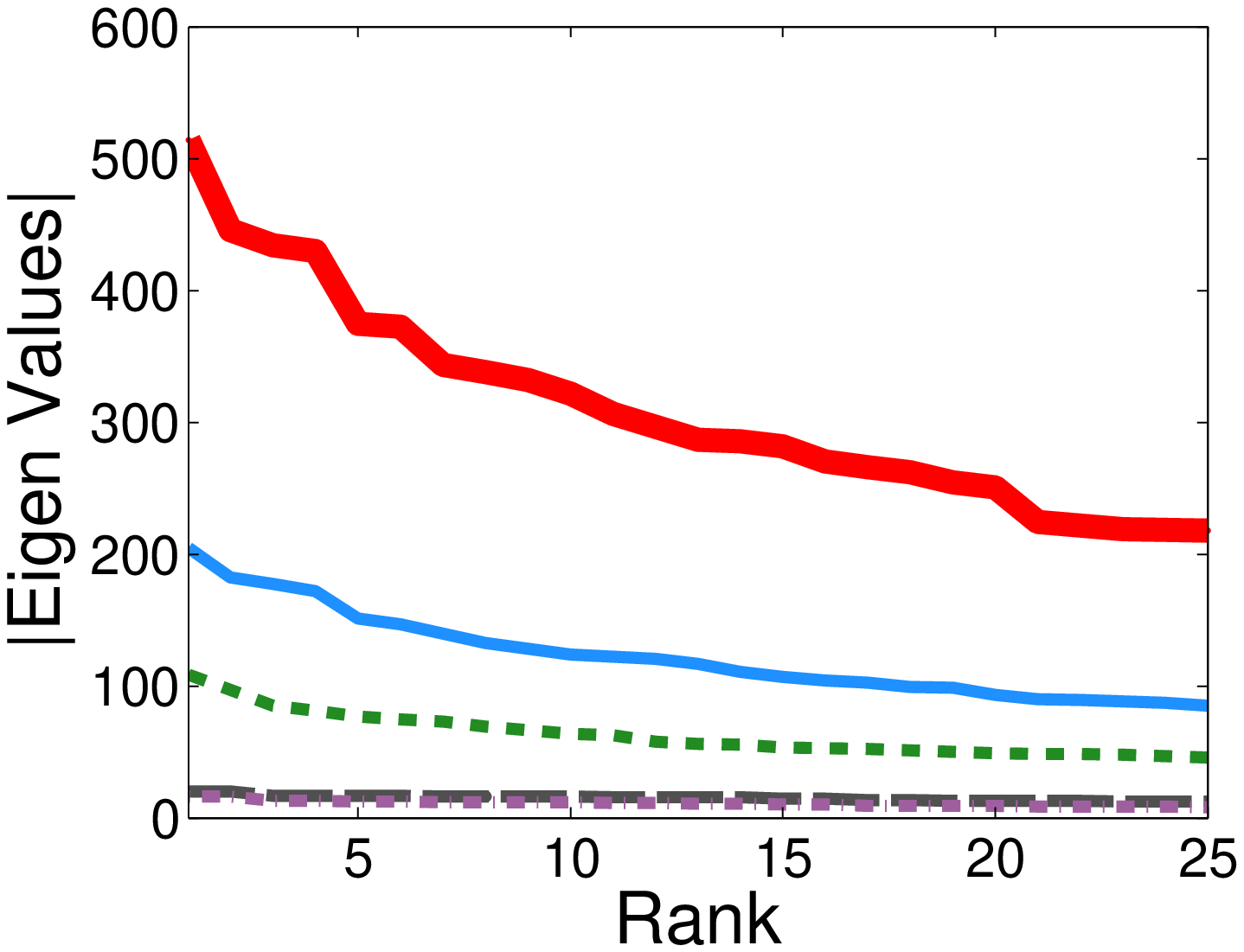}}
\hspace{-5.mm}
\subfigure{\label{fig:socjor cc dist}\includegraphics[width=0.33\linewidth]{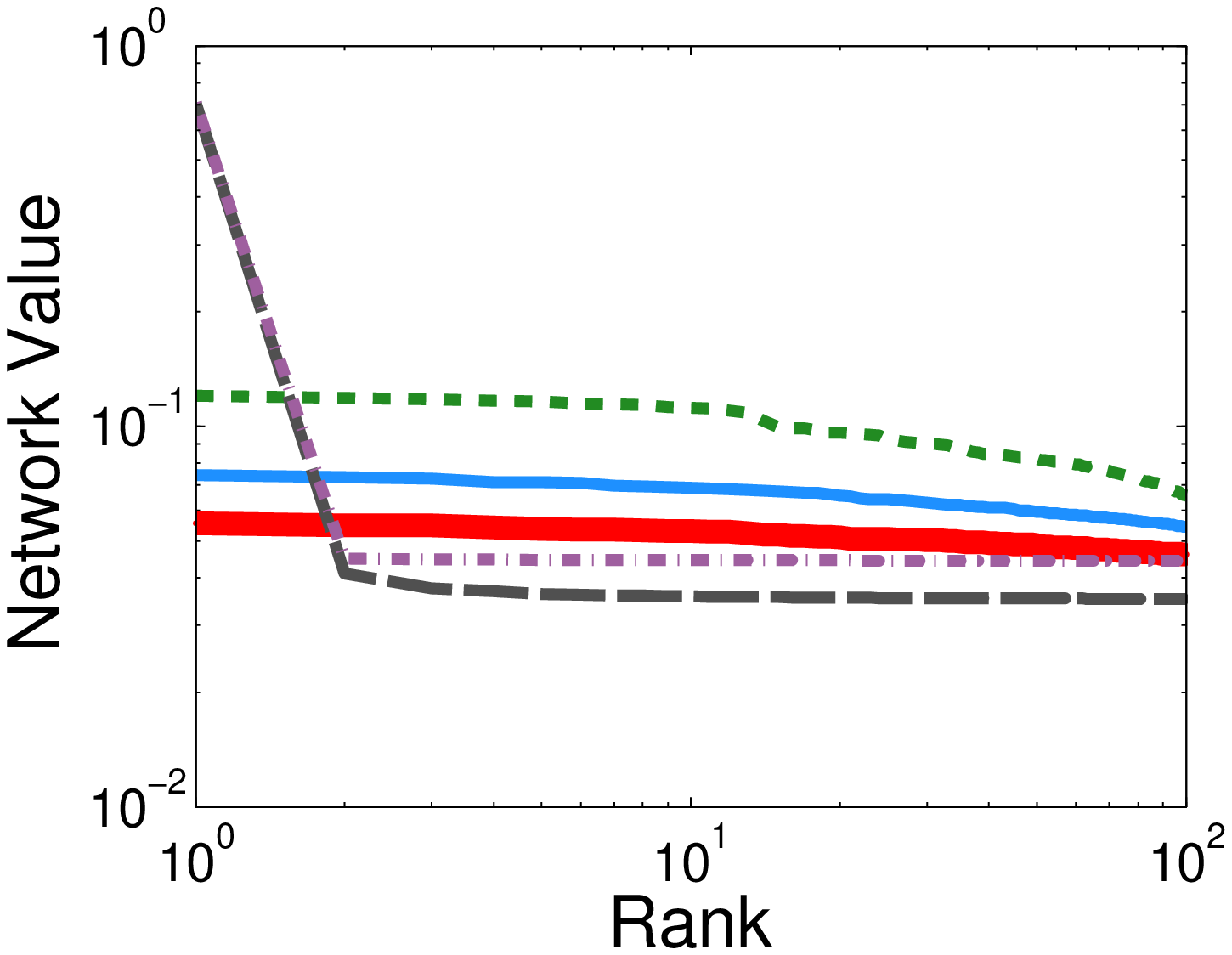}}
\caption{\textsc{LiveJournal} Graph}
\label{fig:stream_dist_comp_socjor}
\vspace{-3.mm}
\end{figure}